\documentclass[11pt]{article}
\usepackage{graphicx} %

\usepackage{mathpazo}
\usepackage{pdfpages}
\usepackage{bm}
\usepackage[margin=1in]{geometry}
\usepackage{amsmath,amssymb,mathtools}

\usepackage[utf8]{inputenc}
\usepackage{color,amsfonts,fullpage,amsthm,mathrsfs,fancyhdr,verbatim,framed,hyperref,xspace,mdframed,tabularx,booktabs,enumitem,comment,float}
\usepackage{cleveref}
\usepackage[dvipsnames]{xcolor}
\usepackage{graphicx}
\usepackage{amsthm}
\usepackage{mathpazo}
\usepackage{fec} 
\usepackage{soul}
\usepackage{authblk}

\usepackage[compat=0.6]{yquant}

\newcommand{\dd}{\,\mathrm{d}}
\newcommand{\dvol}{\,\dd\textup{vol}_g}

\DeclareMathOperator{\diver}{div}
\DeclareMathOperator{\Hess}{Hess}
\DeclareMathOperator{\Ric}{Ric}

\newcommand{\Simp}{\mathcal{S}}
\newtheorem{theorem}{Theorem}
\numberwithin{theorem}{section}
\newtheorem{lemma}[theorem]{Lemma}

\newtheorem{proposition}[theorem]{Proposition}

\newtheorem{corollary}[theorem]{Corollary}

\newtheorem{fact}[theorem]{Fact}

\newtheorem{definition}[theorem]{Definition}

\usepackage{tikz}
\usetikzlibrary{arrows.meta,positioning,fit,calc,backgrounds}

\newcommand{\Ran}{\operatorname{Ran}}
\newcommand{\ip}[2]{\left\langle #1,#2\right\rangle}
\newcount\Comments
\newcommand{\kibitz}[2]{\ifnum\Comments=1\textcolor{#1}{#2}\fi}

\newcommand{\vol}{\operatorname{vol}}

\newcommand{\ketbra}[2]{\ket{#1}\!\bra{#2}}

\DeclareMathOperator{\Tr}{Tr}

\DeclareMathOperator{\supp}{supp}

\definecolor{lightteal}{rgb}{0.5, 0.8, 0.8}
\definecolor{spectralrevision}{RGB}{255,0,0}

\usepackage[style=alphabetic, maxbibnames=99, backref=true, backrefstyle=three]{biblatex}
\bibliography{biblio}

\title{Quantum Simulation on Riemannian Manifolds}
\author[1]{Dylan Herman\thanks{\texttt{dylan.a.herman@jpmchase.com}}}
\author[1]{Jacob Watkins}
\author[1]{Guneykan Ozgul}
\author[1]{Jiayu Shen}
\author[1]{\\ Brandon Augustino}
\author[1]{Junhyung Lyle Kim}
\author[1]{Shouvanik Chakrabarti\thanks{\texttt{shouvanik.chakrabarti@jpmchase.com}}}
\affil[1]{Global Technology Applied Research, JPMorganChase, New York, NY 10001, USA}
\begin{document}

\maketitle

\begin{abstract}
    We investigate algorithms for the quantum simulation of the Schrödinger equation on a Riemannian manifold, where the kinetic operator is defined by the Laplace--Beltrami operator corresponding to the metric. Our first algorithms are based on a global spectral method based on the identification of an efficient transform to the eigenbasis of the Laplace--Beltrami operator. We use this method to provide explicit, efficient, quantum simulation algorithms for the Riemannian Schrödinger equation on tori and spheres with their standard metrics, simplices with the Wright--Fisher metric, truncated positive orthants and their invertible affine images with the log-barrier Hessian metric, and $\ell_p$ balls with a metric induced by the Duffy map. Our second algorithm is based on a coherent simulation of local spectral methods on multiple charts, and is in principle applicable to any compact manifold. We first analyze this algorithm in the continuum and derive conditions under which a polynomial spectral cutoff suffices. We also provide a discretization analysis of a polynomial spectral cutoff for tensor-products of constant-dimensional manifolds. Finally, we consider applications of these methods to optimization and physical simulation. For optimization, we provide results including a generalization and convergence analysis of Quantum Hamiltonian Descent for geodesically convex functions that leads to explicit algorithms on the sphere and simplex, and a Riemannian generalization of the Real-Space Adiabatic Algorithm. For physical simulation, we show that our algorithms can simulate certain spatially discretized field theories, including a variant of the nonlinear sigma model.
\end{abstract}

\clearpage
\tableofcontents
\newpage

\section{Introduction}

Quantum simulation is the paradigmatic application of quantum computers and was the primary impetus for the interest in this form of computing~\cite{feynman1981simulating}. The study of explicit algorithms for quantum simulation was initiated by Lloyd~\cite{lloyd1996universal}, and over time many powerful quantum algorithms have been developed for the simulation of appropriately specified Hamiltonians that act on a discrete space, using techniques that include product formulae~\cite{lloyd1996universal,childs2021theory}, linear combinations of unitaries~\cite{childs2012hamiltonian,berry2015optimal}, or qubitization and quantum signal processing for block-encoded Hamiltonians~\cite{low2017optimal,low2019qubitization,gilyen2019quantum}.

There has also been significant interest in simulating Schr\"odinger dynamics in real space. This problem takes the schematic form
\begin{equation}
	i\partial_t\psi(t)=H(t)\psi(t), \qquad H(t)=-a(t)\Delta+b(t)f,
\end{equation}
on a Hilbert space such as $L^2(\Omega,\dd x)$, where $\Omega$ is a bounded subset of $\R{d}$. Simulating such systems requires appropriate discrete representations of the continuum operators. This continuum-to-circuit passage must be an explicit part of the algorithmic analysis in real space, since the choice of discretization method, and the resulting grid spacing, basis cutoff, smoothing procedure, quadrature rule, and aliasing analysis can all affect the number of qubits, and the query and gate complexity. Kivlichan et al. bounded real-space simulation costs using finite-difference discretizations~\cite{kivlichan2017bounding}, Childs et al.~\cite{childs2022quantumsim} gave an explicit Fourier pseudospectral framework for real-space dynamics, and Chakrabarti et al. gave a refined analysis of this method and concrete estimates for Lipschitz-continuous potentials~\cite{chakrabarti2025speedups}.

In addition to the natural physics applications, another motivation for real-space simulation comes from quantum algorithms for continuous optimization that use quantum dynamics as the algorithmic foundation, with the objective function encoded in the potential. 
A systematic study of such algorithms was initiated by Leng et al. under the name ``Quantum Hamiltonian Descent"~\cite{leng2023quantumhamiltoniandescent}. Subsequently, quantum dynamical algorithms have been explored for linear programming~\cite{augustino2023quantum}, convex optimization~\cite{catli2025exponentially,leng2025quantumhamiltoniandescentnonsmooth,chakrabarti2025speedups, gribling2025selfconcordantschrodingeroperatorsspectral}, and non-convex optimization~\cite{Liu_2023,leng2023quantum,chen2025quantumlangevin,leng2025sub,herman2025mechanisms}. It is notable that in several regimes, the strongest known quantum speedups are achieved by dynamical methods rather than by more traditional quantum optimization primitives. Examples include the best-known speedups for high-dimensional stochastic convex optimization~\cite{chakrabarti2025speedups}, worst-case query speedups for constructed nonconvex optimization instances~\cite{leng2025sub}, and speedups over known classical algorithms for specific explicit families of nonconvex functions~\cite{herman2025mechanisms}.

An important generalization is that of quantum simulation over a Riemannian manifold $(\mathcal M,g)$. In this setting, the corresponding kinetic operator is not the standard Laplacian but rather the Laplace--Beltrami operator, expressed in local coordinates as,
\begin{equation}
	\Delta_g u
	=\rho^{-1}\partial_i\!\left(\rho g^{ij}\partial_j u\right), \qquad \text{where }\rho(x)=\sqrt{\det(g_{ij}(x))},
\end{equation}
where repeated coordinate indices are summed according to the Einstein summation convention. The target for simulation is now the Riemannian Schr\"odinger equation
\begin{equation} \label{eq:Riemannian-SE}
	i\partial_t\psi(t)=H_g(t)\psi(t),
	\qquad
	H_g(t)=-a(t)\Delta_g+b(t)f,
	\qquad
	\psi(t)\in L^2(\mathcal M,\dd\operatorname{vol}_g)
\end{equation}
where $\dd\operatorname{vol}_g(x) = \rho(x)\dd x$. One recovers the Euclidean case by setting $g = I$ in local coordinates.

Why should we care about simulating such dynamics? Physics provides a direct motivation. A particle subject to regular holonomic constraints, or a rigid rotor, has a smooth Riemannian configuration space. Its Hamiltonian uses the corresponding Laplace--Beltrami operator for the kinetic term, potentially supplemented by a scalar-curvature term from the quantization prescription or by an extrinsic-curvature potential generated when the constraint is realized through strong confinement~\cite{dewitt1957dynamical,dacosta1981constrained}. These elementary examples are usually low-dimensional. For the high-dimensional setting where speedups from quantum computing are most important, the more relevant examples arise from spatially discretized field theories. A standard nonlinear sigma model with target $T$ has lattice configuration space $T^{|\Lambda|}$; its kinetic term is a sum of target-space Laplacians, while the spatial-gradient and on-site terms act as a multiplication potential. This includes the $O(N)$ models with $T=S^{N-1}$ and the $\mathbb{C}P^{N-1}$ models with the Fubini--Study metric~\cite{burgess2000goldstone,alexandru2019sigma,singh2022qubit,eichenherr1978sigma,dadda1978expandable}. Hamiltonian lattice gauge theory with compact gauge group $\mathcal{G}$ similarly has kinematic configuration space $\mathcal{G}^{|\mathcal{E}|}$, with one factor for each lattice link; the electric term is a sum of group Laplacians, while the magnetic plaquette term is a multiplication potential~\cite{kogut1975hamiltonian,banuls2020simulating}. In both cases, the configuration-space dimension grows with the number of lattice sites or links even when each factor has fixed dimension. On any fixed spatial lattice, these Hamiltonians are therefore exact instances of the Riemannian Schr\"odinger equation above.

A second motivation for non-Euclidean geometry comes from classical optimization, where geometry can be part of both the problem and the algorithm. Constrained optimization is a central example. When equality constraints define a smooth feasible manifold, Riemannian methods use tangent directions and retractions to preserve feasibility. Spheres, Stiefel and Grassmann manifolds, and fixed-rank matrix manifolds are typical cases~\cite{boumal2023introduction}. More broadly, mirror descent~\cite{nemirovskii1983problem, bubeck2015convex} replaces Euclidean geometry with the Bregman geometry of a chosen mirror map, which can be selected to reflect problem structure; negative entropy on the probability simplex is a canonical example~\cite{beck2003mirror}. In interior-point methods, the geometry induced by the Hessian of a \textit{self-concordant barrier} is more fundamental still: it determines the local norm in which the Newton direction is defined, its unit ball is a Dikin ellipsoid contained in the feasible region, and self-concordance controls how quickly this geometry changes from one point to the next. These properties allow Newton steps to track the central path and underlie the polynomial iteration bounds for path-following methods~\cite{nesterov1994interior, renegar2001mathematical}.

Non-Euclidean geometry plays many important roles beyond enforcing constraints. Natural-gradient and related methods use a position-dependent metric as a preconditioner to reduce sensitivity to parametrization and anisotropic scaling~\cite{amari1998natural}; when the objective is geodesically convex, the chosen geometry can additionally support global first-order convergence guarantees~\cite{zhang2016geodesic}. Thus non-Euclidean geometry can encode feasibility, exploit symmetry, and improve conditioning. Geometry also plays an algorithmic role in sampling, which in turn underlies simulated-annealing methods for convex optimization and algorithms for volume computation~\cite{abernethy2016faster, badenbroek2022simulated}. Riemannian Hamiltonian Monte Carlo adapts its trajectories to the local scaling and correlation structure of a target distribution, and has led to improved worst-case mixing bounds and faster volume algorithms for polytopes~\cite{girolami2011riemann,lee2022geodesic,lee2018riemannianhmc}, as well as practical gains on high-dimensional, ill-conditioned problems~\cite{kook2022sampling}. 

The quantum-dynamical speedups for convex and nonconvex optimization highlighted earlier concern unconstrained problems: the dynamics uses a Euclidean kinetic operator and encodes the objective in the potential~\cite{chakrabarti2025speedups,leng2025sub,herman2025mechanisms}. To extend this approach to constrained optimization while preserving feasibility by construction, the kinetic evolution must also respect the feasible set. For a smooth feasible manifold $(\mathcal M,g)$, this means replacing $-\Delta$ by $-\Delta_g$; for an inequality-constrained domain, a barrier-induced metric can similarly encode the geometry of its interior. Gribling, Apers, Nieuwboer, and Walter~\cite{gribling2025selfconcordantschrodingeroperatorsspectral} make this connection explicit for a self-concordant barrier $\Phi$: they equip the feasible domain with the Hessian metric $g=\nabla^2\Phi$ and study central-path Schr\"odinger operators built from $-\Delta_g$. They prove spectral-gap bounds without the Euclidean condition-number dependence and, under assumptions that include access to the corresponding Riemannian evolution, use these bounds to formulate a quantum interior-point method. Efficient digital simulation of Riemannian Schr\"odinger dynamics is therefore a concrete primitive for extending quantum-dynamical optimization to constrained problems.

Finally, we note that the simulation of \eqref{eq:Riemannian-SE} is generally understood to be significantly more complicated than that of simulating that standard real-space dynamics. In flat real-space simulation, the Fourier transform gives a global basis in which the kinetic energy is simple, while a position grid gives a representation in which the potential is simple. For a general manifold, such a transform requires an efficiently implementable map to the eigenbasis of the Laplace--Beltrami operator, and simulation requires us to both identify such a map and develop an efficient quantum algorithm for the transform. All the nuances of the discretization analysis in the flat space must also be handled again, and this analysis may in general be more complicated. The situation is even more complex for a general manifold: there may be no single global coordinate chart, no efficiently implementable transform to Laplace--Beltrami eigenfunctions, and no uniform volume form. In such cases, one can hope to use local methods where Euclidean simulation methods apply for each local chart, however this brings additional difficulties: Local charts may overlap and amplitudes on overlaps must be recombined coherently rather than treated as independent classical alternatives. Additionally, the continuum-to-circuit passage causes inconsistencies among different charts. As a consequence, the simulation of Riemannian Schr\"{o}dinger operators has been explicitly left open by papers such as~\cite{gribling2025selfconcordantschrodingeroperatorsspectral}. The main goal of this paper is to make the first substantial algorithmic progress towards resolving this question.

\subsection{Results}

Our main results make a systematic analysis of simulation algorithms for Riemannian Schr\"odinger equations. In~\eqref{eq:Riemannian-SE}, the Hamiltonian contains differential and multiplication operators acting on wavefunction $\psi(t)$ defined continuously on $\mathcal{M}$. A digital algorithm must first choose a finite representation of the wavefunction and a finite matrix representing these operators. The analogous problem on a periodic Euclidean domain is handled by the Fourier pseudospectral method. One truncates the Fourier expansion of the wavefunction, uses that the Laplacian is diagonal in the Fourier basis, and applies the discrete Fourier transform to evaluate the potential on a finite set of quadrature nodes. Since the digital evolution does not need to agree with the continuous evolution, a rigorous pseudospectral method provides an error guarantee for a chosen Fourier cutoff and the quadrature rule that determine the space in which digital evolution takes place and how accurately the potential is represented respectively.

\paragraph{Global Spectral Methods.} On a Riemannian manifold, the pseduospectral method can be extended by forming an analogous basis using the eigenfunctions of the Laplace--Beltrami operator. We call this approach \emph{global spectral method} where one relies on a single global transform encoding the geometry of the manifold. This approach becomes viable when one is be able implement this basis transform and approximate the potential in this transformed space using the quadrature rules. Similar to the Euclidean case, one needs to determine the right cutoff threshold for the spectral degree and the quadrature rules to provide a rigorous simulation guarantee. The main additional issue is that in Euclidean case, multiplication by a function with maximum spectral degree $K$, increases the degree of the wavefunction by at most $K$. This does not hold for general manifolds, however for the manifolds considered in this paper, each such multiplication can increase spectral degree by only a controlled amount due to \emph{slow spectral growth} property.

Using the global spectral method, we give end-to-end digital simulation algorithms with rigorous guarantee for several important settings. These include flat torus, the sphere with its standard metric, and the simplex with the
Wright–Fisher metric. We also obtain algorithms for positive orthants, affine images
of such domains, and $\ell_p$ balls by reducing their chosen metrics to flat periodic
problems through explicit coordinate changes.

\paragraph{General Manifolds and Coherent-Chart Methods.} The need for an efficiently implementable global transform and the explicit global spectral structure limits the applicability of the global spectral method. A compact manifold always has a Laplace–Beltrami eigenbasis, but this fact alone does not provide slow spectral growth, useful quadrature rules, or efficient quantum transforms. To address this limitation, we also introduce the \emph{coherent-chart method} that does not rely on one single global basis transformation. Since a manifold resembles an Euclidean space in  the near vicinity of a point, one can decompose the space into local charts and use the ordinary Fourier expansion similar to the pseudospectral method for each chart. To be able to this, one first needs to provide a localization procedure that allows the wavefunction to be decomposed into local charts that can be simulated efficiently by retaining the relative phases so that it can be recombined into original global state. We give a localization procedure inspired by the well known IMS formula. The Riemannian IMS identity allows to decompose the wavefunction and the Hamiltonian into local charts  and also provide the right correction term to the potential that allows the local operators recombine into right global form. We give simulation guarantees for the resulting algorithm under explicit Sobolev and geometric regularity assumptions that are required to control the overlapping regions between the local charts. The resulting approximation is controlled by a local Fourier cutoff: in each chart, only Fourier modes whose coordinate frequencies below the cutoff are retained. For general compact manifolds, we bound the difference between the exact and truncated continuum evolutions, and show that a cutoff that is polynomial in the dimension suffices at fixed accuracy under explicit geometric and Sobolev assumptions. For tensor products of constant-dimensional manifolds, we extend the analysis to a finite-dimensional discretization, using a global interpolation space to connect the local grids, and prove a corresponding polynomial cutoff under various interpolation and regularity assumptions.

We highlight that the two sets of simulation results differ in the scope of their primary contribution. In the case of global spectral methods, the simulation algorithm is clear as it is an extension of the pseudospectral algorithm used in flat space. Our main contribution there is to obtain explicit end-to-end algorithms with rigorous guarantees and resource estimates, for particular well-studied manifolds of application-specific interest. On the other hand for general manifolds, to the best of our knowledge, there did not exist an algorithmic framework that is conducive to analysis. Our contribution is to propose such a framework and determine conditions under which the truncation requirements (which are one of the main bottlenecks in the analysis of simulation algorithms) are benign.

Table~\ref{tab:simulation-results-summary} summarizes the scope and status of the results. The following subsections give a more detailed overview of the two simulation methods. Finally, Section~\ref{sec:results-applications} gives an overview of our study of applications of quantum simulation on Riemanninan manifolds, and discusses how our algorithms fit into these applications.

\begin{table}[htbp]
    \centering
    \small
    \renewcommand{\arraystretch}{1.25}
    \begin{tabularx}{\textwidth}{>{\raggedright\arraybackslash}p{0.21\textwidth}>{\raggedright\arraybackslash}p{0.32\textwidth}>{\raggedright\arraybackslash}X}
    \toprule
    \textbf{Configuration space} & \textbf{Metric} & \textbf{Guarantee} \\
    \midrule
    Flat torus $\mathbb T^d$ & Euclidean metric $g_{ij}=\delta_{ij}$ & Efficient digital simulation (Theorem~\ref{thm:torus-digital-simulation}). \\
    Truncated positive orthant and its invertible affine images & $g=A^{\mathsf T}\diag ((Ax-c)_i^{-2})A$, with $A=I$ and $c=0$ for the orthant & Efficient digital simulation (Theorem~\ref{thm:orthant-affine-digital-simulation}). \\
    Coordinate-truncated $\ell_p$ ball and hemisphere ($p\geq2$) & Duffy Metric \eqref{eq:lp-flat-duffy-metric} & Efficient digital simulation (Theorem~\ref{thm:lp-duffy-digital-simulation}). \\
    Sphere $\mathbb S^d$ & Standard round metric & Efficient digital simulation (Theorem~\ref{thm:spherical-digital-simulation}). \\
    Simplex $\Simp_d$ & Wright--Fisher metric $g^{ij}(x)=x_i(\delta_{ij}-x_j)$ & Efficient simulation (Theorem~\ref{thm:simplex-digital-simulation}). \\
    \midrule
    General compact manifolds & General smooth metric $g$ & Simulation on continuum with polynomial spectral cutoff (Theorem~\ref{thm:coherent-chart-continuum-error} and Corollary~\ref{cor:poly_cutoff}). \\
    Products of constant-dimensional manifolds & Product metric $g=\bigoplus_\nu g_\nu$ & Digital simulation with polynomial spectral cutoff (Theorem~\ref{thm:tensor-local-main}). \\
    \bottomrule
    \end{tabularx}
    \caption{Summary of our simulation results.}
    \label{tab:simulation-results-summary}
\end{table}

\subsubsection{Global spectral methods}

The global spectral method represents the wavefunction by its coefficients in finitely many eigenfunctions of the Laplace--Beltrami operator on the whole manifold. We introduce a binning function for the eigenfunctions $L : \mathcal{I} \rightarrow \mathbb{N}$. We define the $\ell$-th level to be
\begin{equation*}
    \mathcal H_\ell
    \coloneqq \operatorname{span}\{Y_{\alpha}: L({\alpha}) = \ell\},
    \qquad
    \mathcal{V}_N \coloneqq \bigoplus_{\ell=0}^N\mathcal{H}_{\ell}.
\end{equation*}
Let $\{{Y}_{\alpha}\}_{\alpha \in \mathcal{I}}$ be an orthonormal basis of eigenfunctions for $-\Delta_g$. Consider a function $L : \mathcal{I} \rightarrow \mathbb{N}$ that bins the eigenfunctions into spaces $\mathcal H_{\ell} \coloneqq  \textup{span}\{{Y}_{\alpha
} : L(\alpha) = \ell\}$, with $\mathcal V_N=\mathcal H_0\oplus\cdots\oplus\mathcal H_N$. The kinetic term is diagonal in this basis and it multiplies each coefficient with a phase. The potential instead acts by pointwise multiplication of $f$. We apply it by transforming the wavefunction to weighted values on a quadrature grid, querying $f$ at the grid points, and transforming back. The method therefore requires a cutoff $N$ for which the evolution restricted to $\mathcal V_N$ approximates the continuum evolution, and a quantum circuit that maps coefficients in $\mathcal V_N$ to weighted function values on the grid.
For example,  $Y_{k}(x) =e^{2\pi i k\cdot x}, L(k) = \lVert k \rVert_{\infty}$ corresponds to the Fourier spectral method.

This approach has two distinct sources of approximation error. The first comes from restricting the wavefunction to $\mathcal V_N$. Multiplication by the potential generally produces components outside $\mathcal V_N$, even when the initial state has low degree. The second comes from the quadrature rule. A digital implementation replaces the integrals that define the matrix elements of the potential by finite weighted sums over the grid. If the grid cannot resolve a component produced by the potential, that component is mistaken for a combination of retained eigenfunctions. This error is called \emph{aliasing error}, and previous analyses of real-space simulation algorithms control aliasing through high-order regularity bounds on the evolving wavefunction~\cite{chakrabarti2025speedups}. For black-box Lipschitz potentials, the grid resolution obtained from such bounds can grow exponentially with the dimension.

Our analysis instead replaces the potential by a function $f_K\in\mathcal V_K$ that approximates $f$ uniformly. This replacement appears only in the analysis and our algorithm still queries $f$. The the evolutions generated by $f$ and by $f_K$ differ by an amount proportional to $\norm{f-f_K}_\infty$. The analysis then uses slow spectral growth of the eigenfunctions:
\begin{equation}
    \mathcal H_{\ell_1}\mathcal H_{\ell_2}
    \subseteq \mathcal V_{\ell_1+\ell_2}.
    \label{eq:results-slow-spectral-growth}
\end{equation}

This condition states the product of eigenfunctions $Y_{\alpha}, Y_{\beta}$ with $\alpha \in L^{-1}(\ell_1)$ and $\beta \in L^{-1}(\ell_2)$ has no component $Y_{\gamma}$ with  $L(\gamma) > \ell_1+\ell_2$. Consequently, each multiplication by $f_K$ raises the $L$-index of the wavefunction by at most $K$. If the initial state lies in $\mathcal V_L$, the term of the interaction-picture Dyson series with $q$ applications of $f_K$ lies in $\mathcal V_{L+qK}$, and the norms of these terms decrease with $q$. Choosing the cutoff $N$ large enough to contain the terms of low order therefore bounds the truncation error without estimating derivatives of the evolving wavefunction.

Slow spectral growth also determines how many quadrature points are needed. A matrix element of $f_K$ between two eigenfunctions in $\mathcal V_N$ is the integral of a product of three functions of known degree, so the integrand lies in $\mathcal V_{2N+K}$. A quadrature rule that integrates exactly every function in $\mathcal V_{2N+K}$ therefore reproduces the matrix of $f_K$ on $\mathcal V_N$ without aliasing error. This approach is called \textit{oversampling}, because it uses more points than are needed to represent a state in $\mathcal V_N$. The total error then consists of the uniform approximation error $\norm{f-f_K}_\infty$ and the contribution of the terms of high order in the Dyson series; See Corollary~\ref{cor:oversampled-quadrature-error}.

We first give explicit constructions for torus, sphere and simplex. On the torus, averaging over a finite Fourier kernel approximates the potential, and a padded quantum Fourier transform changes representations. On the sphere, evolving the potential for a short time under the heat equation suppresses high spherical harmonics while changing a Lipschitz function only slightly. The recursive structure of spherical harmonics then reduces the spectral transform to a sequence of one-dimensional polynomial transforms. On the simplex, averaging over multinomial samples produces a finite-degree polynomial approximation, while a recursive factorization of the global polynomial eigenfunctions provides the corresponding transform. The sphere and simplex transforms have polynomial rather than polylogarithmic dependence on the one-dimensional cutoff, but the discretization theorems keep that cutoff polynomial in the relevant problem parameters; see Theorems~\ref{thm:constructive-sphere-complexity}, \ref{thm:spherical_harm_total_disc_error}, \ref{thm:constructive-simplex-complexity}, and~\ref{thm:simplex_wright_fisher_total_disc_error}.

The second group of examples consists of constrained domains from optimization. The positive orthant encodes positivity constraints, its invertible affine images encode systems of $d$ linearly independent linear inequalities, and $\ell_p$ balls encode norm constraints. On these domains, an explicit change of coordinates makes the metric Euclidean. On the positive orthant, logarithmic coordinates transform the Laplace--Beltrami operator of the log-barrier Hessian metric into the Euclidean Laplacian, so simulation on a truncated orthant reduces to simulation on a box. On a truncated $\ell_p$ domain, the Duffy map, a triangular change of coordinates, gives the same reduction for a metric that we define through this map.

The global results therefore use two kinds of structure. On the torus, sphere, and simplex, the eigenfunctions satisfy slow spectral growth, the potential has explicit uniform approximations by finitely many eigenfunctions, and the quantum transforms are built from explicit quadrature rules. On the orthant and $\ell_p$ domains, a change of coordinates makes the metric Euclidean.

\subsubsection{Coherent-chart methods}
\label{sec:results-coherent-chart}
The coherent-chart method applies when no tractable transform to a global eigenbasis of $-\Delta_g$ is available. Instead of expanding the wavefunction in global eigenfunctions, we use ordinary Fourier expansions in local coordinate boxes. To avoid discontinuities when a local function is extended periodically across the boundary of its box, we localize the wavefunction with a smooth quadratic partition of unity $\{\chi_\ell\}_\ell$. Each $\chi_\ell$ is supported away from the boundary of the corresponding box, and $\sum_\ell\chi_\ell^2=1$. The localized function $\chi_\ell u$ therefore vanishes near the boundary and has a periodic Fourier expansion in that chart.

Because coordinate charts overlap, the same global amplitude can appear in several localized functions. These functions must remain as components of one quantum state so that their amplitudes interfere correctly when they are mapped back to the manifold. In the $\ell$th chart, we first multiply by $\chi_\ell$, then express the result in the coordinates $y=\kappa_\ell(x)$, and finally multiply by $\rho_\ell(y)^{1/2}$, where $\rho_\ell(y)\dd y$ is the Riemannian volume measure in those coordinates. The last operation converts the local weighted $L^2$ norm to the ordinary Lebesgue $L^2$ norm. These operations define the isometry
\begin{equation*}
    (\mathcal T u)_\ell(y)=\rho_\ell(y)^{1/2}\chi_\ell(\kappa_\ell^{-1}(y))u(\kappa_\ell^{-1}(y)).
\end{equation*}
The adjoint $\mathcal T^\dagger$ maps the local functions back to the manifold and adds their amplitudes on chart overlaps. Thus $\mathcal T$ provides a coherent local representation of one global wavefunction.

The main obstacle is that localizing the wavefunction also changes its kinetic energy. If one applies $-\Delta_g$ separately to the functions $\chi_\ell u$ and then recombines them, the derivatives act on both $u$ and $\chi_\ell$. The resulting extra terms prevent the uncorrected local operators from reproducing $-\Delta_g$. The Riemannian IMS identity computes their total contribution exactly:
\begin{equation*}
    -\Delta_g=\sum_\ell\chi_\ell(-\Delta_g)\chi_\ell-W_\mathrm{IMS},\qquad W_\mathrm{IMS}=\sum_\ell\norm{\dd\chi_\ell}_{g^{-1}}^2.
\end{equation*}
Here $\norm{\dd\chi_\ell}_{g^{-1}}^2=\norm{\nabla\chi_\ell}_g^2$ measures how rapidly the cutoff changes across the manifold. The potential commutes with the cutoff functions, so no analogous correction is needed for the potential term. We subtract the full IMS potential $W_\mathrm{IMS}$ from every local kinetic block. After converting the local blocks to flat measure, their corrected direct sum $\widetilde{\boldsymbol H}^{\oplus}(t)$ satisfies
\begin{equation*}
    H(t)=\mathcal T^\dagger\widetilde{\boldsymbol H}^{\oplus}(t)\mathcal T.
\end{equation*}
The IMS correction therefore turns the collection of local differential operators into an exact representation of the global Riemannian Hamiltonian before any Fourier truncation is introduced.

A spectral simulation replaces an infinite expansion by finitely many modes, and the cutoff determines both the approximation error and the resources needed to represent the truncated operator. An algorithm that retains coordinate frequencies through $N$ requires $\mathcal O(d\log N)$ qubits for the local Fourier labels. The largest retained derivatives also grow with $N$, so the normalization and simulation cost of the local kinetic blocks depend on the same parameter. A useful convergence result must therefore show that the target accuracy can be reached without requiring the cutoff to grow exponentially with the dimension.

We analyze the cutoff in two stages. First, for a general compact manifold, we give conditions ensuring that the energy cutoff required to truncate the continuum model to local Fourier modes grows at most polynomially with the manifold dimension. Second, for products of fixed-dimensional manifolds, we give conditions ensuring that the retained continuum modes admit a finite-dimensional discretization with polynomial local resolution.

\paragraph{Polynomial cutoff in the continuum.} We begin with an analysis of the algorithm in the continuum. In each chart, the projection $P_N$ retains the Fourier modes with $\norm{k}_\infty\leq N$ and discards all higher coordinate frequencies. Applying this projection before and after each local operator restricts the input to the retained modes and removes higher frequencies created by the variable coefficients.

Theorem~\ref{thm:coherent-chart-continuum-error} quantifies how well the truncated local operators approximate the exact evolution. While we refer to the theorem for the precise regularity conditions, the error is controlled when the evolving wavefunction has sufficiently many square-integrable derivatives, and when the change of coordinates, metric coefficients, volume densities, partition of unity, potential, and IMS correction have bounded derivatives through the same order. These conditions ensure that the high-frequency components discarded by $P_N$ decrease as $N$ grows. If the number of controlled derivatives is proportional to $d$ and the corresponding derivative bounds grow no faster than $\operatorname{poly}(d)^d$, Corollary~\ref{cor:poly_cutoff} shows that $N=\operatorname{poly}(d)$ suffices for a fixed accuracy.

\paragraph{Polynomial cutoff after discretization.} A finite-dimensional implementation introduces a second problem: grids belonging to overlapping charts generally represent different points of the manifold. For a product of constant-dimensional manifolds $\mathcal{M}=\prod_{\nu=1}^m\mathcal{M}_\nu$, we connect the local grids through finite source spaces defined separately on the factors. Choose a finite basis $\{L_q^{(\nu)}\}_q$ for the source space on each factor. Products of these basis functions form a basis for the global source space, so their coefficients define one global function. Evaluation maps send these global coefficients to the values of that function on each chart grid. A local discrete Fourier transform converts the sampled values to Fourier coefficients, the local Hamiltonian acts in the chart representation, and the adjoint evaluation maps return the result to the global coefficients. Because the source bases and evaluation maps factor across the manifolds $\mathcal{M}_\nu$, these operations can be applied factor by factor on constant-dimensional (for an appropriately chosen source interpolation basis) registers. The potential may still couple all factors through its value at a product-grid point.

Theorem~\ref{thm:tensor-local-main} analyzes the resulting finite-dimensional evolution. The result assumes that the geometry of each constant-dimensional factor remains uniformly smooth as the number of factors grows. It also requires source spaces that approximate sufficiently smooth functions without introducing an exponentially larger resolution, and it requires the exact and intermediate wavefunctions in the error analysis to remain sufficiently differentiable. Under the precise bounds stated in the theorem, with constants $C_0$ and $\beta$ independent of $d$ and $N$, satisfy
\begin{equation*}
    \sup_{0\leq t\leq T}\norm{u(t)-u_{R,N}(t)}_2\leq\left(\frac{C_0d^\beta}{N}\right)^d.
\end{equation*}
Consequently, $N=\lceil C_0d^\beta\epsilon^{-1/d}\rceil$ achieves error at most $\epsilon$.

\subsubsection{Applications}
\label{sec:results-applications}

Section~\ref{sec:applications} studies physical and optimization problems can be expressed as evolution under a Riemannian Schr\"odinger operator. For these applications, we give details that illustrate the uplift that can be obtained by quantum simulation on quantum manifolds. For many of the highlighted settings, the geometry and boundary in question is compatible with one of the end-to-end simulation algorithm given in this paper. However, we also describe some more general cases.

Most of the applications below should not be interpreted as complete end-to-end quantum speedups for the surrounding task. Firstly, in addition to a simulation procedure, an end-to-end algorithm might also require the analysis of state preparation, measurement procedure, spectral gaps, or other convergence guarantees. We therefore distinguish applications that identify Riemannian Schr\"odinger evolution as a required subroutine, from settings where the end-to-end algorithm follows naturally. Secondly, we do not make an effort in this paper to optimize the polynomially scaling resources required by our simulation algorithms (a notable exception is our analysis for simulation on the torus which is the most efficient that we have found in the literature.) Therefore, even when a natural end-to-end algorithm arises from our results, it does not immediately yield a quantum speedup. Finally, a detailed comparison with classical algorithms is well beyond the scope of this paper.

\paragraph{Physical simulation.} Section~\ref{sec:physics-simulations} shows that several well-studied physical systems can be written as Riemannian Schr\"odinger equations on high-dimensional configuration spaces. The examples include quantum mechanics on curved spaces, nonlinear sigma models with general target manifolds, $\mathbb{C}P^{N-1}$ models, and Hamiltonian lattice gauge theories. After spatial discretization, the field-theory Hamiltonians act on products of the target manifold or gauge group, with a sum of Laplace--Beltrami operators as the kinetic term and the interactions as a multiplication potential. We then apply the sphere algorithm to a spatially discretized $O(N)$ nonlinear sigma model with a nearest-neighbor chordal interaction in Section~\ref{sec:nlsm-spherical-simulation}. Theorem~\ref{thm:on-nlsm-digital-simulation} gives an end-to-end digital simulation with explicit bounds on the state error and on the query, gate, and qubit costs.

\paragraph{Interior-point methods.} Section~\ref{sec:connection-to-IPMs} discusses the quantum path-following method of Gribling, Apers, Nieuwboer, and Walter~\cite{gribling2025selfconcordantschrodingeroperatorsspectral}, which uses Schr\"odinger operators whose kinetic term is the Laplace--Beltrami operator of a self-concordant barrier metric. Their algorithm assumes access to controlled evolution under these operators. For the standard logarithmic barrier on the positive orthant, Theorem~\ref{thm:orthant-affine-digital-simulation} supplies a digital simulation of this evolution after the domain is truncated to a compact box. The result therefore implements one of the central algorithmic primitives required by that method.

\paragraph{Quantum Hamiltonian Descent.} The QHD algorithm~\cite{leng2023quantumhamiltoniandescent} performs optimization by evolving a wavefunction under a time-dependent Schr\"odinger equation whose potential is the objective function. Its Euclidean analysis uses a Lyapunov function to prove convergence for convex potentials. Theorem~\ref{thm:manifold-qhd} in Section~\ref{sec:riemannian-qhd} extends this analysis to Riemannian manifolds and geodesically convex potentials. The sphere and simplex specializations in Theorems~\ref{thm:sphere-qhd} and~\ref{thm:simplex-qhd} combine with our digital simulation algorithms to give end-to-end optimization algorithms for geodesically convex functions on those manifolds.

\paragraph{Real-Space Adiabatic Algorithm.} Herman et al.~\cite{herman2025mechanisms} introduced the Real-Space Adiabatic Algorithm algorithm to study quantum speedups for structured nonconvex optimization problems in Euclidean space. The algorithm begins from the ground state of a kinetic operator and adiabatically turns on the objective potential, so that the final ground state concentrates near the minimizers. In~\cite{herman2025mechanisms}, the authors identify mechanisms by which this algorithm offers large quantum speedups over known classical algorithms. Section~\ref{sec:rsaa-riemannian} defines the corresponding Riemannian algorithm by replacing the Euclidean Laplacian with $-\Delta_g$. Theorem~\ref{thm:riemannian-real-space-adiabatic} gives the adiabatic guarantee, and Theorem~\ref{cor:digital-riemannian-rsaa} gives a digital implementation on a periodic Duffy $\ell_p$ domain. The Duffy map makes this adiabatic path unitarily equivalent to the corresponding path on a Euclidean box, so the mechanisms identified by Herman et al. carry over to these constrained problems.

\subsection{Related work}
\label{sec:related-work}
\paragraph{Real-space simulation on flat domains.}
Digital simulation of the time-dependent Schr\"odinger equation on a position grid goes back to Wiesner and Zalka~\cite{wiesner1996simulations,zalka1998simulating}. Kassal et al.\ used the grid-based split-operator approach, which alternates between position and momentum representations via the quantum Fourier transform, to simulate chemical dynamics~\cite{kassal2008polynomial}. First-quantized plane-wave algorithms later achieved sublinear scaling in the basis size~\cite{babbush2019quantum} and explicit fault-tolerant resource estimates for chemistry~\cite{su2021fault}. The analyses discussed in the introduction~\cite{kivlichan2017bounding,childs2022quantumsim,chakrabarti2025speedups} all bound the discretization error through the regularity of the evolving wavefunction, either assumed or, in the black-box Lipschitz setting, proved. Our global spectral method takes a different approach: aliasing is removed exactly for a finite-mode comparison potential, so no high-order regularity bound on the evolution is needed. On the flat torus, this yields a register of $\mathcal{O}(d\log Q)$ qubits with $Q=\operatorname{poly}(d,1/\epsilon)$ (Theorem~\ref{thm:improved_torus_discretization}). 

\paragraph{Quantum simulation on curved configuration spaces in physics.}
Several physical models already have curved configuration spaces, and their quantum simulation has been studied directly. For $O(N)$ nonlinear sigma models, proposals include an analog laboratory simulator for the $O(3)$ model~\cite{schutzhold2005quantum}, a digital scheme that replaces the target sphere by the fuzzy sphere while preserving the exact $O(3)$ symmetry~\cite{alexandru2019sigma}, and qubit regularizations with few qubits per site that reproduce the relevant critical behavior~\cite{singh2019qubit,singh2022qubit}; Hamiltonian lattice gauge theories raise the same questions for compact gauge groups~\cite{banuls2020simulating}. These works construct model-specific finite Hilbert spaces and justify them through universality arguments supported by numerics, rather than through bounds on the error relative to the continuum dynamics. On the contrary, we provide a general construction that applies to any compact manifold with the required structure, together with explicit bounds on the error relative to the continuum Schr\"odinger evolution with a black-box potential.

\paragraph{Quantum dynamical optimization with geometry.}
Beyond the works mentioned in the introduction, Quantum Hamiltonian Descent now includes stochastic~\cite{peng2025stochastic} and gradient-augmented \cite{leng2025gradient} variants, as well as fault-tolerant resource estimates~\cite{liu2026encoding}. In all of these, the kinetic term is Euclidean. Constraints have so far been handled outside the dynamics, for instance by embedding QHD in an augmented-Lagrangian loop that solves a sequence of unconstrained subproblems~\cite{li2025qhdalm,wu2026alqhd}. Two works put geometry into the kinetic term: Gribling et al.~\cite{gribling2025selfconcordantschrodingeroperatorsspectral} assume access to the Riemannian evolution, and Abe and Nagai~\cite{abe2026quantum} assume sparse access to a chosen discretization of the Laplace--Beltrami operator; neither constructs these objects or bounds the resulting error. Our algorithms supply this step for the manifolds in Table~\ref{tab:simulation-results-summary}; Section~\ref{sec:applications} discusses the consequences for optimization in more detail.

To our knowledge, the global spectral methods of this paper are the first digital quantum algorithms with end-to-end discretization error bounds for Schr\"odinger dynamics with a black-box potential on non-flat configuration spaces, and the coherent-chart method extends the continuum convergence analysis to general compact manifolds under explicit geometric and regularity assumptions.

\section{Preliminaries and Notation}
\label{sec:preliminaries}

    Riemannian geometry is a rich topic of mathematics that we cannot hope to cover in detail here. See~\cite{lee2018introduction} for an excellent introduction. We will merely establish some basic notation and results relevant to this work. Throughout this work, the natural numbers $\mathbb{N}$ are defined with $0$ included.

    A manifold is a Hausdorff, second countable topological space $\mathcal M$, which is locally Euclidean, in the sense that there exists an \emph{atlas} $\{(U_\alpha, \kappa_\alpha)\}_{\alpha \in \mathcal{I}}$, where $\{U_\alpha\}_{\alpha\in\mathcal{I}}$ is an open cover, and 
    \begin{equation}
        \kappa_\alpha:U_\alpha \rightarrow \Omega_\alpha \subseteq \mathbb{R}^d
    \end{equation}
    is a homeomorphism defining local coordinates. Here $d \in \mathbb{Z}_+$ is the manifold dimension. $\mathcal M$ is, more specifically, a \emph{smooth} manifold when the transition maps
    \begin{equation}
        \tau_{\alpha\beta}=\kappa_\alpha\circ\kappa_\beta^{-1}.
    \end{equation}
    are diffeomorphisms whenever $U_\alpha \cap U_\beta$ is not empty. For finite-dimensional smooth manifolds, it is known that there exists a smooth atlas containing at most $d + 1$ charts~\cite{solecki1974finite}. From computational purposes, assuming a finite set of charts is natural and perhaps necessary. In this work, \emph{all atlases are finite}, and one can take $\mathcal{I} = [n_\mathrm{ch}]$ for some $n_\mathrm{ch} \in \mathbb{Z}_+$. Smoothness allows for the construction of \emph{tangent spaces} $T_p \mathcal M$ at each point $x\in \mathcal M$, and the disjoint union of these $d$-dimensional vector spaces defines the tangent bundle $T\mathcal M$.     
    
    A \emph{Riemannian} manifold $(\mathcal M, g)$ is a smooth manifold $\mathcal M$ endowed with a metric $g$. This metric is a smooth, positive definite symmetric $(0,2)$-tensor field, which defines an inner product at each $p \in \mathcal M$ via
    \begin{equation}
        \langle w,v\rangle_g \coloneqq g_p(w,v), \qquad w,v \in T_p \mathcal M.
    \end{equation}
    Local coordinates computations via the atlas provide a convenient setting for calculation, and are indicated by use of indices with appropriate subscript or superscript. We denote the metric in coordinates as $g_{ij}$ and inverse metric is given by $g^{ij}$. Einstein summation notation is used for brevity.
    
    As the name suggests, the metric provides a local notion of distance, and encodes geometrical properties of the space such as curvature, volume, and shortest paths between points. Let us now define those relevant to this work. Given a smooth vector field $v$, define the scalar field $\abs{v}_g \coloneqq \sqrt{g(v,v)}$. Let $\abs{g} \coloneqq \det (g_{ij})$. In local coordinates $x = (x^1,\ldots, x^d)$, the natural, coordinate-invariant volume measure on $\mathcal M$ is then
    \begin{equation}
        \dvol(x) \coloneqq \rho(x) dx, \qquad \rho(x) \coloneqq \sqrt{\abs{g(x)}}
    \end{equation}
    where $dx\equiv dx^1\ldots dx^d$ is the volume form. This can be used to define the Hilbert space of wavefunctions on $\mathcal M$.
    \begin{equation}
        \mathcal{H} = L^2(\mathcal M, \dvol) = \left\{\psi:\mathcal M \rightarrow \mathbb{C} \, \middle\vert \, \int_\mathcal M \abs{\psi(x)}^2 \dvol(x) < \infty\right\}
    \end{equation}
    Because $\mathcal M$ is second countable, $\mathcal{H}$ is separable~\cite{folland1999real}, thus admitting a countable orthonormal basis. For brevity, we may simply write $L^2$ or $L^2(\mathcal M)$. The inner product $\langle \psi, \phi\rangle$ and norm $\norm{\psi}$ are then defined in the usual manner on $\mathcal{H}$. 
    The distance $d_g$ between two points $x, y$ in (a connected component of) $\mathcal M$ is
    \begin{equation}
        d_g(x,y) \coloneqq \inf_{\gamma \in C(x,y)} \int_0^1|\dot\sigma(u)|_g\, du
        \label{eq:riemannian-distance-length}
    \end{equation}
    where $C(x,y)$ is the set of smooth curves from $x$ to $y$ parametrized on $[0,1]$, and $\dot{\sigma}(u) \in T_{\sigma(u)} \mathcal M$ denotes a derivative. 
    
    Let $f, v$ be a smooth scalar and vector fields, respectively. Riemannian generalizations of the gradient and divergence are given, in coordinates, by
    \begin{equation}
        \nabla^i f = g^{ij}\partial_j f, \qquad \nabla \cdot v = \frac{1}{\sqrt{\abs{g}}} \partial_i\left(\sqrt{\abs{g}} v^i\right).
    \end{equation}
    The natural generalization of the Laplacian to Riemannian manifolds, sometimes termed the \allowbreak \emph{Laplace--Beltrami} operator, is then simply the divergence of the gradient. Explicitly, 
    \begin{equation}
    \label{eq:LB}
        \Delta_g f
        =\rho^{-1}\partial_i\left(\rho g^{ij}\partial_j f\right).
    \end{equation}
    When $\mathcal M$ is compact, $\Delta_g$ is negative semi-definite and its eigenfunctions form a smooth orthogonal basis. The spectrum of $-\Delta_g$ is a discrete sequence tending to infinity
    \begin{equation*}
        0=\lambda_0 < \lambda_1 < \ldots < \lambda_n \rightarrow +\infty.
    \end{equation*}
    The associated eigenspaces are finite dimensional. See \cite{rosenberg1997laplacian} for more detail. 
    
    Let $\exp_x:T_x\mathcal M\to \mathcal M$ be the exponential map, which sends a tangent vector $v\in T_x\mathcal M$ to the point reached at time $1$ by the geodesic starting at $x$ with initial velocity $v$.  Wherever $\exp_x$ is one-to-one along the minimizing geodesic from $x$ to $y$, the Riemannian logarithm $\log_x: U_x \rightarrow T_x \mathcal M$ is defined as its local inverse.
    \begin{equation}
        \log_x(y)=v \quad \iff \quad \exp_x(v)=y
    \end{equation}
    Equivalently, if $\gamma \in C(x,y)$ is the unique minimizing geodesic from $x$ to $y$, then
    \begin{equation}
        \log_x(y)=\dot\gamma(0), \qquad
        |\log_x(y)|_g=d_g(x,y).
    \end{equation}
    In short, $\log_x(y)$ naturally generalizes the idea of the vector $y - x$ in Euclidean space, which points from $x$ to $y$ and whose length is the distance. The corresponding unit direction is $\log_x(y)/d_g(x,y)$ when $x\neq y$. We emphasize that this logarithm function is defined only under the provided assumptions of unique geodesics. A particular setting where this occurs is for geodesically convex sets.
    \begin{definition}[Geodesic convexity]
        A subset $\mathcal{C}$ of a Riemannian manifold $(\mathcal M,g)$ is geodesically convex if for each $p,q \in S$, there exists a unique minimizing geodesic segment from $p$ to $q$, whose image lies entirely in $\mathcal{C}$. 
    \end{definition}
    \noindent Geodesically convex sets always exists: take, for instance, sufficiently small geodesic balls. In continuous optimization, convexity of a function is a useful and often practical assumption that allows for efficient gradient-based algorithms. See~\cite{bubeck2015convex} or other standard sources for background and basic definitions. A natural generalization to manifolds is as follows.
    \begin{definition}[Geodesically convex functions] \label{def:geodesic-convex-function}
        For $\mathcal{C}$ geodesically convex, a function $f:\mathcal{C}\to\mathbb{R}$ is geodesically convex if, for every unit-velocity geodesic arc $\gamma$ contained in $\mathcal{C}$, the composition $f\circ\gamma$ is convex as a real function.
    \end{definition}
    
    \begin{proposition}[Geodesic convexity, differentiable function] \label{lem:geodesic-convexity}
        For $\mathcal{C}$ geodesically convex, a differentiable function $f:\mathcal{C}\to\mathbb{R}$ is geodesically convex if and only if for every $x, y \in \mathcal{C}$ connected by a geodesic:
        \begin{align*}
              f(y)\geq f(x)+\ip{\nabla f(x)}{\log_x(y)}_g
        \end{align*}
    \end{proposition}
    \noindent We say that a function $f:\mathcal M\rightarrow \mathbb{R}$ is \emph{geodesically $G$-Lipschitz} if it is $G$-Lipschitz continuous with respect to $d_g$. 

    A useful identity for bounding the difference in two Hamiltonian evolutions is given by Duhamel's formula.
    \begin{fact}[Duhamel's formula] \label{fact:duhamels}
        Let $H(t)$ and $\widetilde H(t)$ be self-adjoint families on a Hilbert space
        with unitary propagators, and suppose both evolutions start from $u(0)$. If $u$ and $\widetilde u$ solve the two Schr\"odinger equations with initial state $u(0)$ then
        \begin{align}
             \norm{u(t)-\widetilde u(t)}
             \leq \int_0^t
             \norm{[H(s)-\widetilde H(s)]u(s)}\dd s.
             \label{eq:general-Duhamel-generator-bound}
        \end{align}
    \end{fact}

\section{Global Spectral Methods}
\label{sec:global_spectral_methods}
Spectral methods approximate the solution of a differential equation using a
finite-dimensional space spanned by globally supported basis functions.
The basis is chosen to reflect the domain, boundary conditions, and
differential operator; standard examples include Fourier modes on periodic
domains and orthogonal polynomial systems, such as Chebyshev or Legendre
polynomials, on bounded intervals. Galerkin methods impose the equation by
projection, whereas collocation and pseudospectral methods evaluate some
terms on a quadrature grid. For a periodic Euclidean Schr\"odinger equation,
the natural specialization is the Fourier pseudospectral method: a finite
Fourier series represents the wavefunction, the kinetic operator is diagonal
in the Fourier basis, and the potential is evaluated in the dual position
representation. This specialization was used for real-space quantum
simulation in~\cite{childs2022quantumsim}, and a rigorous end-to-end
analysis for black-box Lipschitz potentials was given in~\cite{chakrabarti2025speedups}.

On a compact Riemannian manifold, the natural generalization is to replace
Fourier modes by global eigenfunctions of the nonnegative
Laplace--Beltrami operator. One retains a finite spectral subspace, applies
the kinetic evolution diagonally in its coefficient basis, and evaluates
the potential by mapping to weighted values on a quadrature grid and back.
We call this construction the \emph{global spectral method}.

This formal replacement does not by itself give an efficient digital algorithm. One must control both leakage from the the finite-dimensional space of retained eigenmodes (i.e. Galerkin method) under multiplication by the potential and aliasing when its matrix elements of the projected Hamiltonian are replaced by finite quadrature (i.e. collocation method). In the periodic Euclidean setting, \cite{chakrabarti2025speedups} controls dynamical aliasing for black-box Lipschitz potentials using mollification and high-order Sobolev bounds on the evolving wavefunction. Since the derivative order grows with the dimension and the mollified potential is not band limited, the resulting sufficient grid resolution can be exponential in the dimension. This does not yield a polynomial implementation of the spherical and simplex constructions below, whose transform costs are polynomial in the one-dimensional quadrature resolution. 

For each of the manifolds that we consider, we introduce a uniformly close, exactly band-limited proxy for the potential only in the analysis. Thus for finite time, if the initial wave function has finite bandwidth, then the Galerkin evolution  can be shown to be close to a truncate evolution with finite-bandwidth. Since the wavefunction now only have support on a finite band, we can use Gauss quadrature to exactly evaluated the Hamiltonian matrix elements, by sufficient oversampling. The framework below formalizes these analytic and implementation requirements before treating the torus, sphere, and simplex.

Consider finite-dimensional spaces spanned by eigenfunctions of the nonnegative Laplace--Beltrami operator $-\Delta_g$ on $L^2(M)$. Let
$ \{Y_{\alpha}\}_{\alpha \in \mathcal{I}}$ %
be an orthonormal eigenbasis, so that
\begin{equation}
 -\Delta_gY_{\alpha}=\lambda_{\alpha}Y_{\alpha},
 \label{eq:filtered-Laplacian-eigenbasis}
\end{equation}
where it leave it implicit that we allow for degeneracy, for example $\lambda_{\alpha'} = \lambda_{\alpha}$ for $\alpha \neq \alpha'$. We introduce a binning function for the eigenfunctions $L : \mathcal{I} \rightarrow \mathbb{N}$. Define the $\ell$-the level to be
\begin{equation*}
    \mathcal H_\ell
    \coloneqq \operatorname{span}\{Y_{\alpha}: L({\alpha}) = \ell\},
    \qquad
    \mathcal{V}_N \coloneqq \bigoplus_{\ell=0}^N\mathcal{H}_{\ell}.
\end{equation*}
We assume that every $\mathcal V_N$ is finite-dimensional and that their
union is dense in $L^2(\mathcal{M})$. Denote by $P_N$ the orthogonal projector onto
$\mathcal V_N$. For example in the case of Fourier modes $e^{2\pi i k \cdot x}$, the levels are determined by $\lVert k \rVert_{\infty}$.

We will use the following notion of a $Q$-exact quadrature rule for $(\mathcal{M}, g, \Delta_g)$. Such a rule will be required by our algorithm and explictly constructed for our example manifolds.
\begin{definition}[$Q$-exact quadrature]
\label{def:q_exact_quad}
Using the notation introduced above. We say that  $\{(x_j,w_j)\}_{j=1}^{J_Q}$, with points $x_j \in \mathcal{M}$ and associated weights $w_j > 0 $ form a $Q$-exact quadrature rule if  
\begin{equation*}
    \langle \phi, \psi\rangle_{w_Q} \coloneqq \sum_{j=1}^{J_{Q}} w_j \overline{\phi}(x_j)\psi(x_j)
\end{equation*}
is equal to the standard inner product $\langle\phi, \psi\rangle$ on the subspace $\mathcal{V}_Q$. 
\end{definition}

For example, for the symmetric Fourier filtration, $Q$-exactness requires
the scalar quadrature to integrate every Fourier mode with
$\lVert k\rVert_{\infty}\leq 2Q$ exactly. The symmetric grid with
$2Q+1$ points per coordinate has this property.

Fix $N\leq Q$ and consider a computational basis consisting of elements
$\ket{\alpha}$ for $L(\alpha) \leq N$, used to represent $\mathcal{V}_N$
in the spectral basis. Then, the linear map
\begin{equation}
    W\ket{\alpha} = \ket{Y_{\alpha}} \coloneqq \sum_{j=1}^{J_Q}\sqrt{w_j}\,Y_{\alpha}(x_j)\ket{j},
    \qquad L(\alpha) \leq N
    \label{eq:global-weighted-spectral-transform}
\end{equation}
is an isometry. As a guiding example, on the flat torus with standard
Laplacian, choose the symmetric $(2Q+1)$-point grid in each coordinate.
Then $J_Q=(2Q+1)^d$, and the weighted evaluation isometry is a padded
inverse $(2Q+1)$-point QFT. The odd grid length prevents the endpoint
modes $Q$ and $-Q$ from aliasing.

With these ingredients, and from a spectral point of view the reader may anticipate the natural discretized Hamiltonian to consider. For the potential, let $\hat{f} \coloneqq \sum_j f(x_j) \ketbra{j}{j}$. Define the diagonalized Laplace--Beltrami operator mode by mode as
\begin{equation} \label{eq:discretized-kinetic-operator}
    D_N \coloneqq \sum_{\alpha : L(\alpha) \leq N} \lambda_{\alpha}|\alpha\rangle\langle \alpha|.
\end{equation}
Then the discrete Hamiltonian we will simulate is 
\begin{equation}
    \hat{H}(t) \coloneqq a(t)D_N+b(t)W^\dagger\hat fW.
    \label{eq:discretized-Hamiltonian}
\end{equation}

With access to $f$ provided via, say, standard binary oracles, one arrives at setting where methods of finite-dimensional time dependent Hamiltonian simulation can be applied. In particular, we will consider an interaction picture method in the frame of $U_0$, the propagator for $a(t)D_N$. Additionally, in our setting $W$ will likely be only an isometry, due to the number of grid points $J$ being chosen to be larger than $N$. This will not be a problem as the underlying digital simulation algorithm only needs  access to a block-encoding  for $W^{\dagger}\hat{f}W$.

The particular implementation of $W, W^\dagger$, determination quadrature, etc. are necessarily case-by-case, and not generally efficient. For several cases of interest where these transformations are tractible and provide polynomial-time algorithms, are first considered starting in subsection~\ref{subsec:flat-torus}. Before that, we first consider a general error analysis which informs the required overhead needed to achieve a given accuracy with respect to the true continuous dynamics.

\subsection{General framework}

Consider the ideal Schr\"odinger evolution
\begin{align*}
    i\partial_t u(t) =  [-a(t)\Delta_g + b(t)f]u(t),
\end{align*}
for initial data $u(0) \in L^2(\mathcal{M})$. The Galerkin evolution corresponding to $P_N$ is
\begin{align*}
    i\partial_t u_N(t) =  -a(t)\Delta_gu_N(t)+ b(t)P_N(fu_N(t)), \quad u_N(0) = u(0)
\end{align*}
which continuously projects the evolution back into $\mathcal{V}_N$ even as $f$ induces transitions outside $\mathcal{V}_N$. However, even this approximation to the actual evolution may not be possible to implement in practice. This is because it may not be possible to exactly compute matrix elements of the Hamiltonian, for example $P_N f P_N$, using a finite grid if $f$ does not lie in $\mathcal{V}_N$. This due to the aliasing phenomenon, which is a result of the quadrature grid being unable to resolve components associated with eigenfunctions outside $\mathcal{V}_N$. Thus with numerical quadrature the actual evolution corresponds to
\begin{align*}
     i\partial_t u_{N,Q} = -a(t)\Delta_g u_{N,Q} + b(t)I_{N,Q}\bigl(fu_{N,Q}\bigr), \qquad u_{N,Q}(0)=u(0),
\end{align*}
where
\begin{equation}
    I_{N,Q} \psi \coloneqq \sum_{\alpha \in \mathbb{N}~|~L(\alpha) \leq N} \langle Y_{\alpha}, \psi\rangle_{w_Q} Y_{\alpha}.
    \label{eq:quadrature-projection}
\end{equation}
The following derives a direct connection between the above evolution and the one in \eqref{eq:discretized-Hamiltonian}.
\begin{proposition}
\label{prop:quadrature-Hamiltonian-intertwining}
Let $n_N=\dim\mathcal V_N$ and let
\begin{equation*}
    S_N:\mathbb C^{n_N} \rightarrow \mathcal V_N,
    \qquad S_N\ket{\alpha}=Y_{\alpha},
\end{equation*}
be the coefficient-to-function unitary. Let $W$ be the  isometry in \eqref{eq:global-weighted-spectral-transform},
formed using the same $Q$-exact quadrature rule as $I_{N,Q}$, and let
$\hat H(t)$ be defined by \eqref{eq:discretized-Hamiltonian}. Then
\begin{equation}
    S_N\hat H(t)S_N^\dagger v
    =-a(t)\Delta_gv+b(t)I_{N,Q}(fv),
    \qquad v\in\mathcal V_N.
    \label{eq:quadrature-Hamiltonian-generator-intertwining}
\end{equation}
Consequently, if $ u_{N,Q}$ solves
\eqref{eq:quadrature-evolution}, then
\begin{equation*}
    c_{N,Q}(t)\coloneqq S_N^\dagger u_{N,Q}(t)
\end{equation*}
solves
\begin{equation}
    i\partial_tc_{N,Q}(t)=\hat H(t)c_{N,Q}(t).
    \label{eq:quadrature-Hamiltonian-coefficient-evolution}
\end{equation}
\end{proposition}

\begin{proof}
For $v=S_Nc\in\mathcal V_N$, the definition of $D_N$ gives
$S_ND_Nc=-\Delta_gv$. Moreover, for every 
\begin{align*}
    \bra{\alpha}W^\dagger\hat fWc\rangle
    &=\sum_{j=1}^{J_Q}w_j\overline{Y_{\alpha}(x_j)}
        f(x_j)v(x_j)\\
    &=\langle Y_{\alpha},fv\rangle_{w_Q}.
\end{align*}
Applying $S_N$ and summing over $\ell,m$ gives
$S_NW^\dagger\hat fWS_N^\dagger v=I_{N,Q}(fv)$, which proves
\eqref{eq:quadrature-Hamiltonian-generator-intertwining}. Applying
$S_N^\dagger$ to \eqref{eq:quadrature-evolution} proves
\eqref{eq:quadrature-Hamiltonian-coefficient-evolution}.
\end{proof}
Because of the above relation,
we can then pass a block-encoding of the Hamiltonian \eqref{eq:discretized-Hamiltonian} to a standard time-dependent Hamiltonian simulation routine to approximate the evolution of the coefficients $\langle {Y}_{\alpha}, \psi\rangle_{w_{Q}}$. Suppose that $U_f$ is a
$\Lambda_f$-normalized block encoding of the sampled potential,
\begin{equation}
 (\bra{0^a}_A\otimes I_R)U_f(\ket{0^a}_A\otimes I_R)
 =\frac{\hat f}{\Lambda_f},
 \qquad \Lambda_f\geq\norm f_\infty.
 \label{eq:global-sampled-potential-block-encoding}
\end{equation}

Since $W$ is an isometry, there exists a unitary dilation $U_W$  acts on a
quadrature register $R$ and coefficient register $C$ according to
\begin{equation}
 U_W\bigl(\ket0_R\ket\psi_C\bigr)
 =\bigl(W\ket\psi\bigr)_R\ket0_C
 \label{eq:global-evaluation-unitary-dilation}
\end{equation}
for every valid coefficient state. Then
\begin{equation}
 U_V\coloneqq (I_A\otimes U_W^\dagger)(U_f\otimes I_C)
       (I_A\otimes U_W)
 \label{eq:global-compressed-potential-block-unitary}
\end{equation}
is a block encoding of the compressed potential on coefficient space:
\begin{equation}
 (\bra{0^a}_A\bra0_R\otimes I_C)U_V
 (\ket{0^a}_A\ket0_R\otimes I_C)
 =\frac{W^\dagger\hat fW}{\Lambda_f}.
 \label{eq:global-compressed-potential-signal-block}
\end{equation}
Such a block-encoded can be used by interaction picture simulation methods \cite{low2019interaction}. The evolution under $D_{N}$ is diagonal and thus can be fast forwarded. For each of the manifolds we consider, will need to show bound the cost of building this access.

In comparison to what \cite{childs2022quantumsim, chakrabarti2025speedups}, who took $Q = N$ for the Euclidean spectral method, we will choose $Q$ in a different manner. Particularly, our chose of $Q$ will come from analying how the support of a band-limited wavefunction evolves over time given a band-limited potential. The band-limited potential will be a sufficiently close proxy for the true one. Thus a sufficiently large $Q$ can make the Galerkin and quadrature-based evolutions equal if both are started from the same initial bounded-bandwidth wavefunction. In what follows, we will prove this first in the abstract setting under a set of sufficient technical conditions. Then in later sections we show that manifolds that we consider satisfy those conditions.

The first condition  we require the eigenfunction filtration
$\{\mathcal V_N\}_{N\geq0}$ to satisfy what we call \emph{slow spectral growth}. For
subspaces of pointwise-defined functions, write
\begin{equation*}
 \mathcal V_K\mathcal V_N
 \coloneqq \operatorname{span}\{uv:u\in\mathcal V_K,\ v\in\mathcal V_N\}.
\end{equation*}
We assume that the filtration is invariant under pointwise complex conjugation and that
\begin{align}
\label{eqn:eigfunc_mul_exp_prop}
    \overline{\mathcal V_N}=\mathcal V_N,
    \qquad
    \mathcal V_K\mathcal V_N\subseteq\mathcal V_{K+N},
    \qquad K,N\in\mathbb N,
\end{align}
where the overline denotes pointwise complex conjugation. Equivalently, it
is enough to require
$\mathcal H_k\mathcal H_n\subseteq\mathcal V_{k+n}$ together with
conjugation invariance. Since every level $\mathcal{V}_N$ is spanned by eigenfunctions
of $-\Delta_g$, the kinetic propagator preserves each $\mathcal V_N$ even
when a level contains several distinct eigenvalues. In our applications, the relevant levels
are Fourier degree $\|k\|_\infty$ on the flat torus, spherical-harmonic
degree on $\mathbb S^d$, and total Jacobi degree on the simplex.

The second requirement is a \emph{smoothing operation} that, for a bounded function $f$ and target accuracy $\epsilon>0$, produces a band limit $K$ and a function $f_K\in\mathcal V_K$ such that $\norm{f-f_K}_\infty\leq\epsilon$. This is to introduce the aformentioend finite-bandwidth proxy for the potential. Then suppose
$u$ and $u_K$ solve the Schr\"odinger equations with potentials $f$ and
$f_K$, respectively, from a common normalized initial state
$u(0)\in\mathcal V_{K_0}$. Write
$B_1(t)\coloneqq\int_0^t\abs{b(s)}\dd s$ and $B_1\coloneqq B_1(T)$.
Duhamel's formula gives
\begin{equation*}
    \norm{u(t)-u_K(t)}
    \leq B_1(t)\norm{f-f_K}_\infty.
\end{equation*}

The last requirement is a \emph{transform condition}. For each exactness degree $Q$, let $J_Q$ be the total number of quadrature
nodes. Associate to the construction a local resolution parameter
$R_Q\geq2$ satisfying
\begin{equation*}
    \log J_Q=\mathcal O(d\log R_Q).
\end{equation*}
For a product rule with coordinate node counts
$R_{Q,1},\ldots,R_{Q,d}$, one may take
$R_Q\coloneqq \max_iR_{Q,i}$.
For every $N\leq Q$, we require the weighted evaluation isometry
\eqref{eq:global-weighted-spectral-transform} and its adjoint to be
implementable to operator error $\epsilon_\mathrm{tr}$ using
\begin{equation*}
 \poly\!\left(d,N,R_Q,\log(1/\epsilon_\mathrm{tr})\right)
\end{equation*}
elementary gates and
\begin{equation*}
 \mathcal O\!\left(
 \log J_Q+\log\dim\mathcal V_N
 +\polylog(dNR_Q/\epsilon_\mathrm{tr})
 \right)
\end{equation*}
qubits.  For isotropic product grids,
$R_Q=\Theta(J_Q^{1/d})$.  The torus, sphere, and simplex constructions
below satisfy this condition with $R_Q=\Theta(Q)$.

This condition is deliberately weaker than the fast-transform regime
available in special cases.  Fourier transforms, and  Hermite
transforms \cite{jain2026efficient} in settings admitting fast quantum implementations, can have
gate complexity polynomial in
$d$, $\log R_Q$, and $\log(1/\epsilon_\mathrm{tr})$.  We allow polynomial
dependence on $R_Q$ itself so that the framework also includes the
spherical and simplex transforms constructed below, while the required
number of qubits remains logarithmic in the total node count.

Even in this weaker form, the transform condition is not automatic.
For generic polynomial systems on nonrectangular domains, no uniform
construction with gate complexity polynomial in
$d$, $N$, $R_Q$, and $\log(1/\epsilon_\mathrm{tr})$ is known.
The sphere and simplex satisfy the condition because their eigenfunctions
admit recursive factorizations into controlled one-dimensional polynomial
transforms. Furthermore, for our analysis, the transform needs to load the eigenfunctions over a $Q$-exact quadrature  grid (Definition \eqref{def:q_exact_quad}) for arbitrary $Q$. This is critical so that sufficient oversampling can make discrete sums equal integrals, i.e. pseudospectral becomes Galerkin. It may be possible to obtain more efficient transforms that load the eigenfunctions over a completely different grid. 

With the three main conditions---slow spectral growth, a smoothing operation, and the transform condition---in place, we are ready show how these can be used to (1) substantially simplify the analysis of global spectral methods and (2) lead to resource-efficient discretizations.

For a band-limited $f_K$, we first bound the Galerkin error via a Dyson series approach.
\begin{lemma}[Galerkin error from Dyson series] \label{lem:galerkin-tail}
Suppose $f_K \in \mathcal{V}_K$. Consider the evolution
\begin{align*}
    i\partial_t u_K(t) =  [-a(t)\Delta_g + b(t)f_K]u_K(t), \quad u_K(0) = u_0
\end{align*}
and its Galerkin approximation
\begin{align*}
    i\partial_t u_{K,N}(t) =  P_N[-a(t)\Delta_g + b(t)f_K]P_Nu_{K,N}(t), \quad u_{K,N}(0) = u_0,
\end{align*}
with $u_0 \in \mathcal{V}_{K_0}$. If $N\geq K_0 +(q-1)K$ for a positive integer $q\geq 2e B_1(t) \norm{f_K}_\infty$, then
    \begin{equation} \label{eq:galerkin-tail}
        \norm{u_K(t)-u_{K,N}(t)} \leq \frac{4}{2^q}.
    \end{equation}
\end{lemma}

\begin{proof}
    Let $U_0(t,s)$ be the kinetic propagator generated by $-a(t)\Delta_g$. We compare the two evolutions in the interaction picture $U_0$. Let
    \begin{equation*}
        v_K(t) \coloneqq U_0(t,0)^\dagger u_K(t), \qquad v_{K,N}(t) \coloneqq U_0(t,0)^\dagger u_{K,N}(t).
    \end{equation*}
    Define the interaction-picture potential operator
    \begin{equation*}
        f_{K,I}(t) \coloneqq U_0(t,0)^\dagger f_K U_0(t,0).
    \end{equation*}
   Because $U_0(t,0)$ is diagonal in the eigenbasis of $-\Delta_g$, $f_{K,I}(t)$ has the same spectral bandwidth as multiplication by $f_K$. The exact and Galerkin interaction-picture equations are, respectively,
    \begin{align*}
        i\partial_t v_K(t) = b(t)f_{K,I}(t)v_K(t),\qquad i\partial_t v_{K,N}(t) = b(t) P_N f_{K,I}(t) v_{K,N}(t).
    \end{align*}
    
    Both initial values equal $u(0)$ because $u(0)\in\mathcal V_N$.  Their Dyson expansions are
    \begin{align}
    \label{eq:Galerkin-two-Dyson-series}
         v_K(t)=\sum_{n=0}^\infty\mathcal D_n(t)u(0),\qquad v_{K,N}(t)=\sum_{n=0}^\infty\mathcal D_{n,N}(t)u(0),
    \end{align}
    which converge absolutely in spectral norm. For $n\geq1$,
    \begin{align}
        \mathcal D_n(t)u(0)
        &\coloneqq (-i)^n\int_{0<t_n<\cdots<t_1<t}
        \left[\prod_{j=1}^nb(t_j)\right]
        f_{K,I}(t_1)\cdots f_{K,I}(t_n)u(0)
        \dd t_n\cdots\dd t_1,
        \label{eq:Galerkin-continuum-Dyson-term}\\
        \mathcal D_{n,N}(t)u(0)
        &\coloneqq (-i)^n\int_{0<t_n<\cdots<t_1<t}
        \left[\prod_{j=1}^nb(t_j)\right]
        P_Nf_{K,I}(t_1)P_N\cdots
        P_Nf_{K,I}(t_n)P_Nu(0)
        \dd t_n\cdots\dd t_1.
        \label{eq:Galerkin-finite-Dyson-term}
    \end{align}
    while $\mathcal{D}_0 = \mathcal{D}_{0,N} = I$. Consider a fixed term $\mathcal{D}$ in either sum.
    
    We first prove equality of the low-order terms.  Fix $n < q$.  For
    every intermediate index $0\leq j\leq n$, the cutoff assumption gives
    \begin{equation}
         K_0 + jK\leq K_0 + (q-1)K \leq N.
         \label{eq:Galerkin-low-order-inside-cutoff}
    \end{equation}
    After the first $j$ applications of $f_{K,I}$ to a state on the right, either integrand has degree at most $K_0 + jK$. For such states, $P_N$ acts trivially. This implies
    \begin{equation}
         \mathcal D_{n,N}(t)u(0)=\mathcal D_n(t)u(0),
         \qquad 0\leq n\leq q-1.
         \label{eq:Galerkin-low-order-Dyson-equality}
    \end{equation}
    
    Next, we bound the Dyson terms for $n \geq q$. By properties of projectors and unitarity,
    \begin{equation}
        \norm{P_Nf_{K,I}(s)P_N} \leq \norm{f_{K,I}(s)} = \norm{f_K}_\infty.
    \end{equation}
    Thus, by the triangle inequality and the bounds above, both terms satisfy
    \begin{align}
        \max\left\{\norm{\mathcal D_{n,N}(t)u(0)},
        \norm{\mathcal D_n(t)u(0)}\right\} \leq\norm{f_K}_\infty^n
        \int_{0<t_n<\cdots<t_1<t}
        \prod_{j=1}^n\abs{b(t_j)}\dd t_n\cdots\dd t_1.
    \label{eq:Galerkin-two-term-norm-before-simplex}
    \end{align}
    Moreover,
    \begin{equation} \label{eq:Galerkin-ordered-simplex-volume}
        \int_{0<t_n<\cdots<t_1<t} \prod_{j=1}^n\abs{b(t_j)}\dd t_n\cdots\dd t_1 = \frac1{n!}\left(\int_0^t\abs{b(s)}\dd s\right)^n = \frac{1}{n!} B_1(t)^n.
    \end{equation}
    Altogether, this gives the common estimate
    \begin{equation}
        \norm{\mathcal D_n(t)u(0)}, \norm{\mathcal D_{n,N}(t)u(0)} \leq\frac{B_1(t)^n \norm{f_K}_\infty^n}{n!}.
        \label{eq:Galerkin-common-Dyson-bound}
    \end{equation}
    This proves absolute convergence for both series with respect to spectral norm.
    
    Subtracting the series, terms of order below
    $q$ cancel exactly. The remaining difference can be bounded via the triangle inequality.
    \begin{align}
        \norm{v_K(t)-v_{K,N}(t)} &\leq\sum_{n=q}^\infty\norm{\mathcal D_n(t)u(0)} + \sum_{n=q}^\infty\norm{\mathcal D_{n,N}(t)u(0)} \leq 2 \sum_{n=q}^\infty\frac{B^n}{n!}.
        \label{eq:Galerkin-two-tail-bound}
    \end{align}
    where $B \coloneqq B_1(t) \norm{f_K}_\infty$. The remainder term of the exponential can be bounded using elementary arguments. A Stirling bound gives
    \begin{equation*}
     \frac{B^q}{q!}\leq\left(\frac{eB}{q}\right)^q
     \leq2^{-q}
    \end{equation*}
    when $q\geq2eB$.  For $n\geq q$, consecutive terms satisfy
    \begin{equation*}
     \frac{B^{n+1}/(n+1)!}{B^n/n!}=\frac{B}{n+1}
     \leq\frac{B}{q+1}\leq\frac12.
    \end{equation*}
    Therefore
    \begin{equation*}
    \sum_{n=q}^\infty\frac{B^n}{n!}
     \leq2^{-q}\sum_{j=0}^\infty2^{-j}=2^{1-q}.
    \end{equation*}
    Plugging this into Eq.~\eqref{eq:Galerkin-two-tail-bound}, and using unitarity of $U_0$, gives the result.
\end{proof}

In practice, we will be interested in the $N$-level Galerkin evolution for the \emph{exact} potential $f$, which we assume accessed is via oracle. As such, the error bound above for truncated $f_K$ serves as a useful tool for extracting the true Galerkin error, as given below. The error between the evolutions with potential $f$ and $f_K$ can be bounded via the previously mentioned Duhamel's principle argument.
\begin{theorem}[Exact-potential Galerkin error]
\label{thm:over_samp_total_error}
Assume the slow spectral growth condition
\eqref{eqn:eigfunc_mul_exp_prop}.  Let $f\in L^\infty(M)$ be real
valued, let $f_K\in\mathcal V_K$ be real valued, and let $u$, $u_K$,
and $u_{K,N}$ be the evolutions defined above from a common normalized
initial state $u(0)\in\mathcal V_{K_0}$. Let $u_N(t)\in\mathcal V_N$ solve
the level-$N$ Galerkin equation with the exact potential $f$,
\begin{equation}
    i\partial_tu_N
    =P_N\left[-a(t)\Delta_g+b(t)f\right]P_Nu_N,
    \qquad u_N(0)=u(0).
    \label{eq:exact-potential-galerkin}
\end{equation}
If $N\geq K_0 + (q-1)K$ for $q \geq 2eB_1(t)\lVert f_K \rVert_{\infty}$,  then
\begin{equation}
\label{eqn:galerkin-truncation-error}
    \norm{u(t)-u_N(t)} \leq 2 B_1(t)\norm{f-f_K}_\infty+\frac{4}{2^q}.
\end{equation}
\end{theorem}
\begin{proof}
Fact~\ref{fact:duhamels} and Lemma~\ref{lem:galerkin-tail} first give
\begin{equation}
    \norm{u(t)-u_{K,N}(t)}
    \leq B_1(t)\norm{f-f_K}_\infty+\frac{4}{2^q}.
    \label{eq:bandlimited-galerkin-bridge}
\end{equation}
For every $v\in\mathcal V_N$, Bessel's inequality gives
\begin{align*}
    \norm{P_N\bigl((f-f_K)v\bigr)}^2
    &=\sum_{\alpha \in \mathbb{N}~|~L(\alpha) \leq N} 
      \abs{\langle Y_{\alpha},(f-f_K)v\rangle}^2\\
    &\leq\norm{(f-f_K)v}^2
    \leq\norm{f-f_K}_\infty^2\norm v^2.
\end{align*}
Thus the generators of $u_N$ and $u_{K,N}$ differ in norm by at most
$\abs{b(t)}\norm{f-f_K}_\infty$. A second application of Duhamel's
formula yields
\begin{equation*}
    \norm{u_N(t)-u_{K,N}(t)}
    \leq B_1(t) \norm{f-f_K}_\infty.
\end{equation*}
Combining this estimate with \eqref{eq:bandlimited-galerkin-bridge}
proves \eqref{eqn:galerkin-truncation-error}.
\end{proof}

We will now move to analyzing the error associated with the quadrature-based (collocation) method. As mentioned earlier, the
continuous-space projector $P_N$ is replaced with the quadrature-based projection $I_{N,Q}$ given by \eqref{eq:quadrature-projection}. The discrepancy between $I_{N,Q}$ and
$P_N$ is referred to as \emph{aliasing}. More generally, for
$\psi\in\mathcal V_R$,
\begin{equation}
    I_{N,Q}\psi=P_N\psi
    \qquad\text{whenever}\qquad Q\geq \max(N, R).
    \label{eq:quadrature-equals-projection}
\end{equation}
Indeed, for $\langle Y_{\alpha}, \psi\rangle$ with $L(\alpha) \leq N$, both $\psi$ and $Y_{\alpha}$
belong to $\mathcal V_Q$, so the identity follows directly from
$Q$-exactness of the discrete inner product and \eqref{eq:quadrature-projection}.

This has the immediate implication.
\begin{proposition}[Exact quadrature implementation of the Galerkin projector]
\label{prop:Galerkin-quadrature}
Suppose the filtration $\{\mathcal V_N\}_{N\geq0}$ satisfies slow spectral growth,
$f_K\in\mathcal V_K$, and $I_{N,Q}$ is constructed from a $Q$-exact
quadrature rule. If $Q\geq N+K$, then
\begin{equation}
    I_{N,Q}\big|_{\mathcal V_{N+K}}
    =P_N\big|_{\mathcal V_{N+K}}.
    \label{eq:quadrature-implements-galerkin-projector}
\end{equation}
In particular, for every $v\in\mathcal V_N$,
\begin{equation} \label{eq:exact-dealiasing}
    I_{N,Q}(f_K v)=P_N(f_Kv).
\end{equation}
\end{proposition}
\begin{proof}
Equation~\eqref{eq:quadrature-implements-galerkin-projector} follows
from \eqref{eq:quadrature-equals-projection}. Slow spectral growth gives
$f_Kv\in\mathcal V_{N+K}$, which proves \eqref{eq:exact-dealiasing}.
\end{proof}

We now use this exactness to pass from the continuum Galerkin estimate to the quadrature evolution.
\begin{corollary}[Oversampled quadrature error] \label{cor:oversampled-quadrature-error}
    Under the assumptions of Theorem~\ref{thm:over_samp_total_error}, let
    $I_{N,Q}$ be constructed from a $Q$-exact quadrature rule with
    $Q\geq N+K$, and let $ u_{N,Q}(t)\in\mathcal V_N$ solve
    \begin{equation}
        i\partial_t u_{N,Q}
        =-a(t)\Delta_g u_{N,Q}
         +b(t)I_{N,Q}\bigl(f u_{N,Q}\bigr),
        \qquad
        \widetilde u_{N,Q}(0)=u(0),
        \label{eq:quadrature-evolution}
    \end{equation}
    where $f : \mathcal{M} \rightarrow \mathbb{R}$ is pointwise bounded.
    Then for all $t\in [0,T]$
    \begin{equation}
    \label{eqn:oversampling_method_error}
        \norm{u(t)- u_{N,Q}(t)}
        \leq2B_1(t)\norm{f-f_K}_\infty+2^{2-q}.
    \end{equation}
    Moreover, let $h : \mathcal{M} \rightarrow \mathbb{R}$ be pointwise bounded and suppose that a real
    $h_K\in\mathcal V_K$ is given. Then
    \begin{align}
    \label{eqn:observable-gap}
    &
    \left|\sum_{j=1}^{J_Q}w_jh(x_j)
        \abs{ u_{N,Q}(t,x_j)}^2
        -\int_Mh(x)\abs{u(t,x)}^2\dvol(x)\right|\nonumber\\
    &\qquad\leq
        4B_1(t)\norm{h_K}_\infty\norm{f-f_K}_\infty
        +2^{3-q}\norm{h_K}_\infty
        +2\norm{h-h_K}_\infty.
    \end{align}
\end{corollary}
\begin{proof}
    Proposition~\ref{prop:Galerkin-quadrature} identifies the quadrature evolution with potential $f_K$ exactly with $u_{K,N}$. The same Duhamel comparison, with $I_{N,Q}$ in place of $P_N$, proves
    \eqref{eqn:oversampling_method_error}.
    
    Finally, $Q\geq N+K$ and slow spectral growth imply
    \begin{equation*}
        \sum_{j=1}^{J_Q}w_jh_K(x_j)
            \abs{ u_{N,Q}(t,x_j)}^2
        =\int_M h_K(x)\abs{ u_{N,Q}(t,x)}^2\dvol(x).
    \end{equation*}
    Both evolutions preserve the norm. Inserting $h_K$ on the discrete and continuous sides, and using
    \begin{equation*}
        \left|\langle v,h_Kv\rangle-\langle z,h_Kz\rangle\right|
        \leq2\norm{h_K}_\infty\norm{v-z}
    \end{equation*}
    for normalized $v$ and $z$, proves \eqref{eqn:observable-gap}.
\end{proof}

\subsection{Flat Torus} \label{subsec:flat-torus}
The flat torus is the Riemannian quotient manifold $\mathbb{T}^d = \mathbb{R}^d/\mathbb{Z}^d$ with Euclidean metric. Concretely, it admits a global parametrization via the set $S=[-1/2,1/2]^d$, with periodic boundary conditions in each coordinate. For the torus, the Laplace--Beltrami operator $\Delta_{g}$ is diagonalized by Fourier modes and there already exists other simulation algorithms for the torus \cite{childs2022quantumsim, chakrabarti2025speedups}. We present an alternative algorithm and analysis that leads to \textit{exponential} improvement in discretization spacing and a quadratic reduction in the number of qubits in terms of $d$. Through a change-of-coordinates, this also leads to simulation algorithms for the positive orthant with the log-barrier Hessian metric, polyhedral log-barrier Hessian metric for full-rank constraint matrix, and $\ell_p$ balls.

We start by presenting a smoothing procedure that takes $f \in L^2(M)$ and maps it to a finite-degree Fourier polynomial with controllable error. For an integer $m\geq2$, define the following Jackson Kernel
\begin{align} 
    A_m(t)&=\left(\frac{\sin(m\pi t)}{\sin(\pi t)}\right)^4,
    \qquad t\in[-1/2,1/2],\label{eq:Jackson-raw}\\
    J_m(t)&=\frac{A_m(t)}{\int_{-1/2}^{1/2}A_m(s)\dd s},\\
    \mathcal J_{m,d}(x)&=\prod_{j=1}^dJ_m(x_j). \label{eq:Jackson-tensor}
\end{align}
At $t=0$ the ratio is defined by continuity.  The kernel $\mathcal{J}_{m,d}$ is a probability distribution on $S$ with i.i.d. distributed components according to $J_m$. It is also a Fourier polynomial of degree $2(m-1)$. 
\begin{lemma}[Jackson-Fourier Smoother]
\label{lem:Jackson-moments}
    Let $f:S\to\mathbb{R}$ be geodesically $G$-Lipschitz continuous on $\mathbb{T}^d$. Then the Jackson-smoothed function
    \begin{equation}
        f_m \coloneqq \mathcal J_{m,d}*_{\mathbb T^d}f = \int_S \mathcal{J}_{m,d}(z)f(x-z) dz
        \label{eq:Jackson-proxy}
    \end{equation}
    satisfies
    \begin{align} \label{eq:Jackson-uniform-error}
        \norm{f-f_m}_\infty\leq\frac{3G\sqrt d}{m},
        \qquad
        \norm{f_m}_\infty\leq\norm f_\infty.
    \end{align}
    and has Fourier support in $[-2(m-1),2(m-1)]^d$.
 \end{lemma}
\begin{proof}
    First, as a continuous function on a compact space $\mathbb T^d$, $f$ is bounded. Note that $\mathcal {J}_{m,d}$ is a probability density on $S$ by construction. Thus, 
    \begin{equation*}
        |f_m(x)|\leq\int_{S}\mathcal J_{m,d}(z)|f(x-z)|\dd z
        \leq\norm f_\infty,
    \end{equation*}
    which proves the second inequality in \eqref{eq:Jackson-uniform-error}.

    Next, we prove that $f_m$ only has support on Fourier modes with frequencies in $[-2(m-1), 2(m-1)]^d$. Define
    \begin{equation*}
        D_m(t) \coloneqq \sum_{r=0}^{m-1}e^{2\pi i r t}.
    \end{equation*}
    For $t \notin \mathbb Z$, by the formula for finite geometric series,
    \begin{equation*}
        D_m(t) = \frac{1-e^{2\pi i mt}}{1-e^{2\pi i t}} = e^{\pi i(m-1)t}
        \frac{\sin(m\pi t)}{\sin(\pi t)}.
    \end{equation*}
    This equality also holds at integer arguments after continuous extension. Consequently,
    \begin{align*}
         \left(\frac{\sin(m\pi t)}{\sin(\pi t)}\right)^2 &= |D_m(t)|^2 = \sum_{r,s=0}^{m-1}e^{2\pi i(r-s)t}\nonumber\\
         &= \sum_{k=-(m-1)}^{m-1}(m-|k|)e^{2\pi i k t}.
    \end{align*}
    Indeed, for fixed $k=r-s$ there are exactly $m-|k|$ pairs $(r,s)$.
    Thus the squared ratio has Fourier support in
    $[-(m-1),m-1]$.  Since $A_m$ is the square of this trigonometric
    polynomial, multiplication convolves its Fourier coefficients and gives
    \begin{equation*}
        \operatorname{supp}\widehat A_m
        \subseteq[-2(m-1),2(m-1)].
    \end{equation*}
    And hence also for $J_m$. Finally, the product definition gives
    \begin{equation*}
        \operatorname{supp}\widehat{\mathcal J_{m,d}}
        \subseteq[-2(m-1),2(m-1)]^d.
        \label{eq:Jackson-tensor-support}
    \end{equation*}
    The convolution theorem for Fourier transforms implies $f_m$ has the same Fourier support.

    Finally, we prove the first inequality in \eqref{eq:Jackson-uniform-error}. For every $x \in S$, the convolution definition and the Lipschitz condition give
    \begin{align}
        |f_m(x)-f(x)|
        &\leq\int_{S}\mathcal {J}_{m,d}(z)
        |f(x-z)-f(x)|\dd z \nonumber\\
        &\leq G \mathbb{E}_{\mathcal {J}_{m,d}}[\lVert Z \rVert],
        \label{eqn:jackson-lipschitz-error-bound}
    \end{align}
    and thus we only need to analyze the moments of $\mathcal{J}_{m, d}$.

    The function
    $x\mapsto\sin x$ is concave on $[0,\pi/2]$, so it lies above the chord
    joining $(0,0)$ and $(\pi/2,1)$.  Hence
    \begin{equation*}
     \sin x\geq\frac{2x}{\pi},
     \qquad 0\leq x\leq\frac\pi2.
    \end{equation*}
    If $|t|\leq1/(2m)$, then $x=m\pi|t|\in[0,\pi/2]$, and oddness of sine
    gives
    \begin{equation*}
        |\sin(m\pi t)| = \sin(m\pi|t|) \geq \frac{2m\pi|t|}{\pi} = 2m|t|.
    \end{equation*}
    The elementary inequality $|\sin y|\leq|y|$ also gives
    \begin{equation*}
     |\sin(\pi t)|\leq\pi|t|.
    \end{equation*}
    For $0<|t|\leq1/(2m)$, division therefore yields
    \begin{equation*}
     \left|\frac{\sin(m\pi t)}{\sin(\pi t)}\right|
     \geq\frac{2m}{\pi}.
    \end{equation*}
    The same bound holds at $t=0$ by continuity, since the ratio tends to $m$
    and $m\geq2m/\pi$.  It follows that
    \begin{equation}
    \label{eq:Jackson-normalization-lower}
        \int_{-1/2}^{1/2}A_m(t)\dd t \geq \int_{-1/(2m)}^{1/(2m)}
        \left(\frac{2m}{\pi}\right)^4\dd t = \frac1m\frac{16m^4}{\pi^4} = \frac{16m^3}{\pi^4}.
    \end{equation}
    From the above, for all $|t|\leq1/2$,
    \begin{align*}
        \lvert \sin(\pi t) \rvert \geq 2 \lvert t \rvert,
    \end{align*}
    so $|t|\leq1/2$
    \begin{equation}
        \left|\frac{\sin(m\pi t)}{\sin(\pi t)}\right|
        \leq\min\left\{m,\frac1{2|t|}\right\}.
        \label{eq:Jackson-pointwise}
    \end{equation}
    
    We now split the moment integrals at $1/m$.  This point lies in
    $[0,1/2]$ because $m\geq2$.  By evenness of $A_m$ and
    \eqref{eq:Jackson-pointwise},
    \begin{equation}
     A_m(t)\leq
     \begin{cases}
      m^4,&0\leq t\leq1/m,\\
      (2t)^{-4}=1/(16t^4),&1/m\leq t\leq1/2.
     \end{cases}
     \label{eq:Jackson-piecewise-A-bound}
    \end{equation}
    Thus the numerator of the second moment satisfies
    \begin{align}
     \int_{-1/2}^{1/2}t^2A_m(t)\dd t
     &=2\int_0^{1/m}t^2A_m(t)\dd t
       +2\int_{1/m}^{1/2}t^2A_m(t)\dd t\nonumber\\
     &\leq2m^4\int_0^{1/m}t^2\dd t
       +\frac{2}{16}\int_{1/m}^{1/2}t^{-2}\dd t\nonumber\\
     &=2m^4\left[\frac{t^3}{3}\right]_0^{1/m}
       +\frac18\left[-\frac1t\right]_{1/m}^{1/2}\nonumber\\
     &=\frac{2m}{3}+\frac{m-2}{8}
     =\frac{19}{24}m-\frac14
     \leq\frac{19}{24}m.
     \label{eq:Jackson-second-moment-numerator-details}
    \end{align}
    Division by \eqref{eq:Jackson-normalization-lower} gives
    \begin{align}
     \int t^2J_m(t)\dd t
     &\leq\frac{19\pi^4}{384m^2}<\frac9{m^2}.
    \end{align}
    Also,
    \begin{equation}
     \mathbb E\norm Z_2
     \leq\left(\mathbb E\norm Z_2^2\right)^{1/2}
     \leq\frac{3\sqrt d}{m},
    \end{equation}
    which combined with \eqref{eqn:jackson-lipschitz-error-bound} proves the first inequality in \eqref{eq:Jackson-uniform-error}
\end{proof}
\noindent 
To summarize, for sufficiently large Fourier degree $m$, any Lipschitz function on $\mathbb T^d$ can be made $\mathcal{O}(\epsilon)$ close to a Fourier polynomial, with error falling as $O(m^{-1})$.

Let $Q\geq N$ be the desired quadrature exactness degree and set
\begin{equation}
 M_Q\coloneqq 2Q+1.
 \label{eq:flat-QFT-length}
\end{equation}
Define the centered $M_Q$-point grid in each coordinate by
\begin{equation}
 \mathcal I_Q\coloneqq \{-Q,\ldots,Q\}^d,
 \qquad
 \mathcal G_Q\coloneqq \left\{\frac{j}{M_Q}:j\in\mathcal I_Q\right\}
 \subset[-1/2,1/2)^d.
 \label{eq:flat-centered-grid}
\end{equation}
The signed label $j\in\{-Q,\ldots,Q\}$ is stored in the standard
residue-class bin $j\bmod M_Q\in\{0,\ldots,M_Q-1\}$. More precisely, for
$r\in\{-Q,\ldots,Q\}$ define the centered computational-basis label by
\begin{equation*}
 \ket r_\mathrm{c}\coloneqq \ket{r\bmod M_Q}_\mathrm{std}.
\end{equation*}
We suppress the subscript $\mathrm{c}$ below.  This is only a signed
interpretation of the standard computational basis.
Let $E_{N,Q}$ denote the corresponding frequency embedding,
\begin{equation}
 E_{N,Q}\ket n=\ket n,
 \qquad n\in\mathcal B_N.
 \label{eq:flat-coherent-frequency-embedding}
\end{equation}
Because $Q\geq N$, every frequency in $\mathcal B_N$ has a distinct
centered label. Define the centered $M_Q$-point QFT by
\begin{equation}
 F_{M_Q,\mathrm{c}}\ket r
 =\frac1{\sqrt{M_Q}}
 \sum_{s=-Q}^{Q}e^{-2\pi i rs/M_Q}\ket s,
 \qquad r\in\{-Q,\ldots,Q\}.
 \label{eq:flat-QFT-convention}
\end{equation}
The phase depends only on the residue classes of $r$ and $s$ modulo
$M_Q$, so $F_{M_Q,\mathrm{c}}$ is the QFT over $\mathbb Z_{M_Q}$ under the
centered labeling above.

For circuit compilation, embed this $M_Q$-dimensional logical register in
$\lceil\log_2M_Q\rceil$ qubits and extend the transform arbitrarily on
the unused computational-basis states. Although $M_Q$ is generally not a
power of two, standard arbitrary-modulus QFT constructions approximate
$F_{M_Q,\mathrm{c}}$ to operator error $\epsilon_\mathrm{tr}/d$ using
$\operatorname{poly}(\log M_Q,\log(d/\epsilon_\mathrm{tr}))$ elementary
gates. The telescoping product inequality then bounds the error of the
$d$-fold transform by $\epsilon_\mathrm{tr}$, without changing the
$\mathcal O(d\log M_Q)$ qubit count.

The inverse padded QFT gives the weighted samples
\begin{equation}
 \bigl((F_{M_Q,\mathrm{c}}^\dagger)^{\otimes d}E_{N,Q}c\bigr)_j
 =\frac1{M_Q^{d/2}}
 \sum_{n\in\mathcal B_N}c_ne^{2\pi i n\cdot j/M_Q},
 \qquad j\in\mathcal I_Q,
 \label{eq:flat-QFT-weighted-samples}
\end{equation}
which is $M_Q^{-d/2}$ times the value of the represented Fourier
polynomial at the node $x_j=j/M_Q$.

For completeness, this  isometry can be embedded into a unitary (via dilation) as follows.  Let a source register $C$ hold the compact label
$k\in\{0,\ldots,2N\}^d$, corresponding to frequency
$n=k-N\mathbf 1$, and initialize a padded register $R$ to $\ket0_R$.
Reversible modular arithmetic implements:
\begin{equation}
 U_\mathrm{emb}\ket0_R\ket k_C
 =\ket{k-N\mathbf 1}_R\ket0_C,
 \label{eq:flat-reversible-label-transfer}
\end{equation}
where the frequency on the right is stored in the centered convention.

Extending this injection register-wise and applying the inverse QFT
gives the unitary
\begin{equation}
 U_W^{\mathbb T}\coloneqq
 \bigl((F_{M_Q,\mathrm{c}}^\dagger)^{\otimes d}\otimes I_C\bigr)
 U_\mathrm{emb}.
 \label{eq:flat-explicit-QFT-dilation}
\end{equation}
For $\ket\psi_C=\sum_{k\in\{0,\ldots,2N\}^d}c_k\ket k_C$, its action is
\begin{align}
 U_W^{\mathbb T}\ket0_R\ket\psi_C
 ={}&\sum_{j\in\mathcal I_Q}
 \left[
    \frac1{M_Q^{d/2}}
  \sum_k c_k
    e^{2\pi i(k-N\mathbf 1)\cdot j/M_Q}
 \right]\ket j_R\ket0_C.
 \label{eq:flat-explicit-dilation-action}
\end{align}
Thus $U_W^{\mathbb T}$ is a unitary dilation of the weighted evaluation
isometry $W$
in \eqref{eq:global-weighted-spectral-transform}.  It uses one
$M_Q$-point QFT in each coordinate together with arithmetic polynomial in
$d$ and $\log M_Q=\mathcal O(\log(Q+1))$.

The centered grid has exactly the quadrature exactness needed by the
general oversampling result.  Indeed, for $p\in\mathbb Z^d$ with
$\lVert p\rVert_\infty<M_Q$,
\begin{equation}
 \frac1{M_Q^d}\sum_{j\in\mathcal I_Q}
 e^{2\pi i p\cdot j/M_Q}
 =\begin{cases}
  1,&p=0,\\
  0,&p\ne0.
 \end{cases}
 \label{eq:flat-centered-character-sum}
\end{equation}
For $\phi,\psi\in\mathcal V_Q$, every frequency in
$\overline\phi\psi$ has infinity norm at most $2Q<M_Q$. Consequently,
the rule with nodes $\mathcal G_Q$ and weights $M_Q^{-d}$ is $Q$-exact
in the inner-product sense of the general framework, and
$J_Q=M_Q^d$. Therefore the hypothesis of
Corollary~\ref{cor:oversampled-quadrature-error} is satisfied whenever
\begin{equation}\label{eq:flat-centered-dealiasing-condition}
    Q\geq N+K.
\end{equation}

Our main contribution here is to improve the number of grid points required by the discrete quantum algorithm. The cost of simulation then follows from plugging this choice of $N$ in \cite[Theorem 4.4]{chakrabarti2025speedups}. However, we will also include a theorem for completeness.

\begin{theorem}[Improved discretization for quantum simulation on the torus]\label{thm:improved_torus_discretization}
Let $g:\mathbb T^d\to\mathbb R$ be a real-valued $L_g$-Lipschitz function, and consider the evolution
\begin{equation}
i\partial_t\Phi(t)=[-a(t)\Delta+b(t)g(x)]\Phi(t),\qquad \Phi(0)\in\mathcal V_{L_0},\qquad \norm{\Phi(0)}=1,\qquad t\in[0,T].
\end{equation}
Let $h:\mathbb T^d\to\mathbb R$ be a nonzero real-valued $L_h$-Lipschitz function. Fix $0<\epsilon<1$ and define $B_1\coloneqq \int_0^T\abs{b(t)}\dd t$. Then there is a quadrature degree $Q$ for which the centered grid $\mathcal G_Q$ in \eqref{eq:flat-centered-grid} contains $M\coloneqq 2Q+1$ nodes per coordinate, with
\begin{equation}\label{eq:torus-theorem-grid-scaling}
M=\mathcal O\!\left(L_0+\left[1+\frac{\sqrt d}{\epsilon}\max\left\{L_gB_1,\frac{L_h}{\norm h_\infty}\right\}\right]\left(1+B_1\norm g_\infty+\log(1/\epsilon)\right)\right).
\end{equation}
The resulting $M^d$-point grid supports an $\mathcal O(d\log M)$-qubit finite-dimensional evolution with initial state
\begin{equation}
\ket{\Psi(0)}=M^{-d/2}\sum_{x_j\in\mathcal G_Q}\Phi(0,x_j)\ket j.
\end{equation}
Let $\mathsf M_h$ denote multiplication by $h$ on $L^2(\mathbb T^d)$, and define the sampled diagonal observable
\begin{equation}
\widehat h_{\mathcal G_Q}\coloneqq \sum_{x_j\in\mathcal G_Q}h(x_j)\ketbra{j}{j}.
\end{equation}
The final grid state satisfies
\begin{equation}\label{eq:torus-theorem-observable-error}
\left|\langle\Psi(T)|\widehat h_{\mathcal G_Q}|\Psi(T)\rangle-\langle\Phi(T)|\mathsf M_h|\Phi(T)\rangle\right|\leq4\norm h_\infty\epsilon.
\end{equation}
\end{theorem}
\begin{proof}
Fix
\begin{align*}
m&\coloneqq \left\lceil\max\left\{2,\frac{6L_gB_1\sqrt d}{\epsilon},\frac{6L_h\sqrt d}{\epsilon\norm h_\infty}\right\}\right\rceil,\\
q&\coloneqq \left\lceil\max\left\{2eB_1\norm g_\infty,3+\log_2(1/\epsilon)\right\}\right\rceil,
\end{align*}
and set
\begin{equation*}
K\coloneqq 2(m-1),\qquad N\coloneqq L_0+(q-1)K.
\end{equation*}
Set $Q\coloneqq N+K$ and $M\coloneqq 2Q+1$. These choices fix all cutoff and grid parameters used below.

Finite Fourier polynomials satisfy the slow spectral growth property \eqref{eqn:eigfunc_mul_exp_prop}. Define
\begin{equation*}
g_K=\mathcal J_{m,d}*_{\mathbb T^d}g,\qquad h_K=\mathcal J_{m,d}*_{\mathbb T^d}h.
\end{equation*}
Lemma~\ref{lem:Jackson-moments} shows that $g_K,h_K\in\mathcal V_K$ and that
\begin{equation*}
\norm{g-g_K}_\infty\leq\frac{3L_g\sqrt d}{m},\qquad \norm{h-h_K}_\infty\leq\frac{3L_h\sqrt d}{m},
\end{equation*}
while positivity and normalization of the Jackson kernel imply
\begin{equation*}
\norm{g_K}_\infty\leq\norm g_\infty,\qquad \norm{h_K}_\infty\leq\norm h_\infty.
\end{equation*}
The fixed value of $m$ therefore controls both approximation errors simultaneously:
\begin{equation*}
2B_1\norm{g-g_K}_\infty\leq\epsilon,\qquad 2\norm{h-h_K}_\infty\leq\norm h_\infty\epsilon.
\end{equation*}
The fixed value of $q$ implies
\begin{equation*}
q\geq2eB_1\norm{g_K}_\infty,\qquad 2^{2-q}\leq\frac\epsilon2,\qquad 2^{3-q}\leq\epsilon.
\end{equation*}

By construction, $Q=N+K$, so
\eqref{eq:flat-centered-dealiasing-condition} holds. Moreover, $M=2Q+1$
makes the centered grid $Q$-exact by
\eqref{eq:flat-centered-character-sum}, and
\begin{equation*}
M=2Q+1=2(N+K)+1=2L_0+2qK+1\leq2L_0+4qm+1.
\end{equation*}
Substituting the fixed values of $m$ and $q$ proves \eqref{eq:torus-theorem-grid-scaling}.

Let $S_N$ be the coefficient-to-function unitary for $\mathcal V_N$ and
let
\begin{equation*}
W_{N,Q}\coloneqq (F_{M,\mathrm{c}}^\dagger)^{\otimes d}E_{N,Q}
\end{equation*}
be the padded inverse QFT in \eqref{eq:flat-QFT-weighted-samples}. Define
\begin{equation*}
\widehat g_{\mathcal G_Q}\coloneqq \sum_{x_j\in\mathcal G_Q}g(x_j)\ketbra{j}{j}.
\end{equation*}
Evolve the coefficient state according to
\begin{equation*}
i\partial_t\ket{c(t)}=\left(a(t)D_N+b(t)W_{\mathbb T,N,Q}^\dagger \widehat g_{\mathcal G_Q}W_{\mathbb T,N,Q}\right)\ket{c(t)},\qquad \ket{c(0)}=S_N^\dagger \Phi(0),
\end{equation*}
and set $\ket{\Psi(t)}\coloneqq W_{\mathbb T,N,Q}\ket{c(t)}$. Since $N\leq Q$ and the
grid is $Q$-exact, the padded inverse QFT is an isometry on the retained
coefficient space. Equation~\eqref{eq:flat-QFT-weighted-samples}
therefore gives the initial grid encoding in the theorem. The coefficient
register and the $M^d$-point grid workspace use $\mathcal O(d\log M)$
qubits.

Apply Corollary~\ref{cor:oversampled-quadrature-error} with $f=g$, $f_K=g_K$, and the observable proxy $h_K$. Its observable estimate gives
\begin{align*}
&\left|\langle\Psi(T)|\widehat h_{\mathcal G_Q}|\Psi(T)\rangle-\langle\Phi(T)|\mathsf M_h|\Phi(T)\rangle\right|\\
&\qquad\leq4B_1\norm{h_K}_\infty\norm{g-g_K}_\infty+2^{3-q}\norm{h_K}_\infty+2\norm{h-h_K}_\infty\\
&\qquad\leq2\norm h_\infty\epsilon+\norm h_\infty\epsilon+\norm h_\infty\epsilon=4\norm h_\infty\epsilon.
\end{align*}
\end{proof}

Theorem~\ref{thm:improved_torus_discretization} should be compared with Theorem 4.8
and Lemma 4.7 of~\cite{chakrabarti2025speedups}, which together provide the discretization guarantees used by its main simulation theorem \cite[Theorem 4.2]{chakrabarti2025speedups}.  The first
of those results proves $L^2$-closeness between the Fourier interpolant of
the collocation evolution and the continuum solution.  Because the
collocation grid also fixes the retained Fourier band, its proof controls
dynamical aliasing through high-order Sobolev norms of the evolving
wavefunction.  Bounding those norms for a Lipschitz potential requires
mollification and a Sobolev-growth estimate, and leads in general to a grid
resolution that is exponential in the dimension.  Here the Jackson proxy
is instead exactly band limited, and oversampling makes its quadrature
potential identical to the Galerkin potential.  The state error is therefore
controlled only by the uniform proxy error and the tail of the Dyson series,
with no Sobolev-norm estimate for the evolved wavefunction, yielding a grid
resolution polynomial in the dimension under the assumptions of the
theorem.

The expectation-value comparison is simplified for the same reason. The general expectation-error lemma of~\cite{chakrabarti2025speedups}
contains, in addition to the state error, a Riemann-summation error involving $\sup\norm{\nabla|\Phi_M|^2}_\infty$ and a Fourier-tail term involving $|\Phi|_{H^1}/M$.  Obtaining an end-to-end guarantee requires bounding these wavefunction-dependent quantities, and the generic estimate
$\sup\norm{\nabla|\Phi_M|^2}_\infty=O(M^{d+1})$ contributes to the large
grid requirement.  In the present analysis, the observable proxy $h_K$ is
band limited and the oversampled rule integrates $h_K|\widetilde\Psi|^2$ exactly  The observable error therefore reduces to the state error and the uniform approximation error $\norm{h-h_K}_\infty$; no separate Riemann-summation or density-gradient estimate is needed. The Hamiltonian-simulation primitive, and hence its oracle-query bound, may still
be taken from \cite[Theorem~4.4]{chakrabarti2025speedups}; the improvement is in the discretization and qubit requirements.

For concreteness, we present an improved version of Theorem \cite[Theorem 4.2]{chakrabarti2025speedups} using our new discretization result. The torus case is equivalent to their periodic boundary conditions case. The case without periodic boundary conditions can be handlded using the same $\delta$ low-leakage evolution assumption \cite[Definition 4.5]{chakrabarti2025speedups}.

{
\begin{theorem}[Digital interaction-picture simulation on the torus]
\label{thm:torus-digital-simulation}
Let $g:\mathbb T^d\to\mathbb R$ be real-valued and
$L_g$-Lipschitz, let $h:\mathbb T^d\to\mathbb R$ be nonzero,
real-valued, and $L_h$-Lipschitz, and consider
\begin{equation*}
 i\partial_t\Phi(t)=[-a(t)\Delta+b(t)g]\Phi(t),
 \qquad \Phi(0)\in\mathcal V_{L_0},
 \qquad \norm{\Phi(0)}=1.
\end{equation*}
Assume $a$ and $b$ are piecewise $C^1$ and $b$ has finitely many sign
intervals. Fix $0<\epsilon<1$, put
$B_1\coloneqq\int_0^T|b(t)|\dd t$, and let
$\Lambda_g\geq\max\{1,\norm g_\infty\}$.

Then there are integers $N\leq Q$ and $M=2Q+1$ such that
\begin{equation*}
 M=\mathcal O\!\left(L_0+
 \left[1+\frac{\sqrt d}{\epsilon}
 \max\left\{L_gB_1,\frac{L_h}{\norm h_\infty}\right\}\right]
 \left[1+B_1\norm g_\infty+\log(1/\epsilon)\right]\right).
\end{equation*}
Writing $\mathcal I_Q=\{-Q,\ldots,Q\}^d$, let $S_N$ be the Fourier
coefficient map and $ W_{\mathbb T,N,Q}$ as above. 
Assume coherent circuits can prepare $S_N^\dagger\Phi(0)$ to error
$\epsilon_{\rm prep}\leq c_0\epsilon$, query $g(j/M)$ through a binary
oracle $O_g$ to error $0<\epsilon_g\leq\Lambda_g$ with
$B_1\epsilon_g\leq c_0\epsilon$ at a cost of $\mathcal C_g$ gates,
and, for every requested $0<\delta<1$, implement a unitary dilation
$\widetilde U_W^{\mathbb T}$ of $W_{\mathbb T,N,Q}$, the schedule
arithmetic, and the Fourier kinetic phases to operator error $\delta$
using $d\,\polylog(dM/\delta)$ gates per use. Here $c_0>0$ is a
sufficiently small universal constant, and state preparation uses
$\mathcal C_{\rm prep}$ gates and
$\mathcal Q_{\rm prep}$ work qubits.

Then there is a quantum circuit producing a normalized coefficient state
$\ket{\widetilde c(T)}$, with success probability at least
$1-\mathcal O(\epsilon)$, such that
\begin{equation}\label{eq:torus-end-to-end-state-error}
 \norm{S_N\ket{\widetilde c(T)}-\Phi(T)}
 =\mathcal O(\epsilon).
\end{equation}
On grid and coefficient registers $R,C$, define
\begin{equation*}
 \ket{\widetilde\Psi_Q(T)}_{RC}
 \coloneqq \widetilde U_W^{\mathbb T}
 \bigl(\ket0_R\ket{\widetilde c(T)}_C\bigr),
 \qquad
 \widehat h_Q\coloneqq
 \left(\sum_{j\in\mathcal I_Q}h(j/M)\ketbra{j}{j}_R\right)\otimes I_C.
\end{equation*}
Then
\begin{equation}\label{eq:torus-end-to-end-observable-error}
 \left|
 \langle\widetilde\Psi_Q(T)|\widehat h_Q
 |\widetilde\Psi_Q(T)\rangle
 -\langle\Phi(T)|h|\Phi(T)\rangle
 \right|=\mathcal O(\norm h_\infty\epsilon).
\end{equation}

Let $\tau_g\coloneqq\Lambda_gB_1$. The circuit makes
\begin{equation}\label{eq:torus-binary-query-complexity}
 N_g=\mathcal O\!\left(
 (1+\tau_g)
 \frac{\log(\mathrm e^{\mathrm e}(1+\tau_g)/\epsilon)}
 {\log\log(\mathrm e^{\mathrm e}(1+\tau_g)/\epsilon)}
 \right)
 =\widetilde{\mathcal O}(1+\tau_g)
\end{equation}
calls to $O_g$ or $O_g^\dagger$ and uses
\begin{equation}\label{eq:torus-end-to-end-gate-complexity}
 \mathcal C_{\rm gates}
 =\mathcal C_{\rm prep}
 +\widetilde{\mathcal O}\!\left((1+\tau_g)\left[
 d+\mathcal C_g
 +\log^2\!\left(\frac{2\Lambda_g}{\epsilon_g}\right)
 \right]\right)
\end{equation}
elementary gates. Here $\widetilde{\mathcal O}$ suppresses logarithmic
factors in $d,M,(1+\tau_g)/\epsilon$ and in the schedule arithmetic.
The total qubit count is
\begin{equation}\label{eq:torus-end-to-end-qubit-complexity}
 \mathcal Q_{\rm prep}+\mathcal O\!\left(
 d\log M+\log\frac{2\Lambda_g}{\epsilon_g}
 +\polylog\frac{dM(1+\tau_g)}{\epsilon}
 \right).
\end{equation}
\end{theorem}

\begin{proof}
Choose
\begin{align*}
 m&=\left\lceil\max\left\{2,
 \frac{6L_gB_1\sqrt d}{\epsilon},
 \frac{6L_h\sqrt d}{\epsilon\norm h_\infty}\right\}\right\rceil,
 &K&=2(m-1),\\
 q&=\left\lceil\max\left\{2eB_1\norm g_\infty,
 3+\log_2(1/\epsilon)\right\}\right\rceil,
 &N&=L_0+(q-1)K,
\end{align*}
and set $Q=N+K$ and $M=2Q+1$. These choices give the grid bound in
the statement. Define
\begin{equation*}
 \widehat g_Q\coloneqq
 \sum_{j\in\mathcal I_Q}g(j/M)\ketbra{j}{j},
 \qquad
 D_N\ket n=4\pi^2\norm n_2^2\ket n,
\end{equation*}
and let the exact coefficient state solve
\begin{equation*}
 i\partial_t\ket{c_{N,Q}(t)}
 =\left[a(t)D_N+b(t)W_{\mathbb T,N,Q}^\dagger
 \widehat g_QW_{\mathbb T,N,Q}\right]\ket{c_{N,Q}(t)},
 \qquad \ket{c_{N,Q}(0)}=S_N^\dagger\Phi(0).
\end{equation*}
The centered character sum shows that this coefficient evolution is the
oversampled Fourier--Galerkin evolution. If
$g_K=\mathcal J_{m,d}*_{\mathbb T^d}g$ and
$h_K=\mathcal J_{m,d}*_{\mathbb T^d}h$, Jackson smoothing and the
Dyson-tail estimate give
\begin{align}
 \sup_{0\leq t\leq T}
 \norm{S_N\ket{c_{N,Q}(t)}-\Phi(t)}
 &\leq2B_1\norm{g-g_K}_\infty+2^{2-q}
 \leq\frac32\epsilon.
 \label{eq:torus-digital-discretization-state-error}
\end{align}
They also give the ideal sampled-observable estimate
\begin{equation}
 \left|
 \langle\Psi_Q(T)|\widehat h_Q|\Psi_Q(T)\rangle
 -\langle\Phi(T)|h|\Phi(T)\rangle
 \right|\leq4\norm h_\infty\epsilon,
 \qquad
 \ket{\Psi_Q(T)}=W_{\mathbb T,N,Q}\ket{c_{N,Q}(T)}.
 \label{eq:torus-digital-discretization-observable-error}
\end{equation}
Choose all padded-QFT, kinetic-phase, and schedule subroutines to
operator error at most
\begin{equation*}
 \delta=\Theta\!\left(
 \frac{\epsilon\,
 \log\log(\mathrm e^{\mathrm e}(1+\tau_g)/\epsilon)}
 {(1+\tau_g)\log(\mathrm e^{\mathrm e}(1+\tau_g)/\epsilon)}
 \right),
\end{equation*}
with a sufficiently small implied constant. The arbitrary-modulus QFT
construction supplies the synthesized dilation
$\widetilde U_W^{\mathbb T}$ assumed in the theorem; write
$U_W^{\mathbb T}$ for its ideal unitary dilation of
$W_{\mathbb T,N,Q}$.

To encode the sampled potential, query $O_g$, reversibly clamp the
returned value to $[-\Lambda_g,\Lambda_g]$, compute a single-qubit
rotation whose clean amplitude is $\widetilde g(x_j)/\Lambda_g$, and
uncompute the value register. This uses two calls to $O_g$ or
$O_g^\dagger$ and
$\mathcal O(\log^2(2\Lambda_g/\epsilon_g)+\log^2(1/\delta))$ gates,
apart from the oracle cost. Conjugating this sampled-value block by the
implemented forward and inverse padded QFTs block encodes
$W_{\mathbb T,N,Q}^\dagger\widehat g_QW_{\mathbb T,N,Q}$ with
unnormalized operator error at most
$\epsilon_g+\mathcal O(\Lambda_g\delta)$.

Move to the kinetic frame generated by $a(t)D_N$ and rescale time using
$\dd\sigma=\Lambda_g|b(t)|\dd t$. The rescaled evolution lasts for
$\tau_g=\Lambda_gB_1$ and has a norm-one interaction Hamiltonian. On
each sign interval of $b$, a time-controlled interaction-picture query
consists of the compressed-potential block surrounded by the controlled
kinetic phases; intervals on which $b=0$ are omitted. The access
assumptions supply the sign of $b$, the inverse time conversion, and the
controlled phases.

The rescaled truncated-Dyson interaction-picture algorithm
\cite{low2019interaction,berry2020time} simulates this evolution to
operator error $\mathcal O(\epsilon)$, with failure probability
$\mathcal O(\epsilon)$, using the number of time-controlled queries in
\eqref{eq:torus-binary-query-complexity}. Returning to the
Schr\"odinger picture requires one final kinetic phase. Each
interaction-picture query contains one sampled-potential block and hence
two binary-oracle calls, one forward padded QFT, one inverse padded QFT,
and two controlled kinetic phases. This proves
\eqref{eq:torus-binary-query-complexity}.

Regard the synthesized transform sandwich as an exact block encoding of
a nearby compressed Hamiltonian. Duhamel's formula contributes
$B_1\epsilon_g+\mathcal O(\tau_g\delta)$ from this Hamiltonian
perturbation, while the hybrid argument over the time-controlled queries
contributes another $\mathcal O(\epsilon)$ from kinetic-phase and
schedule synthesis. By the choice of $\delta$, the truncated-Dyson,
oracle, transform, kinetic, and schedule errors therefore sum to
$\mathcal O(\epsilon)$. Initial-state preparation adds at most
$\epsilon_{\rm prep}$ by unitarity, so
\begin{equation*}
 \norm{\ket{\widetilde c(T)}-\ket{c_{N,Q}(T)}}
 =\mathcal O(\epsilon).
\end{equation*}
Combining this with
\eqref{eq:torus-digital-discretization-state-error} proves
\eqref{eq:torus-end-to-end-state-error}.

At readout, applying $\widetilde U_W^{\mathbb T}$ instead of
$U_W^{\mathbb T}$ adds at most $\delta$ in state norm. The ideal joint
readout state is
\begin{equation*}
 \ket{\Psi_Q(T)}_{RC}
 \coloneqq
 \bigl(W_{\mathbb T,N,Q}\ket{c_{N,Q}(T)}\bigr)_R\ket0_C.
\end{equation*}
Together with the coefficient estimate above, this gives
\begin{equation*}
 \norm{\ket{\widetilde\Psi_Q(T)}_{RC}-\ket{\Psi_Q(T)}_{RC}}
 =\mathcal O(\epsilon).
\end{equation*}
Then
\begin{equation*}
 |\langle v|\widehat h_Q|v\rangle
 -\langle w|\widehat h_Q|w\rangle|
 \leq2\norm h_\infty\norm{\ket v-\ket w}.
\end{equation*}
Combining this inequality with
\eqref{eq:torus-digital-discretization-observable-error} proves
\eqref{eq:torus-end-to-end-observable-error}.

The compiled padded QFT has per-transform cost
$d\,\polylog(dM/\delta)$ at the chosen value of $\delta$. The assumed
schedule and kinetic-phase cost is polylogarithmic at the same tolerance.
Multiplying the per-query cost by
\eqref{eq:torus-binary-query-complexity}, and adding state preparation
and final readout, gives \eqref{eq:torus-end-to-end-gate-complexity}.
The transform registers, binary value register, truncated-Dyson time and
order registers, and reusable schedule and kinetic workspace give
\eqref{eq:torus-end-to-end-qubit-complexity}.
\end{proof}
}

\subsection{Positive orthant (and its invertible affine images) with log-barrier metrics}
\label{sec:polyhedra-simulation}

Here we will consider the setting of simulating the Schr\"odinger equation constrained a $d$-dimensional polyhedron $\mathcal{P}$ defined a by $Ax \leq b$, where $A$ is invertible. The associated metric with be the Hessian metric induced by the log-barrier function
\begin{align*}
    \Phi_{\mathcal{P}}(x) = - \sum_{j=1}^d \log(b_j - [Ax]_{j}). 
\end{align*}
Since $A$ is invertible we can switch to a coordinate system $y = A^{-1}(x + b) $, where the constrain becomes a positivity one on $x$, i.e. remaining in the positive orthant of $\mathbb{R}^d$. 

The standard log barrier function (for the positive othant) is
\begin{align*}
\Phi(x)=-\sum_{j=1}^d\log x_j,\qquad x\in(0,\infty)^d.
\end{align*}
Here, we choose the manifold $\mathcal{M}$ to be the positive orthant $(0,\infty)^d$ and the metric to be $g(x)=\nabla^2\Phi(x)$. The corresponding Laplace--Beltrami operator is
\begin{equation}\label{eq:orthant-log-barrier-Laplacian}
\Delta_g=\sum_{i=1}^d(x_i\partial_i)^2.
\end{equation}
Under the coordinate change $y_i=\log x_i$, the metric becomes Euclidean and \eqref{eq:orthant-log-barrier-Laplacian} becomes the ordinary Laplacian on $\mathbb R^d$. This is due to the property of the log barrier: $D\log(x)^\top D\log(x)=\nabla^2\Phi(x)$.

The operator on the full orthant has continuous spectrum, and an infinite domain cannot be represented by a finite digital register. We consider a truncated orthant for two reasons: (1) the above does not have a point spectrum and (2) because we cannot digitally encode the infinite orthant.  We therefore restrict to $[r_-,r_+]^d$, where $0<r_-<r_+<\infty$. A self-adjoint realization on this compact box remains unbounded but has compact resolvent.

We now derive the form of the Riemannian Schr\"odinger equation after the co-ordinate change and truncation. Set $\mu\coloneqq (\log r_-+\log r_+)/2$ and $\ell\coloneqq \log(r_+/r_-)$. Rescaling the logarithmic box to $S=[-1/2,1/2]^d$ by $y=\mu\mathbf1+\ell w$ we have the Schr\"odinger equation
\begin{equation}\label{eq:orthant-rescaled-evolution}
i\partial_t\widetilde\Psi(t,w)=\left[-\frac{a(t)}{\ell^2}\Delta+b(t)\widetilde f(w)\right]\widetilde\Psi(t,w),\qquad \widetilde f(w)\coloneqq f\!\left(e^{\mu+\ell w_1},\ldots,e^{\mu+\ell w_d}\right).
\end{equation}
If $f$ is $G_f$-Lipschitz on $[r_-,r_+]^d$, then $\widetilde f$ is at most $r_+\ell G_f$-Lipschitz on $S$. The same statement holds for an observable $h$ with Lipschitz constant $G_h$. In order to apply the simulation for the torus, we assume that $\widetilde f$ and $\widetilde h$ have periodic extensions with these Lipschitz bounds.

If $A\in\mathbb R^{d\times d}$ is invertible and $c\in\mathbb R^d$, then the metric for the log barrier 
\begin{align*}
    \Phi_{A,c}(x)\coloneqq -\sum_{j=1}^d\log(Ax-c)_j
\end{align*} is given by
\begin{equation}\label{eq:affine-orthant-log-barrier-metric}
g(x)=\nabla^2\Phi_{A,c}(x)=A^{\mathsf T}\diag \!\left((Ax-c)_1^{-2},\ldots,(Ax-c)_d^{-2}\right)A.
\end{equation}
Through the change of coordinates $z=Ax-c$, with inverse $x=A^{-1}(z+c)$, we arrive back at the orthant case.

Fix $0<r_-<r_+<\infty$ and let $\mathcal P\coloneqq \{x\in\mathbb R^d:r_-\mathbf1\leq Ax-c\leq r_+\mathbf1\}$. Call a continuous function $v$ on $\mathcal P$ face-compatible if $v(A^{-1}(z+c))$ takes the same value at $z_i=r_-$ and $z_i=r_+$ for every $i$ and every fixed choice of the other coordinates. Equivalently, its pullback to $S$ descends to a continuous function on the quotient torus. The face compatibility condition can potentially be alleviated using a similar leakage condition \cite[Definition 4.5]{chakrabarti2025speedups}. The authors used this condition tackle issues that come with the potential not being continuous on the Torus.

Let $\Delta_g^{\mathrm{per}}$ denote the self-adjoint realization obtained by identifying these face pairs in logarithmic affine coordinates.  This operator remains unbounded but has compact resolvent. For $n\in\mathbb Z^d$, define
\begin{equation}\label{eq:polyhedron-log-affine-modes}
Y_n^{\mathcal P}(x)\coloneqq \ell^{-d/2}\exp\!\left(\frac{2\pi i}{\ell}n\cdot\left[\log(Ax-c)-\mu\mathbf1\right]\right),\qquad \mathcal V_N^{\mathcal P}\coloneqq \operatorname{span}\left\{Y_n^{\mathcal P}:\norm n_\infty\leq N\right\}.
\end{equation}

\begin{corollary}[Discretization on a truncated log-barrier polyhedron]\label{cor:polyhedron-log-barrier-discretization}
Equip $\mathcal P$ with the metric $g=\nabla^2\Phi_{A,c}$ in \eqref{eq:affine-orthant-log-barrier-metric}. Let $f:\mathcal P\to\mathbb R$ be bounded, real-valued, face-compatible, and $G_f$-Lipschitz in Euclidean distance. Consider
\begin{equation}\label{eq:orthant-continuum-evolution}
i\partial_t\psi(t)=\left[-a(t)\Delta_g^{\mathrm{per}}+b(t)f\right]\psi(t),\qquad \psi(0)\in\mathcal V_{L_0}^{\mathcal P},\qquad \norm{\psi(0)}_{L^2(\mathcal P,\dd\operatorname{vol}_g)}=1,
\end{equation}
for $t\in[0,T]$. Let $h:\mathcal P\to\mathbb R$ be bounded, nonzero, real-valued, face-compatible, and $G_h$-Lipschitz in Euclidean distance. Fix $0<\epsilon<1$ and define $B_1\coloneqq \int_0^T\abs{b(t)}\dd t$. Then there is a quadrature degree $Q$ and an $M^d$-point grid $\mathcal G_{\mathcal P,Q}\subset\mathcal P$, with $M\coloneqq 2Q+1$ nodes along each transformed coordinate, where
\begin{equation}\label{eq:orthant-grid-resolution}
M=\mathcal O\!\left(L_0+\left[1+\frac{r_+\ell\norm{A^{-1}}_{\mathrm{op}}\sqrt d}{\epsilon}\max\left\{G_fB_1,\frac{G_h}{\norm h_\infty}\right\}\right]\left(1+B_1\norm f_\infty+\log(1/\epsilon)\right)\right).
\end{equation}
This grid supports an $\mathcal O(d\log M)$-qubit finite-dimensional evolution with initial state
\begin{equation}
\ket{\Psi(0)}=\left(\frac{\ell}{M}\right)^{d/2}\sum_{x_j\in\mathcal G_{\mathcal P,Q}}\psi(0,x_j)\ket j.
\end{equation}
Let $\mathsf M_h$ denote multiplication by $h$ on $L^2(\mathcal P,\dd\operatorname{vol}_g)$, and define the sampled diagonal observable
\begin{equation}
\widehat h_{\mathcal P,Q}\coloneqq \sum_{x_j\in\mathcal G_{\mathcal P,Q}}h(x_j)\ketbra{j}{j}.
\end{equation}
The final grid state satisfies
\begin{equation}\label{eq:orthant-observable-guarantee}
\left|\langle\Psi(T)|\widehat h_{\mathcal P,Q}|\Psi(T)\rangle-\langle\psi(T)|\mathsf M_h|\psi(T)\rangle\right|\leq4\norm h_\infty\epsilon.
\end{equation}
The truncated positive orthant is the special case $A=I$ and $c=0$.
\end{corollary}

\begin{proof}
Define
\begin{equation*}
\Theta(w)\coloneqq A^{-1}\!\left(\left(e^{\mu+\ell w_1},\ldots,e^{\mu+\ell w_d}\right)+c\right),\qquad (U_\Theta\psi)(w)\coloneqq \ell^{d/2}\psi(\Theta(w)),\qquad w\in S.
\end{equation*}
It follows from the definition of the metric that $D\Theta(w)^{\mathsf T}g(\Theta(w))D\Theta(w)=\ell^2I$. Hence $U_\Theta$ is unitary and
\begin{equation*}
U_\Theta(-\Delta_g^{\mathrm{per}})U_\Theta^\dagger =-\frac{1}{\ell^2}\Delta.
\end{equation*}
Conjugating \eqref{eq:orthant-continuum-evolution} by $U_\Theta$ produces the torus evolution with kinetic coefficient $a(t)/\ell^2$, potential $\widetilde f\coloneqq f\circ\Theta$, and observable $\widetilde h\coloneqq h\circ\Theta$. The face-compatibility assumptions make $\widetilde f$ and $\widetilde h$ periodic. Since $\norm{D\Theta(w)}_{\mathrm{op}}\leq r_+\ell\norm{A^{-1}}_{\mathrm{op}}$, their periodic extensions are Lipschitz with constants at most $r_+\ell\norm{A^{-1}}_{\mathrm{op}}G_f$ and $r_+\ell\norm{A^{-1}}_{\mathrm{op}}G_h$. Moreover, $U_\Theta\mathcal V_{L_0}^{\mathcal P}=\mathcal V_{L_0}$ by \eqref{eq:polyhedron-log-affine-modes}. Applying Theorem~\ref{thm:improved_torus_discretization} and setting $\mathcal G_{\mathcal P,Q}\coloneqq \Theta(\mathcal G_Q)$ proves \eqref{eq:orthant-grid-resolution} and gives the stated initial state because $(U_\Theta\psi)(w_j)=\ell^{d/2}\psi(\Theta(w_j))$. Finally, $\Theta$ is a bijection up to boundary points of measure zero, so the sup norms and continuum expectation values are preserved.
\end{proof}

We have the following simulation resource estimates for the positive orthant and invertible affine images.
{
\begin{theorem}[Digital simulation on a truncated positive orthant and its invertible affine images]
\label{thm:orthant-affine-digital-simulation}
Let $A\in\mathbb R^{d\times d}$ be invertible, $c\in\mathbb R^d$,
$0<r_-<r_+$, and
\begin{equation*}
 \mathcal P=\{x\in\mathbb R^d:r_-\mathbf1\leq Ax-c\leq r_+\mathbf1\},
 \qquad
 \ell=\log(r_+/r_-),\qquad
 \mu=\tfrac12(\log r_-+\log r_+).
\end{equation*}
Equip $\mathcal P$ with
\begin{equation*}
 g(x)=A^{\mathsf T}\diag\!\left((Ax-c)_1^{-2},\ldots,
 (Ax-c)_d^{-2}\right)A
\end{equation*}
and periodic boundary conditions obtained by identifying opposite faces
in the coordinates $\log(Ax-c)$. Let $f,h:\mathcal P\to\mathbb R$ be
bounded, real-valued, face-compatible, respectively $G_f$- and
$G_h$-Lipschitz in Euclidean distance, with $h\ne0$, and consider
\begin{equation*}
 i\partial_t\psi(t)=\left[-a(t)\Delta_g^{\mathrm{per}}+b(t)f\right]
 \psi(t),\qquad
 \norm{\psi(0)}_{L^2(\mathcal P,\dd\operatorname{vol}_g)}=1.
\end{equation*}
Assume $a,b$ are piecewise $C^1$ and $b$ has finitely many sign
intervals. Fix $0<\epsilon<1$, put
$B_1=\int_0^T|b(t)|\dd t$, and choose
$\Lambda_f\geq\max\{1,\norm f_\infty\}$.

For coefficients indexed by $\norm n_\infty\leq N$, define the
log-affine Fourier reconstruction by
\begin{equation*}
 (\mathcal R_{\mathcal P,N}c)(x)
 \coloneqq\ell^{-d/2}\sum_{\norm n_\infty\leq N}c_n
 \exp\!\left(\frac{2\pi i}{\ell}n\cdot
 [\log(Ax-c)-\mu\mathbf1]\right).
\end{equation*}
Suppose $\psi(0)\in\operatorname{Ran}(\mathcal R_{\mathcal P,L_0})$.
Then there are integers $N\leq Q$ and $M=2Q+1$ such that
\begin{equation*}
 M=\mathcal O\!\left(L_0+
 \left[1+\frac{r_+\ell\norm{A^{-1}}_{\mathrm{op}}\sqrt d}{\epsilon}
 \max\left\{G_fB_1,\frac{G_h}{\norm h_\infty}\right\}\right]
 \left[1+B_1\norm f_\infty+\log(1/\epsilon)\right]\right).
\end{equation*}
Writing $\mathcal I_Q=\{-Q,\ldots,Q\}^d$, set
\begin{equation*}
 x_j\coloneqq A^{-1}\!\left(e^{\mu\mathbf1+\ell j/M}+c\right),
 \qquad
 W_{\mathcal P,N,Q}\ket n\coloneqq M^{-d/2}
 \sum_{j\in\mathcal I_Q}e^{2\pi i n\cdot j/M}\ket j.
\end{equation*}
Assume coherent circuits can prepare
$\mathcal R_{\mathcal P,N}^*\psi(0)$ to error
$\epsilon_{\rm prep}\leq c_0\epsilon$, query $f(x_j)$ through a binary
oracle $O_f$ to error $0<\epsilon_f\leq\Lambda_f$ with
$B_1\epsilon_f\leq c_0\epsilon$ at a cost of $\mathcal C_f$ gates and
$\mathcal Q_f$ work qubits, including reversible evaluation and
uncomputation of $x_j$, and implement a unitary dilation
$\widetilde U_W^{\mathcal P}$ of $W_{\mathcal P,N,Q}$, the schedule
arithmetic, and the log-affine kinetic phases to requested operator error
$0<\eta<1$ using $d\,\polylog(dM/\eta)$ gates per use. Here $c_0>0$ is
a sufficiently small universal constant, while state preparation uses
$\mathcal C_{\rm prep}$ gates and $\mathcal Q_{\rm prep}$ work qubits.

Then there is a quantum circuit producing a normalized coefficient state
$\ket{\widetilde c(T)}$, with success probability at least
$1-\mathcal O(\epsilon)$, such that
\begin{equation}\label{eq:orthant-affine-end-to-end-state-error}
 \norm{\mathcal R_{\mathcal P,N}\ket{\widetilde c(T)}-\psi(T)}_
 {L^2(\mathcal P,\dd\operatorname{vol}_g)}=\mathcal O(\epsilon).
\end{equation}
On grid and coefficient registers $R,C$, define
\begin{equation*}
 \ket{\widetilde\Psi_{\mathcal P,Q}(T)}_{RC}
 \coloneqq\widetilde U_W^{\mathcal P}
 \bigl(\ket0_R\ket{\widetilde c(T)}_C\bigr),
 \qquad
 \widehat h_{\mathcal P,Q}\coloneqq
 \left(\sum_{j\in\mathcal I_Q}h(x_j)\ketbra{j}{j}_R\right)\otimes I_C.
\end{equation*}
Then
\begin{equation}\label{eq:orthant-affine-end-to-end-observable-error}
 \left|
 \langle\widetilde\Psi_{\mathcal P,Q}(T)|\widehat h_{\mathcal P,Q}
 |\widetilde\Psi_{\mathcal P,Q}(T)\rangle
 -\langle\psi(T)|h|\psi(T)\rangle
 \right|=\mathcal O(\norm h_\infty\epsilon).
\end{equation}

Let $\tau_f=\Lambda_fB_1$. The circuit makes
\begin{equation}\label{eq:orthant-affine-binary-query-complexity}
 N_f=\mathcal O\!\left(
 (1+\tau_f)
 \frac{\log(\mathrm e^{\mathrm e}(1+\tau_f)/\epsilon)}
 {\log\log(\mathrm e^{\mathrm e}(1+\tau_f)/\epsilon)}
 \right)
 =\widetilde{\mathcal O}(1+\tau_f)
\end{equation}
calls to $O_f$ or $O_f^\dagger$ and uses
\begin{equation}\label{eq:orthant-affine-end-to-end-gate-complexity}
 \mathcal C_{\rm prep}
 +\widetilde{\mathcal O}\!\left((1+\tau_f)\left[
 d+\mathcal C_f
 +\log^2\!\left(\frac{2\Lambda_f}{\epsilon_f}\right)
 \right]\right)
\end{equation}
elementary gates. The total qubit count is
\begin{equation}\label{eq:orthant-affine-end-to-end-qubit-complexity}
 \mathcal Q_{\rm prep}+\mathcal Q_f+\mathcal O\!\left(
 d\log M+\log\frac{2\Lambda_f}{\epsilon_f}
 +\polylog\frac{dM(1+\tau_f)}{\epsilon}
 \right).
\end{equation}
\end{theorem}

\begin{proof}
Define
\begin{equation*}
 \Theta(w)\coloneqq A^{-1}\!\left(e^{\mu\mathbf1+\ell w}+c\right),
 \qquad
 (\mathcal U\varphi)(w)\coloneqq\ell^{d/2}\varphi(\Theta(w)).
\end{equation*}
Then $\mathcal U:L^2(\mathcal P,\dd\operatorname{vol}_g)
\to L^2(\mathbb T^d)$ is unitary,
\begin{equation*}
 \mathcal U(-\Delta_g^{\mathrm{per}})\mathcal U^*
 =-\frac{1}{\ell^2}\Delta,
\end{equation*}
and the explicit reconstruction in the statement satisfies
\begin{equation*}
 (\mathcal U\mathcal R_{\mathcal P,N}c)(w)
 =\sum_{\norm n_\infty\leq N}c_ne^{2\pi i n\cdot w}.
\end{equation*}
Thus the coefficient labels are the ordinary torus Fourier labels.

The pulled-back potential $\widetilde f=f\circ\Theta$ and observable
$\widetilde h=h\circ\Theta$ are periodic and have Lipschitz constants
at most $r_+\ell\norm{A^{-1}}_{\mathrm{op}}G_f$ and
$r_+\ell\norm{A^{-1}}_{\mathrm{op}}G_h$. Applying
Theorem~\ref{thm:torus-digital-simulation} to
\begin{equation*}
 i\partial_t(\mathcal U\psi)
 =\left[-\frac{a(t)}{\ell^2}\Delta
 +b(t)\widetilde f\right](\mathcal U\psi)
\end{equation*}
gives the stated $N,Q,M$ and all state, query, gate, qubit, and success
probability bounds. A potential query at $j/M$ is implemented by
computing $x_j=A^{-1}(e^{\mu\mathbf1+\ell j/M}+c)$, querying $O_f$,
and uncomputing $x_j$. The kinetic phase is
\begin{equation*}
 \exp\!\left(-\frac{4\pi^2i}{\ell^2}\norm n_2^2
 \int_0^t a(s)\dd s\right).
\end{equation*}

Finally, $W_{\mathcal P,N,Q}$ is the same padded Fourier evaluation
isometry used for the pulled-back torus problem, and
$\widetilde h(j/M)=h(x_j)$. Unitary invariance under $\mathcal U$ gives
the state and observable estimates in the statement.
\end{proof}
}

\subsection{\texorpdfstring{$\ell_p$}{ell-p} balls with a Duffy metric}
\label{sec:global-lp-duffy}
We now consider simulation on $\ell_p$ balls. The cases $p=1$ and $p=2$ have familiar geometric descriptions: each orthant of the $\ell_1$ ball is a simplex, while the $\ell_2$ sphere carries the standard round metric. For $p\notin\{1,2\}$, there is no comparably distinguished metric adapted to the $\ell_p$ geometry with a known tractable eigenbasis. The metric inherited from the ambient Euclidean space is one possible choice, but the corresponding eigenfunctions are not known in closed form in general. We instead choose a metric that becomes Euclidean under the Duffy map. Schr\"odinger evolution on the resulting compact subsets of an $\ell_p$ ball and the upper half of an $\ell_p$ sphere then reduces to the now familiar evolution on the flat torus.

Let $p\geq2$. This assumption makes the maps below at least $C^2$ on every compact coordinate box used in the construction, which provides the regularity needed for the Laplace--Beltrami operators below. The open $\ell_p$ ball and the upper hemisphere of the $\ell_p$ sphere are
\begin{equation}\label{eq:lp-ball-and-sphere}
B_p^d\coloneqq \left\{u\in\mathbb R^d:\sum_{k=1}^d|u_k|^p<1\right\},\qquad \mathbb S_{p,+}^d\coloneqq \left\{u\in\mathbb R^{d+1}:\sum_{k=1}^{d+1}|u_k|^p=1,\ u_{d+1}>0\right\}.
\end{equation}
For $x\in(-1,1)^d$, the Duffy map onto the upper hemisphere and its projection onto the ball are
\begin{align}
[\mathcal D_p^{\mathrm{sph}}(x)]_k&\coloneqq x_k\prod_{j<k}(1-|x_j|^p)^{1/p},\quad 1\leq k\leq d,\qquad [\mathcal D_p^{\mathrm{sph}}(x)]_{d+1}\coloneqq \prod_{j=1}^d(1-|x_j|^p)^{1/p},\label{eq:lp-spherical-duffy-map}\\
\widehat{\mathcal D}_p(x)&\coloneqq \left([\mathcal D_p^{\mathrm{sph}}(x)]_1,\ldots,[\mathcal D_p^{\mathrm{sph}}(x)]_d\right).\label{eq:lp-simulation-projected-duffy-map}
\end{align}
The identity $\sum_{k=1}^{d+1}|[\mathcal D_p^{\mathrm{sph}}(x)]_k|^p=1$ follows by telescoping. The map $\widehat{\mathcal D}_p$ is a bijection from $(-1,1)^d$ onto $B_p^d$, with inverse
\begin{equation}\label{eq:lp-projected-duffy-inverse}
x_1=u_1,\qquad x_k=\frac{u_k}{\left(1-\sum_{j<k}|u_j|^p\right)^{1/p}},\qquad 2\leq k\leq d.
\end{equation}
Since the last coordinate of a point in $\mathbb S_{p,+}^d$ is determined by the first $d$ coordinates, $\mathcal D_p^{\mathrm{sph}}$ is also a bijection onto $\mathbb S_{p,+}^d$.

The derivatives of these maps become unbounded near the boundary of $(-1,1)^d$. Fix $0<\delta<1$, set $r_\delta\coloneqq 1-\delta$ and $C_\delta\coloneqq [-r_\delta,r_\delta]^d$, and restrict the maps to $C_\delta$. Their images are
\begin{equation}\label{eq:lp-truncated-ball-and-cap}
\Omega_{p,\delta}^{\mathrm{ball}}\coloneqq \widehat{\mathcal D}_p(C_\delta),\qquad \Omega_{p,\delta}^{\mathrm{sph}}\coloneqq \mathcal D_p^{\mathrm{sph}}(C_\delta).
\end{equation}
This truncation restricts the coordinates in $C_\delta$ rather than the $\ell_p$ radius. In particular, $\Omega_{p,\delta}^{\mathrm{ball}}$ need not be a smaller concentric $\ell_p$ ball.

We place metrics on the two images by requiring the coordinate maps to be isometries:
\begin{align}
(\widehat{\mathcal D}_p)^*g_{p,\delta}^{\mathrm{ball}}&=I_d,& \dd\mu_{p,\delta}^{\mathrm{ball}}&\coloneqq (\widehat{\mathcal D}_p)_\#\dd x,\notag\\
(\mathcal D_p^{\mathrm{sph}})^*g_{p,\delta}^{\mathrm{sph}}&=I_d,& \dd\mu_{p,\delta}^{\mathrm{sph}}&\coloneqq (\mathcal D_p^{\mathrm{sph}})_\#\dd x.\label{eq:lp-flat-duffy-metric}
\end{align}
The metric induced from the ambient Euclidean space would instead pull back under either map $F$ to $DF^{\mathsf T}DF$. The measures in \eqref{eq:lp-flat-duffy-metric} are the Riemannian volume measures of the metrics defined there. The coordinate maps therefore define unitaries
\begin{align}
U_{p,\delta}&:L^2(\Omega_{p,\delta}^{\mathrm{ball}},\dd\mu_{p,\delta}^{\mathrm{ball}})\to L^2(C_\delta,\dd x),& (U_{p,\delta}\psi)(x)&\coloneqq \psi(\widehat{\mathcal D}_p(x)),\label{eq:lp-duffy-coordinate-unitary}\\
U_{p,\delta}^{\mathrm{sph}}&:L^2(\Omega_{p,\delta}^{\mathrm{sph}},\dd\mu_{p,\delta}^{\mathrm{sph}})\to L^2(C_\delta,\dd x),& (U_{p,\delta}^{\mathrm{sph}}\psi)(x)&\coloneqq \psi(\mathcal D_p^{\mathrm{sph}}(x)).
\end{align}
For the projected image, we abbreviate $\Omega_{p,\delta}\coloneqq \Omega_{p,\delta}^{\mathrm{ball}}$ and $\mu_{p,\delta}\coloneqq \mu_{p,\delta}^{\mathrm{ball}}$.
When opposite faces of $C_\delta$ are identified, the corresponding Laplace--Beltrami operator on $\Omega_{p,\delta}$ is defined by
\begin{equation}\label{eq:lp-periodic-duffy-kinetic-operator}
-\Delta_{\mathcal D_p}^{\mathrm{per}}\coloneqq U_{p,\delta}^\dagger (-\Delta_{C_\delta}^{\mathrm{per}})U_{p,\delta}.
\end{equation}

The next bounds control the Lipschitz constants after the change of variables.
\begin{lemma}[Bounds for the two coordinate maps]\label{lem:lp-coordinate-map-bounds}
On $C_\delta$,
\begin{align}
\norm{D\widehat{\mathcal D}_p(x)}_{\mathrm{op}}&\leq\kappa_{p,\delta}^{\mathrm{ball}},& (\kappa_{p,\delta}^{\mathrm{ball}})^2&\coloneqq d+\frac{d(d-1)}2\frac{r_\delta^{2p}}{(1-r_\delta^p)^{2-2/p}},\label{eq:lp-ball-duffy-lipschitz-constant}\\
\norm{D\mathcal D_p^{\mathrm{sph}}(x)}_{\mathrm{op}}&\leq\kappa_{p,\delta}^{\mathrm{sph}},& (\kappa_{p,\delta}^{\mathrm{sph}})^2&\coloneqq (\kappa_{p,\delta}^{\mathrm{ball}})^2+d\frac{r_\delta^{2p-2}}{(1-r_\delta^p)^{2-2/p}}.\label{eq:lp-cap-duffy-lipschitz-constant}
\end{align}
Consequently, if $v$ is $G$-Lipschitz in the ambient Euclidean distance on either image in \eqref{eq:lp-truncated-ball-and-cap}, then its pullback to $C_\delta$ is $\kappa_{p,\delta}^{\mathrm{ball}}G$-Lipschitz for the ball and $\kappa_{p,\delta}^{\mathrm{sph}}G$-Lipschitz for the upper hemisphere.
\end{lemma}
\begin{proof}
The diagonal entries of $D\widehat{\mathcal D}_p(x)$ have magnitude at most one, and the entries below the diagonal satisfy
\begin{equation*}
\left|\frac{\partial[\widehat{\mathcal D}_p(x)]_k}{\partial x_j}\right|\leq\frac{r_\delta^p}{(1-r_\delta^p)^{1-1/p}},\qquad j<k.
\end{equation*}
The first bound follows from the Frobenius norm, since there are $d$ diagonal entries and $d(d-1)/2$ entries below the diagonal. The first $d$ rows of $D\mathcal D_p^{\mathrm{sph}}(x)$ are the same, while its last row satisfies
\begin{equation*}
\left|\frac{\partial[\mathcal D_p^{\mathrm{sph}}(x)]_{d+1}}{\partial x_j}\right|\leq\frac{r_\delta^{p-1}}{(1-r_\delta^p)^{1-1/p}}.
\end{equation*}
The second Frobenius-norm bound follows by adding these $d$ entries. The Lipschitz estimates follow by integrating the differential bounds along line segments in the convex set $C_\delta$ and then composing with $v$.
\end{proof}
Set $\kappa_{p,\delta}\coloneqq \kappa_{p,\delta}^{\mathrm{ball}}$. We first impose periodic boundary conditions. Choose either $F=\widehat{\mathcal D}_p$ or $F=\mathcal D_p^{\mathrm{sph}}$. For the projected map, set $\Omega=\Omega_{p,\delta}^{\mathrm{ball}}$, $\mu=\mu_{p,\delta}^{\mathrm{ball}}$, $U=U_{p,\delta}$, and $\kappa=\kappa_{p,\delta}^{\mathrm{ball}}$. For the map onto the upper hemisphere, set $\Omega=\Omega_{p,\delta}^{\mathrm{sph}}$, $\mu=\mu_{p,\delta}^{\mathrm{sph}}$, $U=U_{p,\delta}^{\mathrm{sph}}$, and $\kappa=\kappa_{p,\delta}^{\mathrm{sph}}$.
Identifying each pair of opposite faces of $C_\delta$ defines the periodic Laplacian $-\Delta_{C_\delta}^{\mathrm{per}}$. Its counterpart on $\Omega$ is
\begin{equation*}
-\Delta_\Omega^{\mathrm{per}}\coloneqq U^\dagger (-\Delta_{C_\delta}^{\mathrm{per}})U.
\end{equation*}
A continuous function $v$ on $\Omega$ respects this identification when $v\circ F$ takes the same value at $x_i=-r_\delta$ and $x_i=r_\delta$ for every $i$ and every fixed choice of the remaining coordinates.

Rescale $C_\delta$ to $S=[-1/2,1/2]^d$ by defining
\begin{equation*}
\Theta(w)\coloneqq F(2r_\delta w),\qquad (\mathcal U\psi)(w)\coloneqq (2r_\delta)^{d/2}\psi(\Theta(w)).
\end{equation*}
Then $\mathcal U:L^2(\Omega,\dd\mu)\to L^2(\mathbb T^d)$ is unitary and
\begin{equation}\label{eq:lp-duffy-periodic-rescaling}
\mathcal U(-\Delta_\Omega^{\mathrm{per}})\mathcal U^\dagger =-\frac{1}{4r_\delta^2}\Delta.
\end{equation}
The corresponding Fourier modes on $\Omega$ are
\begin{equation}\label{eq:lp-duffy-periodic-modes}
Y_n^\Omega(u)\coloneqq (2r_\delta)^{-d/2}\exp\!\left(\frac{\pi i}{r_\delta}n\cdot F^{-1}(u)\right),\qquad \mathcal V_N^\Omega\coloneqq \operatorname{span}\left\{Y_n^\Omega:\norm n_\infty\leq N\right\}.
\end{equation}
They are orthonormal eigenfunctions of $-\Delta_\Omega^{\mathrm{per}}$, and $\mathcal U\mathcal V_N^\Omega=\mathcal V_N$.

\begin{corollary}[Discretization with periodic boundary conditions]\label{cor:lp-duffy-periodic-discretization}
Use either choice of $F$ and the accompanying notation above. Let $f:\Omega\to\mathbb R$ be bounded, real-valued, and $G_f$-Lipschitz in the ambient Euclidean distance, and assume that $f\circ F$ has matching values on every pair of opposite faces of $C_\delta$. Consider
\begin{equation}\label{eq:lp-duffy-periodic-continuum-evolution}
i\partial_t\psi(t)=\left[-a(t)\Delta_\Omega^{\mathrm{per}}+b(t)f\right]\psi(t),\qquad \psi(0)\in\mathcal V_{L_0}^\Omega,\qquad \norm{\psi(0)}_{L^2(\Omega,\dd\mu)}=1,
\end{equation}
for $t\in[0,T]$. Let $h:\Omega\to\mathbb R$ be bounded, nonzero, real-valued, and $G_h$-Lipschitz in the ambient Euclidean distance, and assume that $h\circ F$ also has matching values on opposite faces. Fix $0<\epsilon<1$ and define $B_1\coloneqq \int_0^T\abs{b(t)}\dd t$. Then there is a quadrature degree $Q$ such that the grid $\mathcal G_{F,Q}\coloneqq \Theta(\mathcal G_Q)\subset\Omega$ contains $M^d$ points, where $M\coloneqq 2Q+1$ and
\begin{equation}\label{eq:lp-duffy-periodic-grid-resolution}
M=\mathcal O\!\left(L_0+\left[1+\frac{2r_\delta\kappa\sqrt d}{\epsilon}\max\left\{G_fB_1,\frac{G_h}{\norm h_\infty}\right\}\right]\left(1+B_1\norm f_\infty+\log(1/\epsilon)\right)\right).
\end{equation}
This grid supports an $\mathcal O(d\log M)$-qubit finite-dimensional evolution with initial state
\begin{equation}\label{eq:lp-duffy-periodic-sampled-initial-state}
\ket{\Psi(0)}=\left(\frac{2r_\delta}{M}\right)^{d/2}\sum_{u_j\in\mathcal G_{F,Q}}\psi(0,u_j)\ket j.
\end{equation}
Let $\mathsf M_h$ denote multiplication by $h$ on $L^2(\Omega,\dd\mu)$, and define
\begin{equation}
\widehat h_{F,Q}\coloneqq \sum_{u_j\in\mathcal G_{F,Q}}h(u_j)\ketbra{j}{j}.
\end{equation}
The final grid state satisfies
\begin{equation}\label{eq:lp-duffy-periodic-observable-guarantee}
\left|\langle\Psi(T)|\widehat h_{F,Q}|\Psi(T)\rangle-\langle\psi(T)|\mathsf M_h|\psi(T)\rangle\right|\leq4\norm h_\infty\epsilon.
\end{equation}
\end{corollary}

\begin{proof}
Conjugating \eqref{eq:lp-duffy-periodic-continuum-evolution} by $\mathcal U$ and using \eqref{eq:lp-duffy-periodic-rescaling} gives a torus evolution with potential $\widetilde f\coloneqq f\circ\Theta$ and observable $\widetilde h\coloneqq h\circ\Theta$. The matching boundary values make these functions continuous on $\mathbb T^d$. Lemma~\ref{lem:lp-coordinate-map-bounds} shows that they are Lipschitz on the fundamental cube with constants at most $2r_\delta\kappa G_f$ and $2r_\delta\kappa G_h$. The same bounds hold in the torus distance: a shortest torus segment can be divided where it crosses a face, and each piece lifts to a segment in the fundamental cube.

Eq.~\eqref{eq:lp-duffy-periodic-modes} shows that $\mathcal U\psi(0)\in\mathcal V_{L_0}$. Theorem~\ref{thm:improved_torus_discretization} now gives \eqref{eq:lp-duffy-periodic-grid-resolution}. At a node $w_j\in\mathcal G_Q$, $(\mathcal U\psi)(w_j)=(2r_\delta)^{d/2}\psi(\Theta(w_j))$, which gives the initial state in \eqref{eq:lp-duffy-periodic-sampled-initial-state}. The coordinate change preserves the sup norms of $f$ and $h$. Since $\mathcal U\mathsf M_h\mathcal U^\dagger $ is multiplication by $\widetilde h$, it also preserves the continuum expectation value. The observable estimate in Theorem~\ref{thm:improved_torus_discretization} therefore becomes \eqref{eq:lp-duffy-periodic-observable-guarantee}.
\end{proof}

Corollary~\ref{cor:lp-duffy-periodic-discretization} applies to both images in \eqref{eq:lp-truncated-ball-and-cap} when the potential and observable have matching values on opposite coordinate faces.

{\
\begin{theorem}[Digital simulation on a Duffy $\ell_p$ domain]
\label{thm:lp-duffy-digital-simulation}
Let $p\geq2$, $0<\delta<1$, $r_\delta=1-\delta$, and
$C_\delta=[-r_\delta,r_\delta]^d$. Let $F:C_\delta\to\Omega$ be either
the projected Duffy map onto the coordinate-truncated $\ell_p$ ball or
the Duffy map onto the coordinate-truncated upper hemisphere. Equip
$\Omega$ with the flat Duffy metric $F^*g=I_d$, Riemannian measure
$\mu=F_\#\dd x$, and periodic boundary conditions induced by identifying
opposite faces of $C_\delta$. Set
\begin{equation*}
 \kappa\coloneqq\sup_{x\in C_\delta}\norm{DF(x)}_{\mathrm{op}}.
\end{equation*}
Let $f,h:\Omega\to\mathbb R$ be bounded, real-valued, respectively
$G_f$- and $G_h$-Lipschitz, with $h\ne0$, and suppose $f\circ F$ and
$h\circ F$ have matching values on opposite faces. Consider
\begin{equation*}
 i\partial_t\psi(t)=\left[-a(t)\Delta_\Omega^{\mathrm{per}}+b(t)f\right]
 \psi(t),\qquad \norm{\psi(0)}_{L^2(\Omega,\dd\mu)}=1.
\end{equation*}
Assume $a,b$ are piecewise $C^1$ and $b$ has finitely many sign
intervals. Fix $0<\epsilon<1$, put
$B_1=\int_0^T|b(t)|\dd t$, and choose
$\Lambda_f\geq\max\{1,\norm f_\infty\}$.

For coefficients indexed by $\norm n_\infty\leq N$, define the Duffy
Fourier reconstruction explicitly by
\begin{equation*}
 (\mathcal R_{F,N}c)(u)
 \coloneqq(2r_\delta)^{-d/2}
 \sum_{\norm n_\infty\leq N}c_n
 \exp\!\left(\frac{\pi i}{r_\delta}n\cdot F^{-1}(u)\right).
\end{equation*}
Suppose $\psi(0)\in\operatorname{Ran}(\mathcal R_{F,L_0})$. Then there
are integers $N\leq Q$ and $M=2Q+1$ such that
\begin{equation*}
 M=\mathcal O\!\left(L_0+
 \left[1+\frac{2r_\delta\kappa\sqrt d}{\epsilon}
 \max\left\{G_fB_1,\frac{G_h}{\norm h_\infty}\right\}\right]
 \left[1+B_1\norm f_\infty+\log(1/\epsilon)\right]\right).
\end{equation*}
Writing $\mathcal I_Q=\{-Q,\ldots,Q\}^d$, set
\begin{equation*}
 u_j\coloneqq F(2r_\delta j/M),\qquad
 W_{F,N,Q}\ket n\coloneqq M^{-d/2}
 \sum_{j\in\mathcal I_Q}e^{2\pi i n\cdot j/M}\ket j.
\end{equation*}
Assume coherent circuits can prepare
$\mathcal R_{F,N}^*\psi(0)$ to error
$\epsilon_{\rm prep}\leq c_0\epsilon$, query $f(u_j)$ through a binary
oracle $O_f$ to error $0<\epsilon_f\leq\Lambda_f$ with
$B_1\epsilon_f\leq c_0\epsilon$ at a cost of $\mathcal C_f$ gates and
$\mathcal Q_f$ work qubits, and implement a unitary dilation
$\widetilde U_W^F$ of $W_{F,N,Q}$, the schedule arithmetic, and the
Fourier kinetic phases to requested error $0<\eta<1$ using
$d\,\polylog(dM/\eta)$ gates per use. Here $c_0>0$ is a sufficiently
small universal constant, while state preparation uses
$\mathcal C_{\rm prep}$ gates and $\mathcal Q_{\rm prep}$ work qubits.

Then there is a quantum circuit producing a normalized coefficient state
$\ket{\widetilde c(T)}$, with success probability at least
$1-\mathcal O(\epsilon)$, such that
\begin{equation}\label{eq:lp-duffy-end-to-end-state-error}
 \norm{\mathcal R_{F,N}\ket{\widetilde c(T)}-\psi(T)}_
 {L^2(\Omega,\dd\mu)}=\mathcal O(\epsilon).
\end{equation}
On grid and coefficient registers $R,C$, define
\begin{equation*}
 \ket{\widetilde\Psi_{F,Q}(T)}_{RC}
 \coloneqq\widetilde U_W^F
 \bigl(\ket0_R\ket{\widetilde c(T)}_C\bigr),
 \qquad
 \widehat h_{F,Q}\coloneqq
 \left(\sum_{j\in\mathcal I_Q}h(u_j)\ketbra{j}{j}_R\right)\otimes I_C.
\end{equation*}
Then
\begin{equation}\label{eq:lp-duffy-end-to-end-observable-error}
 \left|
 \langle\widetilde\Psi_{F,Q}(T)|\widehat h_{F,Q}
 |\widetilde\Psi_{F,Q}(T)\rangle
 -\langle\psi(T)|h|\psi(T)\rangle
 \right|=\mathcal O(\norm h_\infty\epsilon).
\end{equation}

Let $\tau_f\coloneqq\Lambda_fB_1$. The circuit makes
\begin{equation}\label{eq:lp-duffy-binary-query-complexity}
 N_f=\mathcal O\!\left(
 (1+\tau_f)
 \frac{\log(\mathrm e^{\mathrm e}(1+\tau_f)/\epsilon)}
 {\log\log(\mathrm e^{\mathrm e}(1+\tau_f)/\epsilon)}
 \right)
 =\widetilde{\mathcal O}(1+\tau_f)
\end{equation}
calls to $O_f$ or $O_f^\dagger$ and uses
\begin{equation}\label{eq:lp-duffy-end-to-end-gate-complexity}
 \mathcal C_{\rm prep}
 +\widetilde{\mathcal O}\!\left((1+\tau_f)\left[
 d+\mathcal C_f
 +\log^2\!\left(\frac{2\Lambda_f}{\epsilon_f}\right)
 \right]\right)
\end{equation}
elementary gates. The total qubit count is
\begin{equation}\label{eq:lp-duffy-end-to-end-qubit-complexity}
 \mathcal Q_{\rm prep}+\mathcal Q_f+\mathcal O\!\left(
 d\log M+\log\frac{2\Lambda_f}{\epsilon_f}
 +\polylog\frac{dM(1+\tau_f)}{\epsilon}
 \right).
\end{equation}
\end{theorem}

\begin{proof}
Define
\begin{equation*}
 \Theta(w)\coloneqq F(2r_\delta w),\qquad
 (\mathcal U\varphi)(w)\coloneqq
 (2r_\delta)^{d/2}\varphi(\Theta(w)).
\end{equation*}
Then $\mathcal U:L^2(\Omega,\dd\mu)\to L^2(\mathbb T^d)$ is unitary,
\begin{equation*}
 \mathcal U(-\Delta_\Omega^{\mathrm{per}})\mathcal U^*
 =-\frac{1}{4r_\delta^2}\Delta,
\end{equation*}
and the explicit reconstruction in the statement satisfies
\begin{equation*}
 (\mathcal U\mathcal R_{F,N}c)(w)
 =\sum_{\norm n_\infty\leq N}c_ne^{2\pi i n\cdot w}.
\end{equation*}
Thus the coefficient labels are exactly the ordinary torus Fourier
labels.

The pulled-back potential $\widetilde f=f\circ\Theta$ and observable
$\widetilde h=h\circ\Theta$ are periodic and have Lipschitz constants at
most $2r_\delta\kappa G_f$ and $2r_\delta\kappa G_h$. Applying the torus
interaction-picture  construction (Theorem \ref{thm:torus-digital-simulation}) to
\begin{equation*}
 i\partial_t(\mathcal U\psi)
 =\left[-\frac{a(t)}{4r_\delta^2}\Delta
 +b(t)\widetilde f\right](\mathcal U\psi)
\end{equation*}
gives the stated $N,Q,M$ and all state, query, gate, qubit, and success
probability bounds. A potential query at $j/M$ is implemented by
computing $u_j=F(2r_\delta j/M)$, querying $O_f$, and uncomputing $u_j$.
The kinetic phase is
\begin{equation*}
 \exp\!\left(-\frac{\pi^2i}{r_\delta^2}\norm n_2^2
 \int_0^t a(s)\dd s\right).
\end{equation*}

Finally, $W_{F,N,Q}$ is the same padded Fourier evaluation isometry used
for the pulled-back torus problem, and
$\widetilde h(j/M)=h(u_j)$. Unitary invariance under $\mathcal U$ gives
the state and observable estimates in the statement.
\end{proof}
}

\subsection{Sphere with the standard metric} \label{sec:sphere_standard_met}
We consider the sphere (without loss of generality unit sphere) with its natural Euclidean-inherited metric  $g_{a,b} = \delta_{a,b} - \frac{y_ay_b}{r^2}$. This gives the usual spherical Laplacian $\Delta_{\mathbb{S}^d}$. Thus, $\mathcal{H}_\ell$ is taken to be spanned by the degree $\ell$ spherical harmonics. In this subsection, $\{Y_{\ell,m}:1\leq m\leq h_{\ell,d}\}$ denotes an orthonormal basis in $L^2(\mathbb S^d,\dd S_d)$, where $\dd S_d$ is ordinary Riemannian area measure. Then
\begin{align}
    -\Delta_{\mathbb S^d}Y_{\ell,m} &= \lambda_{\ell,d}Y_{\ell,m},\\
    \lambda_{\ell,d}&=\ell(\ell+d-1),\\
    h_{\ell,d}
    &=\binom{\ell+d}{d}-\binom{\ell+d-2}{d}
    =\frac{(2\ell+d-1)(\ell+d-2)!}{\ell!(d-1)!}.
\end{align}
 Let
$\mathcal P_\ell(\mathbb{R}^{d+1})$ be the homogeneous polynomials of degree
$\ell$.  Using the notation Section \ref{sec:global_spectral_methods} the level function $L$ maps $\mathcal{Y}_{\ell, m}$ to $\ell$. The spaces $\mathcal{H}_{\ell}$ and $\mathcal{V}_N$ are defined using the labeling defined by $L$. Specifically, $\mathcal{H}_{\ell}$ is the space of spherical harmonics with degree $\ell$.

The spherical harmonics on the unit sphere satisfy the Fischer
decomposition:
\begin{equation*}
    \mathcal P_\ell = \mathcal H_\ell\oplus \lVert x \rVert^2\mathcal P_{\ell-2} = \mathcal H_\ell\oplus \mathcal P_{\ell-2}.
\end{equation*}
Recursing the above gives $ \mathcal P_\ell \subseteq \mathcal{V}_{\ell}$. Since $\mathcal{H}_{\ell_1} \cdot \mathcal{H}_{\ell_2} \subseteq \mathcal{P}_{\ell_1 + \ell_2}$, the above implies the eigenfunctions of $\Delta_{\mathbb{S}^{d}}$ satisfy the slow spectral growth in \eqref{eqn:eigfunc_mul_exp_prop}.

The natural smoothing procedure for functions on the sphere expressed in the spherical harmonics basis is the multi-dimensional heat kernel. Recall the multi-dimensional heat semigroup:
\begin{equation*}
    F_t=P_tF = e^{t\Delta_{\mathbb S^d}}F = \sum_{\ell,m}e^{-t\lambda_{\ell,d}} \widehat F_{\ell,m}Y_{\ell,m}.
\end{equation*}
The heat-kernel smoothed function satisfies the following guarantees.
\begin{lemma}
\label{lem:heat-kernel-smoothing}
Let $F:\mathbb{S}^d\to\mathbb{R}$ be bounded and geodesically $G$-Lipschitz. Let $F_t$ be $F$ evolved under the heat kernel for time $t > 0$. Then
\begin{equation}
 \norm{F_t-F}_\infty\leq G\min\{\pi,\sqrt{2dt}\}.
 \label{eq:heat-uniform-error}
\end{equation}
Additionally, for $ K\geq\max\left\{d,\sqrt{\frac{d-1}{2t}}\right\}$,
\begin{equation}
 \norm{(I-P_K)F_t}_\infty
 \leq\norm F\,
 \left[\frac{2^{d-1}\Gamma(d/2)}{(d-1)!}\right]^{1/2}t^{-d/4}e^{-tK^2/2}.
 \label{eq:heat-exact-tail}
\end{equation}
\end{lemma}
\begin{proof}
We will start by representing the heat-evolved function $F_t$ as $\mathbb{E}_x[F(X_t)], X_0 = x$ where $X_t$ is the spherical Brownian motion evolved for time $t$ from $0$ with generator $\Delta_{\mathbb{S}^d}$. Fix $x$ and write
\begin{equation}
 r(y)=d_g(x,y)\in[0,\pi].
\end{equation}
Then Jensen's and geodesic Lipschitzness imply that: 
\begin{align*}
     |F_t(x)-F(x)|
 &\leq\mathbb E_x|F(X_t)-F(x)|
 \leq G\mathbb E_xr(X_t).
\end{align*}
Thus, the goal of the first part of the proof will be to show $\mathbb E_xr(X_t) \leq \min\{\pi,\sqrt{2dt}\}.$

Since $r$ is time-independent and $r(x) = 0$, if $r$ were sufficiently regular, then Ito's formula and taking expectation gives 
\begin{align}
\label{eqn:ito}
    \mathbb{E}_{x}[r(X_t)] = \mathbb{E}_x \int_0^{t} (\Delta_{\mathbb{S}} r)(X_s) ds.
\end{align}
Near $x$, we choose the local coordinate system
\begin{equation}
 y(r, \vartheta)=\cos(r)\,x+\sin(r)\,\vartheta, \quad \vartheta \in T_x\mathbb{S}^{d}, \quad \lVert \vartheta \rVert = 1.
 \label{eq:sphere-geodesic-polar-map}
\end{equation}
This gives the metric
\begin{align*}
g=\dd r^2+\sin^2r\,g_{\mathbb S^{d-1}}, \quad
 \sqrt{|g|}
 =\sin^{d-1}r\sqrt{|g_{\mathbb S^{d-1}}|}.
\end{align*}
Thus for a purely $r$-dependent function $f(r(y))$, the radial Laplacian acts as
\begin{align*}
    \Delta_{\mathbb{S}^d}(f \circ r) = \frac{1}{\sin^{d-1}(r)}\frac{\mathrm d}{\mathrm{dr}}\left(\sin^{d-1}(r)f'(r)\right) = f''(r) + (d-1)\cot r f'(r),
\end{align*}
where $r(y)$ computes the $r$ component of $y$.

Taking $f(r)=r$ yields
\begin{equation}
 \Delta r=(d-1)\cot r,
 \qquad 0<r<\pi.
 \label{eq:sphere-distance-Laplacian-first-order}
\end{equation}
Taking instead
$f(r)=r^2$, so that $f'(r)=2r$ and $f''(r)=2$, gives directly
\begin{equation}
 \Delta r^2=2r\Delta r+2\abs{\nabla r}^2
 =2+2(d-1)r\cot r\leq2d,
 \label{eq:sphere-distance-Laplacian}
\end{equation}
where the inequality follows from $r\cot r \leq 1$ on $r\in[-\pi,\pi]$.

We would like to apply \eqref{eqn:ito} $r^2$, however it fails to be differentiable at the antipod $-x$, and thus is not sufficiently regular for It\^o's lemma.
Hence we apply the following smoothing argument and pass to the limit to use It\^o's lemma.

Fix $0<\eta<\pi/4$.  Choose a nonincreasing cutoff
$\chi_\eta\in C^\infty([0,\pi])$ satisfying
\begin{equation}
 0\leq\chi_\eta\leq1,
 \qquad
 \chi_\eta(r)=1\text{ for }0\leq r\leq\pi-2\eta,
 \qquad
 \chi_\eta(r)=0\text{ for }\pi-\eta\leq r\leq\pi.
 \label{eq:Dynkin-radial-cutoff}
\end{equation}
Define
\begin{equation}
 f_\eta(r)=\int_0^r2s\chi_\eta(s)\dd s,
 \qquad
 \Phi_\eta(y)=f_\eta(d_g(x,y)) = f_\eta(r(y)).
 \label{eq:Dynkin-smooth-approximation}
\end{equation}
Near $r=0$, $f_\eta(r)=r^2$, so $\Phi_\eta$ is smooth at $x$ in normal
coordinates.  Near $r=\pi$, $f_\eta$ is constant, so $\Phi_\eta$ is also
smooth at the antipode.  It is smooth elsewhere because $r$ is smooth away
from those two points.  Hence $\Phi_\eta\in C^\infty(\mathbb S^d)$.

The first two derivatives of $f_\eta$ are
\begin{equation}
 f_\eta'(r)=2r\chi_\eta(r)\geq0,
 \qquad
 f_\eta''(r)=2\chi_\eta(r)+2r\chi_\eta'(r)\leq2,
 \label{eq:Dynkin-approximation-derivatives}
\end{equation}
using $\chi_\eta'\leq0$.  For a radial function
$f(r)$ on the sphere, quote the radial Laplacian formula %
\begin{equation}
 \Delta(f(r))=f''(r)+(d-1)\cot r\,f'(r).
 \label{eq:Dynkin-radial-Laplacian}
\end{equation}
On $0<r\leq\pi-2\eta$, $f_\eta(r)=r^2$, so
\eqref{eq:sphere-distance-Laplacian} gives
$\Delta\Phi_\eta\leq2d$.  On the transition region
$r\geq\pi-2\eta>\pi/2$, one has $\cot r\leq0$ and
$f_\eta'(r)\geq0$.  Substituting
\eqref{eq:Dynkin-approximation-derivatives} into
\eqref{eq:Dynkin-radial-Laplacian} therefore gives
\begin{equation}
 \Delta\Phi_\eta
 =f_\eta''(r)+(d-1)\cot r\,f_\eta'(r)
 \leq f_\eta''(r)\leq2\leq2d.
\end{equation}
Thus
\begin{equation}
 \Delta\Phi_\eta(y)\leq2d
 \qquad\text{for every }y\in\mathbb S^d.
 \label{eq:Dynkin-global-generator-bound}
\end{equation}

We now apply Ito's Lemma with $\Phi_\eta$ instead of $r^2$.  The starting value
is $\Phi_\eta(x)=f_\eta(0)=0$ gives:
\begin{align}
 \mathbb E_x[f_\eta(r(X_t))]
 &=\mathbb E_x\int_0^t(\Delta\Phi_\eta)(X_s)\dd s\nonumber\\
 &\leq\int_0^t2d\dd s=2dt.
 \label{eq:Dynkin-approximating-moment}
\end{align}

It remains to remove the approximation. Consider the limit $\eta\downarrow$. From
\eqref{eq:Dynkin-smooth-approximation},
\begin{align}  \label{eq:Dynkin-uniform-approximation}
    0\leq r^2-f_\eta(r) =\int_0^r2s(1-\chi_\eta(s))\dd s \leq\int_{\pi-2\eta}^{\pi}2s\dd s \leq4\pi\eta.
\end{align}
Therefore $f_\eta(r)\to r^2$ uniformly on $[0,\pi]$. Using this in Eq.~\eqref{eq:Dynkin-approximating-moment} and exchanging limit with expctation gives
\begin{equation}
 \mathbb E_x r(X_t)^2
 \leq 2dt.
 \label{eq:sphere-Brownian-second-moment}
\end{equation}

For the second part of the Lemma, we quote its pointwise expansion:
\begin{equation}
 (I-P_K)F_t(x)=
 \sum_{\ell>K}\sum_{m=1}^{h_{\ell,d}}
 e^{-t\lambda_{\ell,d}}\widehat F_{\ell,m}Y_{\ell,m}(x).
\end{equation}

Apply Cauchy--Schwarz to this sum with first factor
$\widehat F_{\ell,m}$ and second factor
$e^{-t\lambda_{\ell,d}}Y_{\ell,m}(x)$.  This gives
\begin{align}
 |(I-P_K)F_t(x)|^2
 &\leq
 \left(\sum_{\ell>K}\sum_m|\widehat F_{\ell,m}|^2\right)
 \left(\sum_{\ell>K}\sum_m
 e^{-2t\lambda_{\ell,d}}|Y_{\ell,m}(x)|^2\right)\nonumber\\
 &\leq\frac{\norm{F}_{L^2(\dd S_d)}^2}{|\mathbb S^d|}
 \sum_{\ell>K}h_{\ell,d}e^{-2t\lambda_{\ell,d}}.
\end{align}
The second line uses Parseval's identity and the
addition formula for area-normalized spherical harmonics,
\begin{equation*}
 \sum_m|Y_{\ell,m}(x)|^2
 =\frac{h_{\ell,d}}{|\mathbb S^d|}.
\end{equation*}
Together with
$\norm F_{L^2(\dd S_d)}\leq|\mathbb S^d|^{1/2}\norm F_\infty$, the
area factors cancel. Taking square roots and the
supremum over $x$ proves:

\begin{align*}
 \norm{(I-P_K)F_t}_\infty
 \leq\norm F_\infty\,
 \mathcal H_d(K,t),
\end{align*}
where
\begin{align}
 \label{eqn:pre_tail_bound}
 \mathcal H_d(K,t)\coloneqq 
 \left[\sum_{\ell=K+1}^\infty
h_{\ell,d}e^{-2t\lambda_{\ell,d}}\right]^{1/2}.
\end{align}

To obtain the closed tail, first quote the multiplicity formula
\begin{equation}
 h_{\ell,d}=\frac{(2\ell+d-1)(\ell+d-2)!}
 {\ell!(d-1)!}.
\end{equation}
When $\ell\geq d$, bound
$2\ell+d-1\leq3\ell$ and each of the $d-2$ factors in
$(\ell+d-2)!/\ell!$ by $2\ell$.  This yields the convenient envelope
$h_{\ell,d}\leq2^d\ell^{d-1}/(d-1)!$.  Also quote
\begin{equation}
 \lambda_{\ell,d}=\ell(\ell+d-1)\geq\ell^2.
\end{equation}
Substitution into \eqref{eqn:pre_tail_bound} gives
\begin{equation}
 \mathcal H_d(K,t)^2
 \leq\frac{2^d}{(d-1)!}
 \sum_{\ell>K}\ell^{d-1}e^{-2t\ell^2}.
 \label{eq:heat-tail-before-integral}
\end{equation}
The derivative of $x^{d-1}e^{-2tx^2}$ has the sign of
$(d-1)-4tx^2$.  Thus the condition
$K\geq\sqrt{(d-1)/(2t)}$ is more than sufficient to make it decreasing for
$x\geq K$.  For $\ell>K$, split the exponential as
\begin{equation}
 e^{-2t\ell^2}
 =e^{-t\ell^2}e^{-t\ell^2}
 \leq e^{-tK^2}e^{-t\ell^2}.
\end{equation}
By an integral bound for the series then gives,
\begin{align}
    \sum_{\ell>K}\ell^{d-1}e^{-2t\ell^2} &\leq e^{-tK^2} \int_0^\infty x^{d-1}e^{-tx^2}\dd x \\
    &= \frac12t^{-d/2}\Gamma(d/2)e^{-tK^2}.
\end{align}
Inserting this into \eqref{eq:heat-tail-before-integral}, the coefficient is
\begin{equation}
 \frac{2^d}{(d-1)!}\cdot\frac12\Gamma(d/2)
 =A_d^2.
\end{equation}
Taking square roots gives
\begin{align*}\mathcal H_d(K,t)\leq \sqrt{\frac{2^d}{(d-1)!}\cdot\frac12\Gamma(d/2)}t^{-d/4}e^{-tK^2/2},
\end{align*} as desired.
\end{proof}

The following corollary using the above to show that the projected heat kernel smoothed function can be made arbitrarily close to the original function, while remaining band-limited. 

\begin{corollary}[Spherical Heat Kernel Smoother]
\label{cor:heat_kernel_smooth_proxy}
 Suppose $b\in L^1(0,T;\mathbb R)$, put
$B_1=\int_0^T|b(t)|\dd t>0$, and let
 $F:\mathbb{S}^d\to\mathbb{R}$ be bounded and geodesically
 $G$-Lipschitz. For $0<\epsilon<1$,
if $t_\epsilon=\frac{\epsilon^2}{72dG^2B_1^2}$ and 
\begin{align*}
    K_{\epsilon} = \Omega\left(dGB_1\log(dG\lVert F \rVert/\epsilon)/\epsilon\right),
\end{align*}
then
\begin{align*}
     B_1\norm{F-P_{K_\epsilon}F_{t_\epsilon}}_\infty \leq\frac\epsilon3.
\end{align*}
\end{corollary}
\begin{proof}
    By the triangle inequality,
    \begin{equation*}
     B_1\norm{F-P_{K_\epsilon}
     F_{t_\epsilon}}_\infty
     \leq B_1\norm{F-F_{t_\epsilon}}_\infty
     +B_1\norm{F_{t_\epsilon}-P_{K_\epsilon}
     F_{t_\epsilon}}_\infty.
    \end{equation*}
    From Eq.~\eqref{eq:heat-uniform-error} if we select $ t_\epsilon\coloneqq \frac{\epsilon^2}{72dG^2B_1^2}$
    \begin{equation*}
        B_1\norm{F-F_{t_\epsilon}}_\infty
        \leq B_1G\sqrt{2dt_\epsilon} \leq \epsilon/6.
    \end{equation*}
    For the truncation term, we use Equation \eqref{eq:heat-exact-tail}
    \begin{align*}
         B_1\norm{F_{t_\epsilon}-P_{K_\epsilon}F_{t_\epsilon}}_\infty \leq \lVert F \rVert B_1 A_dt_\epsilon^{-d/4}
     e^{-t_\epsilon(K_\epsilon)^2/2}.
    \end{align*}
    Thus take 
    \begin{align*}
        \zeta_\epsilon&\coloneqq \frac{\epsilon}{6B_1\max\{\norm F_2,\epsilon\}},\quad 
        A_d =\sqrt{\frac{2^d}{(d-1)!}\cdot\frac12\Gamma(d/2)},\\
        K_\epsilon &= \left\lceil\max\left\{
        d,\sqrt{\frac{d-1}{2t_\epsilon}},
        \sqrt{\frac{2}{t_\epsilon}
        \log^+\!\left(\frac{A_dt_\epsilon^{-d/4}}
        {\zeta_\epsilon}\right)}
        \right\}\right\rceil,
    \end{align*}
    and so 
    \begin{align*}
         B_1\norm{F_{t_\epsilon}-P_{K_\epsilon^\mathrm{heat}} F_{t_\epsilon}}_\infty \leq \epsilon/6.
    \end{align*}
\end{proof}

With the first two conditions (slow spectral expansion and the smoothing operation) of the oversampling method in place, we are now ready to tackle the last one. Next, we present our choice of spherical quadrature along with a sufficient implementation of the quantum spherical harmonic transform, showing that it satisfies the transform condition. An important property enabling the construction of a sufficient quantum spherical harmonic transform is the recursive structure of spherical coordinates. Specifically, 
\begin{equation}
 x=(\cos\theta_j,\sin\theta_j\,\omega),
 \qquad \omega\in\mathbb S^{j-1},
 \qquad 2\leq j\leq d,
 \label{eq:sphere-recursive-coordinates}
\end{equation}
with $z_j=\cos\theta_j$. Let $\dd S_j$ denote ordinary Riemannian area measure on $\mathbb S^j$. It has the recursive decomposition
\begin{equation}
 \dd S_j=(1-z_j^2)^{(j-2)/2}\dd z_j\,\dd S_{j-1},
 \qquad \dd S_1=\dd\varphi,
 \label{eq:sphere-recursive-measure}
\end{equation}

Thus the recursive structure plus coordinate change transforms ordinary spherical area into the product measure
\begin{align*}
    \dd S_d &=\dd\varphi \prod_{j=2}^d(1-z_j^2)^{(j-2)/2}\dd z_j.
\end{align*}
This coordinate transform is exactly the $p=2$ Duffy transform mentioned earlier. The recursive coordinate map accepts every independently
chosen tuple
\begin{equation}
 (z_d,\ldots,z_2,\varphi)\in(-1,1)^{d-1}\times[0,2\pi)
 \label{eq:sphere-product-coordinate-domain}
\end{equation}
and produces a point of $\mathbb S^d$. This follows from induction on
\begin{equation*}
    (z_j,\omega)\longmapsto
    (z_j,\sqrt{1-z_j^2}\,\omega),
    \qquad \omega\in\mathbb S^{j-1}.
\end{equation*}
Use the following explicit chain of harmonic labels:
\begin{equation}
 \ell_d\geq\ell_{d-1}\geq\cdots\geq\ell_2\geq|m|,
 \qquad \ell_1\coloneqq |m|,
 \qquad \ell_d\leq N.
 \label{eq:sphere-chain-labels}
\end{equation}
Define the  increment labels
\begin{equation}
 n_j=\ell_j-\ell_{j-1}\geq0,
 \qquad
 \ell_j=|m|+\sum_{k=2}^jn_k,
 \qquad
 |m|+\sum_{k=2}^dn_k\leq N.
 \label{eq:sphere-increment-labels}
\end{equation}
In these labels, a
normalized harmonic factors as
\begin{equation}
 Y_{\mathbf n,m}(z_d,\ldots,z_2,\varphi)
 =\frac{e^{im\varphi}}{\sqrt{2\pi}}
 \prod_{j=2}^d\phi_{n_j,\ell_{j-1}}^{(j)}(z_j),
 \qquad \mathbf n=(n_2,\ldots,n_d),
 \label{eq:sphere-separated-harmonic}
\end{equation}
where $e^{im\varphi}/\sqrt{2\pi}$ has unit norm in
$L^2([0,2\pi),\dd\varphi)$. Note that the factor at level
$j$ is controlled only by lower labels
$(m,n_2,\ldots,n_{j-1})$.
At level $j$, after fixing the child degree $h=\ell_{j-1}$, the normalized
one-variable factor has the form
\begin{equation}
 \phi_{n,h}^{(j)}(z)
 =c_{n,h,j}(1-z^2)^{h/2}
 C_n^{h+(j-1)/2}(z),
 \qquad n=\ell_j-h,
 \label{eq:sphere-chain-factor}
\end{equation}
where $C_n^\lambda$ is a Gegenbauer polynomial.

Fix a required quadrature exactness degree $Q$. Thus the rule must
reproduce inner products on $\mathcal V_Q$, equivalently scalar
quadrature for spherical harmonics through degree $2Q$. Set
\begin{equation}
 R\coloneqq Q+1,
 \qquad
 M_\varphi\coloneqq 2Q+1.
 \label{eq:sphere-exactness-node-counts}
\end{equation}
To utilize Corollary \ref{cor:oversampled-quadrature-error} and remove
the aliasing error, we require $Q\geq N+K$. For each
$2\leq j\leq d$, choose the $R$-point Gauss--Jacobi rule
\begin{equation}
 \{(z_{j,r},w_{j,r}):0\leq r<R\}
 \label{eq:sphere-fixed-Jacobi-nodes}.
\end{equation}
These nodes depend on $j$ and $R$, but not on a harmonic label $\ell$. The Gauss-Jacobi rule applies to the usual Jacob weight function:
\begin{equation*}
 \int_{-1}^1(1-z^2)^{(j-2)/2}p(z)\dd z
 =\sum_{r=0}^{R-1}w_{j,r}p(z_{j,r}),
 \qquad \deg p\leq2R-1.
\end{equation*}
For the final angular variable, use the centered labels
$s\in\{-Q,\ldots,Q\}$ and uniform nodes
\begin{equation*}
 \varphi_s\coloneqq \frac{2\pi s}{M_\varphi}\pmod{2\pi},
 \qquad \varphi_s\in[0,2\pi).
\end{equation*}
They satisfy
\begin{equation}
 \frac1{M_\varphi}\sum_{s=-Q}^{Q}e^{im\varphi_s}
 =\begin{cases}
  1,&m\equiv0\pmod{M_\varphi},\\
  0,&m\not\equiv0\pmod{M_\varphi}.
 \end{cases}
 \label{eq:sphere-azimuthal-discrete-orthogonality}
\end{equation}
Thus this discrete average equals
$(2\pi)^{-1}\int_0^{2\pi}e^{im\varphi}\dd\varphi$ for every Fourier mode
with $|m|\leq2Q$.

For every
$h\in\{0,\ldots,N\}$, define the isometry
\begin{equation}
 A_{j,h}(r,n)=\sqrt{w_{j,r}}\,
 \phi_{n,h}^{(j)}(z_{j,r}),
 \qquad 0\leq n\leq N-h.
 \label{eq:sphere-local-isometry}
\end{equation}
The matrix has $R$ rows and $N-h+1$ active columns. Classically pad it to an
$R\times R$ unitary $\widetilde A_{j,h}$ by choosing any orthonormal
completion. The entries of $A_{j,h}$ are computed as follows.  Set
\begin{equation}
 \lambda=h+\frac{j-1}{2}.
\end{equation}
Starting from $C_0^\lambda(z)=1$ and
$C_1^\lambda(z)=2\lambda z$, apply the recurrence
\begin{equation}
 (n+1)C_{n+1}^\lambda(z)
 =2(n+\lambda)zC_n^\lambda(z)
 -(n+2\lambda-1)C_{n-1}^\lambda(z).
 \label{eq:Gegenbauer-recurrence}
\end{equation}
The normalization formula is
\begin{equation}
 \int_{-1}^1(1-z^2)^{\lambda-1/2}
 [C_n^\lambda(z)]^2\dd z
 =\frac{\pi2^{1-2\lambda}\Gamma(n+2\lambda)}
 {n!(n+\lambda)\Gamma(\lambda)^2}.
 \label{eq:Gegenbauer-norm}
\end{equation}
Because $(j-2)/2+h=\lambda-1/2$, insert
\eqref{eq:Gegenbauer-norm} into the area measure
\eqref{eq:sphere-recursive-measure}. This gives
\begin{equation}
 c_{n,h,j}
 =\left[
 \frac{\pi2^{1-2\lambda}\Gamma(n+2\lambda)}
 {n!(n+\lambda)\Gamma(\lambda)^2}
 \right]^{-1/2}.
 \label{eq:sphere-factor-normalization}
\end{equation}

The aforementioned isometry therefore acts on a basis state as 
\begin{align}
 \ket{n_d,\ldots,n_2,m}
 \longmapsto{}&
 \sum_{r_d,\ldots,r_2,s}
 \left(\frac1{\sqrt{M_\varphi}}e^{im\varphi_s}
 \right)
 \prod_{j=2}^dA_{j,\ell_{j-1}}(r_j,n_j)
 \nonumber\\
 &\hspace{38mm}\times
 \ket{r_d,\ldots,r_2,s}.
 \label{eq:sphere-target-state-map}
\end{align}
By \eqref{eq:sphere-separated-harmonic}, the amplitude in
\eqref{eq:sphere-target-state-map} is exactly
\begin{equation}
 \sqrt{\frac{2\pi}{M_\varphi}\prod_{j=2}^dw_{j,r_j}}\,
 Y_{\mathbf n,m}(z_{d,r_d},\ldots,z_{2,r_2},\varphi_s).
 \label{eq:sphere-target-weighted-amplitude}
\end{equation}
For fixed $j$ and $h$, the product
$\phi_{n,h}^{(j)}\phi_{n',h}^{(j)}$ is a polynomial of degree at most
$2h+n+n'\leq2N\leq2Q\leq2R-1$. Hence the active columns of every
$A_{j,h}$ are orthonormal under the $R$-point rule, and the resulting
weighted evaluation map is an isometry on $\mathcal V_N$.

The sequential composition of the controlled maps
\begin{equation}
    \widehat{A}_{j,h} : \ket h\ket n \mapsto
    \ket h\sum_{r=0}^{R-1}A_{j,h}(r,n)\ket r,
    \label{eq:sphere-local-transform}
\end{equation}
together with the $M_\varphi$-point QFT on the azimuthal register, realizes the
desired isometry (\eqref{eq:sphere-target-state-map} extended by linearity) on the fixed product grid
\begin{equation}
    \mathcal G_{R,M_\varphi}^{\mathbb S^d} = \left(\prod_{j=2}^d\{z_{j,r}\}_{r\in[0,R)}\right) \times \{\varphi_s\}_{s \in[-Q,Q]}.
    \label{eq:sphere-product-grid}
\end{equation}

The grid has $J_Q=R^{d-1}M_\varphi$ nodes with product weights
\begin{equation}
    \omega_{\mathbf r,s} \coloneqq \frac{2\pi}{M_\varphi} \prod_{j=2}^d w_{j,r_j}.
    \label{eq:sphere-product-grid-weights}
\end{equation}

We use the notation $W_{\mathbb S,N,Q}$ to denote the desired isometry. Specifically for a general input
\begin{equation}
 \ket{\Psi_\mathrm{spec}}
 =\sum_{\mathbf n,m}c_{\mathbf n,m}
 \ket{n_d,\ldots,n_2,m},
 \label{eq:sphere-general-input-state}
 \end{equation}

\begin{equation}
 W_{\mathbb S,N,Q}\ket{\Psi_\mathrm{spec}}=\sum_{r_d,\ldots,r_2,s}
 \sqrt{\frac1{M_\varphi}\prod_{j=2}^dw_{j,r_j}}\,
 \Psi(z_{d,r_d},\ldots,z_{2,r_2},\varphi_s)
 \ket{r_d,\ldots,r_2,s},
 \label{eq:sphere-final-sample-state}
\end{equation}
where $\Psi=\sum c_{\mathbf n,m}Y_{\mathbf n,m}$. This maps basis states to spherical harmonics and hence the inverse computes the spherical harmonic transform.

For the transform-complexity statements below, the quadrature nodes, weights, normalized polynomial values, and orthogonal completions are supplied by exact-real classical preprocessing. The stated transform error therefore refers to quantum circuit synthesis. We also require the following classical preprocessing to construct the $\widehat{A}_{j,h}$ maps. For each pair $(j,h)$ with $2\leq j\leq d$ and $0\leq h\leq N$, perform the following one-time classical preprocessing for the declared values of $(d,N,Q,\epsilon_\mathrm{tr})$.
\begin{enumerate}
    \item Form the $R \times R$ tridiagonal Jacobi matrix for the fixed base
    weight $(1-z^2)^{(j-2)/2}$.  Diagonalize it to obtain the common nodes
    $z_{j,r}$ and positive weights $w_{j,r}$.
    \item Evaluate the normalized factors
    $\phi_{n,h}^{(j)}(z_{j,r})$ for $0\leq n\leq N-h$.  Stable three-term
    evaluation uses \eqref{eq:Gegenbauer-recurrence}; normalization uses
    \eqref{eq:sphere-factor-normalization}.  Assemble the active matrix
    $W_{j,h}$ in \eqref{eq:sphere-local-isometry}.
    \item Verify, to the working classical precision, that
    $A_{j,h}^{\dagger}A_{j,h}=I$.  Complete its columns deterministically to an
    $R\times R$ orthogonal matrix $\widetilde A_{j,h}$, for example by a QR
    completion seeded by the unused computational basis vectors.
    \item Apply QR elimination to $\widetilde A_{j,h}$ and record at most
    $R(R-1)/2$ Givens rotations, each specified by two row indices and one
    angle, together with the residual diagonal signs.  Round every angle to
    $O(\log(d(N+1)R^2/\epsilon_\mathrm{tr}))$ bits.
    \item Store the node and weight tables for each $j$, and store or provide a
    uniform classical circuit that emits the Givens list controlled on $(j,h)$.
    The latter is required for a uniform quantum circuit family.
\end{enumerate}
Tridiagonal eigensolution costs $O(R^2)$ per level, factor evaluation costs $O(NR)$ per controlled block, and dense completion plus Givens elimination
costs $O(R^3)$ per controlled block. Hence the conservative total is $O(d(N+1)R^3)$ arithmetic operations and $O(d(N+1)R^2)$ stored rotation data.

\begin{algorithm}[htbp]
\caption{Inverse Quantum Spherical Harmonic Transform}
\label{alg:sphere-fixed-quadrature-transform}
\begin{minipage}{\linewidth}
    \textbf{Input:} A state in the chain-label basis
    $\ket{n_d,\ldots,n_2,m}$ satisfying \eqref{eq:sphere-increment-labels}, the fixed
    quadrature data \eqref{eq:sphere-fixed-Jacobi-nodes}, and target transform
    error $\epsilon_\mathrm{tr}$.
    
    \begin{enumerate}
    \item Reversibly compute the initial lower degree
    \begin{equation*}
    H=\ell_{d-1}=|m|+\sum_{k=2}^{d-1}n_k
    \end{equation*}
    into a work register. For $j=d,d-1,\ldots,2$, the invariant at the start
    of iteration $j$ is $H=\ell_{j-1}$. Controlled on $H$, apply
    $\widetilde A_{j,H}$ to overwrite $n_j$ by the fixed node index $r_j$.
    If $j>2$, update $H$ reversibly to
    $H-n_{j-1}=\ell_{j-2}$ before $n_{j-1}$ is transformed. After level
    $2$, subtract $|m|$ to clear $H$. Thus each lower label is used while it
    is still spectral and no degree garbage remains.
    \item After level $2$, only the azimuthal spectral label $m$ remains.
    Apply a centered inverse $M_\varphi$-point QFT synthesized to
    operator-norm error at most $\epsilon_\mathrm{tr}/2$,
    \begin{equation*}
    \ket m\longmapsto\frac1{\sqrt{M_\varphi}}
    \sum_{s=-Q}^{Q}e^{im\varphi_s}\ket s.
    \end{equation*}
    With an ideal QFT the state is exactly
    \eqref{eq:sphere-target-state-map}; the synthesized circuit is within
    the allocated error. Every spectral register has been overwritten by a
    node register and no label garbage remains.
    \end{enumerate}
    \textbf{Output:} The weighted sample state whose amplitudes are the spherical
    harmonic expansion evaluated on \eqref{eq:sphere-product-grid}.
\end{minipage}
\end{algorithm}

The centered azimuthal transform is the QFT over
$\mathbb Z_{M_\varphi}$ under signed labels. Since
$M_\varphi=2Q+1$ need not be a power of two, its logical register is
embedded in $\lceil\log_2M_\varphi\rceil$ qubits and the transform is
extended unitarily on unused basis states. An arbitrary-modulus QFT
compiled to error $\epsilon_\mathrm{tr}/2$ uses
$\operatorname{poly}(\log M_\varphi,\log(1/\epsilon_\mathrm{tr}))$
elementary gates, which is included in the transform-error budget below.

To evaluate a function on $\mathbb{S}^{d}$ in rectangular coordinates, we can use the following mapping (Duffy map) from the rectangular grid to spherical grid: reversibly evaluate the associated node coordinates and product
 weight when a Cartesian point oracle or a potential-value oracle requires
 them.  Starting with $p=1$, set recursively
 \begin{align*}
  x_1&=z_{d,r_d},\\
  x_2&=\sqrt{1-z_{d,r_d}^2}\,z_{d-1,r_{d-1}},
 \end{align*}
 and continue through the polar coordinates, ending with the two azimuthal
 components containing $\cos\varphi_s$ and $\sin\varphi_s$.  Multiply the
 stored one-dimensional weights to obtain
 $(2\pi/M_\varphi)\prod_jw_{j,r_j}$. Query the potential and uncompute all
 coordinate arithmetic afterward.

One important qualification is that Algorithm \ref{alg:sphere-fixed-quadrature-transform} is not quantumly efficient in the usual sense. Specifically, the dependence on the number of grid points in a single radial coordinate $R$ is not logarithmic. However, the discretization theorem keeps $R$ polynomial in the stated problem parameters and inverse accuracy; in particular, $R=\poly(d,1/\epsilon)$ when the remaining parameters are held fixed.

\begin{theorem}[Quantum spherical transform]
\label{thm:constructive-sphere-complexity}
    Let $d\geq2$ and let $N,Q\in\mathbb \mathbb N$ satisfy $Q\geq N$. Set $R=Q+1$ and $M_\varphi=2Q+1$. Choose $0<\epsilon_\mathrm{tr}<1$. Let $W_{\mathbb S,N,Q}$ be the isometry defined in \eqref{eq:sphere-final-sample-state}. Then there is a procedure, using classical  $O(d(N+1)R^3)$ arithmetic operations and $O(d(N+1)R^2)$ classical memory, to construct unitary quantum circuit $\widetilde U_W^{\mathbb S}$ satisfying
    \begin{align} \label{eq:sphere-transform-dilation-error}
        \norm{\widetilde U_W^{\mathbb S}-U_W^{\mathbb S}}\leq\epsilon_\mathrm{tr},
    \end{align}
    where 
    \begin{align}
        U_W^{\mathbb S}\bigl(\ket0_{\mathsf R}\ket\psi_{\mathsf C}\bigr)
        &=\bigl(W_{\mathbb S,N,Q}\ket\psi\bigr)_{\mathsf R}
        \ket0_{\mathsf C}.
    \label{eq:sphere-transform-dilation-contract}
    \end{align}
    The unitary $\widetilde U_W^{\mathbb S}$ uses at most
    \begin{equation*}
        O\left(d(N+1)R^2\operatorname{polylog} \frac{d(N+1)R}{\epsilon_\mathrm{tr}} + \operatorname{polylog}\frac{M_\varphi}{\epsilon_\mathrm{tr}}
        \right)
    \end{equation*}
    one- and two-qubit gates and
    \begin{equation}
        O(d\log R+\log M_\varphi + \operatorname{polylog}(d(N+1)R/\epsilon_\mathrm{tr}))
        \label{eq:constructive-sphere-qubits}
    \end{equation}
    qubits.
\end{theorem}
\begin{proof}
    The case $Q = 0$ is trivial, so henceforth assume it is positive.
    For row indices $p<q$, consider the Givens rotation $G_{pq}(\theta)$ which acts as the rotation
    \begin{align}
     G_{pq}(\theta) =
    \begin{pmatrix}
        \cos\theta&-\sin\theta\\
        \sin\theta&\cos\theta
    \end{pmatrix} = R_y(2\theta),
    \end{align}
    in the $(\ket{p}, \ket{q})$ basis, and fixes other basis vectors.  We first describe how this abstract
     two-level operation is synthesized on qubits.  Put
    \begin{equation}
     n_R=\lceil\log_2R\rceil
    \end{equation}
     and embed the $R$-dimensional matrix as
    \begin{equation}
     \widehat W_{j,h}
     =\widetilde W_{j,h}\oplus I_{2^{n_R}-R}
     \label{eq:Givens-binary-padding}
    \end{equation}
     on an $n_R$-qubit binary index register.  Write $p$ and $q$ as binary
     strings, set $a=p\mathbin{\mathtt{xor}}q$, and choose a bit $k$ for which
    $a_k=1$.  Let $P_{pq}$ be the reversible affine permutation obtained by
     first applying $X$ on every bit where $p$ is one and then, for every
    $i\ne k$ with $a_i=1$, applying a CNOT from bit $k$ to bit $i$.  It obeys
    \begin{equation}
     P_{pq}\ket p=\ket{0^{n_R}},
     \qquad
     P_{pq}\ket q=\ket{e_k},
     \label{eq:Givens-address-permutation}
    \end{equation}
     where $e_k$ is the string whose only nonzero bit is $k$.  Therefore
    \begin{equation}
        G_{pq}(\theta) = P_{pq}^\dagger \left(\Lambda_{r_i=0,\ i\ne k}R_{y,k}(2\theta)\right)P_{pq}.
        \label{eq:Givens-qubit-synthesis}
    \end{equation}
    where $R_y,k$ is $R_y$ on qubit $k$. The middle gate rotates qubit $k$ only when all other index bits are zero. The permutation is bijective, so every other binary address, including every padded address at least $R$, is fixed after the conjugation.

    To apply this rotation only in the block indexed by $h_0$, compute the predicate $[h=h_0]$ and add it as a control on the middle multi-controlled rotation in \eqref{eq:Givens-qubit-synthesis}. The surrounding permutation need not be controlled by $P_{pq}^\dagger P_{pq}=I$. If
    \begin{equation}
        n_h=\lceil\log_2(N+1)\rceil,
    \end{equation}
    the middle rotation has at most $n_R-1+n_h$ controls.  Open controls are
    converted to ordinary controls with $X$ gates.  With
    $O(n_R+n_h)$ clean work qubits, their conjunction can be computed and
    uncomputed using $O(n_R+n_h)$ Toffoli gates, between which one applies a
    singly controlled $R_y(2\theta)$.  The latter has a constant-size
    decomposition into CNOTs and single-qubit rotations.  Consequently, one
    $h$-controlled two-level rotation costs
    \begin{equation}
        O\left(\log R+\log(N+1)+\log\frac1\delta\right)
    \label{eq:Givens-elementary-gate-cost}
    \end{equation}
    Clifford, Toffoli, and synthesized single-qubit gates at operator error
    $\delta$.  Standard ancilla-free decompositions give a quadratic
    polylogarithmic overhead instead and do not change the stated asymptotic
    bound.

    QR elimination reduces one $R\times R$ orthogonal matrix using at most $R(R-1)/2$ such rotations and leaves a diagonal matrix $D=\diag(\pm1)$.  Reconstructing the matrix uses the inverse
    rotations in reverse elimination order, followed by the recorded sign
    correction.  Each negative entry of $D$ is an addressed phase flip and has
    the same polylogarithmic binary-address cost; all $O(R)$ possible sign
    corrections are lower order than the $O(R^2)$ rotations.  This also covers
    an orthogonal completion of determinant $-1$, which cannot be represented
    by determinant-one Givens rotations alone.
    
    There are $d-1$ angular levels and at most $N+1$ control values $h$ per
    level.  Each padded $R$-dimensional unitary uses $O(R^2)$ Givens rotations.
    Multiplying the three counts gives
    \begin{equation}
     L_\mathrm{rot}\leq(d-1)(N+1)R(R-1)/2=O(d(N+1)R^2).
    \end{equation}
    For exact factors $U_k$ and synthesized factors $\widetilde U_k$, quote the
    telescoping product inequality
    \begin{equation}
     \norm{U_L\cdots U_1-\widetilde U_L\cdots\widetilde U_1}
     \leq\sum_{k=1}^L\norm{U_k-\widetilde U_k}.
    \end{equation}
    Use it by setting each local error to
    $\delta=\epsilon_\mathrm{tr}/(2L_\mathrm{rot})$. The total error from the
    controlled radial blocks is then at most $\epsilon_\mathrm{tr}/2$.
    Computing the initial degree accumulator and updating it once per level
    costs $O(d\log(N+1))$ reversible arithmetic gates. Synthesize the centered
    QFT to error at most $\epsilon_\mathrm{tr}/2$ using
    $\operatorname{polylog}(M_\varphi/\epsilon_\mathrm{tr})$ gates. The
    triangle inequality gives total transform error at most
    $\epsilon_\mathrm{tr}$. Substituting the value of $\delta$ into
    \eqref{eq:Givens-elementary-gate-cost} gives the displayed gate bound.
    Binary node registers, the degree accumulator, the control-conjunction
    workspace, and reversible arithmetic give the qubit bound. Dense
    orthogonal completion and decomposition cost at most $O(R^3)$ per
    controlled block classically.

It remains to relate the in-place transform to the
dilation convention in
\eqref{eq:global-evaluation-unitary-dilation}. Pad each coefficient-label
register to the corresponding node-register width. A reversible label
transfer $U_\mathrm{tr}^{\mathbb S}$ maps
\begin{equation*}
 \ket0_{\mathsf R}\ket{\mathbf n,m}_{\mathsf C}
 \longmapsto
 \ket{\mathbf n,m}_{\mathsf R}\ket0_{\mathsf C}
\end{equation*}
on the valid label subspace and is extended arbitrarily to a permutation
of all padded basis states. Applying the ideal in-place spherical
transform on $\mathsf R$ after this transfer gives
$U_W^{\mathbb S}$ in
\eqref{eq:sphere-transform-dilation-contract}. Replacing the ideal
transform by the synthesized circuit gives
$\widetilde U_W^{\mathbb S}$; unitary invariance of the operator norm
and the transform estimate already proved imply
\eqref{eq:sphere-transform-dilation-error}. The transfer consists only of
reversible relabelling and swaps on
$O(d\log R+\log M_\varphi)$ qubits, so it is lower order than the stated
gate bound. 
\end{proof}

\begin{theorem}[Sphere Global Spectral Method Discretization]
\label{thm:spherical_harm_total_disc_error}
    Let $d\geq2$, $L_0\in\mathbb N$, $T>0$, and $\epsilon\in(0,1)$. Let $a,b\in L^1(0,T;\mathbb R)$, put
    $B_1\coloneqq \int_0^T|b(t)|\dd t$, and consider
    \begin{equation*}
     i\partial_t\Phi(t)
     =[-a(t)\Delta_{\mathbb S^d}+b(t)f]\Phi(t),
     \qquad
     \Phi(0)\in\mathcal V_{L_0},
     \qquad
     \norm{\Phi(0)}=1.
    \end{equation*}
    Suppose $f:\mathbb S^d\to\mathbb R$ is bounded and geodesically
    $G$-Lipschitz. Fix a bounded, nonzero, geodesically
    $G_h$-Lipschitz observable $h:\mathbb S^d\to\mathbb R$.
    
    Then there exist spectral and quadrature exactness degrees
    $N,Q\in\mathbb Z_{\geq0}$, each bounded by
    \begin{equation*}
     \poly(d,L_0,B_1,G,G_h,\norm f_\infty,\norm h_\infty,
     1/\epsilon,1/\norm h_\infty),
    \end{equation*}
    with $Q\geq N$. Set $R\coloneqq Q+1$ and $M_\varphi\coloneqq 2Q+1$, let
    $W_{\mathbb S,N,Q}$ be the weighted evaluation isometry on
    $\mathcal G_{R,M_\varphi}^{\mathbb S^d}$, and put
    \begin{equation*}
     \widehat f_Q\coloneqq \sum_{\mathbf r,s}f(x_{\mathbf r,s})
     \ketbra{\mathbf r,s}{\mathbf r,s}.
    \end{equation*}
    Let the exact finite-dimensional coefficient state solve
    \begin{equation}
     i\partial_t|c_{N,Q}(t)\rangle
     =\left[a(t)D_N+b(t)W_{\mathbb S,N,Q}^\dagger
     \widehat f_QW_{\mathbb S,N,Q}\right]|c_{N,Q}(t)\rangle,
     \qquad
     |c_{N,Q}(0)\rangle=S_N^\dagger\Phi(0),
     \label{eq:sphere-exact-coefficient-evolution}
    \end{equation}
    and set
    \begin{equation*}
     |\Psi_Q(t)\rangle\coloneqq W_{\mathbb S,N,Q}|c_{N,Q}(t)\rangle,
     \qquad
     \widehat h_Q\coloneqq \sum_{\mathbf r,s}h(x_{\mathbf r,s})
     \ketbra{\mathbf r,s}{\mathbf r,s}.
    \end{equation*}
    Then
    \begin{align}
     \sup_{0\leq t\leq T}
     \norm{S_N|c_{N,Q}(t)\rangle-\Phi(t)}
     &=\mathcal O(\epsilon),
     \label{eq:sphere-discretization-state-error}\\
     \left|
     \langle\Psi_Q(T)|\widehat h_Q|\Psi_Q(T)\rangle
     -\langle\Phi(T),h\Phi(T)\rangle
     \right|
     &=\mathcal O(\norm h_\infty\epsilon).
     \label{eq:sphere-discretization-observable-error}
    \end{align}
    The coefficient and quadrature representations require
    $\mathcal O(d\log(Q+1))$ data qubits.
\end{theorem}

\begin{proof}
Suppose first that $B_1>0$. The Fischer decomposition gives slow
spectral growth. Apply the heat-kernel smoothing construction to $f$ and
$h$ and choose a common $K$ such that real functions
$f_K,h_K\in\mathcal V_K$ satisfy
\begin{equation*}
 B_1\norm{f-f_K}_\infty=\mathcal O(\epsilon),
 \qquad
 \norm{h-h_K}_\infty
 =\mathcal O(\norm h_\infty\epsilon).
\end{equation*}
Choose
\begin{equation*}
 q\coloneqq \left\lceil\max\left\{
 2eB_1\norm{f_K}_\infty,
 3+\log_2(1/\epsilon)
 \right\}\right\rceil,
 \qquad
 N\coloneqq L_0+(q-1)K,
 \qquad
 Q\coloneqq N+K.
\end{equation*}
The heat-kernel estimates above give the asserted polynomial bounds on
$N$ and $Q$. The quadrature \eqref{eq:sphere-product-grid} $R=Q+1$ and
$M_\varphi=2Q+1$ is $Q$-exact, so Corollary
\ref{cor:oversampled-quadrature-error} gives both
\eqref{eq:sphere-discretization-state-error} and
\eqref{eq:sphere-discretization-observable-error}. Proposition
\ref{prop:quadrature-Hamiltonian-intertwining} identifies its quadrature
evolution with \eqref{eq:sphere-exact-coefficient-evolution}.

If $B_1=0$, the kinetic evolution remains in $\mathcal V_{L_0}$. Smooth
only $h$, obtaining $h_K\in\mathcal V_K$ with
$\norm{h-h_K}_\infty=\mathcal O(\norm h_\infty\epsilon)$, and take
$N=L_0$ and $Q=N+K$. The state error is zero. If
$v=S_Nc_{N,Q}(t)$, slow spectral growth gives
$h_Kv\in\mathcal V_{N+K}=\mathcal V_Q$, while
$v\in\mathcal V_N\subseteq\mathcal V_Q$. Thus $Q$-exactness of the
discrete inner product makes the discrete and continuous $h_K$
expectations identical. Inserting $h_K$ on both sides proves
\eqref{eq:sphere-discretization-observable-error}. The chain-label and
product-grid registers use $\mathcal O(d\log(Q+1))$ qubits.
\end{proof}

\begin{theorem}[Digital interaction-picture simulation on the sphere]
\label{thm:spherical-digital-simulation}
Assume the hypotheses of
Theorem~\ref{thm:spherical_harm_total_disc_error} with target accuracy
$0<\epsilon<1$, and let $N,Q$,
$|c_{N,Q}(t)\rangle$, $|\Psi_Q(t)\rangle$, and $\widehat h_Q$ be the
objects furnished there. Assume additionally that $a$ and $b$ are
piecewise $C^1$ and that $b$ has finitely many sign intervals. Put
$B_1\coloneqq \int_0^T|b(t)|\dd t$, and let
$\Lambda_f\geq\max\{1,\norm f_\infty\}$.

Assume the following coherent access for a sufficiently small universal
constant $c_0>0$. An initial-state circuit using
$\mathcal C_\mathrm{prep}$ gates and $\mathcal Q_\mathrm{prep}$ work qubits
prepares a normalized state
$|\widetilde c_0\rangle$ satisfying
\begin{equation*}
 \norm{|\widetilde c_0\rangle-|c_{N,Q}(0)\rangle}
 \leq\epsilon_\mathrm{prep}\leq c_0\epsilon.
\end{equation*}
The quadrature data and compiled spherical transforms of
Theorem~\ref{thm:constructive-sphere-complexity} are available. At any
spherical quadrature node $x_{\mathbf r,s}$, a binary oracle $O_f$ and
its inverse evaluate a number $\widetilde f(x_{\mathbf r,s})$ satisfying
\begin{equation*}
 |\widetilde f(x_{\mathbf r,s})-f(x_{\mathbf r,s})|
 \leq\epsilon_f,
 \qquad 0<\epsilon_f\leq\Lambda_f,
 \qquad
 B_1\epsilon_f\leq c_0\epsilon,
\end{equation*}
at a cost of $\mathcal C_f$ gates per call, including reversible
evaluation and uncomputation of the spherical node coordinates. Finally,
for every requested
$0<\delta<1$, controlled access to the schedules, their required
integrals and inverse time reparameterization, and the spherical kinetic
phases is available to operator error $\delta$ at a cost
$\polylog(d,N,1/\delta)$ gates per use.

Then there is a quantum circuit producing a normalized coefficient state
$|\widetilde c_{N,Q}(T)\rangle$, with success probability at least
$1-\mathcal O(\epsilon)$, such that
\begin{equation}
 \norm{S_N|\widetilde c_{N,Q}(T)\rangle-\Phi(T)}
 =\mathcal O(\epsilon).
 \label{eq:sphere-end-to-end-state-error}
\end{equation}
Let $R$ and $C$ denote the quadrature and coefficient registers. After
the simulation, append $|0\rangle_R$ and define the actual readout output
by
\begin{equation*}
 |\widetilde\Psi_Q(T)\rangle_{RC}
 \coloneqq \widetilde U_W^{\mathbb S}
 \bigl(|0\rangle_R|\widetilde c_{N,Q}(T)\rangle_C\bigr),
\end{equation*}
where $\widetilde U_W^{\mathbb S}$ is the implemented spherical-transform
dilation from Theorem~\ref{thm:constructive-sphere-complexity}. Define
\begin{equation*}
 \widehat h_Q
 \coloneqq \left(\sum_{\mathbf r,s}h(x_{\mathbf r,s})
 \ketbra{\mathbf r,s}{\mathbf r,s}_R\right)\otimes I_C,
\end{equation*}
with the quadrature-register operator set to zero on unused padded node
addresses.
Then
\begin{equation}
 \left|
 \langle\widetilde\Psi_Q(T)|\widehat h_Q
 |\widetilde\Psi_Q(T)\rangle
 -\langle\Phi(T),h\Phi(T)\rangle
 \right|
 =\mathcal O(\norm h_\infty\epsilon).
 \label{eq:sphere-end-to-end-observable-error}
\end{equation}
Let $\tau_f\coloneqq \Lambda_fB_1$. The circuit makes
\begin{equation}
 N_f=\mathcal O\!\left(
 (1+\tau_f)
 \frac{\log(\mathrm e^{\mathrm e}(1+\tau_f)/\epsilon)}
 {\log\log(\mathrm e^{\mathrm e}(1+\tau_f)/\epsilon)}
 \right)
 =\widetilde{\mathcal O}(1+\tau_f)
 \label{eq:sphere-binary-query-complexity}
\end{equation}
calls to $O_f$ or $O_f^\dagger$ and uses
\begin{align}
 \mathcal C_\mathrm{gates}
 ={}&\mathcal C_\mathrm{prep}
 +\widetilde{\mathcal O}\!\left((1+\tau_f)\left[
 d(N+1)(Q+1)^2
 +\mathcal C_f
 +\log^2\!\left(\frac{2\Lambda_f}{\epsilon_f}\right)
 \right]\right)
 \label{eq:sphere-end-to-end-gate-complexity}
\end{align}
elementary gates. Here $\widetilde{\mathcal O}$ suppresses logarithmic
factors in $d,N,Q,(1+\tau_f)/\epsilon$ and in the schedule
arithmetic.
The total qubit count is
\begin{equation}
 \mathcal Q_\mathrm{prep}+\mathcal O\!\left(
 d\log(Q+1)+\log\frac{2\Lambda_f}{\epsilon_f}
 +\polylog\frac{d(N+1)(Q+1)(1+\tau_f)}{\epsilon}
 \right).
 \label{eq:sphere-end-to-end-qubit-complexity}
\end{equation}
\end{theorem}

\begin{proof}
By Theorem~\ref{thm:spherical_harm_total_disc_error}, the target exact
coefficient evolution is \eqref{eq:sphere-exact-coefficient-evolution}.
Choose all spherical-transform, kinetic-phase, and schedule subroutines
to operator error at most
\begin{equation*}
 \delta=\Theta\!\left(
 \frac{\epsilon\,
 \log\log(\mathrm e^{\mathrm e}(1+\tau_f)/\epsilon)}
 {(1+\tau_f)\log(\mathrm e^{\mathrm e}(1+\tau_f)/\epsilon)}
 \right),
\end{equation*}
with a sufficiently small implied constant. Theorem
\ref{thm:constructive-sphere-complexity} then supplies a synthesized
dilation $\widetilde U_W^{\mathbb S}$ satisfying
$\norm{\widetilde U_W^{\mathbb S}-U_W^{\mathbb S}}\leq\delta$.
To encode the sampled potential,
query $O_f$, reversibly clamp the returned value to
$[-\Lambda_f,\Lambda_f]$, compute a single-qubit rotation whose clean
amplitude is $\widetilde f(x_{\mathbf r,s})/\Lambda_f$, and uncompute the
value register. This uses two calls to $O_f$ or $O_f^\dagger$ and
$\mathcal O(\log^2(2\Lambda_f/\epsilon_f)+\log^2(1/\delta))$ gates,
apart from the oracle cost. Conjugating this sampled-value block by the
implemented forward and inverse spherical transforms block encodes the
compressed potential
$W_{\mathbb S,N,Q}^\dagger\widehat f_QW_{\mathbb S,N,Q}$ with
unnormalized operator error at most
$\epsilon_f+\mathcal O(\Lambda_f\delta)$.

Move to the kinetic frame generated by $a(t)D_N$ and rescale time using
$\dd\sigma=\Lambda_f|b(t)|\dd t$. The rescaled evolution lasts for
$\tau_f=\Lambda_fB_1$ and has a norm-one interaction Hamiltonian. On
each sign interval of $b$, a time-controlled interaction-picture query
consists of the compressed-potential block surrounded by the controlled
kinetic phases; intervals on which $b=0$ are omitted. The access
assumptions supply the sign of $b$, the inverse time conversion, and the
controlled phases.

The rescaled truncated-Dyson interaction-picture algorithm
\cite{low2019interaction,berry2020time} simulates this evolution to
operator error $\mathcal O(\epsilon)$, with failure probability
$\mathcal O(\epsilon)$, using the number of time-controlled queries in
\eqref{eq:sphere-binary-query-complexity}. Returning to the
Schr\"odinger picture requires one final kinetic phase. Each
interaction-picture query contains one sampled-potential block and hence
two binary-oracle calls, one forward spherical transform, one inverse
spherical transform, and two controlled kinetic phases. This proves
\eqref{eq:sphere-binary-query-complexity}.

Regard the synthesized transform sandwich as an exact block encoding of
a nearby compressed Hamiltonian. Duhamel's formula contributes
$B_1\epsilon_f+\mathcal O(\tau_f\delta)$ from this Hamiltonian
perturbation, while the hybrid argument over the time-controlled queries
contributes another $\mathcal O(\epsilon)$ from kinetic-phase and
schedule synthesis. By the choice of $\delta$, the truncated-Dyson,
oracle, transform, kinetic, and schedule errors therefore sum to
$\mathcal O(\epsilon)$. Initial-state preparation adds at most
$\epsilon_\mathrm{prep}$ by unitarity, so
\begin{equation*}
 \norm{|\widetilde c_{N,Q}(T)\rangle-|c_{N,Q}(T)\rangle}
 =\mathcal O(\epsilon).
\end{equation*}
Combining this estimate with
\eqref{eq:sphere-discretization-state-error} proves
\eqref{eq:sphere-end-to-end-state-error}.

At readout, applying $\widetilde U_W^{\mathbb S}$ instead of
$U_W^{\mathbb S}$ adds at most $\delta$ in state norm. The ideal joint
readout state is
\begin{equation*}
 |\Psi_Q(T)\rangle_{RC}
 \coloneqq \bigl(W_{\mathbb S,N,Q}|c_{N,Q}(T)\rangle\bigr)_R|0\rangle_C.
\end{equation*}
Together with the coefficient estimate above, this gives
\begin{equation*}
 \norm{|\widetilde\Psi_Q(T)\rangle_{RC}
 -|\Psi_Q(T)\rangle_{RC}}
 =\mathcal O(\epsilon).
\end{equation*}
Then
\begin{equation*}
 |\langle v|\widehat h_Q|v\rangle
 -\langle w|\widehat h_Q|w\rangle|
 \leq2\norm h_\infty\norm{|v\rangle-|w\rangle}.
\end{equation*}
Combining this inequality with
\eqref{eq:sphere-discretization-observable-error} proves
\eqref{eq:sphere-end-to-end-observable-error}.

Finally, Theorem~\ref{thm:constructive-sphere-complexity} gives the
per-transform cost
$\widetilde{\mathcal O}(d(N+1)(Q+1)^2)$ at the chosen value of $\delta$.
The assumed schedule and kinetic-phase cost is polylogarithmic at the same
tolerance. Multiplying the per-query cost by
the query bound \eqref{eq:sphere-binary-query-complexity}, and adding
state preparation and final readout, gives
\eqref{eq:sphere-end-to-end-gate-complexity}. The transform registers,
binary value register, truncated-Dyson time and order registers, and
reusable schedule and kinetic workspace give
\eqref{eq:sphere-end-to-end-qubit-complexity}.
\end{proof}

\subsection{Simplex with Wright--Fisher metric}
\label{sec:global_spec_simplex}

Let
\begin{equation}
    \Simp_d=\left\{x\in\mathbb{R}_{\geq0}^d: \sum_{i=1}^dx_i\leq1\right\}, \qquad x_{d+1}=1-\sum_{i=1}^dx_i,
    \label{eq:simplex}
\end{equation}
be the $d$-dimensional simplex. Equip $\Simp_d$ with the Wright--Fisher metric, whose inverse metric coordinates $(x_1,\ldots,x_d)$ is
\begin{equation}
 g^{ij}(x)=x_i(\delta_{ij}-x_j).
 \label{eq:simplex-WF-metric}
\end{equation}
The corresponding Laplace--Beltrami operator is
\begin{equation} \label{eq:simplex-WF}
\Delta_{\Simp_d} = \sum_{i,j=1}^dx_i(\delta_{ij}-x_j)\partial_i\partial_j + \frac12\sum_{i=1}^d\left(1-(d+1)x_i\right)\partial_i.
\end{equation}
The eigenfunctions of $ \Delta_{\Simp_d}$ are the multidimensional Jacobi polynomials:
\begin{align}
    &\mathcal P_{\mathbf n}(x) = \prod_{i=1}^d \left[ s_i^{n_i} P_{n_i}^{(-1/2,\,A_{i+1}+2N_{i+1}-1)} \left(1-\frac{2x_i}{s_i}\right)\right]\label{eq:simplex-nested-Jacobi-barycentric-form}, \quad s_i\coloneqq \sum_{r=i}^{d+1} x_r\\
    & -\Delta_g\mathcal P_{\mathbf n} = |\mathbf n|\left(|\mathbf n|+\frac{d-1}{2}\right) \mathcal P_{\mathbf n}
\end{align}
where
\begin{equation}
 A_i\coloneqq \sum_{r=i}^{d+1}\frac12=\frac{d+2-i}{2},
 \qquad
 N_i\coloneqq\sum_{r=i}^{d}n_r,
 \qquad
 N_{d+1}\coloneqq0,
 \label{eq:simplex-suffix-sums}
\end{equation}
and $P_n^{(\alpha, \beta)}$ is the $n$-th Jacobi polynomial with parameters $\alpha$ and $\beta$.
For the simplex case, the labeling of the eigenfunctions is, like the sphere case, done according to the total degree $L(\mathcal{P}_{\mathbf{n}}) = \lVert n \rVert_1$. Hence $\mathcal{H}_{\ell}$ refers to space of $\mathcal{P}_{\mathbf{n}}$ satisfying $\lVert n \rVert_1 = \ell$.

\begin{lemma}
\label{lem:jacobi_inclusion_lemma}
Let $ \mathrm{Poly}_{\leq K}(\Simp_d)$ be the space of polynomials on the simplex of degree at most $K$. Then for all $K$, \begin{align*}
    \mathcal{V}_K =  \mathrm{Poly}_{\leq K}(\Simp_d).
\end{align*}
\end{lemma}
\begin{proof}

Consider the multivariable Jacobi polynomial in barycentric coordinates \eqref{eq:simplex-nested-Jacobi-barycentric-form}. A single one-variable Jacobi polynomial factor is a polynomial of
degree $n_i$ in $x_i/s_i$. Consequently,
each bracket in \eqref{eq:simplex-nested-Jacobi-barycentric-form} is a sum
of terms
\begin{equation}
 x_i^q s_i^{n_i-q},
 \qquad 0\leq q\leq n_i,
 \label{eq:simplex-Jacobi-factor-polynomial-terms}
\end{equation}
and is a barycentric polynomial of degree $n_i$.  Their product has total
degree at most $|\mathbf n|$.  In particular,
\begin{equation*}
 \lVert\mathbf n\rVert_1\leq K \quad\implies\quad \mathcal P_{\mathbf n}\in\mathrm{Poly}_{\leq K}(\Simp_d).
\end{equation*}
Taking spans gives the first, nontrivial inclusion
\begin{equation*}
 \mathrm{Jac}_{\leq K}(\Simp_d)
 \subseteq\mathrm{Poly}_{\leq K}(\Simp_d).
\end{equation*}

Conversely, the nonzero polynomials
$\{\mathcal P_{\mathbf n}: \lVert \mathbf n \rVert\leq K\}$ are mutually orthogonal in
$L^2(\Simp_d,\dvol)$ and hence linearly independent. Note that
\begin{align*}
    \#\{\mathbf n\in\mathbb N^d:\lVert\mathbf n\rVert_1\leq K\}
 =\binom{K+d}{d}
 =\dim\mathrm{Poly}_{\leq K}(\Simp_d).
\end{align*}
Hence
\begin{align*}
    \mathrm{Poly}_{\leq K}(\Simp_d)
 =\mathrm{Jac}_{\leq K}(\Simp_d).
\end{align*}

\end{proof}
The above makes it apparent that the Jacobi polynomials satisfy  the slow expansion property.

For the simplex Laplace--Beltrami operator in the Wright--Fisher metric, the appropriate smoothing operation is averaging the function over the multinomial distribution, also called Bernstein smoothing.
\begin{lemma}[Bernstein smoothing on the simplex]
\label{lem:simplex-Bernstein-smoothing}
Let $F:\Delta_d\to\mathbb{R}$ be bounded and $G$-Lipschitz in the Euclidean
distance on barycentric vectors. For an integer $K\geq1$, let
$Y\sim\operatorname{Multinomial}
 (K;x_1,\ldots,x_{d+1})$
and define
\begin{equation}
\label{eq:simplex-Bernstein}
 (\mathcal B_KF)(x)=\mathbb E\left[F(Y/K)\right], x \in \Delta_d
\end{equation}
Then the smoothed function satisfies  $\mathcal B_KF\in\mathcal V_K, \norm{\mathcal B_KF}_\infty
 \leq\norm F_\infty$, and
\begin{equation}
 \norm{F-\mathcal B_KF}_\infty
 \leq\frac{G}{\sqrt K}.
 \label{eq:simplex-Bernstein-error}
\end{equation}
\end{lemma}

\begin{proof}
For a multi-index
$\nu=(\nu_1,\ldots,\nu_{d+1})\in\mathbb N^{d+1}$ with
$|\nu|=K$, the multinomial mass function is
\begin{equation}
 p_{K,\nu}(x)
 \coloneqq\Pr(Y=\nu)
 =\frac{K!}{\nu_1!\cdots\nu_{d+1}!}
 \prod_{i=1}^{d+1}x_i^{\nu_i}.
 \label{eq:simplex-Bernstein-weights}
\end{equation}
Therefore the expectation in \eqref{eq:simplex-Bernstein} is
\begin{equation}
 (\mathcal B_KF)(x)
 =\sum_{|\nu|=K}F(\nu/K)p_{K,\nu}(x).
 \label{eq:simplex-Bernstein-sum}
\end{equation}

Taking absolute values in
\eqref{eq:simplex-Bernstein-sum} gives
\begin{equation}
 |\mathcal B_KF(x)|
 \leq\sum_{|\nu|=K}p_{K,\nu}(x)|F(\nu/K)|
 \leq\norm F_\infty.
\end{equation}
Taking the supremum over $x$ proves the desired infinity norm bound.

We next prove the approximation estimate.  Consider $x \in \Delta_d$. Let $\{ Z^{(j)}\}_{j=1}^{K}$ be independent and identically distributed as
\begin{equation*}
 \Pr(Z^{(j)}=e_i)=x_i.
\end{equation*}
Then define
\begin{equation*}
 Y=Z^{(1)}+\cdots+Z^{(K)},
\end{equation*}
Also,
\begin{equation*}
\mathbb{E} Z_i^{(r)}=x_i \implies  \mathbb{E} Y_i=Kx_i
\end{equation*}
In particular, $Y/K-x$ has mean zero and a sum of independent random variables, so
\begin{align}
 \mathbb E\norm{Y/K-x}_2^2
 &=\frac1{K^2}\sum_{i=1}^{d+1}\operatorname{Var}(Y_i)\nonumber\\
 &=\frac1K\sum_{i=1}^{d+1}(x_i-x_i^2)\nonumber\\
 &=\frac{1-\sum_{i=1}^{d+1}x_i^2}{K}.
 \label{eq:simplex-Bernstein-trace-calculation}
\end{align}
Cauchy--Schwarz applied to
$\sum_i x_i=1$ gives
\begin{equation*}
 1\leq(d+1)\sum_{i=1}^{d+1}x_i^2 \implies  1-\sum_{i=1}^{d+1}x_i^2\leq\frac{d}{d+1} < 1,
\end{equation*}
so
\begin{align*}
     \mathbb E\norm{Y/K-x}_2^2 \leq \frac{1}{K}.
\end{align*}
Finally, Lipschitzness in Euclidean norm and the above give
\begin{align*}
 |F(x)-\mathcal B_KF(x)|
 &\leq\mathbb E|F(x)-F(Y/K)|\nonumber\\
 &\leq G\mathbb E\norm{Y/K-x}_2\nonumber\\
 &\leq G\left(\mathbb E\norm{Y/K-x}_2^2\right)^{1/2}\\
 &\leq \frac{G}{\sqrt{K}}.
\end{align*}
Taking supremum leads to desired error bound.

It remains to verify $\mathcal B_KF\in\mathcal V_K$.  Every
$p_{K,\nu}$ in \eqref{eq:simplex-Bernstein-sum} is a homogeneous
barycentric monomial of degree $K$.
On the simplex, substitute
$x_{d+1}=1-\sum_{i=1}^dx_i$.  Expanding the final factor gives
\begin{equation*}
 x_1^{\nu_1}\cdots x_d^{\nu_d}
 \left(1-\sum_{i=1}^dx_i\right)^{\nu_{d+1}}.
\end{equation*}
Every term in this expansion has total degree between
$\sum_{i=1}^d\nu_i$ and
\begin{equation*}
 \sum_{i=1}^d\nu_i+\nu_{d+1}=K.
\end{equation*}
From the above calculation, we already know
\begin{equation*}
 p_{K,\nu}\in\mathrm{Poly}_{\leq K}(\Delta_d) \implies \mathcal{B}_KF(x) \in \mathrm{Poly}_{\leq K}(\Delta_d).
\end{equation*}
Thus with Lemma \ref{lem:jacobi_inclusion_lemma} $\mathcal{B}_KF \in \mathcal{V}_K$.
\end{proof}

Recall the Simplex Duffy map $\mathcal{D}_{\mathcal{S}_d}$ \eqref{eq:lp-spherical-duffy-map}, which takes $y \in [0, 1)^{d}$ to $x = \mathcal{D}_{\mathcal{S}_d}(y) \in \Simp_d$. The Jacobi polynomial transform as follows:
\begin{align*}
     &\widehat{\mathcal P}_{\mathbf n}(y) \coloneqq \mathcal{P}_{\mathbf{n}}\circ \mathcal{D}_{\mathcal{S}_d}(y)
 =\prod_{i=1}^d
 (1-y_i)^{N_{i+1}}P_{n_i}^{(-1/2,\,A_{i+1}+2N_{i+1}-1)}(1-2y_i).
\end{align*}
Similar to the case of the sphere in the previous section, the above provides a recursive factorization of the multi-dimensional eignefunctions. For each
$1\leq j\leq d$, choose the same $R$-point Gauss--Jacobi rule
\begin{equation}
 \{(y_{j,r},w_{j,r}):0\leq r<R\}
 \label{eq:simplex-fixed-Jacobi-nodes}.
\end{equation}
Fix a $Q$-exact quadrature \eqref{def:q_exact_quad}  set
\begin{equation}
 R\coloneqq Q+1,
 \qquad 2R-1\geq2Q,
 \label{eq:simplex-exactness-node-count}
\end{equation}
with $Q\geq N$.

Again following what was done in Section \ref{sec:sphere_standard_met} but with the Jacobi labels replacing the spherical harmonic labels.
For
$h=N_{i+1}$, define
\begin{equation}
 \psi_{n,h}^{(i)}(y)
 =c_{n,h,i}(1-y)^h
 P_n^{(-1/2,A_{i+1}+2h-1)}(1-2y),
 \label{eq:simplex-local-factor}
\end{equation}
and 
\begin{equation}
 A_{i,h}^\Delta(r,n)
 =\sqrt{w_{i,r}}\,\psi_{n,h}^{(i)}(y_{i,r}),
 \qquad 0\leq n\leq N-h.
 \label{eq:simplex-local-isometry}
\end{equation}
This matrix has $R$ rows and $N-h+1$ active columns.  Since
$R\geq N+1$, we complete the isometry to form a real
$R\times R$ orthogonal matrix $\widetilde A_{i,h}^\Delta$. The entries of $A_{i,h}^\Delta$  are computed by via the Jacobi recurrence relation.

 The 
cubature has $R^d$ nodes and after applying the Duffy map to go back to the simplex:
\begin{equation}
 \mathcal X_R^\Delta
 \coloneqq\left\{x_{\mathbf r}:0\leq r_i<R\right\}
 \subset\Simp_d
 \label{eq:simplex-physical-product-grid}
\end{equation}
and product weights
\begin{equation}
 w_{\mathbf r}=\prod_{i=1}^dw_{i,r_i}.
 \label{eq:simplex-product-grid-weight}
\end{equation}
In Duffy coordinates the grid is \begin{equation}
 \mathcal G_R^\Delta
 =\prod_{i=1}^d\{y_{i,r}:0\leq r<R\},
 \label{eq:simplex-fixed-product-grid}
\end{equation}

The transform has same preprocessing step as quantum spherical harmonic transform, i.e. compute the $A_{i,h}^{\Delta}$ and embed them in unitaries.
\begin{algorithm}[htbp]
\caption{Quantum Jacobi Transform over Simplex}
\label{alg:simplex-fixed-quadrature-transform}
\begin{minipage}{\linewidth}
\textbf{Input:} A total-degree label state
$\ket{n_1,\ldots,n_d}$ with $\sum_i n_i\leq N$, embedded in
$\lceil\log_2R\rceil$-qubit coordinate registers, the fixed quadrature data
\eqref{eq:simplex-fixed-Jacobi-nodes}, and target transform error
$\epsilon_\mathrm{tr}$.
\begin{enumerate}
 \item Reversibly compute the initial suffix
 \begin{equation*}
  H=N_2=\sum_{j=2}^dn_j
 \end{equation*}
 in a work register.  For $i=1,2,\ldots,d$, the invariant at the start of
 iteration $i$ is $H=N_{i+1}$.  Controlled on $H$, apply
 $\widetilde W_{i,H}^\Delta$ to overwrite $n_i$ by the common node index
 $r_i$.  If $i<d$, replace $H$ reversibly by
 $H-n_{i+1}=N_{i+2}$ before $n_{i+1}$ is transformed.  After iteration
 $d-1$ the accumulator is zero, so the final block has $h=0$ and no suffix
 garbage remains.  This increasing order is essential because every label
 subtracted from $H$ is still spectral at the moment it is used.
 \item When the potential oracle needs the
 physical point, initialize a remainder register $s=1$ and compute
 \begin{equation*}
  x_i=s\,y_{i,r_i},
  \qquad
  s\leftarrow s(1-y_{i,r_i}),
  \qquad 1\leq i\leq d.
 \end{equation*}
 At the end set $x_{d+1}=s$.  If required, multiply the stored
 one-dimensional weights to obtain $\prod_iw_{i,r_i}$.  Query the potential
 and uncompute the Duffy-coordinate and weight arithmetic afterward.
\end{enumerate}
\textbf{Output:} The weighted sample state on
$\mathcal G_R^\Delta$, equivalently on the physical grid
$\mathcal X_R^\Delta$.
\end{minipage}
\end{algorithm}

For a general input
\begin{equation}
 \ket{\Psi_\Delta}
 =\sum_{|\mathbf n|\leq N}c_{\mathbf n}
 \ket{n_1,\ldots,n_d},
\end{equation}
the composition of the above operations performs the inverse Jacobi transform $W_{\Simp_d,N,Q}$: 
\begin{equation}
 W_{\Simp_d,N,Q}|\Psi_{\Delta}\rangle = \sum_{r_1,\ldots,r_d}
 \sqrt{\prod_{i=1}^dw_{i,r_i}}\,
 \left[\sum_{|\mathbf n|\leq N}c_{\mathbf n}
 \prod_{i=1}^d\psi_{n_i,N_{i+1}}^{(i)}(y_i)\right]
 \ket{r_1,\ldots,r_d}.
 \label{eq:simplex-final-sample-state}
\end{equation}

\begin{theorem}[Quantum simplex transform]
\label{thm:constructive-simplex-complexity}
    Let $d\geq1$ and let $N,Q\in\mathbb N$ satisfy $Q\geq N$, and set
    $R=Q+1$. Choose $0<\epsilon_\mathrm{tr}<1$. Let
    $W_{\Delta,N,Q}$ be the isometry defined by
    \eqref{eq:simplex-final-sample-state}. Then there is a procedure using
    $O(d(N+1)R^3)$ classical arithmetic operations and
    $O(d(N+1)R^2)$ classical storage that constructs a unitary
    $\widetilde U_W^\Delta$ satisfying
    \begin{align}
     \norm{\widetilde U_W^\Delta-U_W^\Delta}
     &\leq\epsilon_\mathrm{tr},
     \label{eq:simplex-transform-dilation-error}
    \end{align}
    where
    \begin{align}
     U_W^\Delta\bigl(\ket0_{\mathsf R}\ket\psi_{\mathsf C}\bigr)
     &=\bigl(W_{\Delta,N,Q}\ket\psi\bigr)_{\mathsf R}
         \ket0_{\mathsf C}.
     \label{eq:simplex-transform-dilation-contract}
    \end{align}
    The unitary $\widetilde U_W^\Delta$ uses at most
    \begin{equation}
     O\left(
     d(N+1)R^2\operatorname{polylog}
     \frac{d(N+1)R}{\epsilon_\mathrm{tr}}
     \right)
     \label{eq:constructive-simplex-gates}
    \end{equation}
    one- and two-qubit gates and
    \begin{equation}
     O\left(
     d\log R+\operatorname{polylog}
     \frac{d(N+1)R}{\epsilon_\mathrm{tr}}
     \right)
     \label{eq:constructive-simplex-qubits}
    \end{equation}
    qubits.
\end{theorem}
\begin{proof}
The proof is basically the same as Theorem \ref{thm:constructive-sphere-complexity}, except the azimuthal/QFT part is excluded.

\end{proof}

\begin{theorem}[Simplex Global Spectral Method Discretization]
\label{thm:simplex_wright_fisher_total_disc_error}
Let $d\geq1$, $L_0\in\mathbb N$, $T>0$, and
$0<\epsilon<1$. Let $a,b\in L^1(0,T;\mathbb R)$, put
$B_1\coloneqq\int_0^T|b(t)|\dd t$, and consider
\begin{equation*}
 i\partial_t\Phi(t)=[-a(t)\Delta_g+b(t)f]\Phi(t),
 \qquad
 \Phi(0)\in\mathcal V_{L_0},
 \qquad
 \norm{\Phi(0)}=1.
\end{equation*}
Suppose $f:\Delta_d\to\mathbb R$ is bounded and $G$-Lipschitz in the
Euclidean distance on barycentric vectors. Fix a bounded, nonzero,
$G_h$-Lipschitz observable $h:\Delta_d\to\mathbb R$ in the same metric.

Then there exist spectral and quadrature exactness degrees
$N,Q\in\mathbb N$, each bounded by
\begin{equation*}
 \poly(d,L_0,B_1,G,G_h,\norm f_\infty,\norm h_\infty,
 1/\epsilon,1/\norm h_\infty),
\end{equation*}
with $Q\geq N$. Set $R\coloneqq Q+1$, let $W_{\Delta,N,Q}$ be the weighted
evaluation isometry on $\mathcal X_R^\Delta$, and put
\begin{equation*}
 \widehat f_Q\coloneqq\sum_{\mathbf r}f(x_{\mathbf r})
 \ketbra{\mathbf r}{\mathbf r}.
\end{equation*}
Let the exact finite-dimensional coefficient state solve
\begin{equation}
 i\partial_t|c_{N,Q}(t)\rangle
 =\left[a(t)D_N+b(t)W_{\Delta,N,Q}^\dagger
 \widehat f_QW_{\Delta,N,Q}\right]|c_{N,Q}(t)\rangle,
 \qquad
 |c_{N,Q}(0)\rangle=S_N^\dagger\Phi(0),
 \label{eq:simplex-exact-coefficient-evolution}
\end{equation}
and set
\begin{equation*}
 |\Psi_Q(t)\rangle\coloneqq W_{\Delta,N,Q}|c_{N,Q}(t)\rangle,
 \qquad
 \widehat h_Q\coloneqq\sum_{\mathbf r}h(x_{\mathbf r})
 \ketbra{\mathbf r}{\mathbf r}.
\end{equation*}
Then
\begin{align}
 \sup_{0\leq t\leq T}
 \norm{S_N|c_{N,Q}(t)\rangle-\Phi(t)}
 &=\mathcal O(\epsilon),
 \label{eq:simplex-discretization-state-error}\\
 \left|
 \langle\Psi_Q(T)|\widehat h_Q|\Psi_Q(T)\rangle
 -\langle\Phi(T),h\Phi(T)\rangle
 \right|
 &=\mathcal O(\norm h_\infty\epsilon).
 \label{eq:simplex-discretization-observable-error}
\end{align}
The coefficient and quadrature representations require
$\mathcal O(d\log(Q+1))$ data qubits.
\end{theorem}

\begin{proof}
As mentioned near the beggining of this section, the multivariate Jacobi
eigenfunctions satisfy the slow spectral growth property
\eqref{eqn:eigfunc_mul_exp_prop} due to Lemma \ref{lem:jacobi_inclusion_lemma}.  Define
\begin{equation*}
    f_K\coloneqq\mathcal B_Kf,
    \qquad
    h_K\coloneqq\mathcal B_Kh.
\end{equation*}
By Lemma~\ref{lem:simplex-Bernstein-smoothing},
$f_K,h_K\in\mathcal V_K$ and
\begin{equation*}
    \norm{f_K}_\infty\leq\norm f_\infty,
    \qquad
    \norm{f-f_K}_\infty\leq\frac{G}{\sqrt K},
    \qquad
    \norm{h-h_K}_\infty\leq\frac{G_h}{\sqrt K}.
\end{equation*}
Since the Bernstein operator is a positive contraction,
$\norm{h_K}_\infty\leq\norm h_\infty$.
Choose an integer $K$ sufficiently large that
\begin{equation*}
    \frac{B_1G}{\sqrt K}=\mathcal O(\epsilon),
    \qquad
    \frac{G_h}{\sqrt K}
        =\mathcal O(\norm h_\infty\epsilon),
\end{equation*}
and choose the integer
\begin{equation*}
    q\coloneqq\left\lceil
    \max\left\{
        2eB_1\norm{f_K}_\infty,\,
        3+\log_2(1/\epsilon)
    \right\}\right\rceil.
\end{equation*}
Set $N=L_0+(q-1)K$ and $Q=N+K$. As mentioned earlier, the simplex quadrature \eqref{eq:simplex-fixed-product-grid} is
$Q$-exact. Corollary~\ref{cor:oversampled-quadrature-error}, applied
with $f_K$ and $h_K$ as above, defines the continuum quadrature evolution
$\widetilde u_{N,Q}(t)\in\mathcal V_N$ and gives
\begin{align*}
    \sup_{0\leq t\leq T}
    \norm{\Phi(t)-\widetilde u_{N,Q}(t)}
    &\leq
    2B_1\norm{f-f_K}_\infty+2^{2-q}\\
    &\leq
    \frac{2B_1G}{\sqrt K}+2^{2-q}
     =\mathcal O(\epsilon).
\end{align*}
By
Proposition~\ref{prop:quadrature-Hamiltonian-intertwining}, the
coefficient state in the theorem is
\begin{equation*}
 |c_{N,Q}(t)\rangle=S_N^\dagger\widetilde u_{N,Q}(t),
\end{equation*}
and it evolves according to \eqref{eq:simplex-exact-coefficient-evolution}.
From the same corollary, the observable estimate \eqref{eqn:observable-gap}
similarly gives
\begin{align*}
    &\left|
        \langle\Psi_Q(T)|\widehat h_Q|\Psi_Q(T)\rangle
        -\langle\Phi(T)|h|\Phi(T)\rangle
    \right|\\
    &\qquad\leq
    4B_1\norm{h_K}_\infty\norm{f-f_K}_\infty
    +2^{3-q}\norm{h_K}_\infty
    +2\norm{h-h_K}_\infty\\
    &\qquad\leq
    \frac{4B_1G}{\sqrt K}\norm h_\infty
    +2^{3-q}\norm h_\infty
    +\frac{2G_h}{\sqrt K}\\
    &\qquad=\mathcal O(\norm h_\infty\epsilon).
\end{align*}
The grid and coefficient registers use $\mathcal O(d\log(Q+1))$ qubits.
\end{proof}

\begin{theorem}[Digital interaction-picture simulation on the simplex]
\label{thm:simplex-digital-simulation}
Assume the hypotheses of
Theorem~\ref{thm:simplex_wright_fisher_total_disc_error} with target
accuracy $0<\epsilon<1$, and let $N,Q$, $|c_{N,Q}(t)\rangle$,
$|\Psi_Q(t)\rangle$, and $\widehat h_Q$ be the objects furnished there.
Assume additionally that $a$ and $b$ are piecewise $C^1$ and that $b$
has finitely many sign intervals. Put
$B_1\coloneqq\int_0^T|b(t)|\dd t$, and let
$\Lambda_f\geq\max\{1,\norm f_\infty\}$.

Assume coherent access as follows, for a sufficiently small universal
constant $c_0>0$. An initial-state circuit using
$\mathcal C_\mathrm{prep}$ gates and $\mathcal Q_\mathrm{prep}$ work qubits
prepares a normalized $|\widetilde c_0\rangle$ with
\begin{equation*}
 \norm{|\widetilde c_0\rangle-|c_{N,Q}(0)\rangle}
 \leq\epsilon_\mathrm{prep}\leq c_0\epsilon.
\end{equation*}
The quadrature data and compiled Jacobi transforms from
Theorem~\ref{thm:constructive-simplex-complexity} are available. At each
simplex quadrature node, a binary oracle $O_f$ and its inverse evaluate
$\widetilde f(x_{\mathbf r})$ satisfying
\begin{equation*}
 |\widetilde f(x_{\mathbf r})-f(x_{\mathbf r})|
 \leq\epsilon_f,
 \qquad 0<\epsilon_f\leq\Lambda_f,
 \qquad B_1\epsilon_f\leq c_0\epsilon,
\end{equation*}
at a cost of $\mathcal C_f$ gates and $\mathcal Q_f$ work qubits per
call, including reversible Duffy coordinate evaluation and uncomputation.
For every $0<\delta<1$,
controlled schedule access and the Wright--Fisher kinetic phases are
available to operator error $\delta$ using
$\polylog(d,N,1/\delta)$ gates per use.

Then there is a quantum circuit producing a normalized coefficient state
$|\widetilde c_{N,Q}(T)\rangle$, with success probability at least
$1-\mathcal O(\epsilon)$, such that
\begin{equation}
 \norm{S_N|\widetilde c_{N,Q}(T)\rangle-\Phi(T)}
 =\mathcal O(\epsilon).
 \label{eq:simplex-end-to-end-state-error}
\end{equation}
Let $R$ and $C$ denote the quadrature and coefficient registers and define
\begin{equation*}
 |\widetilde\Psi_Q(T)\rangle_{RC}
 \coloneqq\widetilde U_W^\Delta
 \bigl(|0\rangle_R|\widetilde c_{N,Q}(T)\rangle_C\bigr),
\end{equation*}
where $\widetilde U_W^\Delta$ is the implemented Jacobi-transform
dilation. Extend the quadrature observable to the joint register by
\begin{equation*}
 \widehat h_Q
 \coloneqq\left(\sum_{\mathbf r}h(x_{\mathbf r})
 \ketbra{\mathbf r}{\mathbf r}_R\right)\otimes I_C,
\end{equation*}
setting it to zero on unused padded node addresses. Then
\begin{equation}
 \left|
 \langle\widetilde\Psi_Q(T)|\widehat h_Q
 |\widetilde\Psi_Q(T)\rangle
 -\langle\Phi(T),h\Phi(T)\rangle
 \right|
 =\mathcal O(\norm h_\infty\epsilon).
 \label{eq:simplex-end-to-end-observable-error}
\end{equation}
These guarantees include binary-oracle, interaction-picture,
kinetic-phase, and Jacobi-transform synthesis errors.

Let $\tau_f\coloneqq\Lambda_fB_1$. The circuit makes
\begin{equation}
 N_f=\mathcal O\!\left(
 (1+\tau_f)
 \frac{\log(\mathrm e^{\mathrm e}(1+\tau_f)/\epsilon)}
 {\log\log(\mathrm e^{\mathrm e}(1+\tau_f)/\epsilon)}
 \right)
 =\widetilde{\mathcal O}(1+\tau_f)
 \label{eq:simplex-binary-query-complexity}
\end{equation}
calls to $O_f$ or $O_f^\dagger$ and uses
\begin{equation}
 \mathcal C_\mathrm{prep}
 +\widetilde{\mathcal O}\!\left((1+\tau_f)\left[
 d(N+1)(Q+1)^2+\mathcal C_f
 +\log^2\!\left(\frac{2\Lambda_f}{\epsilon_f}\right)
 \right]\right)
 \label{eq:simplex-end-to-end-gate-complexity}
\end{equation}
elementary gates. The total qubit count is
\begin{equation}
 \mathcal Q_\mathrm{prep}+\mathcal Q_f+\mathcal O\!\left(
 d\log(Q+1)+\log\frac{2\Lambda_f}{\epsilon_f}
 +\polylog\frac{d(N+1)(Q+1)(1+\tau_f)}{\epsilon}
 \right).
 \label{eq:simplex-end-to-end-qubit-complexity}
\end{equation}
\end{theorem}

\begin{proof}
By Theorem~\ref{thm:simplex_wright_fisher_total_disc_error}, the exact
target evolution is \eqref{eq:simplex-exact-coefficient-evolution}.
Choose the Jacobi transforms, kinetic phases, and schedule subroutines to
operator error
\begin{equation*}
 \delta=\Theta\!\left(
 \frac{\epsilon\,
 \log\log(\mathrm e^{\mathrm e}(1+\tau_f)/\epsilon)}
 {(1+\tau_f)\log(\mathrm e^{\mathrm e}(1+\tau_f)/\epsilon)}
 \right)
\end{equation*}
with a sufficiently small implied constant. Theorem
\ref{thm:constructive-simplex-complexity} supplies
$\widetilde U_W^\Delta$ with
$\norm{\widetilde U_W^\Delta-U_W^\Delta}\leq\delta$.

Query $O_f$, reversibly clamp the returned value to
$[-\Lambda_f,\Lambda_f]$, compute a one-qubit amplitude
$\widetilde f(x_{\mathbf r})/\Lambda_f$, and uncompute. This uses two
oracle calls and
$\mathcal O(\log^2(2\Lambda_f/\epsilon_f)+\log^2(1/\delta))$ gates,
apart from the oracle cost. Conjugating this block by the implemented
forward and inverse Jacobi transforms block encodes
$W_{\Delta,N,Q}^\dagger\widehat f_QW_{\Delta,N,Q}$ with unnormalized
operator error at most $\epsilon_f+\mathcal O(\Lambda_f\delta)$.

Move to the kinetic frame generated by $a(t)D_N$ and rescale time by
$\dd\sigma=\Lambda_f|b(t)|\dd t$. The rescaled interaction evolution has
duration $\tau_f$ and norm at most one. Each time-controlled query uses
the compressed-potential block surrounded by controlled Wright--Fisher
kinetic phases. The rescaled truncated-Dyson algorithm
\cite{low2019interaction,berry2020time} achieves operator error and
failure probability $\mathcal O(\epsilon)$ with the query count in
\eqref{eq:simplex-binary-query-complexity}. Each query contains two
binary-oracle calls, one forward Jacobi transform, one inverse Jacobi
transform, and two controlled kinetic phases.

Duhamel's formula contributes
$B_1\epsilon_f+\mathcal O(\tau_f\delta)$ from the nearby compressed
Hamiltonian, and the hybrid argument over the time-controlled queries
contributes $\mathcal O(\epsilon)$ from schedule and kinetic synthesis.
Together with truncated-Dyson and initial-state errors, this gives
\begin{equation*}
 \norm{|\widetilde c_{N,Q}(T)\rangle-|c_{N,Q}(T)\rangle}
 =\mathcal O(\epsilon).
\end{equation*}
Equation~\eqref{eq:simplex-discretization-state-error} proves
\eqref{eq:simplex-end-to-end-state-error}. At readout, the implemented
Jacobi dilation adds at most $\delta$ in state norm. Combining the
resulting joint-state estimate with
\begin{equation*}
 |\langle v|\widehat h_Q|v\rangle
 -\langle w|\widehat h_Q|w\rangle|
 \leq2\norm h_\infty\norm{|v\rangle-|w\rangle}
\end{equation*}
and \eqref{eq:simplex-discretization-observable-error} proves
\eqref{eq:simplex-end-to-end-observable-error}.

Finally, Theorem~\ref{thm:constructive-simplex-complexity} gives
per-transform cost
$\widetilde{\mathcal O}(d(N+1)(Q+1)^2)$ at tolerance $\delta$.
Multiplying by \eqref{eq:simplex-binary-query-complexity} and adding
preparation, readout, oracle arithmetic, and reusable Dyson/schedule
workspace proves the gate and qubit bounds.
\end{proof}

\section{Simulation on General Riemannian Manifolds -- Coherent Chart Spectral Methods}
\label{sec:coherent-chart-methods}

This section develops a more generic simulation algorithm for Reimannian Schr\"odinger equations by applying local Euclidean spectral methods in a
finite atlas and coherently recombining the local solutions. These models require an extended access model that provides more information than what was required by the global spectral methods.  This so we can properly localize the global wavefunction on each coordinate chart and apply the local spectral method. Critical to this is the notion of subordinate partition of unity.

The collection of smooth, compactly supported, real-valued functions on some open set $U\subseteq M$ is denoted $C_c^\infty(U)$.
\begin{definition}[Partitions of Unity] \label{def:partition-of-unity}
    We say a collection $\{\chi_{\alpha} \in C_c^\infty(M)\}_{\alpha \in \mathcal{I}}$ is a partition of unity if
    \begin{align} \label{eq:sum-to-one-POU}
        \sum_{\alpha \in \mathcal{I}}\chi_{\alpha}^2(x) = 1, \quad \forall x \in \mathcal{M}.
    \end{align}
\end{definition}
\noindent Taking the square in Eq.~\eqref{eq:sum-to-one-POU} is nonstandard, but is convenient for $L^2$ spaces. A partition of unity $\{\chi_\alpha\}$, is \emph{subordinate} to an atlas $A = \{(U_\ell,\kappa_\ell)\}_{\ell\in[n_\mathrm{ch}]}$ if the support of each partition $\chi_\alpha$ is contained in some $U_\ell$. For a PoU subordinate to a finite atlas, one can always define a new finite PoU $\{\chi_\ell\}$ indexed with the chart. When we speak of subordinate PoUs we will always assume this finitely-indexed version.

For a function $\psi \in L^2(M)$, the $L^2$ norm can be expressed in terms of the subordinate partition of unity as
\begin{align*}
    \lVert \psi \rVert^2 = \sum_{\ell = 1}^{[n_\mathrm{ch}]}\int_{\Omega_{\ell}} \lvert  \chi_{\ell}(\kappa_{\ell}^{-1}(y))\psi_{[\ell]}(y)\rvert^2\rho_{[\ell]}(y)dy.
\end{align*}

The method requires access to a particular kind of atlas, which always exists for compact manifolds. Let $S=[-1/2,1/2]^d$. A finite atlas
$A=\{(U_\ell,\kappa_\ell)\}_{\ell\in\mathcal I}$ is \emph{box-adapted}
if $S\subset\Omega_\ell\coloneqq\kappa_\ell(U_\ell)$ for every $\ell$ and the
smaller coordinate neighborhoods
$\{\kappa_\ell^{-1}(\operatorname{int}S)\}_{\ell\in\mathcal I}$ cover
$M$. Such a finite refinement exists on a compact manifold: first choose
relatively compact coordinate neighborhoods that cover $\mathcal{M}$, and then
restrict and affinely rescale their coordinate maps.

\begin{definition}[Box-subordinate PoU] \label{def:box-subordinate-PoU}
    Let $A$ be a box-adapted atlas. A real quadratic PoU $\{\chi_\ell\}_{\ell\in\mathcal I}$ is
    \emph{box-subordinate} to $A$ if
    \begin{equation*}
        \sum_{\ell\in\mathcal I}\chi_\ell^2=1,
        \qquad
        \overline{\kappa_\ell(\supp\chi_\ell)}
        \subseteq\operatorname{int}S
        \quad\text{for every }\ell\in\mathcal I.
    \end{equation*}
    Equivalently, the support in each chart lies a positive distance from
    $\partial S$. Such a PoU is obtained from smooth bump functions subordinate to the cover $\{\kappa_\ell^{-1}(\operatorname{int}S)\}_\ell$ by normalizing their pointwise Euclidean square sum.
\end{definition}
\noindent The support of each bump function being a positive distance from the bound of $S$ ensure a chart-dependent collar of
$\partial S$ on which every localized representative
$(\chi_\ell\psi)\circ\kappa_\ell^{-1}$ vanishes almost everywhere. Its
zero extension therefore defines a periodic $L^2$ function without a
boundary jump; for Sobolev inputs, the same statement holds weakly through
the available derivative order. The rectangular domain then permits
standard Fourier or Chebyshev discretizations.

The cutoff functions $\chi_\ell$ define the local Hilbert spaces
\begin{equation}
    \mathcal H_\ell=L^2(U_\ell,\dd\vol_g),
    \qquad
    \mathcal H_\oplus=\bigoplus_{\ell\in\mathcal I}\mathcal H_\ell.
\end{equation}
The associated localization isometry mapping a function from $L^2(\mathcal{M})$ to the local charts is defined as follows.
\begin{definition}[Localization map]
Define
\begin{equation}
\label{eq:J}
    J:L^2(M)\rightarrow\mathcal H_\oplus,
    \qquad
    (Ju)_\ell=\chi_\ell u|_{U_\ell}.
\end{equation}
Its adjoint $J^\dagger $ extends local sections by zero and recombines them:
\begin{equation}
\label{eq:Jstar}
    J^\dagger (u_\ell)_\ell=\sum_\ell\chi_\ell u_\ell.
\end{equation}
\end{definition}
The map $J$ is an isometry due to 
the partition of unity (PoU) :
\begin{equation}
    \norm{Ju}_{\mathcal H_\oplus}^2
    =\sum_\ell\int_M\chi_\ell^2|u|^2\dd\vol_g
    =\norm u_{L^2_g}^2,
\end{equation}
where the last equality is definitional. Moreover, $J^\dagger J=I$, and $P=JJ^\dagger $ is the
orthogonal projector in $\mathcal H_\oplus$ onto the physically consistent subspace
$\mathcal H_\mathrm{phys}=\Ran J$. The direct sum $\mathcal H_\oplus$ is a
redundant representation space, whereas $\mathcal H_\mathrm{phys}$ is its
physical subspace. A general tuple $(u_\ell)_\ell$ need not represent a
global function on $\mathcal{M}$. 

It is often preferable to move every local function to flat Lebesgue
space:
\begin{align*}
    u_{\ell}(x) \mapsto \rho_\ell(x)^{1/2}u_{\ell}(x),
\end{align*}
which also involves conjugating the Hamiltonian by $\rho^{1/2}$ as well.

For $\psi\in L^2(M)$, the localized state is
\begin{equation}
Ju=\sum_\ell|\ell\rangle\otimes|\chi_\ell u\rangle.
\end{equation}
Applying $J^\dagger $ adds overlapping amplitudes introducing interference across overlapping charts. This is the main mechanism that enables recombining the results of chart-local evolutions. However, directly conjugating the Hamiltonian $H$ with $J$ does not result in the correct dynamics.  Particularly, 
$J^\dagger (\bigoplus_\ell H)J=\sum_\ell\chi_\ell H\chi_\ell$, whose kinetic
part differs from $-\Delta_g$ because derivatives act on the cutoffs.
The potential term commutes with every $\chi_\ell$. The Riemannian IMS
identity supplies the required kinetic correction.

\begin{fact}[Riemannian IMS identity]
\label{thm:IMS}
For $u$ in the kinetic form domain,
\begin{equation}
\label{eq:IMS}
    \int_{\mathcal{M}}\lVert \nabla u \rVert_{g}^2 \dvol = \sum_\ell\mathfrak \int_\mathcal{M} \lVert \nabla (\chi_\ell u) \rVert_{g}^2 \dvol - \int_M W_\mathrm{IMS}|u|^2\dd\vol_g,
\end{equation}
where
\begin{equation}
\label{eq:WIMS}
    W_\mathrm{IMS}(x)
    =\sum_\ell\norm{\dd\chi_\ell(x)}_{g^{-1}}^2.
\end{equation}
Equivalently (assuming equality means using the above quadratic form version), then
\begin{equation}
\label{eq:operator-IMS}
    -\Delta_g
    =\sum_\ell\chi_\ell(-\Delta_g)\chi_\ell-W_\mathrm{IMS}.
\end{equation}
\end{fact}

\begin{proof}
    Expand
    \begin{equation}
        \dd(\chi_\ell u)=\chi_\ell\dd u+u\dd\chi_\ell.
    \end{equation}
    After summing over $\ell$, the $|\dd u|^2$ term gives
    $\sum_\ell\chi_\ell^2=1$, while the cross term vanishes because
    $\sum_\ell\chi_\ell\dd\chi_\ell=\frac12\dd\sum_\ell\chi_\ell^2=0$.
    The remaining cutoff-gradient term is \eqref{eq:WIMS}.
\end{proof}
The IMS identity is standard in the analysis of
Schr\"odinger operators and has recently been used in spectral-gap
estimates~\cite{herman2025mechanisms,gribling2025selfconcordantschrodingeroperatorsspectral}. Here it instead provides an exact compressed representation of the Riemannian Hamiltonian.

Put $H=-\Delta_g$.  For each chart, let $H^0_\ell$ denote the local
realization of the differential expression $-\Delta_g$ acting on the
already-localized input $v_\ell=\chi_\ell u$.  Define
\begin{equation}
 K^0\coloneqq\bigoplus_\ell H^0_\ell,
 \qquad
 M_{W_\mathrm{IMS}}^\oplus
 \coloneqq\bigoplus_\ell M_{W_\mathrm{IMS}|_{U_\ell}},
 \qquad
 B\coloneqq K^0-M_{W_\mathrm{IMS}}^\oplus.
 \label{eq:continuum-B-definition-early}
\end{equation}
Thus every chart block contains the pullback of the full IMS potential
$W_\mathrm{IMS}=\sum_m\norm{\dd\chi_m}_{g^{-1}}^2$, i.e. the operator $M_{W_\mathrm{IMS}|_{U_\ell}}$ multiplies by $W_\mathrm{IMS}$.

Combining these definitions gives
\begin{equation}
 u
 \xmapsto{\ J\ }
 (\chi_\ell u)_\ell
 \xmapsto{\ B\ }
 \left(H(\chi_\ell u)-W_\mathrm{IMS}\chi_\ell u\right)_\ell
 \xmapsto{\ J^\dagger \ }
 Hu.
 \label{eq:continuum-JBJ-three-steps}
\end{equation}
Indeed, substituting the explicit adjoint gives
\begin{align}
 J^\dagger BJu
 &=\sum_\ell\chi_\ell H(\chi_\ell u)
     -W_\mathrm{IMS}\sum_\ell\chi_\ell^2u\nonumber\\
 &=\left(H+W_\mathrm{IMS}\right)u-W_\mathrm{IMS}u
 =Hu,
 \label{eq:continuum-JstarBJ-explicit-calculation}
\end{align}
where the second line uses the operator IMS identity
\eqref{eq:operator-IMS} and the quadratic partition of unity.  Consequently,
\begin{equation} \label{eq:continuum-JstarBJ-early}
    J^\dagger BJ=H=-\Delta_g
\end{equation}
as an identity of quadratic forms.  The negative full IMS potential is
exactly what removes the localization error introduced by differentiating
the cutoffs.

Let $B_N$ denote $B$ with finite energy cutoff. Then if $U_{B_N}$ block-encodes $B_N/\alpha_{B_N}$, and
$U_J$ is an exact unitary extension of the isometry $J$, then
\begin{equation}
 U_{\mathrm{src},N}\coloneqq U_J^\dagger U_{B_N}U_J
 \label{eq:continuum-formal-unitary-sandwich}
\end{equation}
block encodes
\begin{equation}
 J^\dagger B_NJ/\alpha_{B_N}.
 \label{eq:continuum-formal-JstarBJ-block}
\end{equation}
The explicitly constructed localization oracle below instead has signal
block $J/\alpha_J$ and therefore contributes a factor $\alpha_J^2$ to the
normalization when used on both sides of a sandwich.

We now present the assumed quantum access model, which begins with a chart-labelled potential oracle.
\begin{definition}[Potential over Manifold Oracle] {Let $f : \mathcal{M} \rightarrow \mathbb{R}$ and $\{(U_\ell,\kappa_\ell)\}_{\ell\in\mathcal I}$ be a finite atlas. An $\epsilon$-accurate binary oracle for $f$ is an operation $\mathcal O_f^{(\mathcal{M})}$ satisfying}
\begin{align*}
\mathcal{O}_f^{(\mathcal{M})}|\ell\rangle|y\rangle|z\rangle \mapsto |\ell\rangle|y\rangle|z \oplus \widetilde{f}(\kappa_{\ell}^{-1}(y))\rangle, \quad \ell \in \mathcal{I}, y \in {\Omega}_{\ell}
\end{align*}
where $\lvert {f}(\kappa_{\ell}^{-1}(y)) - \widetilde{f}(\kappa_{\ell}^{-1}(y))\rvert < \epsilon$.
\end{definition}
Note that this local-coordinates presentation is consistent as if $y_{\ell}, y_{\ell'} \in \mathbb{R}$ encode the same point lying in ${U}_{\ell} \cap {U}_{\ell'}$, then $\kappa_{\ell}^{-1}(y_{\ell}) = \kappa_{\ell'}^{-1}(y_{\ell'})$.

{We also require additional oracles provide the coordinate maps,
metric components, and local volume densities.}
\begin{definition}[Manifold Geometry Oracles] For an atlas $\{(U_{\ell}, \kappa_{\ell})\}_{\ell \in \mathcal{I}}$, the manifold geometry oracles are chart-labelled coordinate oracles
\begin{align*}
&\mathcal{O}_{\kappa}^{(\mathcal{M})}|\ell\rangle|x\rangle \mapsto |\ell\rangle|\kappa_{\ell}(x)\rangle\\
&\mathcal{O}_{g}^{(\mathcal{M})}|\ell\rangle|i\rangle|j\rangle|y\rangle|0\rangle
 \mapsto |\ell\rangle|i\rangle|j\rangle|y\rangle
 |g_{[\ell]}^{ij}(y)\rangle,\\
&\mathcal{O}_{\rho}^{(\mathcal{M})}|\ell\rangle|y\rangle|0\rangle
 \mapsto |\ell\rangle|y\rangle|\rho_{[\ell]}(y)\rangle,
\end{align*}
where $y\in\Omega_\ell$ and
$\rho_{[\ell]}(y)=\sqrt{\det(g_{[\ell],ij}(y))}$. 

\end{definition}

\begin{definition}[Subordinate Atlas Oracle] Let $\{ \chi_{\ell} \}_{\ell \in \mathcal{I}}$ be a PoU subordinate to a locally-finite atlas $\{(U_\ell,\kappa_\ell)\}_{\ell\in\mathcal I}$ on $(M,g)$. The triplet $(\mathcal{O}^{(\mathcal{M})}_{\chi}, \mathcal{O}^{(\mathcal{M})}_{\mathsf{ch}_1}, \mathcal{O}^{(\mathcal{M})}_{\mathsf{ch}_2})$ form a subordinate atlas oracle for $\mathcal{M}$ if ~$\forall x \in \mathcal{M}$:
\begin{align*}
&\mathcal{O}^{(\mathcal{M})}_{\chi}|\ell\rangle|x\rangle|0\rangle \mapsto |\ell\rangle|x\rangle|\chi_{\ell}(x)\rangle\\
&\mathcal{O}^{(\mathcal{M})}_{\mathsf{ch}_1} |x\rangle |0\rangle \mapsto |x\rangle |N_{\mathsf{ch}}(x)\rangle\\
&\mathcal{O}^{(\mathcal{M})}_{\mathsf{ch}_2} |x\rangle |j\rangle\mapsto |x\rangle |\ell_j\rangle,
\end{align*}
where $N_{\mathsf{ch}}(x)$ is the number of overlapping charts containing $x$, $\ell_j \in \mathcal{I}$ indicates the $j \in [N_{\mathsf{ch}}(x)]$  chart containing $x$. 
\end{definition}

{
Let $\alpha_J=\sqrt{\sup_{x\in M}N_{\mathsf{ch}}(x)}$. Using the
subordinate-atlas oracles, prepare a uniform superposition over the charts
containing $x$, query $\chi_{\ell_r}(x)$, and apply a controlled rotation
whose clean amplitude is
$\sqrt{N_{\mathsf{ch}}(x)}\chi_{\ell_r}(x)/\alpha_J$. This amplitude has
absolute value at most one. After uncomputing the auxiliary registers the result is
\begin{equation*}
 |x\rangle|0\rangle
 \longmapsto
 |x\rangle\frac{1}{\alpha_J}
 \sum_{r=1}^{N_{\mathsf{ch}}(x)}
 \chi_{\ell_r}(x)|\ell_r\rangle|0\rangle+|\perp_x\rangle.
\end{equation*}
Since $\chi_\ell(x)=0$ outside $U_\ell$, the resulting unitary $U_J$ block encodes
\begin{equation}
 (I\otimes\langle0|)U_J(I\otimes|0\rangle)
 =\frac{J}{\alpha_J}.
 \label{eq:localization-oracle-signal-block}
\end{equation}
Thus $U_J$ is an $\alpha_J$-normalized block encoding of $J$.}

The IMS-corrected operator is block diagonal before compression by $J^\dagger $ and
$J$.  Its input in block $\ell$ is already the localized function
$v_\ell=(J\psi)_\ell=\chi_\ell\psi$; no second factor of $\chi_\ell$ is
inserted inside the differential expression. Write
\begin{align}
    &f_{[\ell]}(y)\coloneqq f(\kappa_\ell^{-1}(y)),\qquad
    {W_{[\ell]}(y)\coloneqq W_\mathrm{IMS}(\kappa_\ell^{-1}(y)),} \notag \\
    &C_{[\ell]}^{ij}(y)\coloneqq \rho_{[\ell]}(y)g_{[\ell]}^{ij}(y),\qquad
    V_\ell(t,y)\coloneqq b(t)f_{[\ell]}(y)-a(t)W_\ell(y).
    \label{eq:local-coefficient-definitions}
\end{align}
The full local block corresponding to the geometric Hamiltonian is 
\begin{equation} \label{eq:local-IMS-Hamiltonian}
    H_\ell(t)v = -a(t)\rho_{[\ell]}^{-1}\partial_i \left(C_{[\ell]}^{ij}\partial_jv\right) + V_\ell(t)v.
\end{equation}
With these definitions,
$J^\dagger (\bigoplus_\ell H_\ell(t))J=H(t)$ (under taking quadratic forms).

To pass from $L^2(\Omega_\ell,\rho_\ell\dd y)$ to flat Lebesgue space, use
the unitary
\begin{equation}
 U_{\rho,\ell}v=\rho_{[\ell]}^{1/2}v,
 \qquad U_{\rho,\ell}^{-1}q=\rho_{[\ell]}^{-1/2}q.
\end{equation}
{Put $h_{[\ell]}\coloneqq \rho_{[\ell]}^{-1/2}$ and
write $M_f$ for multiplication by $f$. The conjugated operator is}
\begin{align}
 \widetilde H_\ell(t)
 &\coloneqq U_{\rho,\ell}H_\ell(t)U_{\rho,\ell}^{-1}\nonumber\\
 &=-a(t)M_{h_\ell}\partial_i
             M_{C_\ell^{ij}}\partial_jM_{h_\ell}
         +M_{V_\ell(t)}.
 \label{eq:local-half-density-Hamiltonian}
\end{align}
{
Fix real smooth periodic extensions of $h_\ell$, $C_\ell$, $f_\ell$, and
$W_\ell$ from a neighborhood of the localized support to $\mathbb T^d$.
Assume the extended $h_\ell$ is bounded away from zero, the extended
matrix $C_\ell$ is real symmetric and uniformly positive definite, and
the extended potentials are bounded. The kinetic quadratic form
\begin{equation*}
 \mathfrak k_\ell[q]
 \coloneqq \sum_{i,j=1}^d
 \left\langle\partial_i(h_\ell q),
 C_\ell^{ij}\partial_j(h_\ell q)\right\rangle_{L^2(\mathbb T^d)},
 \qquad q\in H^1(\mathbb T^d),
\end{equation*}
is closed and nonnegative. The local kinetic operator is defined as its
Friedrichs operator. Under the displayed smoothness and ellipticity
assumptions its domain is $H^2(\mathbb T^d)$, and on that domain it equals
$M_{h_\ell}\sum_{i,j}\partial_i^\dagger M_{C_\ell^{ij}}
\partial_jM_{h_\ell}$, with $\partial_i^\dagger=-\partial_i$.
{For real $a(t),b(t)$, the full local operator is
self-adjoint on $H^2(\mathbb T^d)$ whenever $a(t)\neq0$, by bounded
perturbation of the kinetic operator. If $a(t)=0$, it is the bounded
multiplication operator $b(t)f_\ell$ on its maximal domain
$L^2(\mathbb T^d)$. In either case, its quadratic-form expression is
well-defined on the common form domain $H^1(\mathbb T^d)$.}
Half-density conjugation leaves this multiplication potential unchanged.}

Within each coordinate chart, the local operator is spectrally truncated using a Fourier--Galerkin method. Let
\begin{align*}
    e_{k}(y) = e^{-i2\pi k \cdot y}.
\end{align*}
{Opposite faces of $S$ are identified, thereby
imposing artificial periodic boundary conditions, and the local Fourier
unitary is defined by}
\begin{align*}
 &(U_{\mathsf{FT},\ell}q)_k
 \coloneqq \int_{S}
  e_{k}(y) q(y)\dd y,
 \qquad k\in\mathbb Z^d
 \label{eq:local-continuum-Fourier-transform}\\
 &(U_{\mathsf{FT},\ell}q)_k \implies \widehat{q} \in \ell^2(\mathbb{Z}^d).
\end{align*}
{The support condition in
Definition~\ref{def:box-subordinate-PoU} makes each localized input
$\psi_{[\ell]}\chi_{[\ell]}$ vanish in a collar of the box boundary.
Consequently, its periodic extension has no boundary jump, and the local
half-density admits the Fourier representation}
\begin{align}
    {[U_{\rho,\ell}(\chi_{[\ell]}\psi_{[\ell]})](y)
    = \sum_{k \in \mathbb{Z}^d}
    (U_{\mathsf{FT},\ell}U_{\rho,\ell}\chi_{[\ell]}\psi_{[\ell]})_k
    \overline{e_{k}(y)},\quad\text{with equality in }L^2(\mathbb T^d).}
\end{align}

A general tuple
$(q_\ell)_{\ell\in\mathcal I}$ with
$q_\ell\in L^2(S,\dd y)$ need not arise from a single
$\psi\in L^2(M)$, because its representatives may be inconsistent on
chart overlaps. The synthesis map $J^\dagger $ recombines an arbitrary tuple,
whereas $JJ^\dagger$ is the orthogonal projection onto the consistent subspace$\mathcal H_\mathrm{phys}$.

With the fixed periodic extensions and
self-adjoint realizations specified above, define
\begin{equation}
    \Lambda_{\ell,j}\coloneqq \diag _{k\in\mathbb Z^d} \left(i2\pi k_j\right), \qquad \widehat M_c\coloneqq U_{\mathsf{FT},\ell}M_cU_{\mathsf{FT},\ell}^\dagger.
\end{equation}
Then the Fourier-space operator is
\begin{equation}
    \widehat H_\ell(t) = a(t)\widehat M_{h_\ell} \left(\sum_{i,j=1}^d \Lambda_{\ell,i}^\dagger\widehat M_{C_\ell^{ij}} \Lambda_{\ell,j}\right) \widehat M_{h_\ell} + \widehat M_{V_\ell(t)}.
\label{eq:local-Fourier-Hamiltonian}
\end{equation}
This is precisely $U_{\mathsf{FT},\ell}\widetilde H_\ell(t)
U_{\mathsf{FT},\ell}^\dagger$ and is self-adjoint on the Fourier image of the
periodic $H^2$ domain.  In particular, the two derivative multipliers occur
as $\Lambda_{\ell,i}^\dagger$ and $\Lambda_{\ell,j}$, and both exterior density
factors are $h_\ell=\rho_\ell^{-1/2}$. 

The Galerkin spectral error is analyzed chart by chart for each $\widetilde{H}_\ell(t)$.
Let $\mathcal A=\{\overline e_k:k\in\mathbb Z^d\}$,
$\mathcal A_N=\{\overline e_k:\norm{k}_\infty\leq N\}$, and let
$P_N$ be the orthogonal projection of $L^2(\mathbb T^d)$ onto
$\mathcal V_N\coloneqq \operatorname{span}\mathcal A_N$.

Identify the redundant chart space with its flat half-density realization
and write
{
\begin{align*}
 \norm w_{H_\mathrm{box}^r(S)}^2
 &\coloneqq \sum_{\abs\alpha\leq r}
     \norm{\partial^\alpha w}_{L^2(S)}^2,\\
 \norm w_{H_\mathrm{per}^r(\mathbb T^d)}^2
 &\coloneqq \sum_{k\in\mathbb Z^d}
     (1+4\pi^2\norm{k}_2^2)^r\abs{\widehat w(k)}^2.
\end{align*}
The first is the ordinary Sobolev norm on the box. The second is the
periodic Fourier Sobolev norm. In the Fourier estimates below, the
shorthand $H^r(S)$ means $H_\mathrm{per}^r(\mathbb T^d)$; arbitrary chart
pullbacks are measured in $H_\mathrm{box}^r(S)$ until localization gives a
periodic extension.}
\begin{align}
 \mathcal H_\oplus
 &\coloneqq \bigoplus_{\ell\in\mathcal I}L^2(S,\dd y),
 &\norm q_{H_\oplus^r}^2
 &{\coloneqq \sum_{\ell\in\mathcal I}
     \norm{q_\ell}_{H_\mathrm{per}^r(\mathbb T^d)}^2,}
 \label{eq:continuum-coherent-direct-sum-Sobolev}\\
 (\mathcal T u)_\ell(y)
 &\coloneqq \rho_\ell(y)^{1/2}\chi_\ell(\kappa_\ell^{-1}(y))
     u(\kappa_\ell^{-1}(y)).
 \label{eq:continuum-coherent-flat-analysis-map}
\end{align}

{For an integer $s\geq0$, define the full covariant Sobolev norm by}
\begin{equation}
 \norm v_{H_g^s(\mathcal{M})}^2
 \coloneqq \sum_{m=0}^s\int_M
     \abs{\nabla^m v(x)}_{g^{\otimes m}}^2\dd\vol_g(x),
 \label{eq:continuum-coherent-covariant-Sobolev-norm}
\end{equation}
where $\nabla^0v=v$, $\nabla v=\dd v$, and $\nabla^m v$ is the $m$th
covariant derivative of the scalar field.  For $m\geq1$, its pointwise
tensor norm is
\begin{equation}
 \abs{\nabla^m v}_{g^{\otimes m}}^2
 \coloneqq g^{i_1j_1}\cdots g^{i_mj_m}
 (\nabla^m v)_{i_1\cdots i_m}
 \overline{(\nabla^m v)_{j_1\cdots j_m}}.
 \label{eq:continuum-coherent-covariant-tensor-norm}
\end{equation}
The argument $(\mathcal M)$ is omitted when the
underlying manifold is clear; on $U_\ell$ the same definition is used with
the integral restricted to $U_\ell$.

Let $\widetilde{\mathcal{H}} : = L^2(S, \dd y)$, and $\widetilde{\mathcal{H}}_{\oplus} \coloneqq  \widetilde{\mathcal{H}}^{\oplus \lvert \mathcal{I}\rvert}$. The quadratic partition of unity and the half-density factor make
$\mathcal T:L_g^2(M)\to\widetilde{\mathcal H}_\oplus$ an isometry. We choose a
common spectral cutoff $N$ for every chart, with
\begin{equation*}
    P_N^\oplus\coloneqq \bigoplus_{\ell\in\mathcal I}P_N.
\end{equation*}

With
\begin{equation} \label{eq:continuum-coherent-operator-and-projector}
    \widetilde{\boldsymbol H}^{\oplus}(t)
    \coloneqq \bigoplus_{\ell\in\mathcal I}\widetilde H_\ell(t),
\end{equation}
the IMS identity gives the exact continuum compression
\begin{equation}
 H(t)=\mathcal T^\dagger\widetilde{\boldsymbol H}^{\oplus}(t)\mathcal T.
 \label{eq:continuum-coherent-exact-compression}
\end{equation}
The ideal continuum Fourier--Galerkin approximation considered in this
subsection is therefore
\begin{equation}
    H_N^\mathrm{cFG}(t)
    \coloneqq \mathcal T^\dagger P_N^\oplus\widetilde{\boldsymbol H}^{\oplus}(t) P_N^\oplus\mathcal T.
    \label{eq:continuum-coherent-Fourier-Galerkin-generator}
\end{equation}
It is self-adjoint.  Notice that the evolving state is still a physical
state in $L_g^2(\mathcal{M}, \dvol)$; the finite direct-sum Fourier space is internal to the
generator. 

{This above Fourier--Galerkin construction is what we refer to as 
the coherent-chart method. The following analysis gives conditions under
which its cutoff $N$ is polynomial in $d$.}

\subsection{Analysis of the coherent chart algorithm in continuous space}
\label{sec:analysis_cont_coherent_chart}
Although
\eqref{eq:continuum-coherent-Fourier-Galerkin-generator} has finite Fourier
cutoff, it remains a continuum operator because the projections are exact.
This subsection bounds the cutoff in that model; the subsequent subsection
introduces a grid and works in a digital model.

{Let $u$ and $u_N^\mathrm{cFG}$ denote the exact and
continuum Fourier--Galerkin evolutions:}
\begin{align}
 i\partial_tu(t)&=H(t)u(t),
 &u(0)&=u_0,\\
 i\partial_tu_N^\mathrm{cFG}(t)&=H_N^\mathrm{cFG}(t)u_N^\mathrm{cFG}(t),
 &u_N^\mathrm{cFG}(0)&=u_0.
 \label{eq:continuum-coherent-exact-and-Galerkin-evolutions}
\end{align}

{The following product estimate controls the
Sobolev norms required in the Galerkin analysis.}
\begin{lemma} \label{lem:leibniz_sobolev_bound}
    Let $r\geq0$ be an integer. If
    $a\in W^{r,\infty}(\mathbb T^d)$ and
    $v\in H_\mathrm{per}^r(\mathbb T^d)$, then
    \begin{equation}
        \norm{av}_{H_\mathrm{per}^r(\mathbb T^d)}
        \leq L_{d,r}\norm a_{W^{r,\infty}(S)}
        \norm v_{H_\mathrm{per}^r(\mathbb T^d)}.
    \label{eq:continuum-coherent-explicit-product-rule},
    \end{equation}
    If instead $v\in H_\mathrm{box}^r(S)$ and
    $\overline{\supp} a\operatorname{int}S$, then the zero extension of $av$
    belongs to $H_\mathrm{per}^r(\mathbb T^d)$ and satisfies
    \begin{equation}\label{eq:continuum-coherent-localized-product-rule}
        \norm{av}_{H_\mathrm{per}^r(\mathbb T^d)} \leq L_{d,r}\norm a_{W^{r,\infty}(S)} \norm v_{H_\mathrm{box}^r(S)}.
    \end{equation}
    where
    \begin{equation*}\label{eq:continuum-coherent-Leibniz-constant}
        L_{d,r} \coloneqq \left[r!\binom{d+r}{r}\sum_{m=0}^r4^m\binom{d+m-1}{m}\right]^{1/2}.
    \end{equation*}
\end{lemma}
\begin{proof}
For a multi-index
$\alpha\in\mathbb N^d$, write
$|\alpha|=\alpha_1+\cdots+\alpha_d$ and define
\begin{equation}
 \norm w_{\mathscr D^r(S)}^2
 \coloneqq \sum_{|\alpha|\leq r}\norm{\partial^\alpha w}_2^2.
 \label{eq:continuum-coherent-full-multi-index-norm}
\end{equation}
Expanding the Fourier weight by the binomial and multinomial theorems
gives
\begin{align}
 \norm w_{H^r(S)}^2
 &=\sum_{k\in\mathbb Z^d}\sum_{m=0}^r\binom rm
     \sum_{|\alpha|=m}\binom m\alpha
     (2\pi)^{2m}k^{2\alpha}\abs{\widehat w(k)}^2\nonumber\\
 &\leq r!\sum_{|\alpha|\leq r}\norm{\partial^\alpha w}_2^2
 =r!\norm w_{\mathscr D^r(S)}^2,
 \label{eq:continuum-coherent-Fourier-to-multi-index-norm}
\end{align}
because
$\binom rm\binom m\alpha
=r!/[\alpha!(r-m)!]\leq r!$.
Conversely, for every $|\alpha|\leq r$,
\begin{equation*}
 (2\pi)^{2|\alpha|}k^{2\alpha}
 \leq\left(1+4\pi^2\norm k_2^2\right)^r.
\end{equation*}
The stars-and-bars count gives
\begin{align*}
 \#\{\alpha\in\mathbb N^d:|\alpha|\leq r\}
 =\binom{d+r}{r}.
\end{align*}
The preceding pointwise frequency bound and
this count therefore imply
\begin{equation}
 \norm w_{\mathscr D^r(S)}^2
 \leq\binom{d+r}{r}\norm w_{H^r(S)}^2.
 \label{eq:continuum-coherent-multi-index-to-Fourier-norm}
\end{equation}
The weak
multi-index Leibniz rule gives
\begin{equation*}
 \partial^\alpha(av)
 =\sum_{\beta\leq\alpha}\binom\alpha\beta
     (\partial^\beta a)\partial^{\alpha-\beta}v.
\end{equation*}
Since
$\sum_{\beta\leq\alpha}\binom\alpha\beta=2^{|\alpha|}$,
the triangle inequality implies
\begin{equation}
 \norm{\partial^\alpha(av)}_2
 \leq2^{|\alpha|}\norm a_{W^{r,\infty}(S)}
             \norm v_{\mathscr D^r(S)}.
 \label{eq:continuum-coherent-one-multi-index-product-bound}
\end{equation}
Summing its square over $|\alpha|\leq r$ and grouping by
$m=|\alpha|$ yields
\begin{equation}
 \norm{av}_{\mathscr D^r(S)}^2
 \leq\left[\sum_{m=0}^r4^m\binom{d+m-1}{m}\right]
             \norm a_{W^{r,\infty}(S)}^2
             \norm v_{\mathscr D^r(S)}^2.
 \label{eq:continuum-coherent-multi-index-product-sum}
\end{equation}
Combining
\eqref{eq:continuum-coherent-Fourier-to-multi-index-norm},
\eqref{eq:continuum-coherent-multi-index-to-Fourier-norm}, and
\eqref{eq:continuum-coherent-multi-index-product-sum} proves
{\eqref{eq:continuum-coherent-explicit-product-rule}.
For the localized statement, the weak Leibniz rule gives the same
multi-index estimate on $S$. Since $a$ has compact support in the interior,
$av$ has zero trace through order $r-1$ (normal derivatives on the boundary to order $r-1$ vanish) and so its zero extension belongs to
$H_\mathrm{per}^r(\mathbb T^d)$. Applying
\eqref{eq:continuum-coherent-Fourier-to-multi-index-norm} to that extension
and using
$\norm v_{\mathscr D^r(S)}=\norm v_{H_\mathrm{box}^r(S)}$ proves
\eqref{eq:continuum-coherent-localized-product-rule}.}
\end{proof}

The following is one of the main results of this section, which bounds the Galerkin method error in terms of manifold properties and the evolved-state's Sobolev norm.
\begin{theorem}[Coherent Charts Method Galerkin Error]
\label{thm:coherent-chart-continuum-error}
    Let $s\geq3$ be an integer, let
    $a,b\in L^1([0,T];\mathbb R)$, and let $f$ be real-valued. Assume that
    the exact and Galerkin generators defined above are self-adjoint and
    generate unitary propagators on a common domain (when evaluated via inner products). Suppose the fixed
    periodic coefficient extensions have the regularity appearing below and
    that the exact solution satisfies
    \begin{equation}
        \mathcal U_s(T)\coloneqq \sup_{0\leq t\leq T}\norm{u(t)}_{H_g^s}<\infty,
        \label{eq:continuum-coherent-localization-Hs-bound}
    \end{equation}   
    Define the ordinary-box chart distortion constant
    \begin{equation}
     C_{\kappa,\ell,s}(g)
     \coloneqq \sup_{0\neq z\in H_g^s(U_\ell)}
     \frac{\norm{z\circ\kappa_\ell^{-1}}_{H_\mathrm{box}^s(S)}}
        {\norm z_{H_g^s(U_\ell)}}
     \label{eq:continuum-coherent-chart-distortion}
    \end{equation}
    and assume these constants are finite. Then
    \begin{align}
        \sup_{0\leq t\leq T}\norm{u(t)-u_N^\mathrm{cFG}(t)}_2
        \leq C_{\mathcal T,s}^\mathrm{geom}\mathcal U_s(T)\Big[{}&
        A_1\left(\mathfrak K_s^\mathrm{geom}
            +\mathfrak K_2^\mathrm{geom}\right)
         (1+4\pi^2(N+1)^2)^{-(s-2)/2}\\
        &+\mathfrak V_s^\mathrm{geom}(T)
         (1+4\pi^2(N+1)^2)^{-s/2}\Big]
        \label{eq:continuum-coherent-fully-explicit-Fourier-bound},
    \end{align}
    where 
    \begin{align*}
    &C_{\mathcal T,s}^\mathrm{geom} : =\sup_{\ell \in \mathcal{I}}L_{d,s}^2 \sqrt{\lvert \mathcal{I}\rvert} \lVert \rho_{[\ell]}^{1/2}\rVert_{W^{s,\infty}(S)}\lVert \chi_{[\ell]}\rVert_{W^{s,\infty}(S)} C_{\kappa, \ell, s}(g)\\
    &\mathfrak K_2^\mathrm{geom}: =\sup_{\ell \in \mathcal{I}}6d^2\norm{h_{[\ell]}}_{W^{2,\infty}}^2
         \max_{i,j}\norm{C_\ell^{ij}}_{W^{1,\infty}}\\
        &{\mathfrak K_s^\mathrm{geom}\coloneqq \sup_{\ell \in \mathcal{I}}d^2L_{d,s}L_{d,s-1}L_{d,s-2}
        \norm{h_\ell}_{W^{s,\infty}}^2
        \max_{i,j}\norm{C_\ell^{ij}}_{W^{s-1,\infty}}}\\
         &{\mathfrak V_s^\mathrm{geom}(T)\coloneqq 
         B_1\left(L_{d,s}\sup_{\ell\in\mathcal I}
             \lVert f_{[\ell]} \rVert_{W^{s,\infty}(S)}
             +\sup_{\ell\in\mathcal I}\lVert f_{[\ell]} \rVert_{\infty}\right)}\\
         &\hspace{39mm}{+
         A_1\left(L_{d,s}\sup_{\ell\in\mathcal I}
             \lVert W_{\ell} \rVert_{W^{s,\infty}(S)}
             +\sup_{\ell\in\mathcal I}\lVert W_{\ell} \rVert_{\infty}\right).}
    \end{align*}
    are all also assumed to be finite.
\end{theorem}
\begin{proof}
Set $q(t)\coloneqq \mathcal T u(t)$ and
$E_N\coloneqq (I-P_N^\oplus)\mathcal T$.  The exact generator defect has
the two-sided decomposition
\begin{align}
 H(t)-H_N^\mathrm{cFG}(t)
 ={}&E_N^\dagger\widetilde{\boldsymbol H}^{\oplus}(t)\mathcal T
 +(P_N^\oplus\mathcal T)^\dagger
    \widetilde{\boldsymbol H}^{\oplus}(t)E_N.
 \label{eq:continuum-coherent-two-sided-Fourier-defect}
\end{align}
The first term is the output, or test-side, Fourier tail.  The second is
the trial-side consistency tail on which the second-order operator acts.

From Duhamel's formula,
\begin{align}
    \sup_{0\leq t\leq T}\norm{u(t)-u_N^\mathrm{cFG}(t)}_2
    \leq\int_0^T\Big[{}
    &\norm{(I-P_N^\oplus)
     \widetilde{\boldsymbol H}^{\oplus}(t)q(t)}_\oplus
    \nonumber\\
    &+\norm{\widetilde{\boldsymbol H}^{\oplus}(t)
     (I-P_N^\oplus)q(t)}_\oplus\Big]\dd t.
    \label{eq:continuum-coherent-exact-Fourier-residual}
\end{align}

For $0\leq r\leq s$,
\begin{align*}
 \norm{(I-P_N)q}_{H_\mathrm{per}^r}^2
 &=\sum_{\norm{k}_\infty\geq N+1}
     (1+4\pi^2\norm{k}_2^2)^r\abs{\widehat q(k)}^2\\
 &\leq
 (1+4\pi^2(N+1)^2)^{-(s-r)}
 \sum_{\norm{k}_\infty\geq N+1}
     (1+4\pi^2\norm{k}_2^2)^s\abs{\widehat q(k)}^2\\
 &\leq
 (1+4\pi^2(N+1)^2)^{-(s-r)}
 \norm q_{H_\mathrm{per}^s}^2.
\end{align*}
Consequently,
\begin{equation}
 \norm{(I-P_N^\oplus)q(t)}_{H_\oplus^r}
 \leq(1+4\pi^2(N+1)^2)^{-(s-r)/2}
             \norm{q(t)}_{H_\oplus^s}.
 \label{eq:continuum-coherent-correct-Fourier-tail}
\end{equation}

Expand the local Hamiltonian blocks as
\begin{equation*}
 \widetilde{\boldsymbol H}^{\oplus}(t)
 =a(t)\mathcal K^\oplus+M_{V(t)}^\oplus,
 \qquad
 V_\ell(t)=b(t)f_\ell-a(t)W_\ell,
\end{equation*}
{The two residual terms therefore satisfy}
\begin{align*}
&\norm{(I-P_N^\oplus)
     \widetilde{\boldsymbol H}^{\oplus}(t)q(t)}_\oplus \leq \lvert a(t) \rvert \lVert (I-P_N^\oplus)\mathcal{K}^\oplus q(t)\rVert_{\oplus} + \lVert (I-P_N^\oplus)M_{V(t)}^\oplus q(t)\rVert_{\oplus}\\
&\norm{ \widetilde{\boldsymbol H}^{\oplus}(t)(I-P_N^\oplus)
    q(t)}_\oplus \leq \lvert a(t) \rvert \lVert \mathcal{K}^\oplus(I-P_N^\oplus)q(t)\rVert_{\oplus} + \lVert M_{V(t)}^\oplus (I-P_N^\oplus)q(t)\rVert_{\oplus}.
\end{align*}
For $r\geq2$, let us define the operator norms:
\begin{align*}
 K_{\oplus,r}
 &\coloneqq \norm{\mathcal K^\oplus}_{H_\oplus^r\to H_\oplus^{r-2}},
 &\mathsf V_r(t)
 &\coloneqq \sup_\ell\norm{V_\ell(t)}_{W^{r,\infty}(S)}.
\end{align*}
Then combining the above with Lemma \ref{lem:leibniz_sobolev_bound}
\begin{align*}
 \norm{(I-P_N^\oplus)
     \mathcal K^\oplus q(t)}_\oplus
 &{\leq
 (1+4\pi^2(N+1)^2)^{-(s-2)/2}}
 K_{\oplus,s}\norm{q(t)}_{H_\oplus^s},\\
 \norm{\mathcal K^\oplus
     (I-P_N^\oplus)q(t)}_\oplus
 &{\leq
 (1+4\pi^2(N+1)^2)^{-(s-2)/2}}
 K_{\oplus,2}\norm{q(t)}_{H_\oplus^s},\\
 \norm{(I-P_N^\oplus)M_{V(t)}^\oplus q(t)}_\oplus
 &{\leq
 (1+4\pi^2(N+1)^2)^{-s/2}}
 L_{d,s}\mathsf V_s(t)\norm{q(t)}_{H_\oplus^s},\\
 \norm{M_{V(t)}^\oplus
     (I-P_N^\oplus)q(t)}_\oplus
 &{\leq
 (1+4\pi^2(N+1)^2)^{-s/2}}
 \mathsf V_0(t)\norm{q(t)}_{H_\oplus^s}.
\end{align*}

{The kinetic operator norms are bounded first at
order $r=2$ and then at general order. For the $r=2$ estimate, initially
take}
$u\in C_\mathrm{per}^\infty(S)$.  The local kinetic block is again
\eqref{eq:local-half-density-Hamiltonian}.  Since
$\partial_i$ is anti-Hermitian on the torus, define
\begin{align}
 \mathcal K_\ell u
 &\coloneqq M_{h_\ell}\sum_{i,j=1}^d
     \partial_i^\dagger M_{C_\ell^{ij}}\partial_jM_{h_\ell}u
 \nonumber\\
 &=-h_\ell\sum_{i,j=1}^d
     \partial_i\left(C_\ell^{ij}\partial_j(h_\ell u)\right),
 \qquad
 \mathcal K^\oplus\coloneqq \bigoplus_\ell\mathcal K_\ell.
 \label{eq:continuum-coherent-derived-local-kinetic-block}
\end{align}
The ordinary product rule applied to
\eqref{eq:continuum-coherent-derived-local-kinetic-block} gives
\begin{align}
 \mathcal K_\ell u
 =-\sum_{i,j=1}^d h_\ell\Big[{}
 & (\partial_iC_\ell^{ij})h_\ell\partial_ju
 +(\partial_iC_\ell^{ij})(\partial_jh_\ell)u
 +C_\ell^{ij}h_\ell\partial_i\partial_ju\nonumber\\
 &+C_\ell^{ij}(\partial_ih_\ell)\partial_ju
 +C_\ell^{ij}(\partial_jh_\ell)\partial_iu
 +C_\ell^{ij}(\partial_i\partial_jh_\ell)u
 \Big].
 \label{eq:continuum-coherent-kinetic-six-term-expansion}
\end{align}
Every coefficient multiplying $u$, a first derivative of $u$, or a
second derivative of $u$ has supremum norm at most
\begin{equation}
 \norm{h_\ell}_{W^{2,\infty}}^2
 \max_{i,j}\norm{C_\ell^{ij}}_{W^{1,\infty}}.
 \label{eq:continuum-coherent-six-term-coefficient-bound}
\end{equation}
Moreover, each of
$\norm u_2$, $\norm{\partial_i u}_2$, and
$\norm{\partial_i\partial_j u}_2$ is at most $\norm u_{H^2}$ under the
Fourier Sobolev norm.  The triangle inequality over the six displayed
terms and the $d^2$ index pairs therefore yields
\begin{align}
 \norm{\mathcal K_\ell u}_2
 \leq{}&6d^2\norm{h_\ell}_{W^{2,\infty}}^2
     \max_{i,j}\norm{C_\ell^{ij}}_{W^{1,\infty}}
     \norm u_{H^2}.
 \label{eq:continuum-coherent-derived-local-H2-bound}
\end{align}
Density of $C_\mathrm{per}^\infty(S)$ in $H^2(S)$ extends this estimate to
all $u\in H^2(S)$.  Squaring, summing over charts, and using the coefficient
envelopes then gives
\begin{equation}
 K_{\oplus,2}
 \leq \sup_{\ell \in \mathcal{I}}6d^2\norm{h_\ell}_{W^{2,\infty}}^2
     \max_{i,j}\norm{C_\ell^{ij}}_{W^{1,\infty}}.
 \label{eq:continuum-coherent-derived-direct-sum-H2-bound}
\end{equation}

{For general $r\geq2$, the Fourier Sobolev norm
gives, for every integer $m\geq1$,}
\begin{align}
 \norm{\partial_j z}_{H^{m-1}(S)}^2
 &=\sum_{k\in\mathbb Z^d}
   \left(1+4\pi^2\norm{k}_2^2\right)^{m-1}
   4\pi^2k_j^2\abs{\widehat z(k)}^2\nonumber\\
 &\leq\sum_{k\in\mathbb Z^d}
   \left(1+4\pi^2\norm{k}_2^2\right)^m
   \abs{\widehat z(k)}^2
 =\norm z_{H^m(S)}^2.
 \label{eq:continuum-coherent-derivative-Sobolev-contraction}
\end{align}
Thus both $\partial_j:H^m\to H^{m-1}$ and
$\partial_i^\dagger:H^m\to H^{m-1}$ have norm at most one.

Fix a chart $\ell$, indices $i,j$, and $u\in H^r(S)$, where $r\geq2$.
{Applying Lemma~\ref{lem:leibniz_sobolev_bound}
successively from the rightmost factor gives}
\begin{align*}
 \norm{M_{h_\ell}u}_{H^r}
 &\leq L_{d,r}\norm{h_\ell}_{W^{r,\infty}}
     \norm u_{H^r},\\
 \norm{\partial_jM_{h_\ell}u}_{H^{r-1}}
 &\leq L_{d,r}\norm{h_\ell}_{W^{r,\infty}}
     \norm u_{H^r},\\
 \norm{M_{C_\ell^{ij}}\partial_jM_{h_\ell}u}_{H^{r-1}}
 &\leq L_{d,r-1}L_{d,r}
     \norm{C_\ell^{ij}}_{W^{r-1,\infty}}
     \norm{h_\ell}_{W^{r,\infty}}
     \norm u_{H^r},\\
 \norm{\partial_i^\dagger M_{C_\ell^{ij}}
     \partial_jM_{h_\ell}u}_{H^{r-2}}
 &\leq L_{d,r-1}L_{d,r}
     \norm{C_\ell^{ij}}_{W^{r-1,\infty}}
     \norm{h_\ell}_{W^{r,\infty}}
     \norm u_{H^r}.
\end{align*}
One final application of
\eqref{eq:continuum-coherent-explicit-product-rule}, now at order $r-2$,
gives
\begin{align}
 &\norm{M_{h_\ell}\partial_i^\dagger M_{C_\ell^{ij}}
     \partial_jM_{h_\ell}u}_{H^{r-2}}
 \nonumber\\
 &\quad\leq L_{d,r-2}L_{d,r-1}L_{d,r}
     \norm{h_\ell}_{W^{r-2,\infty}}
     \norm{C_\ell^{ij}}_{W^{r-1,\infty}}
     \norm{h_\ell}_{W^{r,\infty}}
     \norm u_{H^r}.
 \label{eq:continuum-coherent-single-kinetic-term-bound}
\end{align}
The factors $L_{d,r}$, $L_{d,r-1}$, and $L_{d,r-2}$ therefore come from,
respectively, the inner multiplication by $h_\ell$, the middle
multiplication by $C_\ell^{ij}$, and the outer multiplication by
$h_\ell$.  The derivatives only lower the Sobolev order and add no
constant in this Fourier normalization.

Summing
\eqref{eq:continuum-coherent-single-kinetic-term-bound} over the $d^2$
index pairs gives
\begin{align*}
 \norm{\mathcal K_\ell u}_{H^{r-2}}
 \leq{}&d^2L_{d,r}L_{d,r-1}L_{d,r-2}
     \norm{h_\ell}_{W^{r,\infty}}
     \norm{h_\ell}_{W^{r-2,\infty}}\nonumber\\
 &\hspace{29mm}\times
     \max_{i,j}\norm{C_\ell^{ij}}_{W^{r-1,\infty}}
     \norm u_{H^r}.
\end{align*}
Note
$\norm{h_\ell}_{W^{r-2,\infty}}
\leq\norm{h_\ell}_{W^{r,\infty}}$. 
Finally, for $q=(q_\ell)_\ell\in H_\oplus^r$,
\begin{align}
 \norm{\mathcal K^\oplus q}_{H_\oplus^{r-2}}^2
 &=\sum_\ell\norm{\mathcal K_\ell q_\ell}_{H^{r-2}}^2\nonumber\\
 &{\leq
         \left(\sup_{\ell\in\mathcal I}
             \norm{\mathcal K_{\ell}}_{H^r\to H^{r-2}}\right)^2}
     \sum_\ell\norm{q_\ell}_{H^r}^2.
 \label{eq:continuum-coherent-direct-sum-kinetic-bound}
\end{align}
Consequently,
\begin{equation}
 K_{\oplus,r} \leq \sup_{\ell \in \mathcal{I}}d^2L_{d,r}L_{d,r-1}L_{d,r-2}
     \norm{h_\ell}_{W^{r,\infty}}^2
     \max_{i,j}\norm{C_\ell^{ij}}_{W^{r-1,\infty}}
 \label{eq:continuum-coherent-explicit-kinetic-bound}.
\end{equation}

Next bound the potential operator now. For
$r=0,s$, the identity
$V_\ell(t)=b(t)f_\ell-a(t)W_\ell$ and  the triangle
inequality for the $W^{r,\infty}$ norm give, chart by chart,
\begin{align}
 \norm{V_\ell(t)}_{W^{r,\infty}(S)}
 &=\norm{b(t)f_\ell-a(t)W_\ell}_{W^{r,\infty}(S)}\nonumber\\
 &\leq |b(t)|\norm{f_\ell}_{W^{r,\infty}(S)}
     +|a(t)|\norm{W_\ell}_{W^{r,\infty}(S)}.
 \label{eq:continuum-coherent-chartwise-potential-envelope}
\end{align}
Taking the supremum over $\ell$ yields
\begin{equation*}
 {\mathsf V_r(t)
 \leq |b(t)|\sup_{\ell \in \mathcal{I}}
     \lVert f_{\ell} \rVert_{W^{r,\infty}(S)}
 +|a(t)|\sup_{\ell \in \mathcal{I}}
     \lVert W_{\ell} \rVert_{W^{r,\infty}(S)},}
 \qquad r=0,s.
\end{equation*}

{To compare the direct-sum Sobolev norm with the
intrinsic (covariant) Sobolev norm on $\mathcal{M}$, put}
\begin{equation*}
 \vartheta_\ell(y)
 \coloneqq \rho_{[\ell]}(y)^{1/2}\chi_\ell(\kappa_\ell^{-1}(y)).
\end{equation*}

For
$v\in H_g^s(M)$, let $v_\ell\coloneqq v|_{U_\ell}$.  The definition of
$\mathcal T$ gives the chartwise identity
\begin{equation}
 (\mathcal T v)_\ell
 =\vartheta_\ell(v_\ell\circ\kappa_\ell^{-1}).
 \label{eq:continuum-coherent-localization-chartwise-factorization}
\end{equation}
{Applying the localized product estimate
\eqref{eq:continuum-coherent-localized-product-rule} with $r=s$, followed
by the definition of $C_{\kappa,\ell,s}(g)$, yields}
\begin{align}
 {\norm{(\mathcal T v)_\ell}_{H_\mathrm{per}^s(\mathbb T^d)}}
 &\leq L_{d,s}\norm{\vartheta_\ell}_{W^{s,\infty}(S)}
     {\norm{v_\ell\circ\kappa_\ell^{-1}}_{H_\mathrm{box}^s(S)}}\nonumber\\
 &\leq L_{d,s}\norm{\vartheta_\ell}_{W^{s,\infty}(S)}
     C_{\kappa,\ell,s}(g)\norm{v_\ell}_{H_g^s(U_\ell)}.
 \label{eq:continuum-coherent-chartwise-localization-Hs-bound}
\end{align}
Square this inequality and sum over the orthogonal chart blocks.  Since
the covariant Sobolev norm in
\eqref{eq:continuum-coherent-covariant-Sobolev-norm} has a nonnegative
integrand at every derivative order, restriction to $U_\ell$ gives
$\norm{v_\ell}_{H_g^s(U_\ell)}\leq\norm v_{H_g^s(M)}$.  Therefore
\begin{align}
 \norm{\mathcal T v}_{H_\oplus^s}^2
 &=\sum_{\ell\in\mathcal I}
     \norm{(\mathcal T v)_\ell}_{H^s(S)}^2\nonumber\\
 &\leq L_{d,s}^2\sum_{\ell\in\mathcal I}
     \left(\norm{\vartheta_\ell}_{W^{s,\infty}(S)}
     C_{\kappa,\ell,s}(g)\right)^2
     \norm{v_\ell}_{H_g^s(U_\ell)}^2\nonumber\\
 &\leq L_{d,s}^2\left[\sum_{\ell\in\mathcal I}
     \left(\norm{\vartheta_\ell}_{W^{s,\infty}(S)}
     C_{\kappa,\ell,s}(g)\right)^2\right]
     \norm v_{H_g^s(M)}^2.
 \label{eq:continuum-coherent-localization-square-sum}
\end{align}
Taking square roots proves
{the required bound with}
\begin{align}
\lVert \mathcal{T}\rVert_{H_g^{s} \rightarrow H_{\oplus}^s}
&\leq L_{d,s}\left[
 \sum_{\ell\in\mathcal I}
 \left(\norm{\vartheta_\ell}_{W^{s,\infty}(S)}
 C_{\kappa,\ell,s}(g)\right)^2\right]^{1/2} \nonumber\\
 &{\leq
 \sup_{\ell \in \mathcal{I}}L_{d,s}^2
 \sqrt{\lvert \mathcal{I}\rvert}
 \lVert \rho_{[\ell]}^{1/2}\rVert_{W^{s,\infty}(S)}
 \lVert \chi_{[\ell]}\rVert_{W^{s,\infty}(S)}
 C_{\kappa, \ell, s}(g).}
 \label{eq:continuum-coherent-explicit-localization-constant}
\end{align}

\end{proof}

The following quantities state sufficient
conditions for a spectral cutoff $N$ polynomial in $d$:
\begin{align*}
&\mathsf{F} = \sup_{\ell \in \lvert \mathcal{I}\rvert}\max\left(\lVert f_{[\ell]}\rVert_{W^{s,\infty}(S)},\lVert f_{[\ell]}\rVert_{\infty} \right), \quad  \mathsf{W} = \sup_{\ell \in \lvert \mathcal{I}\rvert}\max\left(\lVert W_{[\ell]}\rVert_{W^{s,\infty}(S)}, \lVert W_{[\ell]}\rVert_{\infty} \right),\\
    &\mathsf{H} =\sup_{\ell \in \lvert \mathcal{I}\rvert}\lVert h_{[\ell]}\rVert_{W^{s,\infty}(S)}, \quad
    \mathsf{R} =\sup_{\ell \in \lvert \mathcal{I}\rvert}\lVert \rho_{[\ell]}^{1/2}\rVert_{W^{s,\infty}(S)}, \quad \mathsf{X} = \sup_{\ell \in \lvert \mathcal{I}\rvert}\lVert \chi_{[\ell]}\rVert_{W^{s,\infty}(S)},\\
    &\mathsf{C} =\sup_{\ell \in \lvert \mathcal{I}\rvert}\max_{i,j}\left(\lVert C_{[\ell]}^{ij}\rVert_{W^{s-1,\infty}(S)}\right), \quad
    \mathsf{D} =\max_{\ell} C_{\kappa, \ell, s}(g).
\end{align*}

\begin{corollary}[Polynomial Energy Cutoff]
\label{cor:poly_cutoff}
Suppose the following conditions hold: $\lvert \mathcal{I}\rvert$, $\mathsf{F}, \mathsf{W}, \mathsf{H}, \mathsf{R}, \mathsf{X}, \mathsf{C}, \mathsf{D}$, {$C_{\mathcal T,s}^{\textup{geom}}$}, $\mathcal{U}_s(T), A_1,$ and $B_1$ are all $\mathcal{O}([\textup{poly}(d)]^d)$ when $s = \Theta(d)$.
{Then there is a polynomial $p$ such that choosing
\begin{equation}
 N\geq p(d)\max\left\{1,\epsilon^{-1/(s-2)}\right\}
 \label{eq:continuum-coherent-correct-cutoff-scaling}
\end{equation}
gives}
\begin{align*}
    \sup_{0\leq t\leq T}\norm{u(t)-u_N^\mathrm{cFG}(t)}_2 = \mathcal{O}(\epsilon).
\end{align*}
{In particular, $N=\operatorname{poly}(d)$ for
fixed target accuracy.}
\end{corollary}
\begin{proof}
    {The first term in
    Theorem~\ref{thm:coherent-chart-continuum-error} decays as
    $N^{-(s-2)}$, while the second term decays faster as $N^{-s}$.
    Taking the $(s-2)$th root of the assumed prefactor gives a polynomial
    in $d$ when $s=\Theta(d)$, which proves
    \eqref{eq:continuum-coherent-correct-cutoff-scaling}.}
\end{proof}

{The reason that we considered the continuous model is that discretization introduces an additional obstacle:
overlapping charts need not have intersecting local grids. A global source
interpolant transfers one coefficient representation to all local grids.
For a general high-dimensional manifold this map may be inefficient; the
factorized construction below is conditional on constant-dimensional
product structure and explicit implementation assumptions.}

\subsection{Tensor products of manifolds and a discretized algorithm}

{
This subsection presents a discretized version of the preceding algorithm
for tensor products of constant-dimensional manifolds. The low-order
geometric parameters of the factors are assumed to be bounded uniformly in
the number of factors. Under the approximation assumptions stated below,
the local Fourier cutoff is polynomially bounded for fixed target
accuracy. This cutoff theorem is not, by itself, a polynomial-runtime
theorem: such a result additionally requires polynomial bounds on the
source-space dimensions, block-encoding costs, oracle costs, and integrated
Hamiltonian normalizations.}

{Consider the product setting}
\begin{equation}
    \mathcal{M}=\prod_{\nu=1}^{m}\mathcal{M}_\nu, \qquad \dim \mathcal{M}_\nu=p_\nu\leq p_0=\Ocal(1), \qquad d=\sum_{\nu=1}^{m}p_\nu, \qquad g=\bigoplus_{\nu=1}^{m}g_\nu.
    \label{eq:tensor-local-product-manifold}
\end{equation}
{Then $m\leq d\leq p_0m$, every local
approximation problem has dimension at most $p_0$, and the product metric
makes the kinetic operator a Kronecker sum. The factor geometries are
assumed to satisfy the uniform bounds stated below, independently of
$\nu$ and the total dimension.

For each factor, fix a box-adapted atlas $\{(U^{(\nu)}_a,\kappa^{(\nu)}_a):a\in\mathcal I_\nu\}$ with a box-adapted subordinate POU $\{\chi^{(\nu)}_a\}_{a\in\mathcal I_\nu}$ (Definition \ref{def:box-subordinate-PoU}).
Define the full factorwise IMS potential
\begin{equation}
 W_\nu(x_\nu)
 \coloneqq \sum_{a\in\mathcal I_\nu}
     \norm{\dd\chi_a^{(\nu)}(x_\nu)}_{g_\nu^{-1}}^2.
 \label{eq:tensor-local-factor-full-IMS-potential}
\end{equation}
A product chart is indexed by
$\boldsymbol a=(a_1,\ldots,a_m)$, with
\begin{equation}
 \kappa_{\boldsymbol a}=\prod_{\nu=1}^{m}\kappa^{(\nu)}_{a_\nu},
 \qquad
 \chi_{\boldsymbol a}(x)
 =\prod_{\nu=1}^{m}\chi^{(\nu)}_{a_\nu}(x_\nu).
 \label{eq:tensor-local-product-chart-cutoff}
\end{equation}
The product cutoff is again a quadratic partition.  After grouping chart
and coordinate registers by factor, the localization and half-density maps
factor as
\begin{equation}
 J=\bigotimes_{\nu=1}^{m}J_\nu,
 \qquad
 \mathcal T=\bigotimes_{\nu=1}^{m}\mathcal T_\nu,
 \qquad
 \rho_{\boldsymbol a}(y)
 =\prod_{\nu=1}^{m}\rho^{(\nu)}_{a_\nu}(y_\nu).
 \label{eq:tensor-local-factorized-analysis-map}
\end{equation}

Discretization introduces a challenge that was absent from the continuum:
even if two product charts overlap, their local Fourier grids need not
intersect.  A global source interpolant resolves this mismatch.  For a
general high-dimensional manifold that interpolation may be computationally
expensive; however for \eqref{eq:tensor-local-product-manifold}, it can be assembled
coherently from constant-dimensional factor interpolants.

We do not select a  particular global interpolation basis. However, one such option is compactly supported wavelets. The result instead
assumes stated approximation and implementation properties of the source
spaces and derives a cutoff bound contingent on those properties. The first part of this section will state the algorithm, while the second part is dedicated to the error analysis.

\subsubsection{Digital Coherent-Charts Method}
The aforementioned global interpolation basis defines a Hilbert space of achievable wavefunctions that globally approximate the true wave function via an interpolation from a fix set of points on the manifold. Each component $\nu$ is given a separate interpolating basis, whose span is referred to as the \emph{source space}.

For each $\nu$ Choose an $L^2(\mathcal{M}_\nu)$-orthonormal basis
$\mathcal{Y}_{\nu}\coloneqq \{L^{(\nu)}_{q_\nu}\}_{q_\nu\in\mathcal{X}_\nu}$ of global interpolating functions. For each factor let the finite-dimensional
source space
\begin{equation}
    V^\mathrm{src}_{\nu,R_\nu}
    \subset H^{\sigma_0}(\mathcal{M}_\nu),
    \qquad \sigma_0>\max\{2,p_0/2\},
\end{equation}
be the complex span of $\mathcal{Y}_{\nu}$. By Sobolev embedding, each source function then has a continuous representative, so the point evaluations used below are well-defined. Thus the interpolating basis for the overall manifold $\mathcal{M}$ is just the tensor product of the component global interpolating bases. Each source space  is associated with a parameter $R_{\nu}$ that we call the \emph{resolution parameter}.  The resolution parameter is directly related to the interpolation error, as will be shown later, and the number of basis functions and grid points is usually a function of $R_{\nu}$.

Let $Q_{\nu,R_\nu}$ be the $L^2$-orthogonal
projector onto the corresponding source space. The product source space
and projector are
\begin{equation}\label{eq:tensor-local-product-source-space}
    V_{\boldsymbol R}^\mathrm{src} = \bigotimes_{\nu=1}^{m}V^\mathrm{src}_{\nu,R_\nu}, \qquad Q_{\boldsymbol R}^\mathrm{src} = \bigotimes_{\nu=1}^{m}Q_{\nu,R_\nu}.
\end{equation}
The coefficient space
$\mathcal H_{\nu,R_\nu}^\mathrm{src}=\mathbb C^{\lvert\mathcal Y_\nu\rvert}$
defines the reconstruction maps
\begin{equation}\label{eq:tensor-local-source-reconstruction}
    \mathscr E_{\nu,R_\nu}:\mathcal H_{\nu,R_\nu}^\mathrm{src} \rightarrow V_{\nu,R_\nu}^\mathrm{src}, \qquad \mathscr E_{\boldsymbol R}:\mathcal H_{\boldsymbol R}^\mathrm{src} \rightarrow V_{\boldsymbol R}^\mathrm{src}, \qquad \mathscr E_{\boldsymbol R} = \bigotimes_{\nu=1}^m\mathscr E_{\nu,R_\nu}.
\end{equation}
Specifically, for $|\mathbf{c}\rangle \coloneqq \sum_{\mathbf{q} \in \prod_{\nu} \mathcal{X}_{v}} c_{\mathbf{q}}|\mathbf{q}\rangle \in \mathcal{H}_{\mathbf{R}}^{\textup{src}} = \bigotimes_{\nu = 1}^{m}\mathcal{H}_{\nu, R_{\nu}}^{\textup{src}}$, 
\begin{align*}
    [\mathscr{E}_{\mathbf{R}}|\mathbf{c}\rangle](x) = \sum_{\mathbf{q} \in \prod_{\nu} \mathcal{X}_{v}} c_{\mathbf{q}}L_{q_1}(x_1)\cdots L_{q_m}(x_m) \in V_{\mathbf{R}}^{\textup{src}}.
\end{align*}
{Each factor chart carries an independent uniform
grid $\mathcal G_{a,N_\nu}^{(\nu)}$ and an independent discrete Fourier
transform. For curved manifolds, the pullbacks of two such grids may be
disjoint even when the charts overlap. Evaluation of the common source
basis on every grid removes any requirement that the grids themselves
intersect.}

 In chart $a$ of manifold $\mathcal{M}_{\nu}$, let the grid by defined by
\begin{equation}
 {r\in \mathcal G^{(\nu)}_{a,N_\nu}
 \coloneqq \mathcal B_{\nu,N_\nu}
 \coloneqq \{-N_{\nu}, \dots, N_{\nu} -1\}^{p_\nu},}
 \qquad
 y^{(\nu)}_{a,r}
 \coloneqq \frac{r}{2 N_\nu}
 \label{eq:tensor-local-factor-grid},
\end{equation}
where $x^{(\nu)}_{a,r}
 \coloneqq (\kappa^{(\nu)}_a)^{-1}(y_{a,r}^{(\nu)})$. For each grid point, we define the following matrices
\begin{align}
 (E^{(\nu)}_{a})_{rq}
 &\coloneqq L^{(\nu)}_q(x^{(\nu)}_{a,r}),\nonumber\\
 (D^{(\nu)}_{a})_{rr}
 &\coloneqq (2N_{\nu})^{-p_\nu/2}
 [\rho^{(\nu)}_a(y^{(\nu)}_{a,r})]^{1/2}
 \chi^{(\nu)}_a(x^{(\nu)}_{a,r}),\nonumber\\
 \mathcal R^{(\nu)}_a&\coloneqq D^{(\nu)}_aE^{(\nu)}_a.
 \label{eq:tensor-local-factor-resampling}
\end{align}
For a vector $|c\rangle \in \mathcal{H}_{\nu, R_{\nu}}^{\textup{src}}$, $E_{a}^{(\nu)}$ evaluates $\mathscr{E}_{\nu, R_{\nu}}|c\rangle$ on all grid points $x_{a, r}^{(\nu)}$ associated with chart $a$. The diagonal matrix $D_{a}^{(\nu)}$ does a few things. First, it applies a discretized version of the map $\mathcal{T}$ from the previous section, which for  single chart is just multiplication by $\chi_{a}^{(\nu)}$. Second, it multiplies by $[\rho_{a}^{(\nu)}]^{1/2}$ to go from $L^2(\kappa_{a}^{(\nu)}(U^{(\nu)}_{a}), \rho_{a}^{(\nu)}\dd y)$ to $L^2(\kappa_{a}^{(\nu)}(U^{(\nu)}_{a}), \dd y)$, like done in the previous section. {Finally, it applies the sample normalization $(2N_\nu)^{-p_\nu/2}$. With the unitary tensor-product DFT convention
\begin{equation*}
 (F_{a,N_\nu}^{(\nu)}z)_k
 \coloneqq (2N_\nu)^{-p_\nu/2}
 \sum_{r\in\mathcal B_{\nu,N_\nu}}
 z_r e^{-2\pi i k\cdot r/(2N_\nu)},
 \qquad k\in\mathcal B_{\nu,N_\nu},
\end{equation*}
these normalized samples produce the discrete coefficients defined in
\eqref{eq:tensor-local-factor-discrete-Fourier-coefficients}.}
Thus in total $\mathcal{R}_{a}^{(\nu)}$ takes an element of $\psi \in V_{\nu, R_{\nu}}^{\textup{src}}$ and maps it to a discrete Fourier approximation to the Fourier projection ($P_N$) of $[\rho_{a}^{(\nu)}]^{1/2}\chi_{a}^{(\nu)}\psi$, i.e. the $q$ function from the previous section. Note like in the global spectral methods section, the number of grids points and retained Fourier coefficients can differ, to combat aliasing. Then, taking the direct sum over all charts in single component manifold gives
\begin{equation}
 \mathcal A_\nu\coloneqq \bigoplus_{a\in\mathcal I_\nu}\mathcal R^{(\nu)}_a.
 \label{eq:tensor-local-factor-analysis-map}
\end{equation}
and thus globally
\begin{align*}
    \mathcal{A} : = \bigotimes_{\nu=1}^{m} \mathcal{A}_{\nu}.
\end{align*}
{
This is a discrete-grid version of $\mathcal T$. Let
$F_{a,L}^{(\nu)}$ be the unitary DFT on the $(2L)^{p_\nu}$-point grid and
define the spectral differentiation matrices by
\begin{equation}
 D_{a,L,j}^{(\nu)}
 \coloneqq (F_{a,L}^{(\nu)})^\dagger
     \diag _{k\in\mathcal B_{\nu,L}}(2\pi i k_j)
     F_{a,L}^{(\nu)}.
 \label{eq:tensor-local-factor-differentiation-matrix}
\end{equation}
For a grid size $L$, define the raw chart block
\begin{align}
 B_{a,L}
 \coloneqq M_{h_a^{(\nu)}}
 \sum_{i,j=1}^{p_\nu}(D_{a,L,i}^{(\nu)})^\dagger
 M_{C_a^{(\nu),ij}}D_{a,L,j}^{(\nu)}M_{h_a^{(\nu)}}
 -M_{W_a^{(\nu)}},
 \qquad
 W_a^{(\nu)}\coloneqq W_\nu\circ(\kappa_a^{(\nu)})^{-1}.
 \label{eq:direct-sum-hamiltonian-for-component}
\end{align}
Choose an implementation grid size $N_\mathrm{pad}\geq N$. Let
$Z_{N\to N_\mathrm{pad}}$ embed the retained Fourier coefficients indexed by
$\mathcal B_{\nu,N}$ into those indexed by
$\mathcal B_{\nu,N_\mathrm{pad}}$ by zero padding, and set
\begin{equation}
 \mathcal Z_a^{(\nu)}
 \coloneqq (F_{a,N_\mathrm{pad}}^{(\nu)})^\dagger
     Z_{N\to N_\mathrm{pad}}F_{a,N}^{(\nu)}.
 \label{eq:tensor-local-factor-oversampled-DFT}
\end{equation}
The implemented padded kinetic block is
\begin{equation}
 B_{\nu,N}^\mathrm{pad}
 \coloneqq \bigoplus_{a\in\mathcal I_\nu}
     (\mathcal Z_a^{(\nu)})^\dagger B_{a,N_\mathrm{pad}}
     \mathcal Z_a^{(\nu)}.
 \label{eq:tensor-local-padded-kinetic-block}
\end{equation}
The source-space kinetic operator is therefore defined by
\begin{equation}
 K_{\nu,R_\nu,N}^\mathrm{src}
 \coloneqq \mathcal A_\nu^\dagger B_{\nu,N}^\mathrm{pad}\mathcal A_\nu.
 \label{eq:tensor-local-factor-padded-source-kinetic}
\end{equation}
The purpose of this padding will be to remove one source of aliasing error using a technique from Section \ref{subsec:flat-torus}.}
{Over the full tensor product, define}
\begin{align*}
    K_{\boldsymbol R,\boldsymbol N}^\mathrm{src}
 &\coloneqq \sum_{\nu=1}^{m}
 K_{\nu,R_\nu,N_\nu}^\mathrm{src}\otimes I_{-\nu}.
\end{align*}
For
$\boldsymbol a=(a_1,\ldots,a_m)$ and
$\boldsymbol r=(r_1,\ldots,r_m)$, put
\begin{equation}
 x_{\boldsymbol a,\boldsymbol r}
 \coloneqq \left(x^{(1)}_{a_1,r_1},\ldots,
                    x^{(m)}_{a_m,r_m}\right) \in \prod_{\nu=1}^{m} U_{a_{\nu}}^{(\nu)},
\end{equation}
and define a global potential operator that acts as 
\begin{equation}
 M_{f,\boldsymbol N}^{\oplus}
 \ket{\boldsymbol a,\boldsymbol r}
 \coloneqq f(x_{\boldsymbol a,\boldsymbol r})
     \ket{\boldsymbol a,\boldsymbol r}.
 \label{eq:tensor-local-coupled-grid-potential}
\end{equation}
{Assume access to the binary evaluation oracle}
\begin{equation}
 O_f\ket{\boldsymbol a,\boldsymbol r,z}
 =\ket{\boldsymbol a,\boldsymbol r,
 z\oplus f(x_{\boldsymbol a,\boldsymbol r})},
 \label{eq:tensor-local-coupled-potential-oracle}
\end{equation}
with $\norm f_\infty\leq\Lambda_f$.  Define
\begin{equation}
 M_{f,\boldsymbol R,\boldsymbol N}^\mathrm{src}
 \coloneqq \mathcal A^\dagger M_{f,\boldsymbol N}^{\oplus}\mathcal A.
 \label{eq:tensor-local-source-potential}
\end{equation}
The fully discrete Hamiltonian is the following Hamiltonian on the source coefficient
space:
\begin{equation}
 H_{\boldsymbol R,\boldsymbol N}^\mathrm{src}(t)
 \coloneqq a(t)K_{\boldsymbol R,\boldsymbol N}^\mathrm{src}
     +b(t)M_{f,\boldsymbol R,\boldsymbol N}^\mathrm{src},
 \label{eq:tensor-local-discrete-Hamiltonian}
\end{equation}
which is the discrete analog of \eqref{eq:continuum-coherent-exact-compression}. This is a finite dimensional Hamiltonian acting on $\mathcal{H}_{\mathbf{R}}^{\textup{src}}$, which can be simulated used standard time-dependent Hamiltonian simulation techniques, e.g. \cite{low2019interaction}. However, unlike the geometric spectral methods case, the kinetic part is not necessarily fast-forwardable. Hence the simulation will necessarily grow with the cutoff $\max_{\nu}N_{\nu}$.

{For the block-encoding analysis, define
$\delta_\nu$ as the factorwise near-isometry defect:}
\begin{align*}
    \norm{\mathcal A_\nu^{\dagger}\mathcal A_\nu-I}\leq\delta_\nu.
\end{align*}
If the above is satisfied, we say that $\mathcal{A}_{\nu}$ is a $\delta_{\nu}$-approximate isometry. {It follows only that
$\norm{\mathcal A_\nu}\leq\sqrt{1+\delta_\nu}$; hence any supplied block
encoding may use a normalization
$\alpha_{A,\nu}\leq\sqrt{1+\delta_\nu}$.}
{
The near-isometry estimate controls the possible normalization but does not
construct a circuit. Accordingly, the block-encoding statements below
assume coherent controlled implementations of
$\mathcal A_\nu/\alpha_{A,\nu}$ and its adjoint, to the required arithmetic
precision, with $\alpha_{A,\nu}\leq\sqrt{1+\delta_\nu}$. Their gate and
oracle costs remain explicit implementation assumptions.}

\begin{proposition}[Block encodings of the sandwiched operators]
\label{prop:tensor-local-block-normalization}
    Suppose the implemented padded raw block
    $B_{\nu,N}^\mathrm{pad}$ from
    \eqref{eq:tensor-local-padded-kinetic-block} has an exact controlled
    block encoding with normalization $\alpha_{B,\nu}^\mathrm{pad}$, and that the exact potential-value access supplies a block encoding of $M_{f,N}^{\oplus}/\Lambda_f$. Assume also the
    controlled block encodings of every $\mathcal A_\nu/\alpha_{A,\nu}$ and
    its adjoint described above.
    
    The sandwiched kinetic and potential operators admit block encodings with normalizations
    \begin{align}
     \alpha_K
     &\coloneqq \sum_{\nu=1}^m
     \alpha_{A,\nu}^2\alpha_{B,\nu}^\mathrm{pad},
     &
     \alpha_f
     &\coloneqq \Lambda_f\prod_{\nu=1}^m\alpha_{A,\nu}^2.
     \label{eq:tensor-local-sandwiched-block-normalizations}
    \end{align}
    The factor-$\nu$ kinetic sandwich uses one call each to the encodings of
    $\mathcal A_\nu$, $\mathcal A_\nu^\dagger$, and $B_{\nu,N}^\mathrm{pad}$.
    The potential sandwich uses one potential-oracle block encoding and one
    call to each factor encoding and its adjoint.  Standard LCU then gives a
    block encoding of $H_{R,N}^\mathrm{src}(t)$ with normalization
    \begin{equation*}
     \alpha_H(t)=\abs{a(t)}\alpha_K+\abs{b(t)}\alpha_f.
    \end{equation*}
    
    If $\delta_\mathrm{samp}\leq1$ and
    $\alpha_{B,\nu}^\mathrm{pad}\leq
    \alpha_{B,0}^\mathrm{pad}(N,N_\mathrm{pad})$ uniformly in $\nu$, then
    \begin{equation}
     \alpha_K\leq(m+1)\alpha_{B,0}^\mathrm{pad}(N,N_\mathrm{pad}),
     \qquad
     \alpha_f\leq e\Lambda_f.
     \label{eq:tensor-local-admissible-block-normalizations}
    \end{equation}
    In particular, if the raw padded kinetic encoding has
    $\alpha_{B,0}^\mathrm{pad}=O(N_\mathrm{pad}^2)$ and $N_\mathrm{pad}\leq7N$, then
    $\alpha_K=O(mN^2)$ while $\alpha_f=O(\Lambda_f)$.
\end{proposition}

\begin{proof}
The block-encoding product lemma gives each sandwich normalization, and
factor selection followed by LCU gives $\alpha_K$ and $\alpha_H(t)$.
Finally,
$\alpha_{A,\nu}^2\leq1+\delta_\nu$ and $1+x\leq e^x$ give
\begin{equation*}
 \sum_{\nu=1}^m\alpha_{A,\nu}^2
 \leq m+\delta_\mathrm{samp},
 \qquad
 \prod_{\nu=1}^m\alpha_{A,\nu}^2
 \leq e^{\delta_\mathrm{samp}},
\end{equation*}
which proves the stated bounds.
\end{proof}

For constant factor-chart counts and uniform resolution, the chart and grid
registers use
\begin{equation}
 \Ocal\left(\sum_{\nu=1}^{m}
 [\log\abs{\mathcal I_\nu}+p_\nu\log N]\right)
 =\Ocal(d\log N)
\end{equation}
qubits.  The formal product grid contains $N^{\Theta(d)}$ nodes, as expected
for a full wavefunction discretization.  {Under the
explicit block-encoding and oracle-access assumptions above, the algorithm
need not materialize an $N^d\times N^d$ resampling matrix or loop over all
product nodes: it applies $m$ constant-dimensional resampling and kinetic
primitives coherently to the factor registers.  This conclusion depends on
those access assumptions and does not follow from register factorization
alone.}

The same formal construction applies beyond
product manifolds, but a global interpolation may then be inefficient.
 Showing a complete polynomial gate complexity still hinges on the efficiency of block encoding $\mathcal{A}$, as even those each component manifold is constant dimensional, the number of global interpolation functions and grid points may grow with $d$. Still, we make a stride in this direction by analyzing a setting where the discretization parameter $N_{\nu}$ can be taken to be $\textup{poly}(m)$.

\subsubsection{Discretization error analysis for products of constant dimensional manifolds}

{The digital coherent-chart error on the product
manifold \eqref{eq:tensor-local-product-manifold} is analyzed by comparing
the exact and fully discrete evolutions in four stages: projection onto
the source space, approximation of the geometric
coefficients, approximation on the local grids, and comparison with the
final discrete operator.  Duhamel's formula then combines the four errors together.}

The main result, Theorem~\ref{thm:tensor-local-main}, assumes that the
dimensions, numbers of charts, chart distortions, and the metric, density,
cutoff, and IMS quantities of the factors are bounded independently of the
factor and of $d$.  It also assumes that one common source resolution can
be chosen for all factors, at least as large as $N$ divided by a polynomial
in $d$.  Along this choice, the source-space projection error defined in
\eqref{eqn:src_interpolation_abastract} and the error from representing a
source-space function on a local grid defined in
\eqref{eq:tensor-local-factor-source-interpolation-defect} must decay at
the rates stated in the theorem.  Finally, the derivatives of the
geometric coefficients and the Sobolev norms used in the later estimates
are allowed to grow with $d$, but by no more than $(C d^\beta)^d$; their
precise definitions appear in
\eqref{eq:tensor-local-geometric-regularity-envelope},
\eqref{eq:tensor-local-potential-mixed-norm}, and
Theorem~\ref{thm:tensor-local-main}.  Under these assumptions, each of the
four errors is at most $(C d^\beta/N)^d$.  Thus a resolution polynomial in
$d$ suffices for fixed target accuracy.  No particular construction of the
source spaces is required.

For clarity, use common resolutions
$R_\nu=R$ and $N_\nu=N$ for every factor.  Accordingly, write
$V_R^\mathrm{src}$, $Q_R^\mathrm{src}$, $\mathscr E_R$,
$K_{R,N}^\mathrm{src}$, $M_{f,R,N}^\mathrm{src}$, and
$H_{R,N}^\mathrm{src}$ for the objects previously defined with
$\boldsymbol R=(R,\ldots,R)$ and $\boldsymbol N=(N,\ldots,N)$.

The local Fourier construction does not prescribe the global source
spaces, so their projection errors are retained explicitly below.  We use
the following notation for the factorwise Sobolev norms.  If $\mathscr X$
is a Hilbert space, then $v\in H^s(\mathcal{M}_\nu;\mathscr X)$ means that
$x_\nu\mapsto v(x_\nu)$ is $\mathscr X$-valued and its weak derivatives in
$x_\nu$ through order $s$ are square integrable in the $\mathscr X$ norm. The operator $Q_{\nu,R_\nu}$ acts only on $x_\nu$ and as the identity on
$\mathscr X$. Here one takes
$\mathscr X=H^r(M_{-\nu})$ when estimating the above mixed global $H^r$ error, where $M_{-\nu}\coloneqq \prod_{\mu\neq\nu}M_\mu$. We also use the shorthand $H^s_{x_{\nu}}(H_{-\nu}^r) \coloneqq  H^s(M_{\nu}; H^r(M_{-\nu}))$. Associated with this space is the following (mixed-order) Sobolev norm
\begin{equation}
 \norm v_{\mathcal H^s_{x_\nu}(H^r_{-\nu})}^2
 \coloneqq \sum_{j=0}^{s}\sum_{\ell = 0}^{r}\int_{M} \lvert \nabla_{x_{-\nu}}^{\ell}\nabla_{x_{\nu}}^jv(x)\rvert_{[g_{-\nu}^{-1}]^{\otimes \ell} \otimes [g_{\nu}^{-1}]\otimes j}^2 \dvol(x)
 \label{eq:tensor-local-mixed-Sobolev-norm},
\end{equation}
where $\nabla$ is the covariant derivative and $g_{-\nu} = \bigoplus_{j \neq \nu} g_j$. Also recall the manifold Sobolev norm \eqref{eq:continuum-coherent-covariant-Sobolev-norm}:
\begin{equation}
 \norm v_{H_g^r(M)}^2
 \coloneqq \sum_{m=0}^r\int_M
     \abs{\nabla^m v(x)}_{[g^{-1}]^{\otimes m}}^2\dd\vol_g(x).
\end{equation}

{In addition to source interpolation, the digital
method incurs aliasing from the local DFTs. The analysis separates four
contributions: source projection, Jackson smoothing of the continuum
kinetic operator, local Fourier discretization, and removal of Jackson
smoothing from the discrete operator. For the exact evolution $u(t)$ and
the discrete evolution $u_{R,N}(t)$, the resulting decomposition is}
\begin{align*}
    \lVert u (t) - u_{{R}, {N}}(t) \rVert &\leq \lVert u(t) - u_{{R}}(t)\rVert \label{eqn:source_interpolation_error}\tag{src interp.} \\
    &+\lVert u_{{R}}(t) - u_{{R}, J}(t) \rVert \label{eqn:jackson_smooth_cont_error}\tag{smoothed -- cont.}\\
    &+\lVert u_{{R}, J}(t) - u_{{R}, J, {N}}(t) \rVert \label{eqn:local_fourier_error}\tag{local Fourier}\\
    &+\lVert u_{{R}, J, {N}}(t) - u_{{R}, {N}}(t)\rVert \label{eqn:jackson_smooth_discr_error}\tag{smoothed -- disc.}.
\end{align*}
The assumption is that $u_{{R}, {N}}(0)= u_{{R}, J ,{N}}(0) = u_{{R}}(0) = u(0) = u_0 \in V_{{R}}^{\textup{src}}$.
The rest of the section is divided in results that analyze each of these error. The end then recombines them to provide the main result.

\paragraph*{Source Interpolation Error}
For $r\in\{0,2\}$, define
\begin{equation}
    S_{\nu,r}(R_\nu) \coloneqq \norm{Q_{\nu,R_\nu}\otimes 
    I_{-\nu}}_{H_g^r(M)\to H_g^r(M)},
    \label{eq:tensor-local-individual-projector-stability}
\end{equation}
and
\begin{equation}
 \mathsf S_{\nu,r}^\mathrm{acc}(\boldsymbol R)
 \coloneqq \prod_{\mu<\nu}S_{\mu,r}(R_\mu),
 \qquad
 \mathsf S_{\mathrm{src},r}(\boldsymbol R)
 \coloneqq \max_{1\leq\nu\leq m}\mathsf S_{\nu,r}^\mathrm{acc}(\boldsymbol R),
 \label{eq:tensor-local-accumulated-projector-stability}
\end{equation}
where the empty product equals one.

\begin{lemma}[Product source-projection error]
\label{lem:tensor-local-product-source-projection}
For every $v\in H_g^r(M)$ and $r\in\{0,2\}$,
\begin{align}
 \norm{(I-Q_{\boldsymbol R}^\mathrm{src})v}_{H^r_g}
 &\leq\sum_{\nu=1}^{m}
 \mathsf S_{\nu,r}^\mathrm{acc}(\boldsymbol R)
 \norm{\bigl[I-(Q_{\nu,R_\nu}\otimes I_{-\nu})\bigr]v}_{H_g^r}
 \nonumber\\
 &\leq\mathsf S_{\mathrm{src},r}(\boldsymbol R)
 \sum_{\nu=1}^{m}
 \norm{\bigl[I-(Q_{\nu,R_\nu}\otimes I_{-\nu})\bigr]v}_{H_g^r}.
 \label{eq:tensor-local-product-source-estimate}
\end{align}
\end{lemma}

\begin{proof}
The identity
\begin{equation}
 I-\bigotimes_{\nu=1}^{m}Q_{\nu,R_\nu}
 =\sum_{\nu=1}^{m}
 \left(\bigotimes_{\mu<\nu}Q_{\mu,R_\mu}\right)
 (I-Q_{\nu,R_\nu})
 \left(\bigotimes_{\mu>\nu}I_\mu\right)
 \label{eq:tensor-local-source-telescoping}
\end{equation}
and the triangle inequality give the first bound.  The second follows by
maximizing the accumulated stability factor over $\nu$.
\end{proof}

Since $Q_{\nu,R_\nu}$ is an $L^2(\mathcal{M}_\nu)$-orthogonal projector,
$S_{\nu,0}(R_\nu)\leq1$.  Orthogonality in $L^2$ does not imply $H^2$
contractivity, so $S_{\nu,2}(R_\nu)$ is retained.  If all factor
projectors are contractive in $H_g^r$, then
$\mathsf S_{\mathrm{src},r}=1$.

For $r\in\{0,2\}$, define the stability-weighted sum of the one-factor
projection errors by
\begin{align}
 \varepsilon_r^\mathrm{src}(T,R)
 &\coloneqq {\mathsf S_{\mathrm{src},r}(\boldsymbol R)}
 \sum_{\nu=1}^m\sup_{0\leq t\leq T}
 \norm{\bigl[I-(Q_{\nu,R}\otimes I_{-\nu})\bigr]u(t)}_{H_g^r},
 \qquad \boldsymbol R=(R,\ldots,R).
 \label{eqn:src_interpolation_abastract}
\end{align}

Like in earlier sections define
\begin{equation}
 A_1\coloneqq \int_0^T\abs{a(t)}\dd t,
 \qquad
 B_1\coloneqq \int_0^T\abs{b(t)}\dd t.
\end{equation}

{The \eqref{eqn:source_interpolation_error} is controlled by
\eqref{eqn:src_interpolation_abastract}.}
\begin{lemma}[source interpolation error]
\begin{align}
 \sup_{0\leq t\leq T}
 \norm{u(t)-u_{R}(t)}
 \leq{}&\varepsilon_0^\mathrm{src}(T, {R})
 +A_1\sqrt{d}\varepsilon_2^\mathrm{src}(T, {R})
 +B_1\Lambda_f\varepsilon_0^\mathrm{src}(T, {R}).
 \label{eq:tensor-local-first-dynamical-comparison}
\end{align}    
\end{lemma}
\begin{proof}
Define the source-projection residual and, for $r=0,2$, its error by
\begin{equation}
 e_{ R}(t)\coloneqq (I-Q_{ R}^\mathrm{src})u(t),
 \qquad
 \epsilon_r^\mathrm{src}(T,R)
 \coloneqq \sup_{0\leq t\leq T}\norm{e_{ R}(t)}_{H_g^r}.
 \label{eq:tensor-local-source-error-definition}
\end{equation}
Let $u_{ R}^\mathrm{src}$ denote the source-projected trajectory,
\begin{equation}
 i\partial_tu_{\boldsymbol R}^\mathrm{src}(t)
 =Q_{ R}^\mathrm{src}H(t)
  Q_{ R}^\mathrm{src}u_{ R}^\mathrm{src}(t),
 \qquad
 u_{ R}^\mathrm{src}(0)
 =Q_{ R}^\mathrm{src}u(0).
 \label{eq:tensor-local-source-projected-trajectory}
\end{equation}
Lemma~\ref{lem:tensor-local-product-source-projection} and
\eqref{eqn:src_interpolation_abastract} give
\begin{align}
 \epsilon_r^\mathrm{src}(T, {R})
 &\leq \varepsilon_r^\mathrm{src}(T,R)
 \label{eq:tensor-local-source-error-budget-in-proof}.
\end{align}
For the  dynamical comparison, put
\begin{equation}
 \theta_{ R}(t)
 \coloneqq Q_{ R}^\mathrm{src}u(t)-u_{ R}^\mathrm{src}(t).
\end{equation}
Since both source trajectories start from
$Q_{ R}^\mathrm{src}u(0)$, one has
$\theta_{ R}(0)=0$.  Since $Q = Q_{ R}^\mathrm{src}$ is time independent and
$u=Qu+e_{ R}$, projection of the exact equation gives
\begin{align}
 i\partial_t(Qu)
 &=QH(t)u
 =QH(t)\bigl(Qu+e_{\boldsymbol R}\bigr)\nonumber\\
 &=QH(t)Q(Qu)+QH(t)e_{\boldsymbol R}.
 \label{eq:tensor-local-projected-exact-equation}
\end{align}
Subtracting \eqref{eq:tensor-local-source-projected-trajectory} from
\eqref{eq:tensor-local-projected-exact-equation} gives
\begin{equation}
    i\partial_t\theta_{\boldsymbol R}(t)
    =QH(t)Q
     \theta_{\boldsymbol R}(t)
     +QH(t)e_{\boldsymbol R}(t).
    \label{eq:tensor-local-source-error-equation}
\end{equation}
Variation of constants therefore yields
\begin{equation}
 \theta_{ R}(t)
 =-i\int_0^tU_{ R}^\mathrm{src}(t,s)
 QH(s)e_{ R}(s)\dd s.
 \label{eq:tensor-local-source-Duhamel}
\end{equation}
The propagator is unitary and $Q$ is an $L^2$ contraction.  Consequently,
\eqref{eq:tensor-local-source-Duhamel} gives
\begin{align}
 \norm{\theta_{\boldsymbol R}(t)}_2
 &\leq\int_0^t\norm{QH(s)e_{\boldsymbol R}(s)}\dd s\nonumber\\
 &\leq\int_0^t\left[
    \abs{a(s)}\norm{-\Delta_ge_{ R}(s)}
  +\abs{b(s)}\norm{f e_{ R}(s)}
  \right]\dd s.
 \label{eq:tensor-local-source-forcing-split}
\end{align}

For $v\in H_g^2(M)$, the identity
 $\Delta_gv=\operatorname{tr}_g(\nabla^2v)$ and Cauchy--Schwarz for the
 metric trace give
 \begin{equation}
  \norm{-\Delta_gv}\leq \sqrt{d}\norm v_{H_g^2}.
  \label{eq:tensor-local-global-H2-kinetic-bound}
 \end{equation}
The kinetic estimate
\eqref{eq:tensor-local-global-H2-kinetic-bound} and
$\norm{fv}\leq\Lambda_f\norm v_2$ imply, pointwise in time,
\begin{equation}
 \norm{QH(s)e_{\boldsymbol R}(s)}
 \leq\abs{a(s)}\sqrt{d}\norm{e_{ R}(s)}_{H_g^2}
 +\abs{b(s)}\Lambda_f\norm{e_{ R}(s)}.
 \label{eq:tensor-local-source-forcing-pointwise-bound}
\end{equation}
Taking the defining suprema of the source errors outside the integral and
using the definitions of $A_1$ and $B_1$ yields
\begin{equation}
 \sup_{0\leq t\leq T}\norm{\theta_{\boldsymbol R}(t)}_2
 \leq \sqrt{d}A_1\varepsilon_2^\mathrm{src}(T, {R})
 +B_1\Lambda_f\varepsilon_0^\mathrm{src}(T, {R}).
 \label{eq:tensor-local-source-driven-error-bound}
\end{equation}
Using
\begin{equation}
 u(t)-u_{ R}^\mathrm{src}(t)
 =e_{ R}(t)+\theta_{ R}(t),
 \label{eq:tensor-local-source-total-split}
\end{equation}
and  triangle inequality the result follows.
\end{proof}

\paragraph*{Errors from Jackson Smoothing -- Kinetic} 
{The next estimates control the two Jackson
smoothing terms \eqref{eqn:jackson_smooth_cont_error} and
\eqref{eqn:jackson_smooth_discr_error}.}

To match the notation of Section \ref{subsec:flat-torus}
$m=n_\mathrm{J}$ and replace the dimension $d$ in
\eqref{eq:Jackson-tensor} by the factor dimension $p_\nu$.  Thus, for
$z=(z_1,\ldots,z_{p_\nu})\in S_\nu$, define the factor kernel
\begin{equation}
 \mathcal J_{n_\mathrm{J},p_\nu}(z)
 \coloneqq \prod_{j=1}^{p_\nu}J_{n_\mathrm{J}}(z_j),
 \qquad
 \kappa_\mathrm{J}\coloneqq \kappa_{n_\mathrm{J}}=2(n_\mathrm{J}-1),
 \label{eq:tensor-local-factor-Jackson-kernel}
\end{equation}
where $J_{n_\mathrm{J}}$ is the one-dimensional normalized kernel in
\eqref{eq:Jackson-raw}--\eqref{eq:Jackson-tensor}.  For
$n_\mathrm{J}\geq2$, let $\mathcal S_{\nu,n_\mathrm{J}}$ denote the associated
one-step Jackson--Fourier smoothing operator:
\begin{equation}
 (\mathcal S_{\nu,n_\mathrm{J}}g)(y)
 \coloneqq (\mathcal J_{n_\mathrm{J},p_\nu}
     *_{\mathbb T^{p_\nu}}g)(y)
 \coloneqq \int_{S_\nu}\mathcal J_{n_\mathrm{J},p_\nu}(z)
     g(y-z)\dd z,
 \label{eq:tensor-local-factor-Jackson-smoother}
\end{equation}
where subtraction is modulo $\mathbb Z^{p_\nu}$.  In the notation of
\eqref{eq:Jackson-proxy},
$\mathcal S_{\nu,n_\mathrm{J}}g$ is precisely $g_m$ after the substitutions
$m=n_\mathrm{J}$ and $d=p_\nu$.

{For an integer $r\geq1$, define the higher-order
Jackson smoother}
\begin{equation}
 \begin{aligned}
 \mathfrak J_{\nu,n_\mathrm{J}}^{[r]}
 &\coloneqq I-(I-\mathcal S_{\nu,n_\mathrm{J}})^r,\\
 \mathfrak J_{\nu,n_\mathrm{J}}^{[r]}g
 &=\sum_{j=1}^r(-1)^{j+1}\binom rj
     \mathcal S_{\nu,n_\mathrm{J}}^j g.
 \end{aligned}
 \label{eq:tensor-local-iterated-Jackson-operator}
\end{equation}
Here $I$ is the identity and
$\mathcal S_{\nu,n_\mathrm{J}}^j$ denotes $j$-fold composition of the
one-step smoother in
\eqref{eq:tensor-local-factor-Jackson-smoother}.  In particular,
$\mathfrak J_{\nu,n_\mathrm{J}}^{[1]}g
=\mathcal S_{\nu,n_\mathrm{J}}g$ is exactly the smoother from
Lemma~\ref{lem:Jackson-moments}.
{Here $S=[-1/2,1/2]^d$ denotes the common local
coordinate box.}

\begin{lemma}[High-order Jackson smoother]
\label{lem:tensor-local-high-order-Jackson}
Let $q \geq 0$, $r\geq1$ and $0\leq\ell\leq r$.  If a scalar periodic function $g$
belongs to $W^{q+\ell,\infty}(S)$, then
\begin{equation}
 \norm{g-\mathfrak J_{\nu,n_\mathrm{J}}^{[r]}g}_{W^{q,\infty}}
 \leq2^{r-\ell}
 \left(\frac{c_{p_0}}{n_\mathrm{J}}\right)^\ell
 \norm g_{W^{q+\ell,\infty}},
 \label{eq:tensor-local-high-order-Jackson-bound}
\end{equation}
where $c_{p_0}$ depends only on the factor-dimension bound.  The approximation
$\mathfrak J_{\nu,n_\mathrm{J}}^{[r]}g$ is real when $g$ is real and has
coordinatewise Fourier bandwidth at most
$\kappa_\mathrm{J}$.  The same statements hold entrywise for
matrix fields.
\end{lemma}

\begin{proof}
The first-order input is exactly the Jackson-Fourier smoother from
Lemma~\ref{lem:Jackson-moments}, applied in the factor dimension.  More
precisely, fix an integer $s\geq0$ and
$h\in W^{s+1,\infty}(S_\nu)$.  Take $d=p_\nu$, $m=n_\mathrm{J}$, and apply
that lemma to the weak derivative $f=\partial^\alpha h$, where
$\abs{\alpha}\leq s$.  Since
$\partial^\alpha h\in W^{1,\infty}(S_\nu)$,
\begin{equation}
 \operatorname{Lip}(\partial^\alpha h)
 \leq\norm{\nabla\partial^\alpha h}_{L^\infty}
 \leq\sqrt{p_\nu}\,\norm h_{W^{s+1,\infty}}.
 \label{eq:tensor-local-derivative-Lipschitz-bound}
\end{equation}
Lemma~\ref{lem:Jackson-moments} therefore gives
\begin{align}
 \norm{(I-\mathcal S_{\nu,n_\mathrm{J}})
     \partial^\alpha h}_\infty
 &\leq\frac{3\sqrt{p_\nu}}{n_\mathrm{J}}
     \operatorname{Lip}(\partial^\alpha h)\nonumber\\
 &\leq\frac{3p_\nu}{n_\mathrm{J}}
     \norm h_{W^{s+1,\infty}}.
 \label{eq:tensor-local-Jackson-one-derivative-bound}
\end{align}
Periodic convolution commutes with weak derivatives, so
\begin{equation}
 \partial^\alpha(I-\mathcal S_{\nu,n_\mathrm{J}})h
 =(I-\mathcal S_{\nu,n_\mathrm{J}})\partial^\alpha h.
 \label{eq:tensor-local-Jackson-derivative-commutation}
\end{equation}
Taking the maximum over $\abs{\alpha}\leq s$ proves the first-order
Sobolev estimate
\begin{equation}
 \norm{(I-\mathcal S_{\nu,n_\mathrm{J}})h}_{W^{s,\infty}}
 \leq\frac{c_{p_0}}{n_\mathrm{J}}
 \norm h_{W^{s+1,\infty}},
 \qquad c_{p_0}\coloneqq 3p_0,
 \label{eq:tensor-local-derived-first-order-Jackson-bound}
\end{equation}
under the componentwise maximum convention for $W^{s,\infty}$.

Set $T_{\nu,n_\mathrm{J}}\coloneqq I-\mathcal S_{\nu,n_\mathrm{J}}$.  Iterating
\eqref{eq:tensor-local-derived-first-order-Jackson-bound}, with the
Sobolev order raised by one at each step, gives
\begin{align}
 \norm{T_{\nu,n_\mathrm{J}}^\ell g}_{W^{q,\infty}}
 &\leq\frac{c_{p_0}}{n_\mathrm{J}}
     \norm{T_{\nu,n_\mathrm{J}}^{\ell-1}g}_{W^{q+1,\infty}}
 \leq\cdots\nonumber\\
 &\leq\left(\frac{c_{p_0}}{n_\mathrm{J}}\right)^\ell
     \norm g_{W^{q+\ell,\infty}}.
 \label{eq:tensor-local-iterated-first-order-Jackson-bound}
\end{align}
On the other hand, positivity and unit mass of the Jackson kernel imply
\begin{equation}
 \norm{\mathcal S_{\nu,n_\mathrm{J}}g}_{W^{s,\infty}}
 \leq\norm g_{W^{s,\infty}},
 \qquad
 \norm{T_{\nu,n_\mathrm{J}}g}_{W^{s,\infty}}
 \leq2\norm g_{W^{s,\infty}}
 \label{eq:tensor-local-Jackson-contraction-bound}
\end{equation}
for every integer $s\geq0$.  Since
\begin{equation}
 g-\mathfrak J_{\nu,n_\mathrm{J}}^{[r]}g
 =T_{\nu,n_\mathrm{J}}^r g
 =T_{\nu,n_\mathrm{J}}^{r-\ell}T_{\nu,n_\mathrm{J}}^\ell g,
\end{equation}
combining \eqref{eq:tensor-local-iterated-first-order-Jackson-bound} with
\eqref{eq:tensor-local-Jackson-contraction-bound} yields
\eqref{eq:tensor-local-high-order-Jackson-bound}.

It remains to verify the structural claims.  By
Lemma~\ref{lem:Jackson-moments}, the one-step operator
$\mathcal S_{\nu,n_\mathrm{J}}$ is the real Fourier multiplier with symbol
$\widehat{\mathcal J}_{n_\mathrm{J},p_\nu}(k)$, and this symbol vanishes for
$\norm{k}_\infty>\kappa_\mathrm{J}$.  The multiplier of
$I-(I-\mathcal S_{\nu,n_\mathrm{J}})^r$ is therefore
\begin{equation}
 1-\left(1-\widehat{\mathcal J}_{n_\mathrm{J},p_\nu}(k)\right)^r,
\end{equation}
which vanishes whenever
$\widehat{\mathcal J}_{n_\mathrm{J},p_\nu}(k)=0$.  Hence the iterated
approximation has Fourier support contained in that of the original
Jackson kernel.
Finally, the kernel is real and the iterated approximation is a real
polynomial in its convolution operator, so it preserves real-valuedness.
All arguments apply entrywise to matrix fields.
\end{proof}

For a factor chart $a$, abbreviate
\begin{equation*}
 h_a^{(\nu)}\coloneqq (\rho_a^{(\nu)})^{-1/2},
 \qquad
 C_a^{(\nu),ij}\coloneqq \rho_a^{(\nu)}(g_\nu)_a^{ij},
 \qquad
 W_a^{(\nu)}\coloneqq W_\nu\circ(\kappa_a^{(\nu)})^{-1},
\end{equation*}
Apply the  operator
$\mathfrak J_{\nu,n_\mathrm{J}}^{[d+2]}$ entrywise, and set
\begin{equation*}
 h_{a,\mathrm{J}}^{(\nu)}
 \coloneqq \mathfrak J_{\nu,n_\mathrm{J}}^{[d+2]}h_a^{(\nu)},
 \qquad
 C_{a,\mathrm{J}}^{(\nu),ij}
 \coloneqq \mathfrak J_{\nu,n_\mathrm{J}}^{[d+2]}C_a^{(\nu),ij},
 \qquad
 W_{a,\mathrm{J}}^{(\nu)}
 \coloneqq \mathfrak J_{\nu,n_\mathrm{J}}^{[d+2]}W_a^{(\nu)}.
\end{equation*}

{
Define the IMS-inclusive factor operators
\begin{align}
 \mathcal K_\nu^\oplus
 &\coloneqq \bigoplus_{a\in\mathcal I_\nu}
 \left[
 M_{h_a^{(\nu)}}\sum_{i,j=1}^{p_\nu}
 \partial_i^\dagger M_{C_a^{(\nu),ij}}\partial_jM_{h_a^{(\nu)}}
 -M_{W_a^{(\nu)}}
 \right],\\
 \mathcal K_{\nu,\mathrm{J}}^\oplus
 &\coloneqq \bigoplus_{a\in\mathcal I_\nu}
 \left[
 M_{h_{a,\mathrm{J}}^{(\nu)}}\sum_{i,j=1}^{p_\nu}
 \partial_i^\dagger M_{C_{a,\mathrm{J}}^{(\nu),ij}}
 \partial_jM_{h_{a,\mathrm{J}}^{(\nu)}}
 -M_{W_{a,\mathrm{J}}^{(\nu)}}
 \right].
 \label{eq:tensor-local-IMS-inclusive-continuum-blocks}
\end{align}
For every grid size $L$, define the Jackson-smoothed chart matrix
\begin{equation}
 B_{a,L}^\mathrm{J}
 \coloneqq M_{h_{a,\mathrm{J}}^{(\nu)}}
 \sum_{i,j=1}^{p_\nu}(D_{a,L,i}^{(\nu)})^\dagger
 M_{C_{a,\mathrm{J}}^{(\nu),ij}}D_{a,L,j}^{(\nu)}
 M_{h_{a,\mathrm{J}}^{(\nu)}}-M_{W_{a,\mathrm{J}}^{(\nu)}}.
 \label{eq:tensor-local-Jackson-chart-grid-block}
\end{equation}
All coefficient multipliers in these formulas use the fixed periodic
extensions introduced in the continuum construction (Section \ref{sec:coherent-chart-methods}).

For every integer $r\geq0$, define the factorwise bound on the geometric quantities associated with each component manifold:
\begin{equation}
 \mathcal G_{\nu,r}
 \coloneqq \max_{a\in\mathcal I_\nu}
 \max\left\{
    \norm{h_a^{(\nu)}}_{W^{r,\infty}(S_\nu)},
    \max_{1\leq i,j\leq p_\nu}
        \norm{C_a^{(\nu),ij}}_{W^{r,\infty}(S_\nu)},
    \norm{W_a^{(\nu)}}_{W^{r,\infty}(S_\nu)}
 \right\}.
 \label{eq:tensor-local-geometric-regularity-envelope}
\end{equation}
In particular, $\mathcal G_{\nu,d+2}$ contains exactly the coefficient
norms needed when the order-$(d+2)$ iterated Jackson operator is applied
below.

Let $\mathcal B_{\nu,R_\nu}$ operator be an idealized version of $\mathcal{A}_{\nu}$ that includes global interpolation but does not utilize a grid:
\begin{equation}
 (\mathcal B_{\nu,R_\nu}c)_a(y)
 \coloneqq \rho_a^{(\nu)}(y)^{1/2}
 \chi_a^{(\nu)}((\kappa_a^{(\nu)})^{-1}(y))
 (\mathscr E_{\nu,R_\nu}c)((\kappa_a^{(\nu)})^{-1}(y)).
 \label{eq:tensor-local-factor-continuum-analysis}.
\end{equation}
{The first comparison concerns the
globally-interpolated kinetic operators with raw and Jackson-smoothed
coefficients:}
\begin{align}
\label{eqn:kinetic_op_cont_no_smooth}
     K_\nu^\mathrm{cont}
 =\mathcal B_{\nu,R_\nu}^\dagger\mathcal K_\nu^\oplus
     \mathcal B_{\nu,R_\nu},
\end{align}
and
\begin{align}
\label{eqn:kinetic_op_cont_smooth}
    K_{\nu,\mathrm{J}}^\mathrm{cont} = \mathcal B_{\nu,R_\nu}^\dagger  \mathcal K_{\nu,\mathrm{J}}^\oplus \mathcal B_{\nu,R_\nu},
\end{align}
respectively.

Conversion between
$\mathcal H^2_{\nu,\oplus}(L^2(M_{-\nu}))$ and
$H^2_{x_\nu}(L^2(M_{-\nu}))$ norms is controlled by the factor localization
norm. Set
\begin{equation}
    \mathcal H_{-\nu,R}^\mathrm{src}
    \coloneqq \bigotimes_{\mu\ne\nu}\mathcal H_{\mu,R}^\mathrm{src},
    \qquad
    \mathscr E_{-\nu,R}
    \coloneqq \bigotimes_{\mu\ne\nu}\mathscr E_{\mu,R}:
    \mathcal H_{-\nu,R}^\mathrm{src}
    \rightarrow L^2(M_{-\nu}),
\end{equation}
and define
\begin{equation}
    C_{\mathcal T,\nu,2}^\mathrm{geom} \coloneqq \norm{\mathcal T_\nu}
    _{\mathcal L(H^2_{x_\nu}(L^2(M_{-\nu})),H^2_{\nu,\oplus})}.
    \label{eq:tensor-local-factor-localization-constant}
\end{equation}
Put
\begin{equation}
 \vartheta_a^{(\nu)}(y)
 \coloneqq \rho_a^{(\nu)}(y)^{1/2}
 \chi_a^{(\nu)}((\kappa_a^{(\nu)})^{-1}(y)),
\end{equation}
and let $C_{\kappa,a,2}(g_\nu)$ denote
\eqref{eq:continuum-coherent-chart-distortion} with
$(\ell,d,g)=(a,p_\nu,g_\nu)$.  The continuum localization calculation
\eqref{eq:continuum-coherent-chartwise-localization-Hs-bound}, specialized
to $s=2$ and $d=p_\nu$, gives
\begin{equation}
 C_{\mathcal T,\nu,2}^\mathrm{geom}
 \leq L_{p_\nu,2}
 \left[\sum_{a\in\mathcal I_\nu}
 \left(\norm{\vartheta_a^{(\nu)}}_{W^{2,\infty}}
 C_{\kappa,a,2}(g_\nu)\right)^2\right]^{1/2}.
 \label{eq:tensor-local-factor-localization-explicit-bound}
\end{equation}

Since $\mathscr E_{-\nu,R}$ is an
isometry and
$\mathcal B_{\nu,R}=\mathcal T_\nu\mathscr E_{\nu,R}$, every
$c\in\mathcal H_{R}^\mathrm{src}$ satisfies
\begin{align}
 &\norm{(\mathcal B_{\nu,R}\otimes
 I_{\mathcal H_{-\nu,R}^\mathrm{src}})c}
 _{H^2_{\nu,\oplus}(\mathcal H_{-\nu,R}^\mathrm{src})}
 \nonumber\\
 &\quad=\norm{(\mathcal T_\nu\otimes I_{L^2(M_{-\nu})})
 \mathscr E_{R}c}
 _{H^2_{\nu,\oplus}(L^2(M_{-\nu}))}\nonumber\\
 &\quad\leq C_{\mathcal T,\nu,2}^\mathrm{geom}
 \norm{\mathscr E_{R}c}_{H_{x_{\nu}}^2(L^2(M_{-\nu}))}.
 \label{eq:tensor-local-Hilbert-valued-localization-bound}
\end{align}
{In the preceding notation,
$H_{x_\nu}^2(L^2(M_{-\nu}))=H_{x_\nu}^2(H_{-\nu}^0)$.}

All factor operators and maps in the following lemmas are tensored with
the identity on $\mathcal H_{-\nu,R}^\mathrm{src}$; {this identity factor is from the notation.  The scalar factor estimates extend with
the same constants: expand in an orthonormal basis of
$\mathcal H_{-\nu,R}^\mathrm{src}$, apply the scalar estimate
componentwise, and sum by Parseval and Minkowski.

\begin{lemma}[Continuum Jackson smoothing error]
\label{lem:tensor-local-continuum-Jackson-smoothing-error}
Let
\begin{equation*}
 C_{p_0}^\mathrm{J}\coloneqq 4+3p_0+\frac94p_0^2,
 \qquad
 n_\mathrm{J}\geq
 \max\left\{2,c_{p_0}
     \left(C_{p_0}^\mathrm{J}\mathcal G_{\nu,d+2}\right)^{1/d}\right\}.
\end{equation*}
Then 
$c\in\mathcal H_R^\mathrm{src}$ satisfies
\begin{equation}
 \norm{(K_\nu^\mathrm{cont}-K_{\nu,\mathrm{J}}^\mathrm{cont})c}_2
 \leq C_{\mathcal T,\nu,2}^\mathrm{geom}A_{c,J}
 \left(\frac{c_{p_0}}{n_\mathrm{J}}\right)^d
 \mathcal G_{\nu,d+2}
 \norm{\mathscr E_{R}c}_{H_{x_{\nu}}^2(L^2(M_{-\nu}))}.
 \label{eq:tensor-local-factor-Jackson-consistency-bound},
\end{equation}
where $A_{c,J} = \left(1+6p_0^2\left[
    2(\mathcal G_{\nu,2}+1)(\mathcal G_{\nu,1}+1)
    +(\mathcal G_{\nu,2}+1)^2\right]\right)(4+3p_0+\frac94p_0^2)$.
\end{lemma}

\begin{proof}
The map
$\mathcal B_{\nu,R}=\mathcal T_\nu\mathscr E_{\nu,R}$ is an
$L^2$ isometry.  Hence its adjoint is an $L^2$ contraction, and
\begin{align*}
 &\norm{(K_\nu^\mathrm{cont}-K_{\nu,\mathrm{J}}^\mathrm{cont})c}_2\\
 &\quad=\norm{\mathcal B_{\nu,R}^\dagger
     (\mathcal K_\nu^\oplus-\mathcal K_{\nu,\mathrm{J}}^\oplus)
     \mathcal B_{\nu,R}c}_2\\
 &\quad\leq\varepsilon_{K,\nu}^\mathrm{J}
 \norm{\mathcal B_{\nu,R}c}_{H^2_{\nu,\oplus}}
 \leq C_{\mathcal T,\nu,2}^\mathrm{geom}
 \varepsilon_{K,\nu}^\mathrm{J}
 \norm{\mathscr E_{R}c}_{\mathcal X_\nu^2},
\end{align*}
where the suppressed identity factors are understood and \begin{equation}
 \varepsilon_{K,\nu}^\mathrm{J}
 \coloneqq \norm{\mathcal K_\nu^\oplus-\mathcal K_{\nu,\mathrm{J}}^\oplus}
 _{H^2_{\nu,\oplus}\to L^2_{\nu,\oplus}}.
 \label{eq:tensor-local-Jackson-kinetic-error-definition}
\end{equation} The second inequality is
\eqref{eq:tensor-local-Hilbert-valued-localization-bound}.

{The estimate of
$\varepsilon_{K,\nu}^\mathrm{J}$ follows the local-operator argument in the
proof of Theorem~\ref{thm:coherent-chart-continuum-error}. Define the
coefficient envelopes}
\begin{align}
 \mathsf H_{\nu,2}^\mathrm{env}
 &\coloneqq \max_{a\in\mathcal I_\nu}
     \max\left\{
        \norm{h_a^{(\nu)}}_{W^{2,\infty}},
        \norm{h_{a,\mathrm{J}}^{(\nu)}}_{W^{2,\infty}}
     \right\},\nonumber\\
 \mathsf C_{\nu,1}^\mathrm{env}
 &\coloneqq \max_{a\in\mathcal I_\nu}
     \max_{1\leq i,j\leq p_\nu}
     \max\left\{
        \norm{C_a^{(\nu),ij}}_{W^{1,\infty}},
        \norm{C_{a,\mathrm{J}}^{(\nu),ij}}_{W^{1,\infty}}
     \right\}.
 \label{eq:tensor-local-Jackson-coefficient-envelopes}
\end{align}

For scalar functions $\eta_L,\eta_R$ and a matrix field $A=(A^{ij})$, put
\begin{equation}
 \mathscr T_a^{(\nu)}[\eta_L,A,\eta_R]
 \coloneqq M_{\eta_L}\sum_{i,j=1}^{p_\nu}
     \partial_i^\dagger M_{A^{ij}}\partial_jM_{\eta_R}.
 \label{eq:tensor-local-kinetic-trilinear-form}
\end{equation}
The same six-term product-rule expansion as in
\eqref{eq:continuum-coherent-kinetic-six-term-expansion} gives, for
$u\in H^2(S_\nu)$,
\begin{align}
 \norm{\mathscr T_a^{(\nu)}[\eta_L,A,\eta_R]u}_2
 \leq{}&6p_\nu^2\norm{\eta_L}_\infty
 \left(\max_{i,j}\norm{A^{ij}}_{W^{1,\infty}}\right)
 \norm{\eta_R}_{W^{2,\infty}}\norm u_{H^2}.
 \label{eq:tensor-local-kinetic-trilinear-bound}
\end{align}
Indeed, after writing $\partial_i^\dagger=-\partial_i$, each $(i,j)$ summand
contains two terms involving a derivative of $A^{ij}$ and four involving
$A^{ij}$ itself; the factors multiplying $u$, $\partial_i u$,
$\partial_j u$, or $\partial_i\partial_j u$ are bounded by the three
coefficient norms displayed above.  Summing the six terms and the
$p_\nu^2$ index pairs proves
\eqref{eq:tensor-local-kinetic-trilinear-bound}.

For one chart, the kinetic difference has the exact telescoping expansion
\begin{align}
 &\mathscr T_a^{(\nu)}[h_a^{(\nu)},C_a^{(\nu)},h_a^{(\nu)}]
 -\mathscr T_a^{(\nu)}[h_{a,\mathrm{J}}^{(\nu)},
                C_{a,\mathrm{J}}^{(\nu)},h_{a,\mathrm{J}}^{(\nu)}]\nonumber\\
 &\quad=\mathscr T_a^{(\nu)}[h_a^{(\nu)}-h_{a,\mathrm{J}}^{(\nu)},
                C_a^{(\nu)},h_a^{(\nu)}]\nonumber\\
 &\qquad+\mathscr T_a^{(\nu)}[h_{a,\mathrm{J}}^{(\nu)},
                C_a^{(\nu)}-C_{a,\mathrm{J}}^{(\nu)},h_a^{(\nu)}]\nonumber\\
 &\qquad+\mathscr T_a^{(\nu)}[h_{a,\mathrm{J}}^{(\nu)},
                C_{a,\mathrm{J}}^{(\nu)},h_a^{(\nu)}-h_{a,\mathrm{J}}^{(\nu)}].
 \label{eq:tensor-local-Jackson-kinetic-telescoping}
\end{align}
Applying \eqref{eq:tensor-local-kinetic-trilinear-bound} to these three
terms and bounding the IMS multiplier difference directly gives
\begin{align}
 \varepsilon_{K,\nu}^\mathrm{J}
 \leq{}&6p_\nu^2\left[
    2\mathsf H_{\nu,2}^\mathrm{env}\mathsf C_{\nu,1}^\mathrm{env}
         \max_{a\in\mathcal I_\nu}
         \norm{h_a^{(\nu)}-h_{a,\mathrm{J}}^{(\nu)}}_{W^{2,\infty}}
    +(\mathsf H_{\nu,2}^\mathrm{env})^2
         \max_{\substack{a\in\mathcal I_\nu\\1\leq i,j\leq p_\nu}}
         \norm{C_a^{(\nu),ij}-C_{a,\mathrm{J}}^{(\nu),ij}}
             _{W^{1,\infty}}\right]\nonumber\\
 &\quad+\max_{a\in\mathcal I_\nu}
         \norm{W_a^{(\nu)}-W_{a,\mathrm{J}}^{(\nu)}}_\infty.
 \label{eq:tensor-local-Jackson-kinetic-error-expanded}
\end{align}

Define
\begin{align}
 \Delta_{\nu,\mathrm{J}}
 \coloneqq \max_{a\in\mathcal I_\nu}\Bigl[{}
 &\norm{h_a^{(\nu)}-h_{a,\mathrm{J}}^{(\nu)}}_{W^{2,\infty}}
 +\max_{i,j}\norm{C_a^{(\nu),ij}
             -C_{a,\mathrm{J}}^{(\nu),ij}}_{W^{1,\infty}}\nonumber\\
 &{}+\norm{W_a^{(\nu)}-W_{a,\mathrm{J}}^{(\nu)}}_\infty
 \Bigr],
 \label{eq:tensor-local-Jackson-geometry-defect}
\end{align}
then 
\begin{align*}
     \varepsilon_{K,\nu}^\mathrm{J}
 \leq \left(1+6p_\nu^2\left[
    2\mathsf H_{\nu,2}^\mathrm{env}\mathsf C_{\nu,1}^\mathrm{env}
    +(\mathsf H_{\nu,2}^\mathrm{env})^2\right]\right)\Delta_{\nu,\mathrm{J}}.
\end{align*}

Let $\xi_\mathrm{J}=\frac{c_{p_0}}{n_\mathrm{J}}$.
Using $r=d+2$ in
\eqref{eq:tensor-local-high-order-Jackson-bound}, with
$(q,\ell)=(2,d)$ for $h$, $(q,\ell)=(1,d+1)$ for $C$, and
$(q,\ell)=(0,d+2)$ for $W$, yields
\begin{align}
 \max_{a\in\mathcal I_\nu}
 \norm{h_a^{(\nu)}-h_{a,\mathrm{J}}^{(\nu)}}_{W^{2,\infty}}
 &\leq4\xi_\mathrm{J}^{d}\mathcal G_{\nu,d+2},\nonumber\\
 \max_{\substack{a\in\mathcal I_\nu\\1\leq i,j\leq p_\nu}}
 \norm{C_a^{(\nu),ij}-C_{a,\mathrm{J}}^{(\nu),ij}}_{W^{1,\infty}}
 &\leq2\xi_\mathrm{J}^{d+1}\mathcal G_{\nu,d+2},\nonumber\\
 \max_{a\in\mathcal I_\nu}
 \norm{W_a^{(\nu)}-W_{a,\mathrm{J}}^{(\nu)}}_\infty
 &\leq\xi_\mathrm{J}^{d+2}\mathcal G_{\nu,d+2},
 \label{eq:tensor-local-Jackson-componentwise-geometry-bounds}
\end{align}
{where
\eqref{eq:tensor-local-geometric-regularity-envelope} was used.}
Consequently,
\begin{equation}
 \Delta_{\nu,\mathrm{J}}
 \leq(4+2\xi_\mathrm{J}+\xi_\mathrm{J}^2)
     \xi_\mathrm{J}^{d}\mathcal G_{\nu,d+2}.
\end{equation}
{Since $n_\mathrm{J}\geq2$,
$\xi_\mathrm{J}\leq c_{p_0}/2$; using $c_{p_0}=3p_0$ gives}
\begin{equation}
 \Delta_{\nu,\mathrm{J}}
 \leq(4+3p_0+\frac94p_0^2)
     \left(\frac{c_{p_0}}{n_J}\right)^{d}\mathcal G_{\nu,d+2}.
 \label{eq:tensor-local-Jackson-geometry-bound-before-uniformization}
\end{equation}
The $\mathsf H_{\nu,2}^\mathrm{env}$ and $\mathsf C_{\nu,1}^\mathrm{env}$ themselves can be bounded in terms of the non-smoothed versions plus  error $\Delta_{\nu, J}$:
\begin{equation}
 \mathsf H_{\nu,2}^\mathrm{env}
 \leq\mathcal G_{\nu,2}+\Delta_{\nu,\mathrm{J}},
 \qquad
 \mathsf C_{\nu,1}^\mathrm{env}
 \leq\mathcal G_{\nu,1}+\Delta_{\nu,\mathrm{J}}.
 \label{eq:tensor-local-Jackson-envelope-original-data}
\end{equation}
For example, once $\Delta_{\nu,\mathrm{J}}\leq1$ (which is satisfied by the lemma's asymptotic condition on $n_J$), one has a bound that depends only on the non-smoothed versions:
\begin{equation}
1+6p_\nu^2\left[
    2\mathsf H_{\nu,2}^\mathrm{env}\mathsf C_{\nu,1}^\mathrm{env}
    +(\mathsf H_{\nu,2}^\mathrm{env})^2\right]
 \leq1+6p_0^2\left[
    2(\mathcal G_{\nu,2}+1)(\mathcal G_{\nu,1}+1)
    +(\mathcal G_{\nu,2}+1)^2\right].
 \label{eq:tensor-local-Jackson-coefficient-constant-original-data}
\end{equation}

\end{proof}

{The discrete analogue compares the source-space
operators with raw and Jackson-smoothed padded chart blocks:}
\begin{align}
\label{eqn:kinetic_op_disc_no_smooth}
    {K^{\textup{src}}_{\nu, R, N}
    = \mathcal{A}_{\nu}^{*}B_{\nu,N}^\mathrm{pad}\mathcal{A}_{\nu},}
\end{align}
and
\begin{align}
\label{eqn:kinetic_op_disc_smooth}
    {K^{\textup{src}, J}_{\nu, R, N}
    = \mathcal{A}_{\nu}^{*}B_{\nu,N}^{\mathrm{J},\mathrm{pad}}
      \mathcal{A}_{\nu}.}
\end{align}

\begin{lemma}[Source-space Jackson smoothing error]
\label{lem:tensor-local-source-Jackson-smoothing-error}
    Suppose $\mathcal A_\nu$ is a
    $\delta_\nu$-approximate isometry and $n_\mathrm{J}$ satisfies the explicit
    condition in Lemma~\ref{lem:tensor-local-continuum-Jackson-smoothing-error}.
    Then every $c\in\mathcal H_{R}^\mathrm{src}$ satisfies
    \begin{align}
        \norm{(K_{\nu,R,N}^{\mathrm{src},\mathrm{J}} - K_{\nu,R,N}^\mathrm{src})c}_2 \leq (1+\delta_\nu) A_{d,J} \left(\frac{c_{p_0}}{n_\mathrm{J}}\right)^{d+2} \mathcal G_{\nu,d+2}\norm c_2,
     \label{eq:tensor-local-factor-finite-Jackson-bound}
    \end{align}
    where
    \begin{equation}
        A_{d,J}\coloneqq 3\left(1+3p_0^2(2\pi N_\mathrm{pad})^2 \left(1+\max_{1\leq\mu\leq m}
        \mathcal G_{\mu,0}\right)^2\right).
        \label{eq:tensor-local-defined-discrete-Jackson-constant}
    \end{equation}
\end{lemma}

\begin{proof}
The definitions of the two source matrices give
\begin{align*}
 &\norm{(K_{\nu,R,N}^{\mathrm{src},\mathrm{J}}
                    -K_{\nu,R,N}^\mathrm{src})c}_2\\
 &\quad=\norm{\mathcal A_\nu^\dagger
 {(B_{\nu,N}^{\mathrm{J},\mathrm{pad}}
     -B_{\nu,N}^\mathrm{pad})}\mathcal A_\nu c}_2\\
 &\quad\leq\norm{\mathcal A_\nu}^2
 \varepsilon_{K,\nu}^{\mathrm{J},\mathrm{grid}}\norm c_2,
\end{align*}
where {$
\varepsilon_{K,\nu}^{\mathrm{J},\mathrm{grid}}
\coloneqq \lVert B_{\nu,N}^{\mathrm{J},\mathrm{pad}}-B_{\nu,N}^\mathrm{pad}
\rVert_{2\to2}$.}

The $\delta_{\nu}$-approximate isometry property gives
$\norm{\mathcal A_\nu}^2\leq1+\delta_\nu$.
Define the bounds
\begin{align}
 \mathsf H_{\nu,0}^\mathrm{env}
 &\coloneqq \max_{a\in\mathcal I_\nu}
    \max\left\{
      \norm{h_a^{(\nu)}}_\infty,
      \norm{h_{a,\mathrm{J}}^{(\nu)}}_\infty
    \right\},\nonumber\\
 \mathsf C_{\nu,0}^\mathrm{env}
 &\coloneqq \max_{a\in\mathcal I_\nu}
    \max_{1\leq i,j\leq p_\nu}
    \max\left\{
      \norm{C_a^{(\nu),ij}}_\infty,
      \norm{C_{a,\mathrm{J}}^{(\nu),ij}}_\infty
    \right\}.
 \label{eq:tensor-local-Jackson-grid-coefficient-envelopes}
\end{align}
For scalar grid functions $\eta_L,\eta_R$ and a matrix-valued grid
function $A=(A^{ij})$, define the single-grid discrete trilinear block
\begin{align}
 \widetilde{\mathscr T}_{a,L}^{(\nu)}
      [\eta_L,A,\eta_R]
 &\coloneqq M_{\eta_L}\sum_{i,j=1}^{p_\nu}
     (D_{a,L,i}^{(\nu)})^\dagger M_{A^{ij}}D_{a,L,j}^{(\nu)}
     M_{\eta_R}.
 \label{eq:tensor-local-single-grid-kinetic-trilinear-form}
\end{align}
Exact diagonal evaluation satisfies
$\norm{M_c}_{2\to2}\leq\norm c_\infty$, and
\begin{equation}
 \norm{D_{a,L,i}^{(\nu)}}_{2\to2}\leq2\pi L.
\end{equation}
Hence submultiplicativity gives
\begin{align}
 \norm{\widetilde{\mathscr T}_{a,L}^{(\nu)}
      [\eta_L,A,\eta_R]}_{2\to2}
 \leq{}&p_\nu^2(2\pi L)^2
   \norm{\eta_L}_\infty
   \left(\max_{i,j}\norm{A^{ij}}_\infty\right)
   \norm{\eta_R}_\infty.
 \label{eq:tensor-local-single-grid-kinetic-trilinear-bound}
\end{align}

For one chart, the difference between the kinetic parts of the raw and
Jackson cores has the exact telescoping expansion
\begin{align}
 &\widetilde{\mathscr T}_{a,L}^{(\nu)}
     [h_a^{(\nu)},C_a^{(\nu)},h_a^{(\nu)}]
 -\widetilde{\mathscr T}_{a,L}^{(\nu)}
     [h_{a,\mathrm{J}}^{(\nu)},C_{a,\mathrm{J}}^{(\nu)},
                        h_{a,\mathrm{J}}^{(\nu)}]\nonumber\\
 &\quad=\widetilde{\mathscr T}_{a,L}^{(\nu)}
     [h_a^{(\nu)}-h_{a,\mathrm{J}}^{(\nu)},
      C_a^{(\nu)},h_a^{(\nu)}]\nonumber\\
 &\qquad+\widetilde{\mathscr T}_{a,L}^{(\nu)}
     [h_{a,\mathrm{J}}^{(\nu)},
      C_a^{(\nu)}-C_{a,\mathrm{J}}^{(\nu)},h_a^{(\nu)}]\nonumber\\
 &\qquad+\widetilde{\mathscr T}_{a,L}^{(\nu)}
     [h_{a,\mathrm{J}}^{(\nu)},C_{a,\mathrm{J}}^{(\nu)},
      h_a^{(\nu)}-h_{a,\mathrm{J}}^{(\nu)}].
 \label{eq:tensor-local-Jackson-single-grid-kinetic-telescoping}
\end{align}
The IMS branches differ by
$-M_{W_a^{(\nu)}-W_{a,\mathrm{J}}^{(\nu)}}$.  Applying
\eqref{eq:tensor-local-single-grid-kinetic-trilinear-bound} to the three
terms above and taking the maximum over charts yields
\begin{align}
 \max_{a\in\mathcal I_\nu}
 {\norm{ B_{a,L}- B_{a,L}^\mathrm{J}}_{2\to2}}
 &\leq{}p_\nu^2(2\pi L)^2\Bigg[
  2\mathsf H_{\nu,0}^\mathrm{env}\mathsf C_{\nu,0}^\mathrm{env}
  \max_{a\in\mathcal I_\nu}
     \norm{h_a^{(\nu)}-h_{a,\mathrm{J}}^{(\nu)}}_\infty\nonumber\\
 &\quad
  +(\mathsf H_{\nu,0}^\mathrm{env})^2
  \max_{\substack{a\in\mathcal I_\nu\\1\leq i,j\leq p_\nu}}
     \norm{C_a^{(\nu),ij}-C_{a,\mathrm{J}}^{(\nu),ij}}_\infty
 \Bigg]\nonumber\\
 &\quad+\max_{a\in\mathcal I_\nu}
     \norm{W_a^{(\nu)}-W_{a,\mathrm{J}}^{(\nu)}}_\infty.
 \label{eq:tensor-local-Jackson-single-grid-error-expanded}
\end{align}

Define
\begin{align}
 \Delta^{(0)}_{\nu,\mathrm{J}}
 \coloneqq \max_{a\in\mathcal I_\nu}\Bigl[{}
 &\norm{h_a^{(\nu)}-h_{a,\mathrm{J}}^{(\nu)}}_\infty
 +\max_{i,j}\norm{C_a^{(\nu),ij}
             -C_{a,\mathrm{J}}^{(\nu),ij}}_\infty +\norm{W_a^{(\nu)}-W_{a,\mathrm{J}}^{(\nu)}}_\infty
 \Bigr].
 \label{eq:tensor-local-Jackson-grid-coefficient-defect}
\end{align}
    Consequently,
    \begin{align}
     \max_{a\in\mathcal I_\nu}
     \norm{ B_{a,L}- B_{a,L}^\mathrm{J}}_{2\to2}
     \leq \left(1+p_\nu^2(2\pi L)^2
     \left[2\mathsf H_{\nu,0}^\mathrm{env}
                \mathsf C_{\nu,0}^\mathrm{env}
          +(\mathsf H_{\nu,0}^\mathrm{env})^2\right]\right)
     \Delta^{(0)}_{\nu,\mathrm{J}}.
     \label{eq:tensor-local-Jackson-single-grid-error-bound}
    \end{align}
    
    The bound on $\Delta^{(0)}_{\nu,\mathrm{J}}$
    follows the proof of Lemma~\ref{lem:tensor-local-continuum-Jackson-smoothing-error}, using
    $(q,\ell)=(0,d+2)$ for each of $h$, $C$, and $W$.  Each of the three summands in \eqref{eq:tensor-local-Jackson-grid-coefficient-defect} is at most $\xi_\mathrm{J}^{d+2}\mathcal G_{\nu,d+2}$ (see proof of Lemma \ref{lem:tensor-local-continuum-Jackson-smoothing-error}), so their sum contributes the factor three. Hence,
    \begin{align*}
        \Delta^{(0)}_{\nu,\mathrm{J}}
        &\leq 3\left(\frac{c_{p_0}}{n_\mathrm{J}}\right)^{d+2} \mathcal G_{\nu,d+2}.
    \end{align*}
    Each of the three component maxima is bounded by
    $\Delta^{(0)}_{\nu,\mathrm{J}}$. 

    The smoothing defect $\Delta^{(0)}_{\nu,\mathrm{J}}$ also allows us to bound the smoothed
    coefficient envelopes in terms of the raw coefficients:
    \begin{equation}
     \mathsf H_{\nu,0}^\mathrm{env}
     \leq\mathcal G_{\nu,0}+\Delta^{(0)}_{\nu,\mathrm{J}},
     \qquad
     \mathsf C_{\nu,0}^\mathrm{env}
     \leq\mathcal G_{\nu,0}+\Delta^{(0)}_{\nu,\mathrm{J}}.
     \label{eq:tensor-local-Jackson-grid-envelope-original-data}
    \end{equation}
    Moreover, $\Delta^{(0)}_{\nu,\mathrm{J}}\leq\Delta_{\nu,\mathrm{J}}$.
    Thus, if $\Delta_{\nu,\mathrm{J}}\leq1$, then
    $\Delta^{(0)}_{\nu,\mathrm{J}}\leq1$ as well.  
    
    Taking $L=N_\mathrm{pad}$, the envelope bounds above imply
    \begin{equation*}
     2\mathsf H_{\nu,0}^\mathrm{env}\mathsf C_{\nu,0}^\mathrm{env}
     +\left(\mathsf H_{\nu,0}^\mathrm{env}\right)^2
     \leq3(\mathcal G_{\nu,0}+1)^2.
    \end{equation*}
    Combining this estimate with
    \eqref{eq:tensor-local-padding-nonexpansive-comparison} and the bound on
    $\Delta_{\nu,\mathrm{J}}^{(0)}$ gives
    \begin{equation*}
     \varepsilon_{K,\nu}^{\mathrm{J},\mathrm{grid}}
     \leq A_{d,J}
     \left(\frac{c_{p_0}}{n_\mathrm{J}}\right)^{d+2}
     \mathcal G_{\nu,d+2}.
    \end{equation*}
    Finally, $\norm{\mathcal A_\nu}^2\leq1+\delta_\nu$ proves
    \eqref{eq:tensor-local-factor-finite-Jackson-bound}.
\end{proof}
No Jackson smoothing of the potential is required. It remains to bound the local Fourier contribution
in \eqref{eqn:local_fourier_error}.

\paragraph*{Local Fourier Error -- Kinetic}
{The local Fourier error is the difference between
the Jackson-smoothed operators in
\eqref{eqn:kinetic_op_cont_smooth} and
\eqref{eqn:kinetic_op_disc_smooth}. An intermediate Galerkin operator,
analogous to that of Section~\ref{sec:analysis_cont_coherent_chart},
isolates this error from aliasing.}

Let $F_{a,N_\nu}^{(\nu)}$ denote the unitary DFT on the factor grid.  The version of $\mathcal{A}_{\nu}$ without aliasing is the following
\begin{equation}
 \mathcal A_\nu^\mathrm{G}c
 \coloneqq \bigoplus_{a\in\mathcal I_\nu}
 (F_{a,N_\nu}^{(\nu)})^{\dagger}
 {\left(
 \widehat{(\mathcal B_{\nu,R_\nu}c)_a}(k)
 \right)_{k\in\mathcal B_{\nu,N_\nu}}.}
 \label{eq:tensor-local-factor-exact-analysis-map}
\end{equation}
Aliasing is avoided because $\widehat{\cdot}$ denotes an exact computation of the Fourier coefficients. This allows us to define an analogous version of the Galerkin kinetic operator from the previous Section \ref{sec:analysis_cont_coherent_chart} but with the source interpolation:
{
For each factor chart, let
\begin{equation}
 P_{a,N}^{(\nu)}:L^2(\mathbb T^{p_\nu})
 \longrightarrow
 \operatorname{span}\left\{e^{2\pi i k\cdot y}:
 k\in\mathcal B_{\nu,N}\right\}
 \label{eq:tensor-local-digital-retained-projector}
\end{equation}
be the orthogonal Fourier projector, and set
$P_{\nu,N}^\oplus\coloneqq \bigoplus_{a\in\mathcal I_\nu}P_{a,N}^{(\nu)}$.
This digital projector retains exactly $(2N)^{p_\nu}$ modes and is
distinct from the symmetric continuum projector $P_N$ used in
Theorem~\ref{thm:coherent-chart-continuum-error}.}
\begin{align}
 K_{\nu,R_\nu,N_\nu}^\mathrm{G}
 {\coloneqq  (\mathcal A_\nu^\mathrm{G})^\dagger
 B_{\nu,N_\nu}^\mathrm{G}}
     \mathcal A_\nu^\mathrm{G},    
\end{align}
where {$B_{\nu,N}^\mathrm{G}
\coloneqq P_{\nu,N}^{\oplus}\mathcal K_{\nu,\mathrm{J}}^\oplus
P_{\nu,N}^{\oplus}$.}

{Jackson smoothing permits the factorwise
de-aliasing argument of Section~\ref{subsec:flat-torus}. With sufficient
padding, the Jackson-smoothed padded block
$B_{\nu,N}^{\mathrm{J},\mathrm{pad}}$ equals the exact Jackson Galerkin block
$B_{\nu,N}^\mathrm{G}$.}

{
Use the zero-padding isometries $\mathcal Z_a^{(\nu)}$ already defined in
\eqref{eq:tensor-local-factor-oversampled-DFT}. Relative to the retained
grid, the sampled diagonal matrices and differentiation matrices in the
raw chart block are evaluated on the $N_\mathrm{pad}$ grid. Now the
additional Jackson-smoothed padded discrete kinetic operator by}
{
\begin{equation}
 B_{\nu,N}^{\mathrm{J},\mathrm{pad}}
 \coloneqq {\bigoplus}_{a\in\mathcal I_\nu}
 (\mathcal Z_a^{(\nu)})^{\dagger}
 B_{a,N_\mathrm{pad}}^\mathrm{J}
 \mathcal Z_a^{(\nu)}.
 \label{eq:tensor-local-Jackson-padded-kinetic-block}
\end{equation}
}
Because $\mathcal Z_a^{(\nu)}$ is an isometry and its adjoint is a
contraction,
\begin{align}
 \norm{B_{\nu,N}^\mathrm{pad}
                    -B_{\nu,N}^{\mathrm{J},\mathrm{pad}}}
 _{2\rightarrow 2}
 \leq\max_{a\in\mathcal I_\nu}
 \norm{ B_{a,N_\mathrm{pad}}
     - B_{a,N_\mathrm{pad}}^\mathrm{J}}_{2\to2}
 \label{eq:tensor-local-padding-nonexpansive-comparison}.
\end{align}
Hence Lemma \ref{lem:tensor-local-source-Jackson-smoothing-error} also work with for the padded operators by replacing $N$ with $N_{\textup{pad}}$.

The following shows that for sufficient padding the aliasing error is zero.
\begin{proposition}
\label{prop:tensor-local-exact-dealiasing}
If
\begin{equation}
 N_\mathrm{pad}\geq N+3\kappa_\mathrm{J},
 \label{eq:tensor-local-padding-condition}
\end{equation}
then
\begin{equation}
 B_{\nu,N}^{\mathrm{J},\mathrm{pad}}=B_{\nu,N}^\mathrm{G}.
 \label{eq:tensor-local-exact-dealiasing}
\end{equation}
\end{proposition}

When $n_\mathrm{J}=N$, one has
$\kappa_\mathrm{J}=2(N-1)$, so the condition above becomes
$N_\mathrm{pad}\geq N+6(N-1)$.

\begin{proof}
An input in the truncated Fourier space has support in
$[-N,N-1]^{p_\nu}$.  Reading the kinetic product from right to
left, the two $h$ multipliers and the $C^{ij}$ multiplier enlarge the final
kinetic support to
\begin{equation}
 [-N-3\kappa_\mathrm{J},
     N-1+3\kappa_\mathrm{J}]^{p_\nu};
\end{equation}
the intermediate supports use one and two copies of $\kappa_\mathrm{J}$ in
the same way.  Fourier differentiation does not enlarge support, and the
IMS branch lies in
$[-N-\kappa_\mathrm{J},N-1+\kappa_\mathrm{J}]^{p_\nu}$.  The padded
retained set is
$[-N_\mathrm{pad},N_\mathrm{pad}-1]^{p_\nu}$, so both endpoints fit
under exactly the same condition
\eqref{eq:tensor-local-padding-condition}, no intermediate crosses the
padded frequency boundary.  Circular convolution on the padded grid thus
equals ordinary Fourier convolution at every multiplication.  The final
adjoint zero-padding map performs the left Galerkin projection, proving
\eqref{eq:tensor-local-exact-dealiasing}.
\end{proof}

Thus padding changes the result in Lemma \ref{lem:tensor-local-source-Jackson-smoothing-error} through
$N\rightarrow N_\mathrm{pad}$; in particular, the two differentiation matrices
replace the factor $N^2$ by $N_\mathrm{pad}^2$. In particular, take $n_\mathrm{J}=N$ and the minimal padding
$ L =N_\mathrm{pad}=N+6(N-1)\leq7N$. The two differentiation matrices cost
exactly the factor $O(N^2)$ that converts the coefficient approximation
rate $N^{-(d+2)}$ into the required padded-matrix rate $N^{-d}$.

{Padding removes coefficient-multiplication
aliasing but not all aliasing in the method. The localization map
$\mathcal T$ multiplies by a function of infinite bandwidth, and the
source-space localization coefficients are themselves computed from grid
samples. These effects are controlled jointly through the defect}
\begin{equation}
 \eta_{\nu,r}^\mathrm{src}(R,N)
 \coloneqq \sup_{\norm c=1}
 \norm{(I-I_{\nu,N}^{\oplus})
 \mathcal B_{\nu,R}c}_{H^r_{\nu,\oplus}},
 \qquad
 I_{\nu,N}^{\oplus}\coloneqq 
 \bigoplus_{a\in\mathcal I_\nu}I^{(\nu)}_{a,N},
 \label{eq:tensor-local-factor-source-interpolation-defect}
\end{equation}
where $I_{a, N}^{\nu}$ is the Fourier interpolation operator for the $a$-th chart of the $\nu$-th component manifold. 
 For a continuous periodic
function $v:S\to\mathscr X$, where
$S=[-1/2,1/2]^{p_\nu}$ and $\mathscr X$ is a Hilbert space, the Fourier interpolation operator is defined as follows:
\begin{align}
 \widehat v_{a,N}^\mathrm{disc}(k)
 &\coloneqq \frac{1}{(2N)^{p_\nu}}
 \sum_{r\in\mathcal G^{(\nu)}_{a,N}}
 v(y^{(\nu)}_{a,r})
 e^{-2\pi i k\cdot y^{(\nu)}_{a,r}},
 \qquad k\in [-N, \dots, 0 , \dots N - 1]^{p_{\nu}},
 \label{eq:tensor-local-factor-discrete-Fourier-coefficients}\\
 (I^{(\nu)}_{a,N}v)(y)
 &\coloneqq \sum_{k\in[-N, \dots, 0 , \dots N - 1]^{p_{\nu}}}
 \widehat v_{a,N}^\mathrm{disc}(k)e^{2\pi i k\cdot y},
 \qquad y\in S.
 \label{eq:tensor-local-factor-Fourier-interpolant-definition}
\end{align}
{Thus the coefficients indexed by
$\mathcal B_{\nu,N}=[-N,N-1]^{p_\nu}$ are computed by the
$(2N)^{p_\nu}$-point tensor-product DFT and define the corresponding
Fourier polynomial.}

The following shows that a bound on $\eta_{\nu, r}^{\textup{src}}$ leads to a bound on $\delta_{\nu}$ (how close $\mathcal{A}_{\nu}$ is to an isometry).
\begin{lemma}
\label{lem:tensor-local-sampling-isometry}
For the maps $\mathcal A_\nu$ and $\mathcal B_{\nu,R}$ defined in
\eqref{eq:tensor-local-factor-analysis-map} and
\eqref{eq:tensor-local-factor-continuum-analysis}, set
\begin{equation}
 \delta_\nu\coloneqq 2\eta_{\nu,0}^\mathrm{src}
 +(\eta_{\nu,0}^\mathrm{src})^2.
 \label{eq:tensor-local-factor-delta-definition}
\end{equation}
Then
\begin{equation}
 \norm{\mathcal A_\nu^\dagger\mathcal A_\nu-I}\leq\delta_\nu,
 \qquad
 \delta_\mathrm{samp}\coloneqq \sum_{\nu=1}^{m}\delta_\nu.
 \label{eq:tensor-local-factor-sampling-isometry}
\end{equation}
\end{lemma}

\begin{proof}
For $q_c\coloneqq \mathcal B_{\nu,R}c$, the normalized samples in
$\mathcal A_\nu c$ are the grid values of
$I_{\nu,N}^{\oplus}q_c$.  Discrete Parseval therefore gives
$\norm{\mathcal A_\nu c}_2
=\norm{I_{\nu,N}^{\oplus}q_c}_{L^2_\oplus}$, while continuum
localization gives $\norm{q_c}_{L^2_\oplus}=\norm c_2$.  Hence
\begin{align}
 \abs{\norm{\mathcal A_\nu c}_2^2-\norm c_2^2}
 &\leq
 \norm{I_{\nu,N}^{\oplus}q_c-q_c}_{L^2_\oplus}
 \left(2\norm{q_c}_{L^2_\oplus}
 +\norm{I_{\nu,N}^{\oplus}q_c-q_c}_{L^2_\oplus}\right)\nonumber\\
 &\leq\delta_\nu\norm c_2^2.
\end{align}
Taking the supremum over unit $c$ proves the claim.
\end{proof}

 Expanding the product
of the factor maps gives
\begin{equation}
 \norm{\mathcal A^\dagger\mathcal A-I}
 \leq\prod_{\nu=1}^{m}(1+\delta_\nu)-1
 {\leq e^{\sum_{\nu=1}^m\delta_\nu}-1
 =e^{\delta_\mathrm{samp}}-1,}
 \label{eq:tensor-local-product-sampling-isometry}
\end{equation}

\begin{lemma}[Fourier consistency of the Jackson comparator]
\label{lem:tensor-local-smoothed-Fourier-consistency}
Suppose
the padding condition \eqref{eq:tensor-local-padding-condition} is used. Then every
$c\in\mathcal H_{R}^\mathrm{src}$ satisfies
\begin{align}
 &\norm{(K_{\nu,\mathrm{J}}^\mathrm{cont}
             -K_{\nu,R,N}^{\mathrm{src},\mathrm{J}})c}_2
 \nonumber\\
 &\quad\leq\overline A_{\textup{Fourier}, J}
 \left[3C_{\mathcal T,\nu,2}^\mathrm{geom}
 \eta_{\nu,0}^\mathrm{src}
 \norm{\mathscr E_{R}c}_{H^2_{x_\nu}(L^2(M_{-\nu}))}
 +\left(1+2\sqrt{1+\delta_\nu}\right)
 \eta_{\nu,2}^\mathrm{src}\norm c_2\right],
 \label{eq:tensor-local-smoothed-Fourier-consistency}
\end{align}
where {$\overline A_{\textup{Fourier}, J}
\coloneqq 6p_\nu^2(\mathsf H_{\nu,2}^\mathrm{env})^2
 \mathsf C_{\nu,1}^\mathrm{env}
 +\max_{a\in\mathcal I_\nu}
 \norm{W_{a,\mathrm{J}}^{(\nu)}}_\infty$.}
\end{lemma}

{
Under the explicit condition on $n_\mathrm{J}$ in
Lemma~\ref{lem:tensor-local-continuum-Jackson-smoothing-error}, one has
$\Delta_{\nu,\mathrm{J}}\leq1$. Equations
\eqref{eq:tensor-local-Jackson-envelope-original-data} and
\eqref{eq:tensor-local-Jackson-geometry-defect} therefore give
$\mathsf H_{\nu,2}^\mathrm{env}\leq\mathcal G_{\nu,2}+1$,
$\mathsf C_{\nu,1}^\mathrm{env}\leq\mathcal G_{\nu,1}+1$, and
$\max_a\norm{W_{a,\mathrm{J}}^{(\nu)}}_\infty
\leq\mathcal G_{\nu,0}+1$.}
\begin{proof}
{Fix $\nu$ and abbreviate}
\begin{equation}
 \mathcal B\coloneqq \mathcal B_{\nu,R},
 \qquad P\coloneqq P_{\nu,N}^\oplus,
 \qquad \mathcal K_\mathrm{J}\coloneqq \mathcal K_{\nu,\mathrm{J}}^\oplus.
\end{equation}

Let
\begin{equation}
 E_{\nu,N}
 \coloneqq (I-P_{\nu,N}^\oplus)\mathcal B_{\nu,R}.
\end{equation}
Under the unitary retained-Fourier/grid identification implicit in
\eqref{eq:tensor-local-factor-exact-analysis-map}, the exact Galerkin
compression can be written as
\begin{equation}
 K_{\nu,R,N}^\mathrm{G}
 =\mathcal B_{\nu,R}^\dagger P_{\nu,N}^\oplus
    \mathcal K_{\nu,\mathrm{J}}^\oplus P_{\nu,N}^\oplus
    \mathcal B_{\nu,R}.
 \label{eq:tensor-local-factor-Galerkin-identification}
\end{equation}
The identity
\begin{equation}
 \mathcal K_{\nu,\mathrm{J}}^\oplus
 -P_{\nu,N}^\oplus\mathcal K_{\nu,\mathrm{J}}^\oplus
    P_{\nu,N}^\oplus
 =(I-P_{\nu,N}^\oplus)\mathcal K_{\nu,\mathrm{J}}^\oplus
 +P_{\nu,N}^\oplus\mathcal K_{\nu,\mathrm{J}}^\oplus
    (I-P_{\nu,N}^\oplus),
\end{equation}
shows that the exact continuum and retained compressions satisfy the
two-sided identity
\begin{align}
 K_{\nu,\mathrm{J}}^\mathrm{cont}-K_{\nu,R,N}^\mathrm{G}
 ={}&E_{\nu,N}^\dagger\mathcal K_{\nu,\mathrm{J}}^\oplus
     \mathcal B_{\nu,R}\nonumber\\
 &+(P_{\nu,N}^\oplus\mathcal B_{\nu,R})^\dagger
     \mathcal K_{\nu,\mathrm{J}}^\oplus E_{\nu,N}.
 \label{eq:tensor-local-factor-two-sided-Fourier-defect}
\end{align}
This is exactly the factor-source specialization of the continuum
two-sided Galerkin defect
\eqref{eq:continuum-coherent-two-sided-Fourier-defect}, with
$\mathcal T$ replaced by $\mathcal B_{\nu,R}$ and the continuum
generator replaced by $\mathcal K_{\nu,\mathrm{J}}^\oplus$.
The first summand is the ``output Fourier tail''.  The second is
the ``input Fourier tail'', on which the second-order operator acts.
Because $I_{\nu,N}^\oplus\mathcal B_{\nu,R}c$ belongs to the
retained Fourier space and Fourier projection is the best approximation in
each $H^r$ norm,
\begin{equation}
 \norm{E_{\nu,N}}_{\mathcal H_{\nu,R}^\mathrm{src}
 \to H^r_{\nu,\oplus}}
 \leq\eta_{\nu,r}^\mathrm{src},
 \qquad r=0,2.
 \label{eq:tensor-local-source-Fourier-tail-bound}
\end{equation}

Applying
\eqref{eq:continuum-coherent-derived-local-H2-bound}--
\eqref{eq:continuum-coherent-derived-direct-sum-H2-bound} with
$d$ replaced by $p_\nu$ and with the Jackson coefficients, and then
adding the IMS multiplier, gives
\begin{align}
 \norm{\mathcal K_{\nu,\mathrm{J}}^\oplus}
 _{\mathcal L(H^2_{\nu,\oplus},L^2_{\nu,\oplus})} \leq&6p_\nu^2(\mathsf H_{\nu,2}^\mathrm{env})^2
 \mathsf C_{\nu,1}^\mathrm{env}
 +\max_{a\in\mathcal I_\nu}
 \norm{W_{a,\mathrm{J}}^{(\nu)}}_\infty
 \label{eq:tensor-local-factor-smoothed-operator-coefficient-bound}
\end{align}
The map $\mathcal B=\mathcal T_\nu\mathscr E_{\nu,R}$ is an
$L^2$ isometry, so $\norm{\mathcal Bc}_{L^2_{\nu,\oplus}}=\norm c_2$ and
$\norm{\mathcal B^\dagger}_{L^2_{\nu,\oplus}\to
\mathcal H_{\nu,R}^\mathrm{src}}=1$.

Hence for the output tail part:
\begin{equation}
 \norm{E_{\nu,N}^\dagger\mathcal K_{\nu,\mathrm{J}}^\oplus
     \mathcal B_{\nu,R}c}_2
 \leq \left(6p_\nu^2(\mathsf H_{\nu,2}^\mathrm{env})^2
 \mathsf C_{\nu,1}^\mathrm{env}
 +\max_{a\in\mathcal I_\nu}
 \norm{W_{a,\mathrm{J}}^{(\nu)}}_\infty\right)\eta_{\nu,0}^\mathrm{src}
     \norm{\mathcal B_{\nu,R}c}_{H^2_{\nu,\oplus}}.
 \label{eq:tensor-local-factor-test-tail-bound}
\end{equation}
For the input tail,
$P_{\nu,N}^\oplus\mathcal B_{\nu,R}$ is an $L^2$ contraction,
while $\mathcal K_{\nu,\mathrm{J}}^\oplus$ acts directly on the tail.  Hence
the same operator norm
\eqref{eq:tensor-local-factor-smoothed-operator-coefficient-bound} and the $r=2$
estimate are required and give
\begin{equation}
 \norm{(P_{\nu,N}^\oplus\mathcal B_{\nu,R})^\dagger
     \mathcal K_{\nu,\mathrm{J}}^\oplus E_{\nu,N}c}_2
 \leq \left(6p_\nu^2(\mathsf H_{\nu,2}^\mathrm{env})^2
 \mathsf C_{\nu,1}^\mathrm{env}
 +\max_{a\in\mathcal I_\nu}
 \norm{W_{a,\mathrm{J}}^{(\nu)}}_\infty\right)\eta_{\nu,2}^\mathrm{src}\norm c_2.
 \label{eq:tensor-local-factor-trial-tail-bound}
\end{equation}
Put
\begin{equation*}
 D_\nu\coloneqq \mathcal A_\nu^\mathrm{G}-\mathcal A_\nu.
\end{equation*}
Adding and subtracting
$\mathcal A_\nu^\dagger B_{\nu,N}^\mathrm{G}\mathcal A_\nu$ gives
\begin{align}
 K_{\nu,R,N}^\mathrm{G}
 -\mathcal A_\nu^\dagger B_{\nu,N}^{\mathrm{J},\mathrm{pad}}\mathcal A_\nu
 ={}&D_\nu^\dagger B_{\nu,N}^\mathrm{G}\mathcal A_\nu^\mathrm{G}
 +\mathcal A_\nu^\dagger B_{\nu,N}^\mathrm{G}D_\nu\nonumber\\
 &+\mathcal A_\nu^\dagger
    (B_{\nu,N}^\mathrm{G}-B_{\nu,N}^{\mathrm{J},\mathrm{pad}})
  \mathcal A_\nu.
 \label{eq:tensor-local-factor-pre-dealiasing-defect}
\end{align}
The padding condition gives
$B_{\nu,N}^{\mathrm{J},\mathrm{pad}}=B_{\nu,N}^\mathrm{G}$ exactly, using Proposition~\ref{prop:tensor-local-exact-dealiasing}.
Therefore the last line of
\eqref{eq:tensor-local-factor-pre-dealiasing-defect} vanishes.  After this cancellation,
the remaining smoothed sampling error is
\begin{align}
 K_{\nu,R,N}^\mathrm{G}
 -K_{\nu,R,N}^{\mathrm{src},\mathrm{J}}
 ={}&D_\nu^\dagger B_{\nu,N}^\mathrm{G}\mathcal A_\nu^\mathrm{G}
 +\mathcal A_\nu^\dagger B_{\nu,N}^\mathrm{G}D_\nu
 \label{eq:tensor-local-factor-sampled-kinetic-defect}
\end{align}

Note that $D_\nu c$ is the padded version of 
$(P_{\nu,N}^\oplus-I_{\nu,N}^\oplus)
\mathcal B_{\nu,R}c$.  The triangle inequality and
\eqref{eq:tensor-local-source-Fourier-tail-bound} therefore give
\begin{equation}
 \norm{\mathcal A_\nu^\mathrm{G}-\mathcal A_\nu}
 _{\mathcal H_{\nu,R}^\mathrm{src}
     \to H^r_{\nu,N,\oplus}}
 \leq2\eta_{\nu,r}^\mathrm{src},
 \qquad r=0,2.
 \label{eq:tensor-local-factor-analysis-defects}
\end{equation}
Note that $B_{\nu,N}^\mathrm{G}$ is the padded version of 
$P_{\nu,N}^\oplus\mathcal K_{\nu,\mathrm{J}}^\oplus
P_{\nu,N}^\oplus$. Moreover, Fourier projection is contractive in $H^2$, so
\begin{equation}
 \norm{\mathcal A_\nu^\mathrm{G}c}_{H^2_{\nu,N,\oplus}}
 =\norm{P_{\nu,N}^\oplus\mathcal B_{\nu,R}c}
     _{H^2_{\nu,\oplus}}
 \leq\norm{\mathcal B_{\nu,R}c}_{H^2_{\nu,\oplus}}.
 \label{eq:tensor-local-factor-exact-analysis-H2-bound}
\end{equation}
Using the $r=0$ estimate in the first term of
\eqref{eq:tensor-local-factor-sampled-kinetic-defect} gives
\begin{equation}
 \norm{D_\nu^\dagger B_{\nu,N}^\mathrm{G}
     \mathcal A_\nu^\mathrm{G}c}_2
 \leq2\left(6p_\nu^2(\mathsf H_{\nu,2}^\mathrm{env})^2
 \mathsf C_{\nu,1}^\mathrm{env}
 +\max_{a\in\mathcal I_\nu}
 \norm{W_{a,\mathrm{J}}^{(\nu)}}_\infty\right)\eta_{\nu,0}^\mathrm{src}
     \norm{\mathcal B_{\nu,R}c}_{H^2_{\nu,\oplus}}.
 \label{eq:tensor-local-factor-sampled-test-bound}
\end{equation}
For the other term, Lemma~\ref{lem:tensor-local-sampling-isometry}
implies
$\norm{\mathcal A_\nu}\leq\sqrt{1+\delta_\nu}$.  The $r=2$ estimate in
\eqref{eq:tensor-local-factor-analysis-defects} therefore gives
\begin{equation}
 \norm{\mathcal A_\nu^\dagger B_{\nu,N}^\mathrm{G}D_\nu c}_2
 \leq2\sqrt{1+\delta_\nu}\,\left(6p_\nu^2(\mathsf H_{\nu,2}^\mathrm{env})^2
 \mathsf C_{\nu,1}^\mathrm{env}
 +\max_{a\in\mathcal I_\nu}
 \norm{W_{a,\mathrm{J}}^{(\nu)}}_\infty\right)
     \eta_{\nu,2}^\mathrm{src}\norm c_2.
 \label{eq:tensor-local-factor-sampled-trial-bound}
\end{equation}
Adding
\eqref{eq:tensor-local-factor-test-tail-bound},
\eqref{eq:tensor-local-factor-trial-tail-bound},
\eqref{eq:tensor-local-factor-sampled-test-bound},
and \eqref{eq:tensor-local-factor-sampled-trial-bound} first gives the
same bound with
$3\eta_{\nu,0}^\mathrm{src}
\norm{\mathcal B_{\nu,R}c}_{H^2_{\nu,\oplus}}$.
Applying \eqref{eq:tensor-local-Hilbert-valued-localization-bound} and
then \eqref{eq:tensor-local-factor-smoothed-operator-coefficient-bound} proves
\eqref{eq:tensor-local-smoothed-Fourier-consistency}.

\end{proof}

\paragraph*{Local Fourier Error -- Potential}

\begin{lemma}[Product-grid interpolation]
\label{lem:tensor-local-product-Fourier-interpolation}
 Let
$v=(v_{\boldsymbol a})_{\boldsymbol a}$ have periodic extensions in
$H^{d+2}(\mathbb T^d)$ on every product chart.  Define
\begin{equation}
 \mathcal V_{\nu,s}(v)^2
 \coloneqq \sum_{\boldsymbol a}
     \sum_{\abs{\alpha}\leq s}
     \norm{\partial_{y_\nu}^{\alpha}v_{\boldsymbol a}}_2^2.
\end{equation}
Then
\begin{equation}
 \norm{(I-\mathsf I_{N}^{\oplus})v}_{\mathcal H_\oplus}
 \leq
 {\frac{\left(1+3^{d/2}\sqrt{\zeta(d+4)}\right)
 m^{(d+2)/2}\sqrt{(d+2)!}}{N^{d+2}}}
 \max_{1\leq\nu\leq m}\mathcal V_{\nu,d+2}(v),
 \label{eq:tensor-local-multidimensional-interpolation-bound}
\end{equation}
where $\zeta$ is the Riemann-zeta function.
\end{lemma}

\begin{proof}

 {Set $s=d+2$. Since $s>d/2$,} weighted Cauchy--Schwarz makes the Fourier coefficients
absolutely summable.  The high-order truncation and multidimensional
aliasing proofs of
\cite[Lemmas~C.5--C.6]{chakrabarti2025speedups} therefore apply chart by
chart.  Using the $\ell_\infty$ norm for the cutoff and
summing the squared estimates over product charts gives
\begin{equation}
 \norm{(I-\mathsf I_{N}^{\oplus})v}_{\mathcal H_\oplus}
 \leq\frac{C_\mathrm{md}(d,s)}{N^s}
 \left[\sum_{\boldsymbol a}
 \norm{v_{\boldsymbol a}}_{H^s(\mathbb T^d)}^2\right]^{1/2},
 \label{eq:tensor-local-even-cited-isotropic-bound}
\end{equation}
with
\begin{equation}
 C_\mathrm{md}(d,s)
 \coloneqq 1+\left(
 \sum_{z\in\mathbb Z^d\setminus\{0\}}
         \norm z_\infty^{-2s}\right)^{1/2}.
 \label{eq:tensor-local-multidimensional-aliasing-constant}
\end{equation}

It remains only to convert the full Sobolev norm into the directional norms
in the statement.  For $k=(k_1,\ldots,k_m)$ grouped by factor, put
$b_\nu(k)\coloneqq 1+4\pi^2\norm{k_\nu}_2^2$.  Convexity and the
binomial--multinomial formula give
\begin{align}
 (1+4\pi^2\norm k_2^2)^s
 &\leq m^{s-1}\sum_{\nu=1}^m b_\nu(k)^s,\nonumber\\
 b_\nu(k)^s
 &\leq s!\sum_{\abs{\alpha}\leq s}
 (2\pi)^{2\abs{\alpha}}\abs{k_\nu^\alpha}^2.
 \label{eq:tensor-local-even-directional-frequency-bound}
\end{align}
Multiplying by $\abs{\widehat v_{\boldsymbol a}(k)}^2$, summing over
frequencies and product charts, and applying Parseval yields
\begin{equation}
 \left[\sum_{\boldsymbol a}
 \norm{v_{\boldsymbol a}}_{H^s(\mathbb T^d)}^2\right]^{1/2}
 \leq m^{s/2}\sqrt{s!}\max_\nu\mathcal V_{\nu,s}(v).
 \label{eq:tensor-local-even-directional-to-isotropic-bound}
\end{equation}
Combining
\eqref{eq:tensor-local-even-cited-isotropic-bound} and
\eqref{eq:tensor-local-even-directional-to-isotropic-bound} proves
\eqref{eq:tensor-local-multidimensional-interpolation-bound}.

Finally, the number of lattice points with $\norm z_\infty=r$ is at most
$3^dr^d$.  At $s=d+2$,
\begin{equation}
 \sum_{z\ne0}\norm z_\infty^{-2d-4}
 \leq3^d\sum_{r=1}^\infty r^{-d-4}
 =3^d\zeta(d+4).
\end{equation}
Hence 
\begin{align*}
    C_\mathrm{md}(d, d+2) \leq 1 + 3^{d/2}\sqrt{\zeta(d+4)}.
\end{align*}

\end{proof}

The error for the potential is a bound on the opertors without applying local Fourier interpolation to each $f_{[\ell]}$, $Q_{R}^\mathrm{src}M_f
 Q_{R}^\mathrm{src}$ and the result of applying Fourir interpolation $\mathscr E_{R}
 M_{f,R,N}^\mathrm{src}
 \mathscr E_{R}^{\dagger}$. Specifically, 
for any path $w:[0,T]\to V_R^\mathrm{src}$, define the potential
discretization error by
\begin{align}
 r_f[w](t)&\coloneqq \left[Q_{R}^\mathrm{src}M_f
 Q_{R}^\mathrm{src}
 -\mathscr E_{R}
 M_{f,R,N}^\mathrm{src}
 \mathscr E_{R}^{\dagger}\right]
 w(t),\nonumber\\
 \epsilon_f[w](T)&\coloneqq 
 \sup_{0\leq t\leq T}\norm{r_f[w](t)}_2.
 \label{eq:tensor-local-potential-error-definition}
\end{align}
The following uses the previous interpolation bound to bound 

\begin{proposition}[Potential error from interpolation]
\label{prop:tensor-local-potential-consistency}
Let $w:[0,T]\to V_R^\mathrm{src}$ be a path such that, for every product
chart $\boldsymbol a$ and $t\in[0,T]$, the function
\begin{equation}
 y\longmapsto
 f(\kappa_{\boldsymbol a}^{-1}(y))
 (\mathcal T w(t))_{\boldsymbol a}(y)
 \label{eq:tensor-local-localized-potential-path}
\end{equation}
has a periodic extension in $H^{d+2}(\mathbb T^d)$.  For derivatives in
the factor coordinate $y_\nu$, define
\begin{align}
 \mathcal U_{f,\nu,r}[w](T)
 \coloneqq \sup_{0\leq t\leq T}
 \left[\sum_{\boldsymbol a}
 \sum_{\abs{\alpha}\leq r}
 \norm{\partial_{y_\nu}^{\alpha}
 \left[f(\kappa_{\boldsymbol a}^{-1}(\mathord\cdot))
 (\mathcal T w(t))_{\boldsymbol a}\right]}_2^2
 \right]^{1/2}.
 \label{eq:tensor-local-potential-mixed-norm}
\end{align}
Set
\begin{align}
 \mathsf C_\mathrm{pot}(d,m)
 &\coloneqq \left(1 + 3^{d/2}\sqrt{\zeta(d+4)}\right)m^{(d+2)/2}\sqrt{(d+2)!},\nonumber\\
 \mathsf S_\mathrm{src-int}
 &\coloneqq \max_{1\leq\nu\leq m}
 \prod_{\mu<\nu}\sqrt{1+\delta_\mu}.
 \label{eq:tensor-local-potential-interpolation-coefficients}
\end{align}
Then
\begin{align}
 \epsilon_f[w](T)
 &\leq
 \frac{\mathsf C_\mathrm{pot}(d,m)}{N^{d+2}}
 \max_{1\leq\nu\leq m}\mathcal U_{f,\nu,d+2}[w](T)
 \nonumber\\
 &\quad+\mathsf S_\mathrm{src-int}
 \left(\sum_{\nu=1}^m\eta_{\nu,0}^\mathrm{src}\right)
 \Bigg[
 \Lambda_f\sup_{0\leq t\leq T}\norm{w(t)}_2
 +\frac{\mathsf C_\mathrm{pot}(d,m)}{N^{d+2}}
 \max_{1\leq\nu\leq m}\mathcal U_{f,\nu,d+2}[w](T)
 \Bigg],
 \label{eq:tensor-local-potential-consistency}
\end{align}
where $\eta^{\textup{src}}_{\nu, 0}$ is defined in \eqref{eq:tensor-local-factor-source-interpolation-defect}.
\end{proposition}

\begin{proof}
Fix $t\in[0,T]$ and a unit vector $v\in V_R^\mathrm{src}$.  Since the
interpolants agree with the localized functions at every grid point and
$\mathcal A$ contains the normalized samples of
$\mathcal T\mathscr E_R$, discrete Parseval gives
\begin{equation}
 \left\langle\mathcal A\mathscr E_R^\dagger v,
 M_{f,N}^\oplus\mathcal A\mathscr E_R^\dagger w(t)\right\rangle
 =\left\langle
 \mathsf I_N^\oplus\mathcal Tv,
 \mathsf I_N^\oplus M_f^\oplus\mathcal Tw(t)
 \right\rangle_{\mathcal H_\oplus}.
 \label{eq:tensor-local-exact-grid-inner-product}
\end{equation}
Using $M_f=\mathcal T^\dagger M_f^\oplus\mathcal T$ and subtracting the right
side of \eqref{eq:tensor-local-exact-grid-inner-product}, we first obtain
the two quantities that must be bounded:
\begin{align}
 &\left\langle v,r_f[w](t)\right\rangle
 \nonumber\\
 &\quad={}
 \left\langle\mathcal Tv,
 (I-\mathsf I_N^\oplus)M_f^\oplus\mathcal Tw(t)
 \right\rangle_{\mathcal H_\oplus}
 \nonumber\\
 &\qquad+\left\langle
 (I-\mathsf I_N^\oplus)\mathcal Tv,
 \mathsf I_N^\oplus M_f^\oplus\mathcal Tw(t)
 \right\rangle_{\mathcal H_\oplus}.
 \label{eq:tensor-local-potential-defect-split}
\end{align}

{For the first term,
Lemma~\ref{lem:tensor-local-product-Fourier-interpolation}}
applied to the function in
\eqref{eq:tensor-local-localized-potential-path} gives
\begin{align}
 &\sup_{0\leq t\leq T}
 \norm{(I-\mathsf I_N^\oplus)M_f^\oplus\mathcal Tw(t)}_{\mathcal H_\oplus}
 \nonumber\\
 &\quad\leq
 \frac{\mathsf C_\mathrm{pot}(d,m)}{N^{d+2}}
 \max_{1\leq\nu\leq m}\mathcal U_{f,\nu,d+2}[w](T).
 \label{eq:tensor-local-potential-interpolation-budget}
\end{align}
Because $\mathcal T$ is an isometry and $\norm v_2=1$, this is also an
upper bound for the absolute value of the first inner product in
\eqref{eq:tensor-local-potential-defect-split}.

{For the second term, write
$v=\mathscr E_Rc$. The identities}
$\mathcal T\mathscr E_R=\bigotimes_\nu\mathcal B_{\nu,R}$ and
$\mathsf I_N^\oplus\mathcal T\mathscr E_R
=\bigotimes_\nu(I_{\nu,N}^\oplus\mathcal B_{\nu,R})$ give
\begin{align}
 &(I-\mathsf I_N^\oplus)\mathcal T\mathscr E_R
 \nonumber\\
 &\quad=\sum_{\nu=1}^m
 \left(\bigotimes_{\mu<\nu}
 I_{\mu,N}^\oplus\mathcal B_{\mu,R}\right)
 \otimes\left[(I-I_{\nu,N}^\oplus)\mathcal B_{\nu,R}\right]
 \otimes\left(\bigotimes_{\mu>\nu}\mathcal B_{\mu,R}\right).
 \label{eq:tensor-local-test-interpolation-telescoping}
\end{align}
Each $\mathcal B_{\mu,R}$ is an $L^2$ isometry.  Discrete Parseval and
the $\delta_{\nu}$-approximate isometry property give
\begin{equation}
 \norm{I_{\mu,N}^\oplus\mathcal B_{\mu,R}}_{2\to2}
 =\norm{\mathcal A_\mu}_{2\to2}
 \leq\sqrt{1+\delta_\mu}.
 \label{eq:tensor-local-interpolated-factor-stability}
\end{equation}
Using \eqref{eq:tensor-local-factor-source-interpolation-defect} in the
middle factor of each summand yields
\begin{align} \label{eq:tensor-local-test-interpolation-error-bound}
    \norm{(I-\mathsf I_N^\oplus)\mathcal Tv}_{\mathcal H_\oplus} &\leq \sum_{\nu=1}^m \left(\prod_{\mu<\nu}\sqrt{1+\delta_\mu}\right)
    \eta_{\nu,0}^\mathrm{src} \nonumber\\
    &\leq \mathsf S_\mathrm{src-int} \sum_{\nu=1}^{m}
    \eta_{\nu,0}^\mathrm{src}.
\end{align}
The product in $\mathsf S_\mathrm{src-int}$ contains only factors preceding
$\nu$ because these factors use interpolated synthesis in
\eqref{eq:tensor-local-test-interpolation-telescoping}; following factors
use exact isometric synthesis.

Finally,
\begin{equation}
 \norm{\mathsf I_N^\oplus M_f^\oplus\mathcal Tw(t)}_{\mathcal H_\oplus}
 \leq\Lambda_f\norm{w(t)}_2
 +\norm{(I-\mathsf I_N^\oplus)M_f^\oplus\mathcal Tw(t)}_{\mathcal H_\oplus}.
\end{equation}
Combining this estimate with the two bounds above, taking the supremum over
unit $v$, and then taking the supremum over $t$ proves
\eqref{eq:tensor-local-potential-consistency}.
\end{proof}

\paragraph*{Combining All Errors}

Let $a,b\in L^1([0,T];\mathbb R)$, let $f$ be real with
$\norm f_\infty\leq\Lambda_f$, and let $u$ solve
$i\partial_tu=H(t)u$ with $u(0)=u_0\in V_R^\mathrm{src}$.  Assume that the
exact and comparison generators below are self-adjoint and generate
unitary propagators, that all comparison evolutions start from $u_0$, and
that the source interpolation bound
\eqref{eq:tensor-local-first-dynamical-comparison}, the three consistency
bounds \eqref{eq:tensor-local-factor-Jackson-consistency-bound},
\eqref{eq:tensor-local-smoothed-Fourier-consistency}, and
\eqref{eq:tensor-local-factor-finite-Jackson-bound}, and the potential
bound \eqref{eq:tensor-local-potential-consistency} hold.  Assume also that
the Sobolev suprema displayed below are finite.

Let $u_R$ solve
\begin{equation*}
 i\partial_tu_R(t)=Q_R^\mathrm{src}H(t)Q_R^\mathrm{src}u_R(t),
 \qquad u_R(0)=u_0.
\end{equation*}
On the source coefficient space set
\begin{align*}
 K_{R,J}^\mathrm{cont}
 &\coloneqq \sum_{\nu=1}^mK_{\nu,J}^\mathrm{cont}\otimes I_{-\nu},
 &K_{R,J,N}^\mathrm{src}
 &\coloneqq \sum_{\nu=1}^mK_{\nu,R,N}^{\mathrm{src},J}\otimes I_{-\nu},\\
 M_{f,R}^\mathrm{cont}
 &\coloneqq \mathscr E_R^\dagger M_f\mathscr E_R.
\end{align*}
Define $u_{R,J}\coloneqq \mathscr E_Rc_{R,J}$ and
$u_{R,J,N}\coloneqq \mathscr E_Rc_{R,J,N}$, where $c_{R,J}$ and $c_{R,J,N}$
are generated by
\begin{align*}
 a(t)K_{R,J}^\mathrm{cont}+b(t)M_{f,R}^\mathrm{cont},
 \qquad
 a(t)K_{R,J,N}^\mathrm{src}+b(t)M_{f,R,N}^\mathrm{src},
\end{align*}
respectively, with initial coefficient vector $\mathscr E_R^\dagger u_0$.
Finally, let $u_{R,N}\coloneqq \mathscr E_Rc_{R,N}$, where $c_{R,N}$ is generated
by $H_{R,N}^\mathrm{src}(t)$ with the same initial coefficient vector.

{The preceding estimates define the following
error budgets. The source-interpolation budget is}
\begin{align*}
 \mathfrak E_\mathrm{src}(T,R)
 &\coloneqq {}
 \varepsilon_0^\mathrm{src}(T,R)
 +A_1\sqrt d\,\varepsilon_2^\mathrm{src}(T,R)
 +B_1\Lambda_f\varepsilon_0^\mathrm{src}(T,R).
\end{align*}
and \eqref{eq:tensor-local-first-dynamical-comparison} states that
\begin{equation}
 \sup_{0\leq t\leq T}\norm{u(t)-u_R(t)}_2
 \leq\mathfrak E_\mathrm{src}(T,R).
 \label{eq:tensor-local-combined-source-bound}
\end{equation}

The continuum Jackson-smoothing bound is
\begin{align*}
 \mathfrak E_{J,\mathrm{cont}}(T,R)
 &\coloneqq {}
 A_1\sum_{\nu=1}^m
 C_{\mathcal T,\nu,2}^\mathrm{geom}A_{c,J}
 \left(\frac{c_{p_0}}{n_\mathrm{J}}\right)^d
 \mathcal G_{\nu,d+2}
 \sup_{0\leq t\leq T}
 \norm{u_R(t)}_{H_{x_\nu}^2(L^2(M_{-\nu}))}.
\end{align*}
Lemma~\ref{lem:tensor-local-continuum-Jackson-smoothing-error} and Duhamel's formula give
\begin{equation}
 \sup_{0\leq t\leq T}\norm{u_R(t)-u_{R,J}(t)}_2
 \leq\mathfrak E_{J,\mathrm{cont}}(T,R).
 \label{eq:tensor-local-combined-continuum-Jackson-bound}
\end{equation}

The kinetic part of the local Fourier bound is
\begin{align*}
 \mathfrak E_\mathrm{F}(T,R,N)
 &\coloneqq {}
 A_1\sum_{\nu=1}^m
 \overline A_{\textup{Fourier},J}
 \Bigg[
 3C_{\mathcal T,\nu,2}^\mathrm{geom}\eta_{\nu,0}^\mathrm{src}
 \sup_{0\leq t\leq T}
 \norm{u_{R,J}(t)}_{H_{x_\nu}^2(L^2(M_{-\nu}))}
 \\
 &\hspace{44mm}
 +\left(1+2\sqrt{1+\delta_\nu}\right)
 \eta_{\nu,2}^\mathrm{src}\norm{u_0}_2
 \Bigg].
\end{align*}
The potential bound from
Proposition~\ref{prop:tensor-local-potential-consistency} is
\begin{align*}
 \mathfrak E_f(T,R,N)
 &\coloneqq {}
 \frac{\mathsf C_\mathrm{pot}(d,m)}{N^{d+2}}
 \max_{1\leq\nu\leq m}
 \mathcal U_{f,\nu,d+2}[u_{R,J}](T)
 \\
 &\quad+\mathsf S_\mathrm{src-int}
 \left(\sum_{\nu=1}^m\eta_{\nu,0}^\mathrm{src}\right)
 \Bigg[
 \Lambda_f\norm{u_0}_2
 +\frac{\mathsf C_\mathrm{pot}(d,m)}{N^{d+2}}
 \max_{1\leq\nu\leq m}
 \mathcal U_{f,\nu,d+2}[u_{R,J}](T)
 \Bigg].
\end{align*}
Consequently, Lemma~\ref{lem:tensor-local-smoothed-Fourier-consistency},
Proposition~\ref{prop:tensor-local-potential-consistency}, and Duhamel's
formula give
\begin{equation}
 \sup_{0\leq t\leq T}\norm{u_{R,J}(t)-u_{R,J,N}(t)}_2
 \leq\mathfrak E_\mathrm{F}(T,R,N)+B_1\mathfrak E_f(T,R,N).
 \label{eq:tensor-local-combined-Fourier-bound}
\end{equation}

Finally, the discrete Jackson-smoothing bound is
\begin{align*}
 \mathfrak E_{J,\mathrm{disc}}(T,R,N)
 &\coloneqq {}
 A_1\sum_{\nu=1}^m
 (1+\delta_\nu)A_{d,J}
 \left(\frac{c_{p_0}}{n_\mathrm{J}}\right)^{d+2}
 \mathcal G_{\nu,d+2}\norm{u_0}_2.
\end{align*}
Lemma~\ref{lem:tensor-local-source-Jackson-smoothing-error} with
Duhamel's formula gives
\begin{equation}
 \sup_{0\leq t\leq T}\norm{u_{R,J,N}(t)-u_{R,N}(t)}_2
 \leq\mathfrak E_{J,\mathrm{disc}}(T,R,N).
 \label{eq:tensor-local-combined-discrete-Jackson-bound}
\end{equation}

\begin{proposition}[Dynamical discretization error]
\label{prop:tensor-local-dynamical-error}
Under the assumptions stated above,
\begin{align}
 \sup_{0\leq t\leq T}\norm{u(t)-u_{R,N}(t)}_2
 &\leq\mathfrak E_\mathrm{src}(T,R)
 +\mathfrak E_{J,\mathrm{cont}}(T,R)
 +\mathfrak E_\mathrm{F}(T,R,N)\nonumber\\
 &\quad+B_1\mathfrak E_f(T,R,N)
 +\mathfrak E_{J,\mathrm{disc}}(T,R,N).
 \label{eq:tensor-local-total-evolution-error}
\end{align}
\end{proposition}

\begin{proof}
The four evolutions satisfy
\begin{align}
 u-u_{R,N}
 ={}&(u-u_R)+(u_R-u_{R,J})\nonumber\\
 &+(u_{R,J}-u_{R,J,N})+(u_{R,J,N}-u_{R,N}).
 \label{eq:tensor-local-four-comparison-split}
\end{align}

Combining \eqref{eq:tensor-local-combined-source-bound},
\eqref{eq:tensor-local-combined-continuum-Jackson-bound},
\eqref{eq:tensor-local-combined-Fourier-bound}, and
\eqref{eq:tensor-local-combined-discrete-Jackson-bound} with
\eqref{eq:tensor-local-four-comparison-split} proves
\eqref{eq:tensor-local-total-evolution-error}.
\end{proof}

\begin{theorem}[{Polynomial Cutoff for Discretized Coherent-Chart Methods}]
\label{thm:tensor-local-main}
    Suppose $M=\prod_{\nu=1}^m\mathcal{M}_\nu$ has product metric $g = \bigoplus_{\nu=1}^mg_\nu$, with $p_\nu \coloneqq \dim \mathcal{M}_\nu\leq p_0=O(1)$ and $m\leq d$. Assume that the numbers
    of factor charts, the chart distortions, and the metric, density, cutoff,
    and IMS data through order two are bounded uniformly in $\nu$ and $d$. Assume also that there are constants $C_R, \kappa_1, \kappa_2, c_R,C_*,\beta_*,\gamma_R>0$, independent of $d$, $N$, and $\nu$,
    such that, for every integer $N\geq C_*d^{\beta_*}$, there is an integer
    source resolution $R(N)$, common to all factors, for which the following
    conditions hold.
    \begin{enumerate}
        \item The assumptions of
        Proposition~\ref{prop:tensor-local-dynamical-error} and of the four
        consistency results used there hold.  In particular, the discrete
        evolution starts from the exact source state, and
        {$\delta_\mathrm{samp}
        \coloneqq \sum_{\nu=1}^m\delta_\nu\leq1$.}
        
        \item The chosen source resolution satisfies $R(N)$ satisfies
        $C_RN^{\kappa_1}d^{\kappa_2} \geq R(N)\geq c_Rd^{-\gamma_R}N$, and the source-projection errors with $R(N)$ scaling this way obey
        \begin{equation*}
        \varepsilon_r^\mathrm{src}(T,R(N))
        \leq (C_*d^{\beta_*})^dR(N)^{-(d+2-r)},
        \qquad r\in\{0,2\}.
        \end{equation*}
        Along the same resolution schedule, the source-to-grid errors separately
        satisfy
        \begin{equation*}
        \eta_{\nu,r}^\mathrm{src}(R(N),N)
        \leq (C_*d^{\beta_*})^dN^{-(d+2-r)},
        \qquad 1\leq\nu\leq m,\quad r\in\{0,2\}.
        \end{equation*}
        
        \item Each quantity in the following display is bounded by
        $(C_*d^{\beta_*})^d$:
        \begin{equation*}
        \begin{aligned}
        &A_1,\quad B_1,\quad \Lambda_f,\quad \norm{u_0}_2,\quad
        \max_{1\leq\nu\leq m}\mathcal G_{\nu,d+2},\\
        &
        \max_{\substack{1\leq\nu\leq m\\v\in\{u_R,u_{R,J}\}}}
        \sup_{0\leq t\leq T}
        \norm{v(t)}_{H_{x_\nu}^2(L^2(M_{-\nu}))},\quad
        \max_{1\leq\nu\leq m}
        \mathcal U_{f,\nu,d+2}[u_{R,J}](T).
        \end{aligned}
        \end{equation*}
    \end{enumerate}
    Then there are constants $C_0\geq1$ and $\beta\geq0$, independent of $d$
    and $N$, such that every integer $N\geq C_0d^\beta$ satisfies
    \begin{equation*}
        \sup_{0\leq t\leq T}\norm{u(t)-u_{R,N}(t)}_2
        \leq\left(\frac{C_0d^\beta}{N}\right)^d.
    \end{equation*}
    Consequently, for every $\epsilon\in(0,1]$, it suffices to choose
    \begin{equation*}
        N=\left\lceil C_0d^\beta\epsilon^{-1/d}\right\rceil
    \end{equation*}
    to obtain error at most $\epsilon$.
\end{theorem}

\begin{proof}
Choose the Jackson order $n_\mathrm{J}=N$.  Since its coefficient bandwidth
is $\kappa_\mathrm{J}=2(N-1)$, choose the minimal de-aliasing padding
\begin{equation*}
 N_\mathrm{pad}=N+3\kappa_\mathrm{J}=N+6(N-1)\leq7N.
\end{equation*}
These choices satisfy the padding condition in
Proposition~\ref{prop:tensor-local-exact-dealiasing} and do not change the
asymptotic resolution.

The uniform factor hypothesis and $p_\nu\leq p_0$ imply
\begin{equation*}
 \sup_{1\leq\nu\leq m}
 C_{\mathcal T,\nu,2}^\mathrm{geom}=O(1).
\end{equation*}
The low-order envelopes
$\mathcal G_{\nu,0}$, $\mathcal G_{\nu,1}$, and
$\mathcal G_{\nu,2}$ are also uniformly bounded.  Inspecting their
definitions in the preceding consistency lemmas therefore gives
\begin{equation*}
 \sup_{1\leq\nu\leq m}A_{c,J}=O(1),
 \qquad
 \sup_{1\leq\nu\leq m}\overline A_{\textup{Fourier},J}=O(1).
\end{equation*}
Finally, $N_\mathrm{pad}\leq7N$ and the displayed formula for $A_{d,J}$ give
\begin{equation*}
 \sup_{1\leq\nu\leq m}\frac{A_{d,J}}{N^2}=O(1).
\end{equation*}
These conclusions use only uniform low-order factor data.  They do not
control the high-order envelope $\mathcal G_{\nu,d+2}$.  They also do not
control the comparison-evolution norms.  Those quantities are assumed
separately above.

The lower bound on $R(N)$ gives
\begin{equation*}
    R(N)^{-d}\leq c_R^{-d}d^{\gamma_Rd}N^{-d}.
\end{equation*}
Therefore the two source terms of order $R^{-d}$, as well as the faster
$R^{-(d+2)}$ terms, are bounded by
$(C_1d^{\beta_1}/N)^d$ after enlarging constants independent of $d$ and
$N$.  Hence $\mathfrak E_\mathrm{src}(T,R)$ has this form.

Since $n_\mathrm{J}=N$, the continuum Jackson factor is $N^{-d}$.  The
coefficient and regularity assumptions, together with the sum over at most
$m\leq d$ factors, therefore give the same bound for
$\mathfrak E_{J,\mathrm{cont}}(T,R)$.

For the local Fourier term, $\eta_{\nu,2}^\mathrm{src}=O(N^{-d})$ and
$\eta_{\nu,0}^\mathrm{src}=O(N^{-(d+2)})$.  The factors
 $1+2\sqrt{1+\delta_\nu}$ are uniformly bounded because
{$\delta_\nu\leq\delta_\mathrm{samp}\leq1$.} Thus $\mathfrak E_\mathrm{F}(T,R,N)$ is also bounded by
$(C_2d^{\beta_2}/N)^d$.

For the potential term, $m\leq d$ and the explicit expression for
$\mathsf C_\mathrm{pot}(d,m)$ imply
$\mathsf C_\mathrm{pot}(d,m)\leq(Cd^2)^d$.  Moreover,
\begin{equation*}
    \mathsf S_\mathrm{src-int} = \max_\nu\prod_{\mu<\nu}\sqrt{1+\delta_\mu} \leq \exp\left(\frac12\sum_{\mu=1}^m\delta_\mu\right) \leq e^{1/2}.
\end{equation*}
The first part of $\mathfrak E_f(T,R,N)$ is proportional to $N^{-(d+2)}$, and its second part contains $\sum_\nu\eta_{\nu,0}^\mathrm{src}=O(N^{-(d+2)})$. Both are therefore bounded by $(C_3d^{\beta_3}/N)^d$.

Finally, $A_{d,J}\leq N^2(C_*d^{\beta_*})^d$ by assumption. Multiplying
this by the discrete Jackson factor $N^{-(d+2)}$ gives $N^{-d}$, so
$\mathfrak E_{J,\mathrm{disc}}(T,R,N)$ has the same form. Absorbing the five
factor sums, products of coefficient envelopes, and fixed constants into
one polynomial base gives constants $C_0$ and $\beta$ such that
\begin{equation*}
    \mathfrak E_\mathrm{src} + \mathfrak E_{J,\mathrm{cont}} + \mathfrak E_\mathrm{F} + B_1\mathfrak E_f + \mathfrak E_{J,\mathrm{disc}} \leq \left(\frac{C_0d^\beta}{N}\right)^d.
\end{equation*}
The result follows from Proposition~\ref{prop:tensor-local-dynamical-error}; the displayed choice of $N$ then makes the right-hand side at most $\epsilon$.
\end{proof}

\section{Applications}
\label{sec:applications}

\subsection{Simulations of Physics}
\label{sec:physics-simulations}
The preceding sections develop quantum algorithms for Hamiltonians of the form
\begin{equation}
\label{eq:intro-hamiltonian}
    \hat H(t)=-a(t)\Delta_G+b(t)\mathcal V,
\end{equation}
acting on $L^2(\mathcal M,\dd\operatorname{vol}_G)$, where $(\mathcal M,G)$ is a Riemannian configuration space and $\mathcal V:\mathcal M\to\mathbb R$ is a real potential.
Here, $\mathcal V$ corresponds to the potential $f$ used in the preceding sections. In this section, we use $G$ for the metric on the full physical configuration space and reserve $g$ for the metric on the target space associated with a single site or link.

This section explains why Hamiltonians of the form \eqref{eq:intro-hamiltonian} arise naturally in physics. For a constrained quantum-mechanical system, as discussed in Section~\ref{sec:qm-curved}, $\mathcal M$ is a curved configuration space. For a spatially discretized quantum field theory, it is often a Cartesian product $\mathcal M=T^{|\Lambda|}$ of copies of a fixed-dimensional target manifold $T$. The product metric makes the kinetic operator a sum of Laplace--Beltrami operators on the factors, while spatial-gradient, on-site, and plaquette terms contribute to the potential $\mathcal V$.

\subsubsection{Quantum Mechanics on Curved Spaces}
\label{sec:qm-curved}

The most immediate case is the one that requires no construction at all. A single non-relativistic particle moving on a Riemannian manifold $(\mathcal{M}, G)$ under a potential $\mathcal{V} : \mathcal{M} \to \mathbb{R}$ has the Hamiltonian
\begin{equation}
\label{eq:qm-curved-hamiltonian}
    \hat{H} = -\frac{\hbar^2}{2m} \Delta_G + \mathcal{V}(x) ,
    \qquad x \in \mathcal{M} ,
\end{equation}
acting on $L^2(\mathcal{M})$, where $\Delta_G$ is the Laplace--Beltrami operator of $G$ and $m$ is the mass. This is already the form assumed in the beginning of this section, requiring neither a lattice regularization nor any field-theoretic input. It is also the template of which the remaining sections are instances: each arrives at \eqref{eq:qm-curved-hamiltonian} on a configuration space that is a Cartesian product, through its own choice of factor manifold, metric and potential.

Such Hamiltonians arise whenever a mechanical system is subject to constraints, so that its configuration space is a curved manifold rather than a Euclidean space. The quantization of constrained systems, and the operator-ordering question it raises, are discussed by DeWitt~\cite{dewitt1957dynamical}.

Quantum systems governed by Hamiltonians of the form~\eqref{eq:qm-curved-hamiltonian} arise in several concrete settings. These include particles constrained to prescribed geometries~\cite{da1981quantum}, with geometry-induced potentials in quantum nanostructures providing a prominent example~\cite{da2017quantum}, as well as molecular Hamiltonians~\cite{meyer2002molecular} and shell-shaped atomic gases~\cite{tononi2024shell}. Treatments of the latter two settings place particular emphasis on motion on the sphere $S^2$.
Moreover, manifolds with nontrivial topology may support topological terms, providing physically interesting toy models for more complicated systems. A particle on a circle with such a term has, for example, been studied as a target for quantum simulation in~\cite{shen2022simulating}. Although our present algorithm does not include topological terms, this example suggests a natural direction in which the framework could be extended.

Quantum field theory on a curved spacetime is a different setting, since there the metric is a metric on spacetime rather than on the configuration space of the dynamical variables, and the configuration space remains a space of field configurations. We therefore do not consider it in these notes. The models below may be field theories, but only on a flat Minkowski spacetime.

\subsubsection{Nonlinear Sigma Models}
\label{sec:nlsm}

Nonlinear sigma models (NLSMs) are an important prototypical class of models in physics. They were first introduced in~\cite{gell1960axial} to describe pions and now play a broad role in both high-energy and condensed matter physics (see the review~\cite{burgess2000goldstone} for a general introduction). A prominent particle-physics example is chiral perturbation theory, whose leading-order Lagrangian is that of a nonlinear sigma model with target manifold $\mathrm{SU}(N_f)$~\cite{ecker1989role}. In condensed matter physics, the $O(3)$ nonlinear sigma model, whose target manifold is $S^2$, describes the low-energy behavior of Heisenberg antiferromagnets in one spatial dimension~\cite{haldane1983nonlinear} and two spatial dimensions~\cite{chakravarty1989two}. Reflecting this broad relevance, the $O(N)$ models have also attracted considerable interest as targets for quantum simulation~\cite{schutzhold2005quantum, alexandru2019sigma, singh2019qubit, singh2022qubit}. Their target manifold is the sphere $S^{N-1}$, with the corresponding $O(N)$ symmetry.

The Lagrangian (density) of the Nonlinear Sigma Model for a target manifold $T$ of dimension $d(T)$ on a $(D+1)$-dimensional Minkowski spacetime is
\begin{equation}
    \mathcal{L}(x) = \frac{1}{2} \sum_{\mu, \nu = 0}^{D} \sum_{a, b = 1}^{d(T)} \eta^{\mu\nu} g_{ab}(\Sigma(x)) \partial_\mu \Sigma^a(x) \partial_\nu \Sigma^b(x) - V(\Sigma(x))
    ,
\end{equation}
where $\eta$ is the $(D+1)$-dimensional Minkowski metric with $\eta^{\mu\nu} = \mathrm{diag}(1, -1, -1, \ldots, -1)$. The target manifold $T$ is equipped with a Riemannian metric $g$, the field is a map $\Sigma : \mathbb{R}^{1,D} \to T$ from spacetime into the target, and $\Sigma^a$ ($a = 1, \ldots, d(T)$) denotes the components of $\Sigma$ in a local chart of $T$. Correspondingly, $g_{ab}(\Sigma(x))$ is the metric of $T$ evaluated at the field value $\Sigma(x)$. The potential $V : T \to \mathbb{R}$ is a fixed function on the target.

It is worth emphasizing the change of viewpoint relative to the main analysis of the paper. There, the dynamical variable is a point ``$x$'' of a manifold. Here, the dynamical variable is the field value $\Sigma$, taking values in the manifold $T$, while the spacetime coordinate $x$ merely \emph{labels} which copy of $\Sigma$ we are talking about. This will become concrete in \Cref{sec:nlsm-lattice}, where discretizing space turns the label $x$ into a lattice site and the configuration space into a Cartesian product of copies of $T$, one per site.

We write $x = (t, \bm{x})$ with $t \in \mathbb{R}$ and $\bm{x} \in \mathbb{R}^D$, and split the spacetime index $\mu \in \{0, 1, \ldots, D\}$ accordingly into the time index $0$ and the spatial indices $i \in \{1, \ldots, D\}$. Since $\eta^{\mu\nu}$ is diagonal, the double sum over $\mu, \nu$ can be carried out explicitly, and the Lagrangian density becomes
\begin{equation}
\label{eq:nlsm-lagrangian-split}
    \mathcal{L}
    = \frac{1}{2} \sum_{a, b = 1}^{d(T)} g_{ab}(\Sigma) \, \dot{\Sigma}^a \dot{\Sigma}^b
    - \frac{1}{2} \sum_{i=1}^{D} \sum_{a, b = 1}^{d(T)} g_{ab}(\Sigma) \, \partial_i \Sigma^a \partial_i \Sigma^b
    - V(\Sigma),
\end{equation}
where $\dot{\Sigma}^a \coloneqq \partial_0 \Sigma^a = \partial \Sigma^a / \partial t$, and we have suppressed the evaluation at $x$ throughout.

\paragraph{Legendre transform: the classical Hamiltonian}

The momentum conjugate to $\Sigma^a(\bm{x})$ is
\begin{equation}
    \pi_a(\bm{x}) = \frac{\partial}{\partial \dot{\Sigma}^a(\bm{x})} \mathcal{L}(\bm{x})
    = \sum_{b = 1}^{d(T)} g_{ab}(\Sigma(\bm{x})) \, \dot{\Sigma}^b(\bm{x}) .
\end{equation}
Since $g$ is positive definite, the matrix $\left( g_{ab}(\Sigma) \right)_{a, b = 1}^{d(T)}$ is invertible at every point of $T$. As in standard notation, we write $\left( g^{ab}(\Sigma) \right)_{a, b = 1}^{d(T)}$ for its matrix inverse, i.e.\ the components of the inverse metric, characterized by
\begin{equation}
\label{eq:inverse-metric}
    \sum_{b = 1}^{d(T)} g^{ab}(\Sigma) \, g_{bc}(\Sigma) = \delta^a_{\;c}
    \qquad \text{for all } a, c \in \{1, \ldots, d(T)\},
\end{equation}
with $\delta^a_{\;c}$ the Kronecker delta. Applying $g^{ab}$ to both sides of the previous equation therefore inverts the relation between velocities and momenta,
\begin{equation}
    \dot{\Sigma}^a(\bm{x}) = \sum_{b = 1}^{d(T)} g^{ab}(\Sigma(\bm{x})) \, \pi_b(\bm{x}) .
\end{equation}
Substituting into the standard Legendre transform for the Hamiltonian density $\mathcal{H}$
\begin{equation}
    \mathcal{H} = \sum_{a = 1}^{d(T)} \pi_a \dot{\Sigma}^a - \mathcal{L}
\end{equation}
gives the Hamiltonian $H$
\begin{equation}
\label{eq:nlsm-classical-hamiltonian}
    H = \int_{\mathbb{R}^D} \dd^D \bm{x} \, \mathcal{H} (\bm{x})
    = \int_{\mathbb{R}^D} \dd^D \bm{x} \left[
        \frac{1}{2} \sum_{a, b = 1}^{d(T)} g^{ab}(\Sigma) \pi_a \pi_b
        + \frac{1}{2} \sum_{i=1}^{D} \sum_{a, b = 1}^{d(T)} g_{ab}(\Sigma) \, \partial_i \Sigma^a \partial_i \Sigma^b
        + V(\Sigma)
    \right] ,
\end{equation}
where the first term is the kinetic term, quadratic in the conjugate momentum $\pi$, the second term is the spatial derivative term, involving only $\Sigma$ but not $\pi$, and the last term is the ordinary potential term.

This Hamiltonian still describes a field theory in continuous space $\mathbb{R}^D$. It carries one copy of the target manifold $T$, with its own conjugate momentum, for every point $\bm{x} \in \mathbb{R}^D$, so the configuration space is infinite-dimensional and not yet amenable to the quantum simulation algorithm we propose. We therefore discretize the spatial directions while keeping the time $t$ still continuous.

\paragraph{Spatial lattice regularization}
\label{sec:nlsm-lattice}

To obtain a finite-dimensional configuration space we discretize space on a hypercubic lattice $\Lambda = (a\mathbb{Z}/L\mathbb{Z})^D$ with lattice spacing $a$ and $|\Lambda| = (L/a)^D$ sites, keeping time continuous. A field configuration is then a collection
\begin{equation}
    \Sigma = (\Sigma_{\bm{n}})_{\bm{n} \in \Lambda} \in \mathcal{M} \coloneqq T^{|\Lambda|},
\end{equation}
so the configuration space is a Cartesian product of $|\Lambda|$ copies of the fixed manifold $T$ of dimension $d(T)$.

Replacing $\int \dd^D \bm{x} \mapsto a^D \sum_{\bm{n} \in \Lambda}$ and $\partial_i \Sigma \mapsto (\Sigma_{\bm{n} + a \hat{e}_i} - \Sigma_{\bm{n}})/a$ where $\hat{e}_i$ is the unit vector for direction $i$, the discretized momentum conjugate to $\Sigma^a_{\bm{n}}$ is $p_{\bm{n} a} = a^D \pi_a(\bm{n})$, and \eqref{eq:nlsm-classical-hamiltonian} becomes
\begin{equation}
\label{eq:nlsm-lattice-hamiltonian}
    H_\Lambda
    = \sum_{\bm{n} \in \Lambda} \frac{1}{2 a^D} \sum_{a, b = 1}^{d(T)} g^{ab}(\Sigma_{\bm{n}}) \, p_{\bm{n} a} p_{\bm{n} b}
    \;+\; \mathcal{V}(\Sigma),
\end{equation}
\begin{equation}
\label{eq:nlsm-lattice-potential}
    \mathcal{V}(\Sigma)
    \coloneqq a^{D} \sum_{\bm{n} \in \Lambda} \left[
        \frac{1}{2 a^2} \sum_{i=1}^{D} d_T\!\left(\Sigma_{\bm{n} + a\hat{e}_i}, \Sigma_{\bm{n}}\right)^2
        + V(\Sigma_{\bm{n}})
    \right],
\end{equation}
where we have used the geodesic distance $d_T$ on $T$ as the manifestly chart-independent (and globally well-defined) discretization of $\sum_{a,b=1}^{d(T)} g_{ab} \partial_i \Sigma^a \partial_i \Sigma^b$; for nearby field values $d_T(\Sigma_{\bm{n}+a\hat e_i}, \Sigma_{\bm{n}})^2 = \sum_{a,b=1}^{d(T)} g_{ab}(\Sigma_{\bm{n}}) \delta_i\Sigma^a \delta_i \Sigma^b + O(|\delta_i \Sigma|^3)$ with $\delta_i \Sigma \coloneqq \Sigma_{\bm{n}+a\hat e_i} - \Sigma_{\bm{n}}$, so the two agree in the continuum limit.

\paragraph{Quantization}

Equation~\eqref{eq:nlsm-lattice-hamiltonian} is the Hamiltonian of a single point particle moving on the Riemannian manifold
\begin{equation}
    (\mathcal{M}, G), \qquad
    \mathcal{M} = T^{|\Lambda|}, \qquad
    G \coloneqq a^D \bigoplus_{\bm{n} \in \Lambda} g ,
\end{equation}
i.e.\ $G$ is the product metric on $T^{|\Lambda|}$ with each factor rescaled by $a^D$, subject to the potential $\mathcal{V}$. Indeed, the kinetic term of \eqref{eq:nlsm-lattice-hamiltonian} is $\tfrac{1}{2} \sum_{A, B} G^{AB}(\Sigma) p_A p_B$, where $A = (\bm{n}, a)$ is a composite index running over the $d(T) \, |\Lambda|$ coordinate directions of $\mathcal{M}$.

Quantizing on the Hilbert space $L^2(\mathcal{M})$ with the ordering prescription by DeWitt~\cite{dewitt1957dynamical},
\begin{equation}
\label{eq:nlsm-ordering}
    \frac{1}{2} \sum_{A, B} G^{AB} p_A p_B
    \;\longmapsto\;
    -\frac{\hbar^2}{2} \Delta_G,
    \qquad
    \Delta_G = \frac{1}{\sqrt{\det G}} \sum_{A, B} \partial_A \left( \sqrt{\det G} \, G^{AB} \partial_B \right),
\end{equation}
we obtain the quantum Hamiltonian on the spatial lattice,
\begin{equation}
\label{eq:nlsm-quantum-hamiltonian}
    \hat{H}_\Lambda
    = -\frac{\hbar^2}{2} \Delta_G + \mathcal{V}(\Sigma)
    = -\frac{\hbar^2}{2 a^D} \sum_{\bm{n} \in \Lambda} \Delta_{g}^{(\bm{n})} + \mathcal{V}(\Sigma) \; ,
\end{equation}
where $\Delta_g^{(\bm{n})}$ denotes the Laplace--Beltrami operator of $(T, g)$ acting on the copy of $T$ at site $\bm{n}$, and the second equality uses that the Laplace--Beltrami operator of a product metric is the sum of those of its factors, $\Delta_G = a^{-D}\sum_{\bm{n}} \Delta^{(\bm{n})}_g$.

\subsubsection{\texorpdfstring{$\mathbb{C}P^{N-1}$}{CP^(N-1)} Models}
\label{sec:cpn}

The $\mathbb{C}P^{N-1}$ models~\cite{eichenherr1978sigma, dadda1978expandable} are nonlinear sigma models whose target is complex projective space. In two spacetime dimensions, they are associated with asymptotic freedom and confinement, and possess instantons and a $\theta$-vacuum structure. They are for this reason a long-standing toy model for the corresponding phenomena in four-dimensional gauge theory~\cite{witten1979instantons}.

From our point of view the essential point is that a $\mathbb{C}P^{N-1}$ model is an instance of \Cref{sec:nlsm} with a different choice of $(T, g)$. We therefore only need to identify the target manifold and its metric. The Legendre transform, lattice regularization and quantization of \Cref{sec:nlsm} then naturally apply to the case of $\mathbb{C}P^{N-1}$.

\paragraph{The target manifold and its metric}

Complex projective space is the set of complex lines through the origin of $\mathbb{C}^N$,
\begin{equation}
\label{eq:cpn-definition}
    \mathbb{C}P^{N-1} \coloneqq \left\{ z \in \mathbb{C}^N : \; \sum_{a=1}^{N} |z^a|^2 = 1 \right\} \Big/ \mathrm{U}(1),
    \qquad \text{where } \mathrm{U}(1) \text{ acts by } z \mapsto e^{i \varphi} z, \; \varphi \in \mathbb{R},
\end{equation}
that is, the unit sphere $S^{2N-1} \subset \mathbb{C}^N$ with points differing by an overall phase identified.\footnote{For $N = 2$ one has $\mathbb{C}P^{1} \cong S^2$, so the model coincides with the $O(3)$ nonlinear sigma model of \Cref{sec:nlsm}.} It is a compact manifold of real dimension
\begin{equation}
\label{eq:cpn-dimension}
    d\left(\mathbb{C}P^{N-1}\right) = 2N - 1 - 1 = 2(N-1),
\end{equation}
the two subtractions removing the radial direction and the phase direction respectively. We write $[z]$ for the equivalence class of $z$, and use $z$ itself as homogeneous coordinates, a redundant description in which each point of the manifold corresponds to a whole set of representatives.

The natural metric on $\mathbb{C}P^{N-1}$ is the Fubini--Study metric, which in the terms of homogeneous coordinates normalized to $\abs{z}^2 = 1$ in \eqref{eq:cpn-definition} is
\begin{equation}
\label{eq:cpn-fubini-study}
    \dd s^2
    = \sum_{a, b = 1}^{N} g_{a \bar{b}} dz^a d\bar{z}^b  = \sum_{a=1}^N d z^a d \bar{z}^a - \left(\sum_{a=1}^N \bar{z}^a d z^a\right) \left(\sum_{b=1}^N z^b d \bar{z}^b\right)
\end{equation}
It is invariant under $z \mapsto e^{i\varphi} z$ and therefore descends to a Riemannian metric on $\mathbb{C}P^{N-1}$. Writing $\Sigma^a$ ($a = 1, \ldots, 2(N-1)$) for coordinates in a local chart, \eqref{eq:cpn-fubini-study} defines the components $g_{ab}$ required as input to \Cref{sec:nlsm}.

\paragraph{The continuum Lagrangian}

The physics literature usually presents the model not through a chart on $\mathbb{C}P^{N-1}$ but in the redundant homogeneous coordinates, with the redundancy promoted to a local gauge symmetry. One introduces a non-dynamical $\mathrm{U}(1)$ gauge field $A_\mu$ and writes the Lagrangian density
\begin{equation}
\label{eq:cpn-gauged-lagrangian}
    \mathcal{L} = \frac{1}{g_0^2} \sum_{\mu, \nu = 0}^{D} \sum_{a = 1}^{N} \eta^{\mu\nu} \overline{\left( D_\mu z^a \right)} \left( D_\nu z^a \right),
    \qquad
    D_\mu z^a \coloneqq \partial_\mu z^a + i A_\mu z^a ,
\end{equation}
subject to $\sum_{a=1}^N |z^a|^2 = 1$, with $g_0 > 0$ the coupling constant. Since $A_\mu$ carries no kinetic term, it is an auxiliary field with the equation of motion
\begin{equation}
\label{eq:cpn-auxiliary}
    A_\mu = i \sum_{a=1}^{N} \bar{z}^a \partial_\mu z^a .
\end{equation}
Substituting \eqref{eq:cpn-auxiliary} back into \eqref{eq:cpn-gauged-lagrangian} eliminates $A_\mu$ entirely and produces the Lagrangian with the Fubini--Study metric \eqref{eq:cpn-fubini-study} as 
\begin{equation}
    \mathcal{L} = \frac{1}{g_0^2} \sum_{\mu, \nu = 0}^{D} \sum_{a, b = 1}^{N} \eta^{\mu\nu} g_{a\bar{b}} \partial_\mu z^a \partial_\nu \bar{z}^b 
    .
\end{equation}

\paragraph{The lattice Hamiltonian}

With $T = \mathbb{C}P^{N-1}$ and $g$ the Fubini--Study metric, the construction of \Cref{sec:nlsm} applies unchanged. The configuration space is
\begin{equation}
\label{eq:cpn-configuration-space}
    \mathcal{M} = \left( \mathbb{C}P^{N-1} \right)^{|\Lambda|},
\end{equation}
and \eqref{eq:nlsm-quantum-hamiltonian} becomes
\begin{equation}
\label{eq:cpn-quantum-hamiltonian}
    \hat{H}_\Lambda
    = -\frac{\hbar^2 g_0^2}{4 a^D} \sum_{\bm{n} \in \Lambda} \Delta_{\mathrm{FS}}^{(\bm{n})}
    + \frac{a^{D-2}}{g_0^2} \sum_{\bm{n} \in \Lambda} \sum_{i=1}^{D} d_{\mathbb{C}P^{N-1}}\!\left( \Sigma_{\bm{n} + a \hat{e}_i}, \Sigma_{\bm{n}} \right)^2 ,
\end{equation}
where $\Delta_{\mathrm{FS}}$ is the Laplace--Beltrami operator of the Fubini--Study metric and we have set $V \equiv 0$, the usual choice for this model. The coupling enters asymmetrically because \eqref{eq:cpn-gauged-lagrangian} carries an overall $1/g_0^2$ rather than the $1/2$ of \Cref{sec:nlsm}. (Setting $g_0^2 = 2$ would recover the normalization of \eqref{eq:nlsm-quantum-hamiltonian}.) As in \Cref{sec:nlsm}, this is of the form $-\Delta_G + \mathcal{V}$ on a Cartesian product of copies of a single fixed manifold.

\subsubsection{Lattice Gauge Theories}
\label{sec:lgt}

Lattice gauge theories~\cite{wilson1974confinement} (LGTs) are the standard non-perturbative regularization of gauge theory, and the Hamiltonian formulation of Kogut and Susskind~\cite{kogut1975hamiltonian} is the natural starting point for quantum simulation~\cite{banuls2020simulating}. We keep the gauge group $\mathcal{G}$ an arbitrary compact connected Lie group, so that $\mathcal{G} = \mathrm{SU}(3)$ recovers lattice QCD and $\mathcal{G} = \mathrm{U}(1)$ compact lattice electrodynamics.

On a fixed lattice, the Kogut--Susskind Hamiltonian~\cite{kogut1975hamiltonian} is already of the form $-\Delta_G + \mathcal{V}$ on a Cartesian product of copies of a fixed compact manifold, with no further continuum limit needed to bring it there. The target manifold is the gauge group $\mathcal{G}$ itself, so its dimension is fixed by the theory and does not grow with the lattice. The one structural difference from \Cref{sec:nlsm} is that the degrees of freedom live on the \emph{links} of the lattice rather than on its sites.

We restrict discussions throughout to the gauge degrees of freedom, so that what we describe is more precisely a lattice Yang--Mills theory. This pure-gauge sector already carries much of the essential non-perturbative physics. Moreover, a complete theory such as lattice QCD adds matter fields on the sites, but retains the pure-gauge Hamiltonian below as a summand.

Throughout we fix once and for all a faithful unitary representation of $\mathcal{G}$ and identify group elements with the corresponding unitary matrices, so that $\Tr$ and $\dagger$ below are the ordinary matrix trace and conjugate transpose.

\paragraph{The configuration space}

Let $\Lambda = (a \mathbb{Z} / L \mathbb{Z})^D$ be a spatial lattice as in \Cref{sec:nlsm-lattice}, and let
\begin{equation}
\label{eq:lgt-links}
    \mathcal{E} \coloneqq \left\{ \ell = (\bm{n}, i) : \; \bm{n} \in \Lambda, \; i \in \{1, \ldots, D\} \right\},
    \qquad |\mathcal{E}| = D \, |\Lambda| ,
\end{equation}
denote its set of oriented links, where $(\bm{n}, i)$ is the link from $\bm{n}$ to $\bm{n} + a \hat{e}_i$. A gauge-field configuration assigns to each link a group element,
\begin{equation}
\label{eq:lgt-configuration-space}
    U = \left( U_\ell \right)_{\ell \in \mathcal{E}} \in \mathcal{M} \coloneqq \mathcal{G}^{|\mathcal{E}|} ,
\end{equation}
with the reversed link carrying $U_{-\ell} = U_\ell^{-1} = U_\ell^{\dagger}$. The configuration space is therefore a Cartesian product of $D |\Lambda|$ copies of the single fixed manifold $\mathcal{G}$, of dimension $d(\mathcal{G}) = \dim \mathfrak{g}$, where $\mathfrak{g}$ is the Lie algebra of $\mathcal{G}$. For $\mathcal{G} = \mathrm{SU}(N)$ one has $d(\mathcal{G}) = N^2 - 1$, so $d(\mathcal{G}) = 8$ for lattice QCD.

The potential term of the LGT will be built from the elementary plaquettes of $\Lambda$. For a plaquette $p$ spanning directions $i < j$ at site $\bm{n}$, let
\begin{equation}
\label{eq:lgt-plaquette}
    U_p \coloneqq U_{(\bm{n}, i)} \, U_{(\bm{n} + a \hat{e}_i, \, j)} \, U_{(\bm{n} + a \hat{e}_j, \, i)}^{-1} \, U_{(\bm{n}, j)}^{-1}
\end{equation}
be the ordered product of the four link variables around its boundary, and write $\mathcal{P}$ for the set of plaquettes.

\paragraph{The Kogut--Susskind Hamiltonian and the Laplace--Beltrami operator}

With the temporal gauge and on a lattice with continuous time but finite spatial lattice spacing, the Lagrangian reads~\cite{hamber2008quantum}
\begin{equation}
\label{eq:lgt-lagrangian}
    L = \frac{a}{4 g^2} \sum_{\ell \in \mathcal{E}} \Tr \left( \dot{U}_\ell^\dagger \dot{U}_\ell \right) + \frac{1}{4 a g^2} \sum_{p \in \mathcal{P}} \Tr \left( U_p + U_p^\dagger \right)
\end{equation}
where $a$ is the spatial lattice spacing, and $g > 0$ is the gauge coupling constant.

Then the Kogut--Susskind Hamiltonian~\cite{kogut1975hamiltonian,PhysRevD.15.1128} from the Lagrangian is obtained through the Legendre transform
\begin{equation}
\label{eq:lgt-legendre}
    H = \sum_{\ell \in \mathcal{E}} \left( \dot{U}_\ell^\dagger \frac{\partial L}{\partial \dot{U}_\ell^\dagger} + \dot{U}_\ell \frac{\partial L}{\partial \dot{U}_\ell} \right) - L = \frac{a}{4 g^2} \sum_{\ell \in \mathcal{E}} \Tr \left( \dot{U}_\ell^\dagger \dot{U}_\ell \right) - \frac{1}{4 a g^2} \sum_{p \in \mathcal{P}} \Tr \left( U_p + U_p^\dagger \right)
    .
\end{equation}

To exhibit \eqref{eq:lgt-legendre} as a Laplace--Beltrami operator plus a potential, we introduce on each link a local chart of $\mathcal{G}$, writing $U_\ell = U(q_\ell)$ with $q_\ell = (q_\ell^1, \ldots, q_\ell^{d(\mathcal{G})})$. The same chart is used on every link, since all factors of $\mathcal{M}$ are copies of the same manifold. By the chain rule $\dot{U}_\ell = \sum_{a = 1}^{d(\mathcal{G})} \partial_a U(q_\ell) \, \dot{q}_\ell^{\, a}$, so that
\begin{equation}
\label{eq:lgt-metric}
    \Tr \left( \dot{U}_\ell^\dagger \dot{U}_\ell \right)
    = \sum_{a, b = 1}^{d(\mathcal{G})} g_{ab}(q_\ell) \, \dot{q}_\ell^{\, a} \dot{q}_\ell^{\, b} ,
    \qquad
    g_{ab}(q) \coloneqq \Tr \left( \partial_a U^\dagger(q) \, \partial_b U(q) \right) ,
\end{equation}
which defines a Riemannian metric on $\mathcal{G}$. The momentum conjugate to $q_\ell^{\, a}$ is
\begin{equation}
\label{eq:lgt-momentum}
    \pi_{\ell a} = \frac{\partial L}{\partial \dot{q}_\ell^{\, a}}
    = \frac{a}{2 g^2} \sum_{b = 1}^{d(\mathcal{G})} g_{ab}(q_\ell) \, \dot{q}_\ell^{\, b} ,
    \qquad \text{equivalently} \qquad
    \dot{q}_\ell^{\, a} = \frac{2 g^2}{a} \sum_{b = 1}^{d(\mathcal{G})} g^{ab}(q_\ell) \, \pi_{\ell b} ,
\end{equation}
and substituting into \eqref{eq:lgt-legendre} gives the classical Hamiltonian in canonical form,
\begin{equation}
\label{eq:lgt-classical-hamiltonian}
    H(q, \pi)
    = \frac{g^2}{a} \sum_{\ell \in \mathcal{E}} \sum_{a, b = 1}^{d(\mathcal{G})} g^{ab}(q_\ell) \, \pi_{\ell a} \pi_{\ell b}
    \; - \; \frac{1}{4 a g^2} \sum_{p \in \mathcal{P}} \Tr \left( U_p(q) + U_p^\dagger(q) \right) .
\end{equation}
The first term is the kinetic term of a particle on the product manifold $\mathcal{M} = \mathcal{G}^{|\mathcal{E}|}$ carrying the product metric, and the second term is a potential term that depends on $q$ alone.

Similar to the discussion of \Cref{sec:nlsm}, \eqref{eq:lgt-classical-hamiltonian} is the Hamiltonian of a single point particle moving on the Riemannian manifold
\begin{equation}
\label{eq:lgt-product-metric}
    (\mathcal{M}, G), \qquad
    \mathcal{M} = \mathcal{G}^{|\mathcal{E}|}, \qquad
    G \coloneqq \frac{a}{2 g^2} \bigoplus_{\ell \in \mathcal{E}} g ,
\end{equation}
that is, $G$ is the product metric on $\mathcal{G}^{|\mathcal{E}|}$ with each factor carrying the metric \eqref{eq:lgt-metric} rescaled by $a / (2 g^2)$, subject to the potential $\mathcal{V}$. Indeed, with this choice the kinetic term of \eqref{eq:lgt-classical-hamiltonian} is $\tfrac{1}{2} \sum_{A, B} G^{AB}(q) \, \pi_A \pi_B$, where $A = (\ell, a)$ is a composite index running over the $d(\mathcal{G}) \, |\mathcal{E}|$ coordinate directions of $\mathcal{M}$.

Quantizing on $L^2(\mathcal{M})$ with the same ordering prescription \eqref{eq:nlsm-ordering} therefore gives
\begin{equation}
\label{eq:lgt-quantum-hamiltonian}
    \hat{H}
    = -\frac{\hbar^2}{2} \Delta_G + \mathcal{V}(q)
    = -\frac{g^2 \hbar^2}{a} \sum_{\ell \in \mathcal{E}} \Delta_{\mathcal{G}}^{(\ell)} + \mathcal{V}(q) ,
    \qquad
    \mathcal{V}(q) \coloneqq - \frac{1}{4 a g^2} \sum_{p \in \mathcal{P}} \Tr \left( U_p(q) + U_p^\dagger(q) \right) ,
\end{equation}
where $\Delta_{\mathcal{G}}^{(\ell)}$ is the Laplace--Beltrami operator of $(\mathcal{G}, g)$ acting on the copy of $\mathcal{G}$ at link $\ell$, and the second equality again uses that the Laplace--Beltrami operator of a product metric is the sum of those of its factors. This is the announced form $-\Delta_G + \mathcal{V}$ on a Cartesian product of copies of the single fixed compact manifold $\mathcal{G}$.

\subsubsection{Concrete simulation of the spatially discretized \texorpdfstring{$O(N)$}{O(N)} nonlinear sigma model}
\label{sec:nlsm-spherical-simulation}

We now give a concrete application of the sphere algorithm to a particular lattice realization of the nonlinear sigma model. Let $O(N)$ be the orthogonal group of real $N\times N$ matrices that preserve the Euclidean inner product on $\mathbb R^N$. An $O(N)$ field takes values in the unit sphere $\mathbb S^{N-1}\subset\mathbb R^N$, and the group acts on every field value by the same orthogonal transformation. The round metric and any interaction formed from inner products of field values are invariant under this action.

Fix $N\geq3$, let $\Lambda=(a\mathbb Z/L\mathbb Z)^D$ be a finite periodic spatial lattice, and write $n_\Lambda\coloneqq |\Lambda|$. A field configuration is a tuple $\sigma=(\sigma_{\bm n})_{\bm n\in\Lambda}\in(\mathbb S^{N-1})^{n_\Lambda}$. We use the squared chordal distance in the ambient Euclidean space. Each coordinate function $\sigma^a$ on $\mathbb S^{N-1}$ is a degree-one spherical harmonic, so $1-\sigma\cdot\sigma'$ contains only degrees zero and one in each spin variable. Multiplication by one nearest-neighbor interaction therefore raises the spherical-harmonic degree at either endpoint by at most one. The interaction depends only on the inner product and hence preserves the $O(N)$ symmetry. Since $\frac12\norm{\sigma-\sigma'}_2^2=1-\sigma\cdot\sigma'$ on the unit sphere, the resulting Hamiltonian is
\begin{equation}
H_{\Lambda}^{O(N)}\coloneqq -\frac{\hbar^2}{2a^D}\sum_{\bm n\in\Lambda}\Delta_{\mathbb S^{N-1}}^{(\bm n)}+\mathcal V_\mathrm{ch}(\sigma),\qquad \mathcal V_\mathrm{ch}(\sigma)\coloneqq a^{D-2}\sum_{\bm n\in\Lambda}\sum_{i=1}^D\left(1-\sigma_{\bm n}\cdot\sigma_{\bm n+a\widehat e_i}\right).
\label{eq:on-nlsm-chordal-hamiltonian}
\end{equation}
The chordal interaction agrees with the squared geodesic interaction to leading order for neighboring field values. The special case $N=3$ has target $\mathbb S^2$ and gives the spatially discretized $O(3)$ model used to describe Heisenberg antiferromagnets.

It is natural now to use the quantum algorithm for simulating dynamcis on the sphere. To formalize the setup, let $\mathcal H_\ell(\mathbb S^{N-1})$ denote the degree-$\ell$ spherical harmonics and define
\begin{equation}
\mathscr V_J^\Lambda\coloneqq \bigotimes_{\bm n\in\Lambda}\left(\bigoplus_{\ell=0}^J\mathcal H_\ell(\mathbb S^{N-1})\right)\subset L^2\!\left((\mathbb S^{N-1})^{n_\Lambda}\right).
\label{eq:on-nlsm-product-filtration}
\end{equation}
A function belongs to $\mathscr V_J^\Lambda$ when its spherical-harmonic degree is at most $J$ at every lattice site. Products satisfy
\begin{equation}
\mathscr V_J^\Lambda\mathscr V_K^\Lambda\subseteq\mathscr V_{J+K}^\Lambda.
\label{eq:on-nlsm-product-growth}
\end{equation}
The preceding degree-one observation implies $\mathcal V_\mathrm{ch}\in\mathscr V_1^\Lambda$, and
\begin{equation}
\norm{\mathcal V_\mathrm{ch}}_\infty\leq\Lambda_\mathrm{ch}\coloneqq \max\left\{1,2a^{D-2}Dn_\Lambda\right\}.
\label{eq:on-nlsm-potential-bound}
\end{equation}

Let $\psi_0\in\mathscr V_{J_0}^\Lambda$ be normalized, let $T>0$, and fix $0<\epsilon<1$. Choose
\begin{equation}
q\coloneqq \left\lceil\max\left\{2eT\Lambda_\mathrm{ch},2+\log_2(1/\epsilon)\right\}\right\rceil,\qquad J\coloneqq J_0+q-1,\qquad Q\coloneqq J+1,
\label{eq:on-nlsm-cutoffs}
\end{equation}
and set $R\coloneqq Q+1$ and $M_\varphi\coloneqq 2Q+1$. Let $W_{\mathbb S,J,Q}$ be the single-site weighted evaluation isometry from Theorem~\ref{thm:constructive-sphere-complexity}, and define the product transform and product grid by
\begin{equation}
W_\Lambda\coloneqq \bigotimes_{\bm n\in\Lambda}W_{\mathbb S,J,Q},\qquad \mathcal G_Q^\Lambda\coloneqq \left(\mathcal G_{R,M_\varphi}^{\mathbb S^{N-1}}\right)^{n_\Lambda}.
\label{eq:on-nlsm-product-transform}
\end{equation}
Let $D_{\Lambda,J}\coloneqq \sum_{\bm n\in\Lambda}D_J^{(\bm n)}$ be the diagonal kinetic matrix in the product spherical-harmonic basis and let
\begin{equation}
\widehat{\mathcal V}_\mathrm{ch,Q}\coloneqq \sum_{\boldsymbol x\in\mathcal G_Q^\Lambda}\mathcal V_\mathrm{ch}(\boldsymbol x)\ketbra{\boldsymbol x}{\boldsymbol x},\qquad \widehat H_{\Lambda,J,Q}\coloneqq \frac{\hbar^2}{2a^D}D_{\Lambda,J}+W_\Lambda^\dagger \widehat{\mathcal V}_\mathrm{ch,Q}W_\Lambda.
\label{eq:on-nlsm-discrete-hamiltonian}
\end{equation}

\begin{theorem}[Digital simulation of the spatially discretized $O(N)$ model]
\label{thm:on-nlsm-digital-simulation}
Let the lattice, Hamiltonian, initial state, and cutoffs be as above. Let $S_{\Lambda,J}$ map product spherical-harmonic coefficients to functions in $\mathscr V_J^\Lambda$. Assume that a circuit using $\mathcal C_\mathrm{prep}$ gates and $\mathcal Q_\mathrm{prep}$ work qubits prepares the coefficient state $S_{\Lambda,J}^\dagger \psi_0$ to state error $\mathcal O(\epsilon)$. At a product quadrature node, the potential \eqref{eq:on-nlsm-chordal-hamiltonian} can be evaluated reversibly by computing the $Dn_\Lambda$ nearest-neighbor inner products. Let $\mathcal C_\mathrm{ch}$ denote the gate cost of this evaluation and its uncomputation to absolute error $\epsilon_\mathrm{ch}$, and assume $T\epsilon_\mathrm{ch}=\mathcal O(\epsilon)$.

Then the tensor product of the spherical transforms in Theorem~\ref{thm:constructive-sphere-complexity}, combined with interaction-picture Hamiltonian simulation as in Theorem~\ref{thm:spherical-digital-simulation}, produces a normalized coefficient state $\ket{\widetilde c(T)}$ such that, with success probability at least $1-\mathcal O(\epsilon)$,
\begin{equation}
\norm{S_{\Lambda,J}\ket{\widetilde c(T)}-e^{-iTH_{\Lambda}^{O(N)}}\psi_0}=\mathcal O(\epsilon).
\label{eq:on-nlsm-state-error}
\end{equation}
Put $\tau_\mathrm{ch}\coloneqq T\Lambda_\mathrm{ch}$. The number of calls to the reversible potential-evaluation circuit or its inverse is
\begin{equation}
N_\mathrm{ch}=\mathcal O\!\left((1+\tau_\mathrm{ch})\frac{\log\!\left(\mathrm e^{\mathrm e}(1+\tau_\mathrm{ch})/\epsilon\right)}{\log\log\!\left(\mathrm e^{\mathrm e}(1+\tau_\mathrm{ch})/\epsilon\right)}\right)=\widetilde{\mathcal O}(1+\tau_\mathrm{ch}).
\label{eq:on-nlsm-query-complexity}
\end{equation}
The gate complexity is
\begin{equation}
\mathcal C_\mathrm{gates}=\mathcal C_\mathrm{prep}+\widetilde{\mathcal O}\!\left((1+\tau_\mathrm{ch})\left[n_\Lambda(N-1)(J+1)(Q+1)^2+\mathcal C_\mathrm{ch}+\log^2\!\left(\frac{2\Lambda_\mathrm{ch}}{\epsilon_\mathrm{ch}}\right)\right]\right),
\label{eq:on-nlsm-gate-complexity}
\end{equation}
and the total qubit count is
\begin{equation}
\mathcal Q_\mathrm{prep}+\mathcal O\!\left(n_\Lambda(N-1)\log(Q+1)+\log\frac{2\Lambda_\mathrm{ch}}{\epsilon_\mathrm{ch}}+\operatorname{polylog}\frac{n_\Lambda(N-1)(J+1)(Q+1)(1+\tau_\mathrm{ch})}{\epsilon}\right).
\label{eq:on-nlsm-qubit-complexity}
\end{equation}
For fixed $D$, $N$, $a$, $T$, and $J_0$, these resources are polynomial in $n_\Lambda$ and $1/\epsilon$.
\end{theorem}

\begin{proof}
The kinetic operator is diagonal in the product spherical-harmonic basis and preserves every space $\mathscr V_J^\Lambda$. Equation~\eqref{eq:on-nlsm-product-growth} and $\mathcal V_\mathrm{ch}\in\mathscr V_1^\Lambda$ show that the interaction-picture Dyson terms of orders below $q$ lie in $\mathscr V_J^\Lambda$. The tensor product of the single-site quadrature rules is $Q$-exact for the product filtration. Since $Q=J+1$, Proposition~\ref{prop:Galerkin-quadrature} identifies the quadrature compression of $\mathcal V_\mathrm{ch}$ with its Galerkin compression. Corollary~\ref{cor:oversampled-quadrature-error}, with $f_K=f=\mathcal V_\mathrm{ch}$ and $B_1=T$, bounds the discretization error by $2^{2-q}\leq\epsilon$.

The product transform is implemented by applying the compiled single-site transform at every lattice site. Synthesizing each factor to operator error $\delta/n_\Lambda$ makes the product-transform error at most $\delta$. The kinetic phase is computed from the sum of the single-site eigenvalues $\ell(\ell+N-2)$, and the sampled potential is computed from the nearest-neighbor inner products. The interaction-picture proof of Theorem~\ref{thm:spherical-digital-simulation} therefore applies with one forward and one inverse product transform in each potential query. A product transform invokes $n_\Lambda$ single-site transforms, each with sphere dimension $N-1$ and spectral cutoff $J$, so its gate cost is $\widetilde{\mathcal O}(n_\Lambda(N-1)(J+1)(Q+1)^2)$. Substituting this cost, the potential-evaluation cost $\mathcal C_\mathrm{ch}$, and the normalization $\Lambda_\mathrm{ch}$ into \eqref{eq:sphere-binary-query-complexity}--\eqref{eq:sphere-end-to-end-qubit-complexity} proves \eqref{eq:on-nlsm-state-error}--\eqref{eq:on-nlsm-qubit-complexity}.
\end{proof}

\subsection{Connection to quantum interior-point methods}
\label{sec:connection-to-IPMs}
Let $\Xcal\subseteq\R{d}$ be a convex set with nonempty interior, and consider
the convex optimization problem
\begin{equation}
    \label{e:cvx_opt}
    \min_{x\in\Xcal}\inner{c,x}.
\end{equation}
Given a $\vartheta$-self-concordant barrier
$\Phi:\operatorname{int}(\Xcal)\to\mathbb R$, path-following interior-point
methods consider the barrier objectives
\begin{equation}
    V_\eta(x)\coloneqq \eta\inner{c,x}+\Phi(x),
    \qquad
    x(\eta)\coloneqq \argmin_{x\in\operatorname{int}(\Xcal)}V_\eta(x).
    \label{eq:ipm-central-path}
\end{equation}
As $\eta$ increases, the central path $\{ x(\eta)\}_{\eta > 0}$ approaches the optimal
set of \eqref{e:cvx_opt}.  The Hessian of the barrier defines the local metric
$g_\Phi=\nabla^2\Phi$.  Since the objective is linear, it is also the Hessian
of $V_\eta$, and the Newton step at an interior point $x$ is
\begin{equation}
    -\nabla^2\Phi(x)^{-1}
        \bigl(\eta c+\nabla\Phi(x)\bigr).
    \label{eq:ipm-newton-correction}
\end{equation}
Thus the same local geometry determines the Newton direction, the Dikin
ellipsoid in which a step remains feasible, and the notion of proximity used
to follow the central path.  Self-concordance controls the variation of this
metric between nearby points and is the central regularity property in the
classical path-following analysis~\cite{nesterov1988general,
nesterov1994interior,renegar2001mathematical}.

Gribling, Apers, Nieuwboer, and Walter proposed a quantum analogue of this
path-following picture~\cite{gribling2025selfconcordantschrodingeroperatorsspectral}.
They associate to each value of $\eta$ the Riemannian Schr\"odinger operator
\begin{equation}
    \label{eq:dutch_hamiltonian}
    H^R(\eta)
        \coloneqq -\frac{1}{2}\Delta_\Phi+\gamma^2V_\eta,
\end{equation}
where $\Delta_\Phi$ is the Laplace--Beltrami operator of
$g_\Phi=\nabla^2\Phi$ and $\gamma$ is a semiclassical parameter.  In the
barrier metric, the harmonic approximation near $x(\eta)$ is normalized,
rather than inheriting the Euclidean condition number of
$\nabla^2\Phi(x(\eta))$.  Their analysis makes this intuition quantitative:
\begin{theorem}[{\cite[Theorem 1.2]{gribling2025selfconcordantschrodingeroperatorsspectral}}]
    \label{thm:GANW_gap}
    There is a universal constant $C>0$ such that, for $d\geq2$ and
    $\gamma\geq C(d\log d)^5$, the operators in
    \eqref{eq:dutch_hamiltonian} satisfy, for every $\eta\geq0$,
    \begin{equation}
        \lambda_1(H^R(\eta))-\lambda_0(H^R(\eta))
        \geq\frac{\gamma}{2}.
    \end{equation}
\end{theorem}
In particular, this lower bound is uniform along the central path and has no
dependence on a Euclidean Hessian condition number.  Early quantum interior-point methods used
quantum linear-system algorithms to approximate the Newton direction at each
iteration, and their runtimes consequently depended polynomially on a uniform
upper bound for the condition numbers of the Newton systems encountered along
the central path~\cite{augustino2021quantum,kerenidis2018quantum}.  Such a
bound may be large and must hold over the entire run.  By incorporating the
barrier Hessian into the kinetic geometry, the Riemannian construction removes
this particular dependence from the spectral-gap estimate underlying quantum
path following.  For sufficiently large $\gamma$, the corresponding ground
state is concentrated near $x(\eta)$.
The cited work uses these states to define a discrete quantum central path:
under assumptions including an initial state with sufficient ground-state
overlap, estimates of the ground-state energies, and access to controlled
Riemannian Hamiltonian evolution, rotations about a sequence of neighboring
ground states prepare a state concentrated near an approximate optimizer.

The simulation results of this paper address one of the primitives required
by that framework.  For the standard logarithmic barrier
$\Phi(x)=-\sum_{i=1}^d\log x_i$ on the positive orthant,
Theorem~\ref{thm:orthant-affine-digital-simulation} gives a digital
discretization of the evolution generated by $-\Delta_\Phi$ plus a bounded
Lipschitz potential after restricting to a box $[a,b]^d$ and imposing
Dirichlet boundary conditions.  For any fixed $\eta$, the restriction of
$V_\eta$ to this box is such a potential, so the theorem applies to a
compactly truncated version of $H^R(\eta)$, subject to its stated initial-state
and regularity hypotheses.  This provides a rigorous finite-dimensional
simulation result in a canonical barrier geometry, but it does not by itself
instantiate the complete quantum path-following algorithm.

Several additional steps would be needed to instantiate an end-to-end algorithm.
In particular, one must choose a truncation that works uniformly over the
relevant central-path parameters, control the effect of its artificial
boundary on ground-state concentration, spectral gaps, and neighboring-state
overlaps, and account for controlled simulation, ground-energy
estimation, state preparation, and readout.  The orthant result should be viewed as a digital simulation primitive whose integration into a
fully discrete quantum interior-point method remains an open problem.

\subsection{Riemannian Quantum Hamiltonian Descent}
\label{sec:riemannian-qhd}

In this section, we extend Quantum Hamiltonian Descent~\cite{leng2023quantumhamiltoniandescent} to non-Euclidean geometry. Abe and Nagai recently proposed Riemannian Quantum Hamiltonian Descent at the
operator and path-integral levels, derived semiclassical equations of motion, and numerically studied representative examples~\cite{abe2026quantum}.  They also gave a conditional oracle-level simulation analysis, assuming phase-oracle access to the objective and sparse access to a chosen discretization of the Laplace--Beltrami operator.  Complementing this formulation, we develop rigorous convergence guarantees for geodesically convex objectives, including the relevant boundary terms, and combine them with explicit digital simulation algorithms on the sphere and the Fisher--Rao simplex. These results make the Riemannian QHD algorithm concrete in settings where the required geometric Hamiltonians can be implemented and analyzed explicitly.

\subsubsection{Additional Preliminaries}

We record the differential-geometric identities used in the computation relevant to this subsection.
Here we note $\nabla_V\psi=V(\psi)$ for vector field $V$.
Metric compatibility of the Levi--Civita connection gives the covariant product
rule
\begin{equation}
    X\ip{Y}{Z}_g
    =\ip{\nabla_XY}{Z}_g+\ip{Y}{\nabla_XZ}_g.
    \label{eq:metric-product-rule}
\end{equation}
For a smooth scalar function $u$, its Hessian is the covariant $2$-tensor
\begin{equation}
    \Hess u(X,Y)=\ip{\nabla_X\nabla u}{Y}_g
    =X(Yu)-(\nabla_XY)u,
    \label{eq:hessian-definition}
\end{equation}
and $\Delta_gu$ is its metric trace.  The first equality means that the Hessian
and the covariant derivative of the gradient contain the same information:
\begin{equation}
    \nabla_X\nabla u=(\Hess u(X,\cdot))^\sharp.
    \label{eq:hessian-gradient-connection}
\end{equation}
Here, $\Hess u(X,\cdot)$ is the $1$-form obtained by fixing the first slot and $\sharp$ converts that $1$-form to a vector using the metric. Thus the Hessian
is the Riemannian analogue of the Jacobian matrix of the vector field
$\nabla u$.

The divergence of a covariant $2$-tensor $S$ is the $1$-form
\begin{equation}
    (\diver S)(Y)=\sum_{i=1}^d(\nabla_{e_i}S)(e_i,Y).
    \label{eq:covariant-tensor-divergence-definition}
\end{equation}
At any fixed point $q$, one can choose a local
orthonormal frame $e_1,\ldots,e_d$ that is geodesic at $q$, meaning
$\nabla_{e_i}e_j=0$ at $q$.  At that point,
\begin{equation}
    \Hess u(e_i,e_j)=e_i(e_ju),
    \qquad
    \Delta_gu=\sum_{i=1}^d e_i(e_iu).
    \label{eq:normal-frame-hessian-laplacian}
\end{equation}
Then, \eqref{eq:hessian-gradient-connection} becomes
\begin{equation}
    \nabla_{e_i}\nabla u
    =\sum_{j=1}^d\Hess u(e_i,e_j)e_j.
    \label{eq:hessian-gradient-frame}
\end{equation}
Taking the trace of the Hessian gives the scalar Laplacian
\begin{equation}
    \Delta_gu=\sum_{i=1}^d\Hess u(e_i,e_i).
\end{equation}

From the covariant derivative product rule and setting $v = \nabla_{e_i}\nabla u$ yields
\begin{align*}
 e_i\langle v, e_j\rangle_g = \langle \nabla_{e_{i}}v, e_j\rangle_g = (\nabla_{e_i}\Hess u)(e_i, e_j).
\end{align*}
Hence differentiating \eqref{eq:hessian-gradient-frame} once more in the
$e_i$ direction and then summing gives a vector:
\begin{align}
    \Delta_g^{\mathrm{rough}}\nabla u
    &\coloneqq \sum_{i=1}^d\nabla_{e_i}\nabla_{e_i}\nabla u\notag\\
    &=\sum_{i,j=1}^d
      (\nabla_{e_i}\Hess u)(e_i,e_j)e_j
      \quad\text{at   }q.
    \label{eq:rough-laplacian-gradient-hessian}
\end{align}
In contrast applying the gradient after the Laplacian gives:
\begin{equation}
    \nabla\Delta_gu
    =\sum_{i,j=1}^d
      (\nabla_{e_j}\Hess u)(e_i,e_i)e_j
      \quad\text{at   }q.
    \label{eq:gradient-laplacian-hessian}
\end{equation}
The expressions \eqref{eq:rough-laplacian-gradient-hessian} and
\eqref{eq:gradient-laplacian-hessian} trace different pairs of the three
indices of $\nabla\Hess u$.  Interchanging those indices produces the Ricci
curvature term derived below.  In flat Euclidean space all covariant derivatives
are ordinary partial derivatives, so the indices commute and the two
expressions agree.

The frame removes connection coefficients at $q$, but not their derivatives;
the latter are what produce curvature when third covariant derivatives are
commuted.

With the curvature convention, we obtain the Ricci curvature tensor $R$:
\begin{equation}
    R(X,Y)Z
    =\nabla_X\nabla_YZ-\nabla_Y\nabla_XZ-\nabla_{[X,Y]}Z,
    \label{eq:curvature-convention}
\end{equation}
the vector $R(X,Y)Z$ measures the failure of covariant derivatives of the
vector field $Z$ to commute in the directions $X$ and $Y$. Specifically, it follows that
\begin{equation}
    (\nabla_X\Hess u)(Y,Z)-(\nabla_Y\Hess u)(X,Z)
    =\ip{R(X,Y)\nabla u}{Z}_g.
    \label{eq:hessian-curvature-commutation}
\end{equation}

The Riemannian QHD convergence proof relies heavily on the following lemmas. For symmetric differential expressions $X$ and $Y$ on a common smooth core, set
\begin{equation}
        \mathfrak c_{X,Y}[\psi]
        \coloneqq i\left(
            \langle X\psi,Y\psi\rangle
            -\langle Y\psi,X\psi\rangle
        \right).
        \label{eq:commutator-form-convention}
\end{equation}
This definition requires only $X\psi,Y\psi\in L^2$, not that either
$XY\psi$ or $YX\psi$ belong to an operator domain.  Whenever both
compositions are defined, it equals
$\langle\psi,i[X,Y]\psi\rangle$.  Pointwise differential commutators below
are used only as intermediate identities for evaluating
\eqref{eq:commutator-form-convention}. 

\begin{lemma}
\label{lem:a_boundary_lemma}
Let $V=\nabla\phi$ and
\begin{align*}
    A_V\coloneqq \frac12(V\cdot p+p\cdot V)
    =-i\left(\nabla_V+\frac12\diver V\right).
\end{align*}
For smooth $u,v$ on $\overline\Omega$,
\begin{align*}
    \langle u,A_Vv\rangle_{L^2(\Omega)}
    -\langle A_Vu,v\rangle_{L^2(\Omega)}
    =-i\int_{\partial\Omega}
      \overline u\,v\,\langle V,n\rangle_g\,\dd S_g.
    \end{align*}
\end{lemma}

\begin{proof}

Let $V = \nabla \phi$. First,
\begin{equation}
    (V\cdot p)\psi=-i\nabla_V\psi.
\end{equation}
The dot product $\cdot$ is with respect to the metric $g$. In the reverse ordering, $p$ also differentiates the coefficient $V$.
Invariantly, this ordering is the divergence of the vector field $V\psi$:
\begin{align}
    (p\cdot V)\psi
    &=-i\diver(V\psi)\notag\\
    &=-i\left((\diver V)\psi+\nabla_V\psi\right),
    \label{eq:p-dot-V-action}
\end{align}
where the second line is the divergence product rule.  Averaging the two gives
\begin{equation}
    \frac12(V\cdot p+p\cdot V)\psi
    =-i\left(\nabla_V+\frac12\diver V\right)\psi.
\end{equation}

The divergence term also explains why the symmetrized operator is formally
self-adjoint.  To see the integration-by-parts step explicitly, apply the
divergence theorem to the vector field
\begin{equation}
        W=\overline u\,v\,V.
\end{equation}
The divergence product rule gives
\begin{align}
        \diver W
        &=\diver(\overline u\,v\,V)\notag\\
        &=\overline u\,\nabla_Vv
            +v\,\nabla_V\overline u
            +(\diver V)\overline u\,v.
        \label{eq:weighted-vector-field-divergence}
\end{align}
The Riemannian divergence theorem states that
\begin{equation}
        \int_\Omega\diver W\,\dd\operatorname{vol}_g
        =\int_{\partial\Omega}\ip{W}{n}_g\,\dd S_g
        =\int_{\partial\Omega}
            \overline u\,v\,\ip{V}{n}_g\,\dd S_g,
        \label{eq:divergence-theorem-A-V}
\end{equation}
where $n$ is the outward unit normal.  Substituting
\eqref{eq:weighted-vector-field-divergence} and moving the last two volume
terms to the other side yields the full integration-by-parts identity
\begin{align}
        \int_\Omega\overline u\,\nabla_Vv\,\dd\operatorname{vol}_g
        ={}&-\int_\Omega
            \left(\nabla_V\overline u+(\diver V)\overline u\right)v
            \,\dd\operatorname{vol}_g +\int_{\partial\Omega}
             \overline u\,v\,\ip{V}{n}_g\,\dd S_g.
        \label{eq:directional-derivative-integration-by-parts}
\end{align}
Then
\begin{align*}
      \int_\Omega\overline u\, i\left(-\nabla_V - \frac{1}{2}\textup{div}(V)\right)v\,\dd\operatorname{vol}_g &= \int_\Omega
            i\left(-\nabla_V\overline u-\frac{1}{2}(\diver V)\overline u\right)v
            \,\dd\operatorname{vol}_g\\
            &\quad -i\int_{\partial\Omega}
             \overline u\,v\,\ip{V}{n}_g\,\dd S_g.
\end{align*}

For $V=\nabla\phi$, the identity can
be computed further using $\diver V=\Delta_g\phi$; set
\begin{equation}
    B=\nabla_V+\frac12\Delta_g\phi,
    \qquad A_V=-iB.
\end{equation}
\end{proof}
Then we can use the alternative expression for $A_{\nabla \phi}$ and arrive at the following critical identity.
\begin{lemma}
\label{lem:main_identity}
Let $\phi\in C^4(\overline\Omega)$ be real, $V = \nabla \phi$, $\psi\in C^\infty(\overline\Omega)$, and define
\begin{align*}
& B_{\phi} = \nabla_{V}  + \frac{1}{2}\Delta_g\phi,
    \quad A_{V} = -iB_{\phi}.
\end{align*}
Also define the total boundary functional
\begin{align}
    \mathcal B_\phi[\psi]
    \coloneqq \operatorname{Re}\int_{\partial\Omega}\Bigl[{}
       &-2\overline\psi\,
          \Hess\phi(\nabla\psi,n)
        -(\partial_n\overline\psi)B_\phi\psi +\overline\psi\,\partial_n(B_\phi\psi)
        +(\partial_n\phi)\overline\psi\,K_0\psi
      \Bigr] \,\dd S_g.
    \label{eq:general-virial-boundary-functional}
\end{align}
If
\begin{equation}
    \mathfrak c_\phi[\psi]
    \coloneqq \mathfrak c_{K_0,A_V}[\psi]
    =i\left(\langle K_0\psi,A_V\psi\rangle
       -\langle A_V\psi,K_0\psi\rangle\right),
\end{equation}
then 
\begin{align}
    \mathfrak c_\phi[\psi]
    ={}&2\int_\Omega
      \Hess\phi(\nabla\psi,\nabla\overline\psi)
      \,\dd\operatorname{vol}_g
      -\frac12\int_\Omega(\Delta_g^2\phi)|\psi|^2
      \,\dd\operatorname{vol}_g
      +\mathcal B_\phi[\psi].
    \label{eq:general-virial-form}
\end{align}
\end{lemma}

\begin{proof}
By the definition of $A_V$,
\begin{equation}
    \mathfrak c_\phi[\psi]
    =2\operatorname{Re}\langle K_0\psi,B_\phi\psi\rangle.
    \label{eq:virial-form-real-part}
\end{equation}
We first compute the formal differential commutator in the interior of the domain.  This is
an intermediate identity, not yet the commutator form. Since
$i[K_0,A_V]=[K_0,B_\phi]$ formally, we seek to compute the following:
\begin{align*}
  [-\Delta_g, \nabla_{V}  +\frac{1}{2}\Delta_g\phi]\psi &= -[\Delta_g, \nabla_V]\psi  -\frac{1}{2}[\Delta_g, \Delta_g\phi]\psi, \\
  [\Delta_g, \nabla_V]\psi&=\Delta_g\langle \nabla \phi, \nabla \psi\rangle_g - \langle \nabla \phi, \nabla \Delta_g \psi\rangle_g, \\
  [\Delta_g, \Delta_g\phi]\psi &= \Delta_g((\Delta_g\phi)\psi) - (\Delta_g\phi)(\Delta_g \psi) = (\Delta_g^2\phi)\psi
      +2\ip{\nabla\Delta_g\phi}{\nabla\psi}_g.
\end{align*}

Consider a point $q$ and a local geodesic frame $e_1, \dots, e_d$:
\begin{align*}
    e_i\!\left(e_i\langle\nabla\phi,\nabla\psi\rangle_g\right)
    &=
    e_i\langle\nabla_{e_i}\nabla\phi,\nabla\psi\rangle_g
    +e_i\langle\nabla\phi,\nabla_{e_i}\nabla\psi\rangle_g\\
    &=
    \langle\nabla_{e_i}\nabla_{e_i}\nabla\phi,\nabla\psi\rangle_g
    +2\langle\nabla_{e_i}\nabla\phi,
              \nabla_{e_i}\nabla\psi\rangle_g\\
    &\quad
    +\langle\nabla\phi,
            \nabla_{e_i}\nabla_{e_i}\nabla\psi\rangle_g.
\end{align*}
Then using the definition of Riemannian Hessian in a geodesic frame, the above implies
\begin{align}
     \Delta_g\ip{\nabla\phi}{\nabla\psi}_g
     =\sum_{i=1}^d\Bigl(&
         \ip{\nabla_{e_i}\nabla_{e_i}\nabla\phi}{\nabla\psi}_g
         +2\ip{\nabla_{e_i}\nabla\phi}
                 {\nabla_{e_i}\nabla\psi}_g\notag\\
         &+\ip{\nabla\phi}
                 {\nabla_{e_i}\nabla_{e_i}\nabla\psi}_g
         \Bigr).
     \label{eq:gradient-product-rule-frame}
\end{align}
By definition and the chosen frame, the middle sum is 
\begin{equation}
     \sum_{i=1}^d
         \ip{\nabla_{e_i}\nabla\phi}{\nabla_{e_i}\nabla\psi}_g
     =\sum_{i,j=1}^d
         \Hess\phi(e_i,e_j)\Hess\psi(e_i,e_j)
     =:\Hess\phi:\Hess\psi.
\end{equation}
Consequently,
\begin{align}
    \Delta_g\ip{\nabla\phi}{\nabla\psi}_g
    ={}&\ip{\Delta_g^{\mathrm{rough}}\nabla\phi}{\nabla\psi}_g
       +2\Hess\phi:\Hess\psi +\ip{\nabla\phi}{\Delta_g^{\mathrm{rough}}\nabla\psi}_g.
    \label{eq:gradient-inner-product-laplacian}
\end{align}
 We have the identity
\begin{equation}
    \Delta_g^{\mathrm{rough}}\nabla u
    =\nabla\Delta_gu+\Ric^\sharp(\nabla u),
    \label{eq:bochner-gradient-commutation}
\end{equation}
where $\Ric(Y,\nabla u)=\ip{\Ric^\sharp(\nabla u)}{Y}_g$ with $\Ric^\sharp(Y) \coloneqq \bigl(\Ric(Y,\cdot)\bigr)^\sharp$. Then using the above identity we have
\begin{align*}
    \ip{\Delta_g^{\mathrm{rough}}\nabla\phi}{\nabla\psi}_g + \ip{\nabla\phi}{\Delta_g^{\mathrm{rough}}\nabla\psi}_g &= \langle \nabla \Delta_g \phi, \nabla \psi\rangle_g + \langle \nabla \phi,  \nabla\Delta_g \psi\rangle_g + \ip{\Ric^\sharp(\nabla \phi)}{\nabla \psi}_g\\ &\quad+ \ip{\Ric^\sharp(\nabla \psi)}{\nabla \phi}_g\\
     &=\langle \nabla \Delta_g \phi, \nabla \psi\rangle_g + \langle \nabla \phi,  \nabla\Delta_g \psi\rangle_g + 2\ip{\Ric^\sharp(V)}{\nabla\psi}_g,
\end{align*}
where we used the symmetry of the Ricci tensor.
Thus
\begin{align*}
    \Delta_g\ip{\nabla\phi}{\nabla\psi}_g  - \langle \nabla \phi,  \nabla\Delta_g \psi\rangle_g= 2\Hess\phi:\Hess\psi + \langle \nabla\Delta_g \phi + 2\textup{Ric}^\sharp(\nabla \phi), \nabla\psi\rangle_g.
\end{align*}

So far we have that
\begin{align*}
    i[K, A_V]\psi = -2\Hess\phi:\Hess\psi -2 \langle \nabla\Delta_g \phi + \textup{Ric}^\sharp(\nabla \phi), \nabla\psi\rangle_g - \frac{1}{2}(\Delta_g^2\phi)\psi.
\end{align*}
The last link is the following identity
    \begin{align}
    &\int_{M} \textup{Hess}\phi(\nabla \psi, \nabla \psi) \dvol - \int_{\partial M} \overline{\psi}\textup{Hess}\phi(\nabla \psi, \nu) dS_{g} \notag
    \\&\qquad=\int_{M} -\overline{\psi}\Hess\phi:\Hess\psi \dvol
     -\int_{M}\overline{\psi}\ip{\nabla\Delta_g\phi+\Ric^\sharp(V)}{\nabla\psi}_g \dvol
    \label{eq:hessian-divergence-operator-expanded},
\end{align}
which proves the main result. We complete the proof by deriving the two identities  \eqref{eq:bochner-gradient-commutation} and \eqref{eq:hessian-divergence-operator-expanded} used above.

First we prove \eqref{eq:bochner-gradient-commutation}. Expressions such as $\nabla_Y\Hess u$ are defined using a
vector field near $q$, so we first extend the single tangent vector $Y$ to a
local vector field $\widetilde Y$.  This can be done with
\begin{equation}
    \widetilde Y(q)=Y,
    \qquad
    \nabla_X\widetilde Y(q)=0
    \quad\text{for every} \quad X\in T_qM.
    \label{eq:Y-normal-extension}
\end{equation}
We henceforth write $Y$ for this extension.  Set $X=Z=e_i$ in
\eqref{eq:hessian-curvature-commutation} and sum over $i$. This gives
\begin{align}
    &\sum_{i=1}^d
      \left[(\nabla_{e_i}\Hess u)(e_i,Y)
            -(\nabla_Y\Hess u)(e_i,e_i)\right] =
      \sum_{i=1}^d\ip{R(e_i,Y)\nabla u}{e_i}_g.
    \label{eq:traced-hessian-curvature}
\end{align}
We identify the two terms separately.  At the center of the geodesic frame,
\begin{align}
    \sum_{i=1}^d(\nabla_{e_i}\Hess u)(e_i,Y)
    &=\ip{\Delta_g^{\mathrm{rough}}\nabla u}{Y}_g,
    \label{eq:rough-laplacian-traced-index}\\
    \sum_{i=1}^d(\nabla_Y\Hess u)(e_i,e_i)
    &=Y\left(\sum_{i=1}^d\Hess u(e_i,e_i)\right) =Y(\Delta_gu)
     =\ip{\nabla\Delta_gu}{Y}_g.
    \label{eq:gradient-laplacian-traced-index}
\end{align}
Since the equality holds
for every $Y\in T_qM$, we obtain the desired identity \eqref{eq:bochner-gradient-commutation}.

To prove the identity \eqref{eq:hessian-divergence-operator-expanded}, start with
\begin{equation}
    (\diver\Hess\phi)(Y)
    =\sum_{i=1}^d(\nabla_{e_i}\Hess\phi)(e_i,Y).
    \label{eq:divergence-hessian-as-trace}
\end{equation}
Using the symmetry of the Hessian, \eqref{eq:hessian-curvature-commutation} and a geodesic frame
we get
\begin{align*}
    \sum_{i=1}^d(\nabla_{e_i}\Hess u)(e_i,Y)
    &=\sum_{i=1}^d(\nabla_{e_i}\Hess u)(Y, e_i)\notag\\
    &=\sum_{i=1}^d(\nabla_{Y}\Hess u)(e_i, e_i)+\sum_{i=1}^d\ip{R(e_i,Y)\nabla\phi}{e_i}_g\\
    &=Y\left(\sum_{i=1}^d\Hess\phi(e_i,e_i)\right) +\ip{\Ric^\sharp(\nabla \phi)}{Y}_g\\
    &=\ip{\nabla\Delta_g\phi}{Y}_g + \ip{\Ric^\sharp(\nabla \phi)}{Y}_g.
\end{align*}
Therefore
\begin{equation}
    (\diver\Hess\phi)(\nabla \psi)
    =\ip{\nabla\Delta_g\phi+\Ric^\sharp(V)}{\nabla \psi}_g.
    \label{eq:divergence_hess}
\end{equation}
Consider 
\begin{align*}
    \nabla_{\nabla \psi}\nabla \phi = \sum_{i, j =1}^d \textup{Hess}\phi(e_i, e_j)(e_i\psi)e_j,
\end{align*}
and take the divergence
\begin{align*}
    \textup{div}(\nabla_{\nabla \psi}\nabla \phi) &= \sum_{i,j=1}^d e_j [\textup{Hess}\phi(e_i, e_j)(e_i\psi)] \\ &= \sum_{i,j=1}^d e_j(\textup{Hess}\phi(e_i, e_j)) (e_i\psi) + \sum_{i,j=1}^d (\textup{Hess}\phi(e_i, e_j)) e_j(e_i\psi)\\
    &=\sum_{i,j=1}^d (\nabla_{e_j}\textup{Hess}\phi(e_i, e_j)) (e_i\psi) + \sum_{i,j=1}^d (\textup{Hess}\phi(e_i, e_j)) e_j(e_i\psi)\\
    &=(\textup{div Hess} \phi)(\nabla \psi) + \textup{Hess} \phi : \textup{Hess}\psi,
\end{align*}
where the third equality follows from the frame being geodesic and the covariant derivative product rule.

Combining with \eqref{eq:divergence_hess},
\begin{align*}
     \textup{div}(\nabla_{\nabla \psi}\nabla \phi) = \ip{\nabla\Delta_g\phi+\Ric^\sharp(\nabla \phi)}{\nabla \psi}_g+ \textup{Hess} \phi : \textup{Hess}\psi. 
\end{align*}
Thus on the interior of $\Omega$:
\begin{align*}
    i[K, A_V]\psi = -2 \textup{div}(\nabla_{\nabla \psi}\nabla \phi) - \frac{1}{2}(\Delta_g^2\phi)\psi.
\end{align*}

Integration by parts, without imposing a boundary condition, gives
\begin{align}
        \langle\psi, i[K, A_V]\psi\rangle
        ={}&2\int_\Omega
            \Hess\phi(\nabla\psi,\nabla\overline\psi)
            \,\dd\operatorname{vol}_g
            -\frac12\int_\Omega(\Delta_g^2\phi)|\psi|^2
            \,\dd\operatorname{vol}_g\notag\\
        &-2\int_{\partial\Omega}
            \overline\psi\,\Hess\phi(\nabla\psi,n)\,\dd S_g.
        \label{eq:virial-interior-form}
\end{align}
The following Green's identities compare this interior expression with the commutator
form:
\begin{align}
        \langle\psi,K_0B_\phi\psi\rangle
        -\langle K_0\psi,B_\phi\psi\rangle
        =\int_{\partial\Omega}\Bigl[
             (\partial_n\overline\psi)B_\phi\psi
             -\overline\psi\,\partial_n(B_\phi\psi)
            \Bigr]\,\dd S_g,
        \label{eq:K-green-boundary-form}\\
        \langle\psi,B_\phi K_0\psi\rangle
        +\langle B_\phi\psi,K_0\psi\rangle
        =\int_{\partial\Omega}
            (\partial_n\phi)\overline\psi\,K_0\psi\,\dd S_g.
        \label{eq:B-green-boundary-form}
\end{align}
Subtracting \eqref{eq:B-green-boundary-form} from
\eqref{eq:K-green-boundary-form} and using
\eqref{eq:virial-form-real-part} yields
\begin{equation}
        \mathfrak c_\phi[\psi]
        =\operatorname{Re}
            \langle\psi,i[K, A_V]\psi\rangle
         +\operatorname{Re}\int_{\partial\Omega}\Bigl[
             -(\partial_n\overline\psi)B_\phi\psi
             +\overline\psi\,\partial_n(B_\phi\psi)
             +(\partial_n\phi)\overline\psi\,K_0\psi
            \Bigr]\,\dd S_g.
        \label{eq:virial-form-domain-correction}
\end{equation}
Combining \eqref{eq:virial-interior-form} and
\eqref{eq:virial-form-domain-correction} proves
\eqref{eq:general-virial-form}. 
\end{proof}

Equipped with the above lemma, We can simply extend the identity \eqref{eq:general-virial-form} to the whole domain of some self-adjoint realization of the Laplace--Beltrami operator.

\begin{proposition}[Domain Extension]
\label{prop:graph-closed-virial-form}
Let $K$ be a nonnegative self-adjoint realization of $K_0=-\Delta_g$ with a
smooth operator core
$\mathcal D\subset C^\infty(\overline\Omega)\cap\operatorname{Dom}(K)$
on which $K=K_0$.  Equip $\operatorname{Dom}(K)$ with the graph norm
\begin{equation}
    \|u\|_K^2\coloneqq \|u\|_2^2+\|Ku\|_2^2.
\end{equation}
Assume that the inclusion
$\operatorname{Dom}(K)\hookrightarrow H^1(\Omega)$ and the map
$A_V:\operatorname{Dom}(K)\to L^2(\Omega)$ are graph-norm continuous.
Define, for $u\in\operatorname{Dom}(K)$,
\begin{equation}
    \mathfrak c_\phi^K[u]
    \coloneqq i\left(\langle Ku,A_Vu\rangle
       -\langle A_Vu,Ku\rangle\right).
    \label{eq:closed-virial-commutator-form}
\end{equation}
Then the classical boundary functional
\eqref{eq:general-virial-boundary-functional}, initially defined on
$\mathcal D$, has a unique graph-norm continuous extension to
$\operatorname{Dom}(K)$, namely
\begin{align}
    \overline{\mathcal B}_\phi^K[u]
    \coloneqq {}&\mathfrak c_\phi^K[u]
      -2\int_\Omega
       \Hess\phi(\nabla u,\nabla\overline u)
       \,\dd\operatorname{vol}_g\notag\\
      &+\frac12\int_\Omega
       (\Delta_g^2\phi)|u|^2\,\dd\operatorname{vol}_g.
    \label{eq:closed-boundary-functional}
\end{align}
Consequently, every $u\in\operatorname{Dom}(K)$ satisfies the closed
quadratic-form identity
\begin{align}
    \mathfrak c_\phi^K[u]
    ={}&2\int_\Omega
      \Hess\phi(\nabla u,\nabla\overline u)
      \,\dd\operatorname{vol}_g
      -\frac12\int_\Omega(\Delta_g^2\phi)|u|^2
      \,\dd\operatorname{vol}_g
      +\overline{\mathcal B}_\phi^K[u].
    \label{eq:closed-general-virial-form}
\end{align}
\end{proposition}

\begin{proof}
The assumptions make $u\mapsto A_Vu$ continuous from the graph domain to
$L^2$, so \eqref{eq:closed-virial-commutator-form} is graph-norm
continuous.  The two bulk integrals are graph-norm continuous by the
continuous embedding into $H^1$, the boundedness of $\Hess\phi$, and
$\Delta_g^2\phi\in L^\infty$.  Hence
\eqref{eq:closed-boundary-functional} is graph-norm continuous.  On
$\mathcal D$, Lemma~\ref{lem:main_identity} identifies it with the
classical boundary functional.  Since $\mathcal D$ is an operator core, it
is graph-norm dense in $\operatorname{Dom}(K)$, proving uniqueness and
\eqref{eq:closed-general-virial-form}.
\end{proof}

We will choose to consider the setting of Dirichlet boundary conditions on the operator $-\Delta_g$ over $\overline{\Omega} \subseteq M$. Under these conditions, a few of the boundary terms simplify. For example $A_V$ becomes a symmetric operator on the common core by Lemma \ref{lem:a_boundary_lemma}.
Graph continuity of $A_V$ and graph density of the core extend the symmetry to $\operatorname{Dom}(K)$.

Elliptic regularity makes the graph norm of $K_D$ equivalent to the $H^2$
norm.  Thus $A_V:H^2\to L^2$ and the inclusion into $H^1$ are continuous,
so Proposition~\ref{prop:graph-closed-virial-form} applies.  Moreover, the
normal-trace map $H^2(\Omega)\to H^{1/2}(\partial\Omega)$ is continuous.
Approximating by the smooth Dirichlet operator core in graph norm and using
\eqref{eq:dirichlet-virial-boundary-functional} proves
the extension of \eqref{eq:general-virial-form} to the domain of $K$.

For Dirichlet boundary conditions the boundary gradient is purely normal:
\begin{equation}
        \nabla\psi=(\partial_n\psi)n
        \qquad\text{on }\partial\Omega.
        \label{eq:dirichlet-gradient-normal}
\end{equation}
Since $V=\nabla\phi$ and the zeroth-order term in $B_\phi\psi$ contains the
vanishing factor $\psi$, equation \eqref{eq:dirichlet-gradient-normal} gives
\begin{equation}
        B_\phi\psi
        =\ip{\nabla\phi}{\nabla\psi}_g
            +\frac12(\Delta_g\phi)\psi
        =(\partial_n\phi)\partial_n\psi
        \qquad\text{on }\partial\Omega.
        \label{eq:dirichlet-B-trace}
\end{equation}
Now inspect the four terms in
\eqref{eq:general-virial-boundary-functional}.  The first, third, and fourth
contain the factor $\overline\psi$ and therefore vanish.  The remaining term
is
\begin{align}
        -\operatorname{Re}\int_{\partial\Omega}
            (\partial_n\overline\psi)B_\phi\psi\,\dd S_g
        &=-\int_{\partial\Omega}
            (\partial_n\phi)|\partial_n\psi|^2\,\dd S_g,
\end{align}
where \eqref{eq:dirichlet-B-trace} was used and
$\partial_n\phi$ is real.  This proves
\begin{equation}
    \mathcal B_\phi[\psi]
    =-\int_{\partial\Omega}
      (\partial_n\phi)|\partial_n\psi|^2\,\dd S_g.
    \label{eq:dirichlet-virial-boundary-functional}
\end{equation}

 The
Hamiltonian is
\begin{equation}
    H(t)=c_t\left(\frac{K}{2m_t}+m_t\omega_t^2 f\right),
    \label{eq:manifold-H}
\end{equation}
where $m_t,\omega_t,c_t>0$ and $K = -\Delta_g$ is the Dirichlet operator on $\overline{\Omega} \subset M$.  We assume the schedule obeys
\begin{equation}
    \frac{\dot m_t}{m_t}=\lambda c_t,
    \qquad \lambda>0.
    \label{eq:mass-scaling}
\end{equation}

Define
\begin{equation}
    P_t=\frac{-i}{m_t}\nabla+\lambda V,
\end{equation}
and define its nonnegative quadratic form by
\begin{equation}
    \mathcal T_t[\psi]
    \coloneqq \frac12\|P_t\psi\|_2^2.
\end{equation}
The Lyapunov functional is
\begin{equation}
    \mathcal E_t
    =\mathcal T_t[\psi_t]
     +\omega_t^2\langle f-f_\star\rangle_t,
    \qquad f_\star=f(x_\star).
    \label{eq:manifold-energy}
\end{equation}

{
Expanding $\mathcal{T}_t[\psi]$ gives
\begin{align*}
   \frac{1}{2}\int_{\Omega}
       \left\lvert \frac{-i\nabla \psi}{m_t}+\lambda V\psi\right\rvert_g^2
       \dvol
   ={}&\frac{1}{2m_t^2}\int_{\Omega}|\nabla\psi|_g^2\,\dvol\\
   &+\frac{\lambda}{2m_t}\int_{\Omega}
       \left(\langle-i\nabla\psi,V\psi\rangle_g
       +\langle V\psi,-i\nabla\psi\rangle_g\right)\dvol
     \\&+\frac{\lambda^2}{2}\int_{\Omega}|V|_g^2|\psi|^2\,\dvol.
\end{align*}
In particular,
\begin{equation*}
    \int_{\Omega}\langle V\psi,-i\nabla\psi\rangle_g\,\dvol
    =-i\int_{\Omega}\overline\psi\,\nabla_V\psi\,\dvol.
\end{equation*}
Using the same integration by parts as in
Lemma~\ref{lem:a_boundary_lemma}, together with the symmetry of $A_V$, the
sum of the two cross terms is $2\langle A_V\rangle_\psi$.  Therefore
\begin{equation}
    \mathcal T_t[\psi]
    =\frac{1}{2m_t^2}\langle K\rangle_{\psi}
     +\frac{\lambda}{m_t}\langle A_V\rangle_\psi
     +\frac{\lambda^2}{2}\langle|V|_g^2\rangle_\psi.
    \label{eq:P-expansion-manifold},
\end{equation}
All terms in \eqref{eq:P-expansion-manifold} are real, and nonnegativity
follows directly from the definition of $\mathcal T_t$.
}

\begin{theorem}[QHD convergence with a favorable boundary form]
\label{thm:manifold-qhd}
Let $\Omega\subset M$ be a geodesically convex, compact domain with smooth boundary, let
$x_\star\in\Omega$, and suppose $\overline\Omega$ is contained in a normal
neighborhood of $x_\star$.  Set
\begin{equation}
    \phi=\frac12d_g(\cdot,x_\star)^2,
    \qquad
    C_M=\|\Delta_g^2\phi\|_{L^\infty(\Omega)},
\end{equation}
and suppose $\Hess(\phi)\preceq g$ on $\Omega$.  Let
$f\in W^{1,\infty}(\Omega)$ be real and geodesically convex, minimized at
$x_\star$.  Let $K$ be a Dirichlet realization.  Let normalized
$\psi_0\in\operatorname{Dom}(K)$ and let $\psi_t$ be the corresponding
strong solution of the Schrodinger equation with Hamiltonian
\eqref{eq:manifold-H}, with regularity
\eqref{eq:strong-solution-regularity}.  Assume only along this trajectory that
\begin{equation}
    \overline{\mathcal B}_\phi^K[\psi_t]\leq0.
    \label{eq:trajectory-boundary-sign}
\end{equation}
Let $m_t,\omega_t,c_t$ be positive $C^1$ functions on the time interval.  If
\eqref{eq:mass-scaling} holds and
\begin{equation}
    2\frac{\dot\omega_t}{\omega_t}\leq \lambda c_t,
    \label{eq:omega-scaling}
\end{equation}
then
\begin{equation}
    \mathcal{E}_t
    \leq
    \mathcal{E}_0+\frac{C_M}{8m_0^2}.
    \label{eq:energy-bound}
\end{equation}
Consequently,
\begin{equation}
    \langle f-f_\star\rangle_t
    \leq
    \left(\mathcal E_0+\frac{C_M}{8m_0^2}\right)\omega_t^{-2}.
    \label{eq:function-bound}
\end{equation}
\end{theorem}
\begin{proof}
We compute $\dot{\mathcal E}_t$ in quadratic form.  

{
Let $\psi$ satisfy
\begin{equation}
    \psi\in C([0,T];\operatorname{Dom}(K))
    \cap C^1([0,T];L^2(\Omega)),
    \label{eq:strong-solution-regularity}
\end{equation} and
$i\partial_t\psi_t=H(t)\psi_t$, where $\operatorname{Dom}(K)$ carries its
graph norm.  All commutators below are interpreted as quadratic-form
identities on this trajectory.

Differentiate $\mathcal T_t[\psi_t]$ once, treating it as a
time-dependent quadratic form.  Its explicit derivative is
\begin{equation}
    (\partial_t\mathcal T_t)[\psi_t]
        =-\frac{\dot m_t}{m_t^3}\mathfrak q_K[\psi_t]
         -\lambda\frac{\dot m_t}{m_t^2}\langle A_V\rangle_t.
    \label{eq:P-explicit-derivative}
\end{equation}
The state-evolution part, obtained from \eqref{eq:P-expansion-manifold}, is
\begin{align}
    \dot{\mathcal E}_t
    ={}&(\partial_t\mathcal T_t)[\psi_t]
      +2\dot\omega_t\omega_t\langle f-f_\star\rangle_t\notag\\
     &+\frac{c_t\omega_t^2}{2m_t}
        \left(\mathfrak c_{f,K}[\psi_t]
              +\mathfrak c_{K,f}[\psi_t]\right)\notag\\
     &+\frac{c_t\lambda}{2m_t^2}\mathfrak c_{K,A_V}[\psi_t]
      +c_t\lambda\omega_t^2\mathfrak c_{f,A_V}[\psi_t]
      +\frac{c_t\lambda^2}{4m_t}
        \mathfrak c_{K,|V|_g^2}[\psi_t].
    \label{eq:qhd-energy-derivative}
\end{align}
In particular, the explicit derivative in
\eqref{eq:P-explicit-derivative} occurs exactly once.  The two forms
coupling $K$ and $f$ vanish because
\begin{equation}
    \frac{c_t\omega_t^2}{2m_t}
    \left(
      \mathfrak c_{K,f}[\psi_t]
      +\mathfrak c_{f,K}[\psi_t]
    \right)=0
    \label{eq:K-f-form-cancellation}
\end{equation}
by antisymmetry of \eqref{eq:commutator-form-convention}.

For $f\in W^{1,\infty}(\Omega)$, the relation
$i[f,A_V]=-\ip{\nabla f}{V}_g$ is understood as a quadratic-form identity
with the weak gradient.  It gives
\begin{equation}
    \mathfrak c_{f,A_V}[\psi_t]
    =-\left\langle\ip{\nabla f}{V}_g\right\rangle_t.
    \label{eq:f-A-commutator-form}
\end{equation}
Its contribution is therefore
$-\lambda c_t\omega_t^2
\langle\ip{\nabla f}{V}_g\rangle_t$.  Thus the complete
potential contribution to $\dot{\mathcal E}_t$ is
\begin{equation}
    2\dot\omega_t\omega_t\langle f-f_\star\rangle_t
    -\lambda c_t\omega_t^2\langle\ip{\nabla f}{V}_g\rangle_t.
    \label{eq:potential-part}
\end{equation}
Geodesic convexity along the unique minimizing geodesic from $x$ to
$x_\star$, together with Rademacher differentiability, gives
\[
    f-f_\star\leq \ip{\nabla f}{V}_g
\]
almost everywhere with respect to Riemannian volume.  Hence
\eqref{eq:potential-part} is bounded above by
\begin{equation}
    \omega_t^2\left(2\frac{\dot\omega_t}{\omega_t}-\lambda c_t\right)
    \langle f-f_\star\rangle_t.
    \label{eq:potential-upper}
\end{equation}
}

It remains to analyze the remaining kinetic parts.  Using
\eqref{eq:P-expansion-manifold}, its contribution to $\dot{\mathcal E}_t$ is
\begin{equation}
    \frac{c_t}{2m_t}
    \left(
            \frac{\lambda}{m_t}\mathfrak c_\phi^K[\psi_t]
      +\frac{\lambda^2}{2}
        \mathfrak c_{K,|V|_g^2}[\psi_t]
    \right).
    \label{eq:kinetic-commutator-form}
\end{equation}
The first term is given completely by
\eqref{eq:closed-general-virial-form}.

For a smooth real multiplier $h$, the gradient-form identity
\begin{align}
        \mathfrak q_K[u,v]
        &=\int_\Omega
            \ip{\nabla\overline u}{\nabla v}_g\,\dd\operatorname{vol}_g,
            &&u,v\in\mathcal D,
        \label{eq:admissible-gradient-form}
\end{align}
gives directly
\begin{align*}
    \mathfrak c_{K,h}[\psi]
    &=i\left(\mathfrak q_K[\psi,h\psi]
       -\mathfrak q_K[h\psi,\psi]\right)\\
    &=2\operatorname{Re}\langle A_{\nabla h}\rangle_\psi,
\end{align*}
where
\begin{equation}
    A_{\nabla h}
    =-i\left(\nabla_{\nabla h}+\frac12\Delta_gh\right).
    \label{eq:K-h-commutator-form}
\end{equation}
{For smooth $h$, multiplication by $h$ preserves
the Dirichlet form domain, so this identity extends from $\mathcal D$ to
$\operatorname{Dom}(K)$ by graph-core density (similarly to the domain extension  in Proposition \ref{prop:graph-closed-virial-form}}).
Here we take $h=|V|_g^2$. For the special vector field $V=\nabla\phi$ with
$\phi=\frac12d_g(\cdot,x_\star)^2$, one also has the exact identity
\begin{equation}
    \frac12\nabla |V|_g^2=V.
    \label{eq:W-equals-V}
\end{equation}
Indeed, $|V|_g=d_g(x,x_\star)$ on the normal neighborhood, so
$\frac12|V|_g^2=\phi$ and taking a gradient gives \eqref{eq:W-equals-V}.  

Thus we have
$A_{\nabla h}=2A_V$ by linearity of both the covariant derivative and the
divergence in the vector-field argument.  Here we use symmetry of $A_V$
explicitly: it gives
$\operatorname{Re}\langle A_V\rangle_t=\langle A_V\rangle_t$, and hence
\begin{equation}
    \mathfrak c_{K,|V|_g^2}[\psi_t]
    =4\langle A_V\rangle_t.
    \label{eq:K-V2-commutator-form}
\end{equation}
Consequently, this part of \eqref{eq:kinetic-commutator-form} is
\begin{equation}
    \frac{\lambda^2c_t}{m_t}\langle A_V\rangle_t.
    \label{eq:V2-cross-contribution}
\end{equation}

Since $\dot m_t=\lambda c_tm_t$, the second term in
\eqref{eq:P-explicit-derivative} is
$-\lambda^2c_t\langle A_V\rangle_t/m_t$ and exactly cancels
\eqref{eq:V2-cross-contribution}.  The first term is
\begin{equation}
        -\frac{\lambda c_t}{m_t^2}\mathfrak q_K[\psi_t]
        =-\frac{\lambda c_t}{m_t^2}
            \int_\Omega|\nabla\psi_t|_g^2\,\dd\operatorname{vol}_g,
\end{equation}
by \eqref{eq:admissible-gradient-form}.

Finally, multiplying \eqref{eq:closed-general-virial-form} by
$\lambda c_t/(2m_t^2)$ gives the full commutator contribution
\begin{align}
    \frac{\lambda c_t}{m_t^2}
    \int_\Omega\Hess\phi(\nabla\psi_t,\nabla\overline{\psi_t})
    \,\dd\operatorname{vol}_g
    -\frac{\lambda c_t}{4m_t^2}
            \langle\Delta_g^2\phi\rangle_t
        +\frac{\lambda c_t}{2m_t^2}
            \overline{\mathcal B}_\phi^K[\psi_t].
        \label{eq:scaled-virial-form}
\end{align}

Adding the last two displays leaves precisely
\begin{align}
    \frac{\lambda c_t}{m_t^2}
    \int_\Omega (\Hess\phi-g)(\nabla\psi_t,\nabla\overline{\psi_t})
    \,\dd\operatorname{vol}_g
    -\frac{\lambda c_t}{4m_t^2}\langle\Delta_g^2\phi\rangle_t
        +\frac{\lambda c_t}{2m_t^2}
            \overline{\mathcal B}_\phi^K[\psi_t],
    \label{eq:kinetic-remainder}
\end{align}

Only now do we use the boundary sign:
$\overline{\mathcal B}_\phi^K[\psi_t]\leq0$ by
\eqref{eq:trajectory-boundary-sign}.  By the assumption
$\Hess\phi\preceq g$, the first term in
\eqref{eq:kinetic-remainder} is also nonpositive.
The second term is bounded
above in absolute value by
\begin{equation}
    \frac{\lambda c_t}{4m_t^2}C_M.
\end{equation}

Combining this with \eqref{eq:potential-upper} gives
\begin{equation}
    \dot{\mathcal E}_t
    \leq
    \omega_t^2\left(2\frac{\dot\omega_t}{\omega_t}-\lambda c_t\right)
    \langle f-f_\star\rangle_t
    +\frac{\lambda c_t}{4m_t^2}C_M.
\end{equation}
Under \eqref{eq:omega-scaling}, the first term is nonpositive.  Therefore
\begin{equation}
    \dot{\mathcal E}_t\leq \frac{\lambda c_t}{4m_t^2}C_M.
\end{equation}

Finally, using $\dot m_t/m_t=\lambda c_t$,
\begin{equation}
    \int_0^t\frac{\lambda c_s}{m_s^2}\,\dd s
    =\int_0^t \frac{\dot m_s}{m_s^3}\,\dd s
    =\frac12\left(\frac1{m_0^2}-\frac1{m_t^2}\right)
    \leq \frac1{2m_0^2}.
\end{equation}

This proves \eqref{eq:energy-bound}.  Since the potential part of
$\mathcal E_t$ is
$\omega_t^2\langle f-f_\star\rangle_t$ and the kinetic part is nonnegative,
\eqref{eq:function-bound} follows.
\end{proof}

{
The convergence theorem itself does not fix a Hamiltonian-simulation model.
For the schedule calculation, we assume the number of queries scales similarily to the quanties provided by Theorem \ref{thm:spherical-digital-simulation} and \ref{thm:simplex-digital-simulation}, which are typical for interaction-picture based simulation \cite{low2019interaction}.  Since the coefficient of
$f$ in \eqref{eq:manifold-H} is $c_tm_t\omega_t^2$, the query count is, up
to constant and polylogarithmic factors,
\begin{equation}
        \mathcal Q_f(T)
        =\|f\|_{L^\infty(\Omega)}
            \int_0^T c_t m_t\omega_t^2\,\dd t
        \label{eq:riemannian-potential-oracle-cost}
\end{equation}
queries to a value oracle for $f$.  This expression counts only
potential-oracle queries.  Costs for implementing the kinetic realization
$K$, preparing the initial state, discretizing the manifold, and performing
measurements are additional and are not included in
\eqref{eq:riemannian-potential-oracle-cost}.
}

For fixed normalized $\psi_0$, $m_0$, and $\omega_0$, define the exact
prefactor certified by Theorem~\ref{thm:manifold-qhd} by
\begin{align}
        B_M(\lambda)
        \coloneqq {}&\mathcal E_0(\lambda)+\frac{C_M}{8m_0^2}\notag\\
        ={}&\frac{\mathfrak q_K[\psi_0]}{2m_0^2}
             +\frac{\lambda}{m_0}\langle A_V\rangle_0
             +\frac{\lambda^2}{2}\langle |V|_g^2\rangle_0
             +\omega_0^2\langle f-f_\star\rangle_0
             +\frac{C_M}{8m_0^2}.
        \label{eq:riemannian-schedule-prefactor}
\end{align}
Thus \eqref{eq:function-bound} reads
\begin{equation}
        \langle f-f_\star\rangle_t
        \leq B_M(\lambda)\omega_t^{-2}.
        \label{eq:riemannian-bound-B-form}
\end{equation}

\begin{proposition}[Optimal Riemannian QHD schedule]
\label{prop:riemannian-optimal-schedule}
Fix $m_0,\omega_0,\lambda>0$ and the initial state $\psi_0$.  Among schedules
satisfying \eqref{eq:mass-scaling} and \eqref{eq:omega-scaling}, the
potential-oracle count \eqref{eq:riemannian-potential-oracle-cost} required
to certify
\begin{equation}
        \langle f-f_\star\rangle_T\leq\epsilon
\end{equation}
through \eqref{eq:riemannian-bound-B-form} is minimized by saturating the
frequency-growth constraint.  In operational time, the optimal path is
\begin{equation}
        m(s)=m_0e^{\lambda s},
        \qquad
        \omega(s)=\omega_0e^{\lambda s/2}.
        \label{eq:riemannian-optimal-operational-path}
\end{equation}
A convenient constant-speed representative is
\begin{equation}
        c_t=1,
        \qquad
        m_t=m_0e^{\lambda t},
        \qquad
        \omega_t=\omega_0e^{\lambda t/2}.
        \label{eq:riemannian-optimal-schedule}
\end{equation}
{If $\epsilon<B_M(\lambda)/\omega_0^2$, its stopping
time and potential-oracle counts are}
\begin{align}
        T_\epsilon
        &=\frac{2}{\lambda}
            \log\left(
                \frac{\sqrt{B_M(\lambda)/\epsilon}}{\omega_0}
            \right),
        \label{eq:riemannian-optimal-stopping-time}\\
        {\mathcal Q_{f,\epsilon}(\lambda)}
        &{=\frac{\|f\|_{L^\infty(\Omega)}m_0\omega_0^2}{2\lambda}
            \left[
                \left(
                    \frac{B_M(\lambda)}{\epsilon\omega_0^2}
                \right)^2-1
            \right].}
        \label{eq:riemannian-optimal-query-exact}
\end{align}
Consequently,
\begin{equation}
        \mathcal Q_{f,\epsilon}(\lambda)
        =O\left(
            \frac{\|f\|_{L^\infty(\Omega)}m_0}{\lambda\omega_0^2}
            \left(\frac{B_M(\lambda)}{\epsilon}\right)^2
        \right).
        \label{eq:riemannian-optimal-query-asymptotic}
\end{equation}
If instead $\epsilon\geq B_M(\lambda)/\omega_0^2$, the initial state already
satisfies the certified accuracy and one may take $T_\epsilon=0$.
\end{proposition}

\begin{proof}
Introduce operational time
\begin{equation}
        s(t)=\int_0^t c_u\,\dd u,
        \qquad S=s(T).
        \label{eq:riemannian-operational-time}
\end{equation}
Because $c_t>0$, it is a valid reparametrization.  The schedule constraints
become
\begin{equation}
        m(s)=m_0e^{\lambda s},
        \qquad
        \frac{\dd}{\dd s}\log\omega(s)\leq\frac{\lambda}{2},
        \label{eq:riemannian-operational-constraints}
\end{equation}
while the cost is
\begin{equation}
        {\mathcal Q_f(S)
        =\|f\|_{L^\infty(\Omega)}
            \int_0^S m(s)\omega(s)^2\,\dd s.}
        \label{eq:riemannian-operational-query-cost}
\end{equation}
Thus $c_t$ changes only the physical-time parametrization of an operational
path.

For the nontrivial accuracy regime, put
\begin{equation}
        \omega_\epsilon=\sqrt{B_M(\lambda)/\epsilon}>\omega_0.
\end{equation}
Any admissible path that first reaches $\omega_\epsilon$ at operational time
$S$ obeys, by integrating the second inequality in
\eqref{eq:riemannian-operational-constraints} backward,
\begin{equation}
        \omega(s)\geq
        \omega_\epsilon e^{-\lambda(S-s)/2},
        \qquad
        S\geq\frac{2}{\lambda}
             \log\frac{\omega_\epsilon}{\omega_0}.
        \label{eq:riemannian-backward-frequency-bound}
\end{equation}
Substitution into \eqref{eq:riemannian-operational-query-cost} yields
\begin{equation}
    {\mathcal Q_f(S)
        \geq
    \frac{\|f\|_{L^\infty(\Omega)}m_0\omega_\epsilon^2}{2\lambda}
    \left(e^{\lambda S}-e^{-\lambda S}\right).}
        \label{eq:riemannian-schedule-lower-bound}
\end{equation}
The right-hand side is strictly increasing in $S$.  Its minimum is therefore
attained at the smallest allowed operational time, and equality throughout
requires \eqref{eq:riemannian-optimal-operational-path}.  Integrating the
cost along that path gives
\begin{equation}
        {
        \int_0^{S_\epsilon}m_0\omega_0^2e^{2\lambda s}\,\dd s
        =\frac{m_0\omega_0^2}{2\lambda}
            \left[
                \left(\frac{\omega_\epsilon}{\omega_0}\right)^4-1
            \right],}
\end{equation}
which proves \eqref{eq:riemannian-optimal-stopping-time} and
\eqref{eq:riemannian-optimal-query-exact}.  Any positive $c_t$ tracing the
same operational path has the same potential-oracle count.
\end{proof}

The remaining scalar parameter $\lambda$ can also be optimized.  Write
\begin{equation}
        B_M(\lambda)=a_M\lambda^2+b_M\lambda+q_M,
        \label{eq:riemannian-B-quadratic}
\end{equation}
where
\begin{align}
        a_M&=\frac12\langle|V|_g^2\rangle_0,
        &b_M&=\frac{\langle A_V\rangle_0}{m_0},
        \label{eq:riemannian-schedule-ab}\\
        q_M&=\frac{\mathfrak q_K[\psi_0]}{2m_0^2}
             +\omega_0^2\langle f-f_\star\rangle_0
             +\frac{C_M}{8m_0^2}.
        \label{eq:riemannian-schedule-q}
\end{align}
{
If $a_M>0$ and $q_M>0$, the positive value minimizing the leading
small-$\epsilon$ cost $B_M(\lambda)^2/\lambda$ is
\begin{equation}
        \lambda_\star
    =\frac{\sqrt{b_M^2+12a_Mq_M}-b_M}{6a_M}.
        \label{eq:riemannian-optimal-lambda}
\end{equation}
Indeed, differentiating its logarithm gives
$2\lambda B_M'(\lambda)=B_M(\lambda)$, or
$3a_M\lambda^2+b_M\lambda-q_M=0$, whose positive root is the displayed
value.
}

\subsubsection{Riemannian QHD over the Sphere}

Let $M=S^d$ be the unit round sphere.  Fix $x_\star\in S^d$ and let
\[
    r(x)=d_{S^d}(x,x_\star),
    \qquad
    \phi(x)=\frac12r(x)^2.
\]
Assume
\[
    \Omega\subset B_{S^d}(x_\star,R),
    \qquad 0<R<\pi,
\]
and that $\Omega$ is geodesically convex with smooth boundary.

Away from $x_\star$, the spherical metric in geodesic polar coordinates is
\[
    g=\dd r^2+\sin^2 r\,g_{S^{d-1}}.
\]
For a radial function $h(r)$,
\begin{equation}
    \Hess(h)=h''(r)\,\dd r\otimes\dd r
    +h'(r)\cot r\,(g-\dd r\otimes\dd r).
    \label{eq:sphere-radial-hess}
\end{equation}
Taking $h(r)=\frac12r^2$, so $h'(r)=r$ and $h''(r)=1$, gives
\begin{equation}
    \Hess(\phi)
    =\dd r\otimes\dd r+r\cot r\,(g-\dd r\otimes\dd r).
    \label{eq:sphere-hess-phi}
\end{equation}
For $0<r<\pi$, one has $r\cot r\leq1$.  Therefore
\begin{equation}
    \Hess(\phi)\preceq g
    \qquad\text{on }B_{S^d}(x_\star,R).
\end{equation}
At $r=0$, the same conclusion follows by smooth extension.

If $\Omega$ is geodesically convex and $x_\star\in\Omega$, then the outward
star-shaped condition also holds:
\begin{equation}
    \partial_n\phi=\ip{\nabla\phi}{n}_g\geq0
    \qquad\text{on }\partial\Omega.
\end{equation}
Indeed, the minimizing geodesic from a boundary point $x$ to $x_\star$ points
inward, so $-\log_x(x_\star)=\nabla\phi(x)$ has nonnegative pairing with the
outward normal.

For a radial function $h(r)$ on $S^d$,
\begin{equation}
    \Delta h=h''(r)+(d-1)\cot r\,h'(r).
    \label{eq:sphere-radial-lap}
\end{equation}
Thus
\begin{equation}
    \Delta\phi=1+(d-1)r\cot r.
\end{equation}
Let
\begin{equation}
    A(r)=r\cot r\,\csc^2 r-\cot^2 r
    =\frac{\cot r}{\sin^2 r}\bigl(r-\sin r\cos r\bigr).
    \label{eq:A-r}
\end{equation}
We use the continuous extension $A(0)=2/3$.
A direct differentiation gives
\begin{equation}
    \Delta(r\cot r)=-2-(d-3)A(r),
\end{equation}
and hence
\begin{equation}
    \Delta^2\phi(r)=-(d-1)\left[2+(d-3)A(r)\right].
    \label{eq:sphere-biharmonic}
\end{equation}
Define the radius-only factor
\begin{equation}
    \mathcal A_R=\max_{0\leq r\leq R}|A(r)|.
    \label{eq:A-R}
\end{equation}
Then all dimension dependence of the biharmonic correction is explicit:
\begin{equation}
    \|\Delta_{S^d}^2\phi\|_{L^\infty(B(x_\star,R))}
    \leq (d-1)\left(2+|d-3|\mathcal A_R\right).
    \label{eq:C-d-R}
\end{equation}
In particular, the geometric error is $O_R(d^2)$ as $d\to\infty$.  In the
special dimension $d=3$, the $A(r)$ term cancels exactly and
$\Delta^2\phi=-4$, independently of $R$.

The factor $\mathcal A_R$ itself admits the dimension-free elementary bound
\begin{equation}
    \mathcal A_R\leq
    \begin{cases}
        \displaystyle \frac23\left(\frac{R}{\sin R}\right)^2,
            &0<R\leq\pi/2,\\[6pt]
        \displaystyle \max\left\{\frac{\pi^2}{6},
            R|\cot R|\csc^2 R+\cot^2R\right\},
            &\pi/2<R<\pi.
    \end{cases}
    \label{eq:A-R-elementary}
\end{equation}
Indeed, the first bound follows from the calculation below.  For
$\pi/2\leq r\leq R$, both $|\cot r|$ and $\csc^2r$ are increasing, and the
definition of $A$ gives
$|A(r)|\leq R|\cot R|\csc^2R+\cot^2R$.

For $R\leq\pi/2$, the identity
\[
    r-\sin r\cos r=\int_0^r2\sin^2 s\,\dd s\leq \frac23r^3
\]
and the inequality $\cot r\leq1/r$ show that $A(r)\geq0$ and
\[
    A(r)\leq \frac23\left(\frac{R}{\sin R}\right)^2.
\]
Thus the simpler upper bound
\begin{equation}
    \|\Delta_{S^d}^2\phi\|_{L^\infty(B(x_\star,R))}
    \leq(d-1)\left[
        2+\frac23|d-3|\left(\frac{R}{\sin R}\right)^2
    \right]
\end{equation}
is available in an open hemisphere.  Since $R/\sin R\leq\pi/2$ there, the
coarser dimension-only bound
\begin{equation}
    \|\Delta_{S^d}^2\phi\|_{L^\infty(B(x_\star,R))}
    \leq (d-1)\left[2+\frac{\pi^2}{6}|d-3|\right]
\end{equation}
holds.

Since $V=\nabla\phi=r\partial_r$ and
$\diver V=\Delta\phi=1+(d-1)r\cot r$, the first-order multiplier in this
specialization is
\begin{equation}
    A_r\coloneqq A_V
    =-i\left(
        r\partial_r+\frac12\bigl[1+(d-1)r\cot r\bigr]
      \right),
    \label{eq:sphere-dilation-operator}
\end{equation}
with its smooth extension at $r=0$ understood.

\begin{theorem}[Spherical QHD convergence]
\label{thm:sphere-qhd}
Let $x_\star\in\Omega\subset S^d$, where $\Omega$ is a smooth geodesically
convex domain contained in $B(x_\star,R)$ with $R<\pi$.

Let $f$ be
geodesically convex and in $W^{1,\infty}(\Omega)$, minimized at $x_\star$.
Let
\begin{equation}
    \psi_0\in
    \operatorname{Dom}(-\Delta_{S^d}^D)
    =H^2(\Omega)\cap H_0^1(\Omega),
    \qquad \|\psi_0\|_2=1,
    \label{eq:sphere-strong-initial-domain}
\end{equation}
and let $\psi_t$ be the corresponding strong solution of
\begin{equation}
    i\partial_t\psi_t
    =c_t\left(\frac{-\Delta_{S^d}^D}{2m_t}+m_t\omega_t^2f\right)\psi_t
\end{equation}
with Dirichlet boundary conditions on $\partial\Omega$.  {Assume that
$\psi_t$ has the trajectory regularity in
\eqref{eq:strong-solution-regularity}.}  If

\[ 
    m_t,\omega_t,c_t\in C^1([0,T];\mathbb R_{>0}),
    \qquad
    \frac{\dot m_t}{m_t}=\lambda c_t,
    \qquad
    2\frac{\dot\omega_t}{\omega_t}\leq\lambda c_t,
\]
then
\begin{equation}
    \langle f-f(x_\star)\rangle_t
    \leq
    \left[
        \mathcal E_0
        +\frac{(d-1)(2+|d-3|\mathcal A_R)}{8m_0^2}
    \right]\omega_t^{-2},
    \label{eq:sphere-function-bound}
\end{equation}
where $\mathcal A_R$ is the dimension-free quantity in \eqref{eq:A-R} and
\begin{equation}
    \begin{split}
 \mathcal E_0
    ={}&\frac{\langle p^2\rangle_0}{2m_0^2}
      +\frac{\lambda}{m_0}\langle A_r\rangle_0
      +\frac{\lambda^2}{2}\langle r^2\rangle_0 +\omega_0^2\langle f-f(x_\star)\rangle_0,
    \end{split}
    \label{eq:sphere-initial-energy}
\end{equation}
Thus the explicit geometric additive error in
\eqref{eq:sphere-function-bound} is $O_R(d^2/m_0^2)$.  Equations
\eqref{eq:sphere-dilation-operator} and
\eqref{eq:sphere-initial-energy} display
the remaining dependence carried by the initial state.  If the schedules or
initial-state family vary with $d$, their dependence through $m_0$, $\omega_t$,
and the displayed expectations must of course be included separately.
\end{theorem}

{
\begin{proof}
For the Dirichlet realization
$K_D=-\Delta_{S^d}^D$, elliptic regularity gives
$\operatorname{Dom}(K_D)=H^2(\Omega)\cap H_0^1(\Omega)$.  The graph-norm
extension described after Proposition~\ref{prop:graph-closed-virial-form},
together with \eqref{eq:dirichlet-virial-boundary-functional}, gives
\begin{equation}
    \overline{\mathcal B}_\phi^{K_D}[u]
    =-\int_{\partial\Omega}
       (\partial_n\phi)|\partial_nu|^2\,\dd S_g
    \leq0,
    \qquad u\in\operatorname{Dom}(K_D),
\end{equation}
because geodesic convexity gives $\partial_n\phi\geq0$.  Thus the
trajectory boundary condition \eqref{eq:trajectory-boundary-sign} holds.
The Hessian estimate \eqref{eq:sphere-hess-phi} and the biharmonic bound
\eqref{eq:C-d-R} verify the remaining geometric hypotheses of
Theorem~\ref{thm:manifold-qhd}.  Applying that theorem proves
\eqref{eq:sphere-function-bound}.
\end{proof}
}

\begin{corollary}[Optimal schedule for spherical QHD]
\label{cor:sphere-optimal-schedule}
Assume the potential-oracle model
\eqref{eq:riemannian-potential-oracle-cost}.  Define
\begin{equation}
        \overline C_{d,R}
        \coloneqq (d-1)\left(2+|d-3|\mathcal A_R\right)
\end{equation}
and the spherical prefactor
\begin{align}
        B_{d,R}(\lambda)
        \coloneqq {}&\frac{\langle p^2\rangle_0}{2m_0^2}
             +\frac{\lambda}{m_0}\langle A_r\rangle_0
             +\frac{\lambda^2}{2}\langle r^2\rangle_0
             +\omega_0^2\langle f-f(x_\star)\rangle_0
             +\frac{\overline C_{d,R}}{8m_0^2}.
        \label{eq:sphere-schedule-prefactor}
\end{align}
Then Proposition~\ref{prop:riemannian-optimal-schedule} applies with
$B_M(\lambda)$ replaced by $B_{d,R}(\lambda)$.  In particular, the
constant-speed schedule is
\begin{equation}
        c_t=1,
        \qquad m_t=m_0e^{\lambda t},
        \qquad \omega_t=\omega_0e^{\lambda t/2},
\end{equation}
and, whenever $\epsilon<B_{d,R}(\lambda)/\omega_0^2$,
\begin{align}
        T_\epsilon
        &=\frac2\lambda
            \log\left(
                \frac{\sqrt{B_{d,R}(\lambda)/\epsilon}}{\omega_0}
            \right),\quad
        {\mathcal Q_{f,\epsilon}(\lambda)}
        {=\frac{\|f\|_{L^\infty(\Omega)}m_0\omega_0^2}{2\lambda}
            \left[
                \left(
                    \frac{B_{d,R}(\lambda)}{\epsilon\omega_0^2}
                \right)^2-1
            \right].}
        \label{eq:sphere-optimal-query-exact}
\end{align}
Moreover,
\begin{equation}
        B_{d,R}(\lambda)
        =a_{d,R}\lambda^2+b_{d,R}\lambda+q_{d,R},
\end{equation}
where
\begin{align}
        a_{d,R}&=\frac12\langle r^2\rangle_0,\quad
        b_{d,R}=\frac{\langle A_r\rangle_0}{m_0},\quad
        q_{d,R}=\frac{\langle p^2\rangle_0}{2m_0^2}
             +\omega_0^2\langle f-f(x_\star)\rangle_0
             +\frac{\overline C_{d,R}}{8m_0^2}.
        \label{eq:sphere-schedule-coefficients}
\end{align}
If $a_{d,R}>0$ and $q_{d,R}>0$, the leading small-$\epsilon$ query count is
minimized at
\begin{equation}
        {\lambda_\star^{S^d}
        =\frac{
             \sqrt{b_{d,R}^2+12a_{d,R}q_{d,R}}-b_{d,R}
            }{6a_{d,R}}.}
        \label{eq:sphere-optimal-lambda}
\end{equation}
\end{corollary}

\begin{proof}
For $\phi=r^2/2$, one has $|V|_g^2=r^2$,
$\mathfrak q_K[\psi_0]=\langle p^2\rangle_0$, and $A_V=A_r$ by
\eqref{eq:sphere-dilation-operator}.  Equation~\eqref{eq:C-d-R} gives
$C_M\leq\overline C_{d,R}$.  Substitution into
\eqref{eq:riemannian-schedule-prefactor},
\eqref{eq:riemannian-optimal-query-exact}, and
\eqref{eq:riemannian-optimal-lambda} proves every displayed formula.
\end{proof}

If the geometric term dominates for fixed $R$ and fixed $\lambda$, then
\eqref{eq:riemannian-optimal-query-asymptotic} gives
\begin{equation}
        {\mathcal Q_{f,\epsilon}(\lambda)
        =O_R\left(
             \frac{\|f\|_{L^\infty(\Omega)}}
                        {\lambda\omega_0^2m_0^3}
             \frac{d^4}{\epsilon^2}
            \right).}
        \label{eq:sphere-fixed-lambda-query-scaling}
\end{equation}
More specifically, for an initial-state family satisfying
\begin{equation}
        a_{d,R}=\Theta_R(1),
        \qquad
        b_{d,R}=O_R(d/m_0),
        \qquad
        q_{d,R}=\Theta_R(d^2/m_0^2),
\end{equation}
one has $\lambda_\star^{S^d}=\Theta_R(d/m_0)$ and
\begin{equation}
        {\mathcal Q_{f,\epsilon}(\lambda_\star^{S^d})
        =O_R\left(
             \frac{\|f\|_{L^\infty(\Omega)}}{m_0^2\omega_0^2}
             \frac{d^3}{\epsilon^2}
            \right).}
        \label{eq:sphere-optimized-query-scaling}
\end{equation}
These dimension estimates concern only the potential-oracle model
\eqref{eq:riemannian-potential-oracle-cost}; they do not include kinetic,
discretization, state-preparation, or measurement costs, nor any hidden
dimension dependence of $\|f\|_{L^\infty(\Omega)}$.

\subsubsection{Riemannian QHD over  the Simplex}

We now specialize the Riemannian theorem to the simplex using Laplace--Beltrami operator from Section \ref{sec:global_spec_simplex}: {
\begin{equation}
    \Delta_{\Simp_d}
    =\sum_{i,j=1}^d x_i(\delta_{ij}-x_j)\partial_i\partial_j
     +\frac12\sum_{i=1}^d
        \left(1-(d+1)x_i\right)\partial_i
    \label{eq:qhd-wf-half-generator}
\end{equation}
on $L^2(\Simp_d,\dvol)$, where
\begin{equation}
        \dvol(x)
        =\prod_{i=1}^{d+1}x_i^{-1/2}
            \,\dd x_1\cdots\dd x_d.
    \label{eq:qhd-wf-half-measure}.
\end{equation}
}
Its contravariant metric is
\begin{equation}
    g_{\mathrm{WF}}^{ij}(x)=x_i(\delta_{ij}-x_j),
    \label{eq:qhd-wf-contravariant-metric}
\end{equation}
or, intrinsically for $\sum_{i=1}^{d+1}\xi_i=0$,
\begin{equation}
    |\xi|_{g_{\mathrm{WF}}}^2
    =\sum_{i=1}^{d+1}\frac{\xi_i^2}{x_i}.
    \label{eq:qhd-wf-intrinsic-metric}
\end{equation}
Let
\begin{equation}
    \pi:S_+^d\rightarrow\Delta_d,
    \qquad
    \pi(y)_i=y_i^2,
    \qquad
    S_+^d=\{y\in S^d:y_i\geq0\}.
    \label{eq:qhd-simplex-square-root-map}
\end{equation}
Since $\dd x_i=2y_i\dd y_i$, the pullback metric and the associated
distance are
\begin{align}
    \pi^*g_{\mathrm{WF}}&=4g_{S^d},
    \label{eq:qhd-fisher-sphere-metric-scaling}\\
    d_{\mathrm{WF}}(x,z)
    &=2\arccos\left(\sum_{i=1}^{d+1}\sqrt{x_iz_i}\right).
    \label{eq:qhd-fisher-rao-distance}
\end{align}
{Let $\dd S_d$ denote ordinary spherical area on
$S_+^d$. The metric identity
$\pi^*g_{\mathrm{WF}}=4g_{S^d}$ implies
\begin{equation*}
 \pi^*\dvol=2^d\dd S_d.
\end{equation*}
Therefore}
\begin{equation}
    (Uu)(y)\coloneqq 2^{d/2}u(\pi(y))
    \label{eq:qhd-wf-unitary-lift}
\end{equation}
is unitary from $L^2(\Delta_d,\dvol)$ to
$L^2(S_+^d,\dd S_d)$. The chain rule gives
{
\begin{equation}
    U\Delta_{\Simp_d}U^{-1}=\frac14\Delta_{S^d}.
    \label{eq:qhd-wf-sphere-generator-equivalence}
\end{equation}
}

Let $\widetilde\Omega$ be a smooth geodesically convex domain compactly
contained in the open spherical orthant
$S_{++}^d=\{y\in S^d:y_i>0\}$, choose
$y_\star\in\widetilde\Omega$, and put
\begin{equation}
    \Omega=\pi(\widetilde\Omega),
    \qquad
    x_\star=\pi(y_\star),
    \qquad
    R_\Omega
      =\sup_{y\in\overline{\widetilde\Omega}}
          d_{S^d}(y,y_\star)<\frac\pi2.
    \label{eq:qhd-wf-covering-radius}
\end{equation}
Fix $R$ with $R_\Omega<R\leq\pi/2$.  Compact containment also gives the
positive face margin
\begin{equation}
    \eta_\Omega
    \coloneqq \min_{y\in\overline{\widetilde\Omega}}
       \min_{1\leq i\leq d+1}y_i>0.
    \label{eq:qhd-wf-face-margin}
\end{equation}
We call $f:\Omega\to\mathbb R$ Fisher--Rao geodesically convex when its
lift
\begin{equation}
    F(y)\coloneqq f(\pi(y))
    \label{eq:qhd-wf-lifted-potential}
\end{equation}
is geodesically convex on $\widetilde\Omega$.  This is equivalent to
geodesic convexity for $g_{\mathrm{WF}}$, because the constant factor in
\eqref{eq:qhd-fisher-sphere-metric-scaling} changes lengths but not geodesic
paths.

\begin{theorem}[QHD convergence on the Fisher--Rao simplex]
\label{thm:simplex-qhd}
Let the domain, minimizer, and radius satisfy the preceding assumptions, and
let $f\in W^{1,\infty}(\Omega)$ be real, Fisher--Rao geodesically convex,
and minimized at $x_\star$.  Let $-\Delta_g^D$ be the Dirichlet
realization on $\Omega$ unitarily equivalent under $U$ to
$-\Delta_{S^d}^D/4$ on $\widetilde\Omega$.  For normalized
$u_0\in\operatorname{Dom}(-\Delta_g^D)$, let $u_t$ be the
corresponding strong solution of
{
\begin{equation}
    i\partial_tu_t
    =c_t\left(
    \frac{-\Delta_{\Simp_d}^D}{2m_t}
       +m_t\omega_t^2f
      \right)u_t.
    \label{eq:qhd-wf-half-Hamiltonian}
\end{equation}
}
{Assume that $u_t$ has the trajectory regularity in
\eqref{eq:strong-solution-regularity}.}
If
\begin{equation}
    \frac{\dot m_t}{m_t}=\lambda c_t,
    \qquad
    2\frac{\dot\omega_t}{\omega_t}\leq\lambda c_t,
\end{equation}
then
\begin{equation}
    \langle f-f(x_\star)\rangle_t
    \leq B_{d,R}^{\mathrm{WF}}(\lambda)\omega_t^{-2},
    \label{eq:qhd-wf-half-convergence}
\end{equation}
where
\begin{align}
    B_{d,R}^{\mathrm{WF}}(\lambda)
    \coloneqq {}&\frac{\langle-\Delta_g^D\rangle_{u_0}}{2m_0^2}
       +\frac{\lambda}{m_0}\langle A_r\rangle_{Uu_0}
       +2\lambda^2\langle r^2\rangle_{Uu_0}\notag\\
       &+\omega_0^2\langle f-f(x_\star)\rangle_{u_0}
       +\frac{C_{d,R}^{\mathrm{WF}}}{32m_0^2},
    \label{eq:qhd-wf-schedule-prefactor}\\
    C_{d,R}^{\mathrm{WF}}
    \coloneqq {}&(d-1)\left(2+|d-3|\mathcal A_R\right).
    \label{eq:qhd-wf-geometric-constant}
\end{align}
Here $r=d_{S^d}(\cdot,y_\star)$, and $A_r$ is
\eqref{eq:sphere-dilation-operator} centered at $y_\star$.
\end{theorem}

\begin{proof}
Let
\begin{equation}
    \phi_{\mathrm{WF}}(x)
      =\frac12d_{\mathrm{WF}}(x,x_\star)^2,
    \qquad
    \phi_S(y)=\frac12r(y)^2.
\end{equation}
Equations \eqref{eq:qhd-fisher-sphere-metric-scaling} and
\eqref{eq:qhd-fisher-rao-distance} give
\begin{equation}
    \phi_{\mathrm{WF}}\circ\pi=4\phi_S.
\end{equation}
Because multiplication of a metric by a constant leaves its Levi--Civita
connection unchanged,
\begin{equation}
    \pi^*\Hess_{\mathrm{WF}}\phi_{\mathrm{WF}}
      =4\Hess_{S^d}\phi_S
      \preceq4g_{S^d}=\pi^*g_{\mathrm{WF}}.
\end{equation}
The lifted virial vector field is
\begin{equation}
    \nabla_{4g_{S^d}}(4\phi_S)=\nabla_{g_{S^d}}\phi_S=r\partial_r,
\end{equation}
and consequently
\begin{align}
    U A_{\nabla_{\mathrm{WF}}\phi_{\mathrm{WF}}}U^{-1}&=A_r,
    \label{eq:qhd-wf-virial-lift}\\
    |\nabla_{\mathrm{WF}}\phi_{\mathrm{WF}}|_{g_{\mathrm{WF}}}^2
      \circ\pi&=4r^2.
    \label{eq:qhd-wf-virial-norm-lift}
\end{align}
Moreover, \eqref{eq:qhd-wf-sphere-generator-equivalence} gives
{
\begin{align}
    \mathfrak q_{-\Delta_g^D}[u_0]
      &=\frac14\langle p_{S^d}^2\rangle_{Uu_0},
    \label{eq:qhd-wf-kinetic-lift}\\
    \|\Delta_g^2\phi_{\mathrm{WF}}\|_\infty
      &=\frac14\|\Delta_{S^d}^2\phi_S\|_\infty
      \leq\frac14C_{d,R}^{\mathrm{WF}}.
    \label{eq:qhd-wf-biharmonic-lift}
\end{align}
}
{By the assumed Dirichlet unitary equivalence, $U$
identifies the corresponding Dirichlet form domains.  Applying the
graph-core extension of \eqref{eq:dirichlet-virial-boundary-functional} on
$\widetilde\Omega$ gives the favorable boundary sign because
$\widetilde\Omega$ is geodesically convex.}  Substituting
\eqref{eq:qhd-wf-virial-lift}--\eqref{eq:qhd-wf-biharmonic-lift} into
Theorem~\ref{thm:manifold-qhd} gives
\eqref{eq:qhd-wf-half-convergence} and
\eqref{eq:qhd-wf-schedule-prefactor}.
\end{proof}

\begin{corollary}[Optimal schedule for simplex QHD]
\label{cor:simplex-optimal-schedule}
Assume the potential-oracle model
\eqref{eq:riemannian-potential-oracle-cost}, now with
$\|f\|_{L^\infty(\Omega)}$ and the Wright--Fisher Hamiltonian
\eqref{eq:qhd-wf-half-Hamiltonian}.  For fixed
$m_0,\omega_0,\lambda>0$ and $u_0$, the optimal operational path certified
by Theorem~\ref{thm:simplex-qhd} is
\begin{equation}
    m(s)=m_0e^{\lambda s},
    \qquad
    \omega(s)=\omega_0e^{\lambda s/2}.
\end{equation}
A constant-speed representative is
\begin{equation}
    c_t=1,
    \qquad
    m_t=m_0e^{\lambda t},
    \qquad
    \omega_t=\omega_0e^{\lambda t/2}.
    \label{eq:qhd-wf-optimal-schedule}
\end{equation}
If
$\epsilon<B_{d,R}^{\mathrm{WF}}(\lambda)/\omega_0^2$, then
\begin{align}
    T_\epsilon^{\mathrm{WF}}
    &=\frac2\lambda
      \log\left(
       \frac{\sqrt{B_{d,R}^{\mathrm{WF}}(\lambda)/\epsilon}}{\omega_0}
      \right),
    \label{eq:qhd-wf-optimal-stopping-time}\\
        {\mathcal Q_{f,\epsilon}^{\mathrm{WF}}(\lambda)}
        &{=\frac{\|f\|_{L^\infty(\Omega)}m_0\omega_0^2}{2\lambda}
      \left[
       \left(
        \frac{B_{d,R}^{\mathrm{WF}}(\lambda)}
             {\epsilon\omega_0^2}
             \right)^2-1
            \right].}
    \label{eq:qhd-wf-optimal-query-exact}
\end{align}
Furthermore,
\begin{equation}
    B_{d,R}^{\mathrm{WF}}(\lambda)
      =a_{\mathrm{WF}}\lambda^2
       +b_{\mathrm{WF}}\lambda+q_{\mathrm{WF}},
    \label{eq:qhd-wf-B-quadratic}
\end{equation}
where
\begin{align}
    a_{\mathrm{WF}}&=2\langle r^2\rangle_{Uu_0},
    &b_{\mathrm{WF}}&=\frac{\langle A_r\rangle_{Uu_0}}{m_0},
    \label{eq:qhd-wf-schedule-ab}\\
    q_{\mathrm{WF}}
    &=\frac{\langle-\Delta_g^D\rangle_{u_0}}{2m_0^2}
       +\omega_0^2\langle f-f(x_\star)\rangle_{u_0}
       +\frac{C_{d,R}^{\mathrm{WF}}}{32m_0^2}.
    \label{eq:qhd-wf-schedule-q}
\end{align}
If $a_{\mathrm{WF}}>0$ and $q_{\mathrm{WF}}>0$, the leading
small-$\epsilon$ potential-oracle count is minimized at
\begin{equation}
        {\lambda_\star^{\mathrm{WF}}
    =\frac{
             \sqrt{b_{\mathrm{WF}}^2+12a_{\mathrm{WF}}q_{\mathrm{WF}}}
       -b_{\mathrm{WF}}
            }{6a_{\mathrm{WF}}}.}
    \label{eq:qhd-wf-optimal-lambda}
\end{equation}
\end{corollary}

\begin{proof}
Apply Proposition~\ref{prop:riemannian-optimal-schedule} with the prefactor
$B_M(\lambda)$ replaced by
$B_{d,R}^{\mathrm{WF}}(\lambda)$ from
\eqref{eq:qhd-wf-schedule-prefactor}.  Collecting its powers of $\lambda$
gives \eqref{eq:qhd-wf-schedule-ab}--\eqref{eq:qhd-wf-schedule-q}, and
\eqref{eq:riemannian-optimal-lambda} gives
\eqref{eq:qhd-wf-optimal-lambda}.
\end{proof}

For fixed $R$ and fixed $\lambda$, if the geometric term dominates, then
\begin{equation}
        {\mathcal Q_{f,\epsilon}^{\mathrm{WF}}(\lambda)
        =O_R\left(
      \frac{\|f\|_{L^\infty(\Omega)}}
                     {\lambda\omega_0^2m_0^3}
            \frac{d^4}{\epsilon^2}
        \right).}
    \label{eq:qhd-wf-fixed-lambda-query-scaling}
\end{equation}
If an initial-state family satisfies
\begin{equation}
    a_{\mathrm{WF}}=\Theta_R(1),
    \qquad
    b_{\mathrm{WF}}=O_R(d/m_0),
    \qquad
    q_{\mathrm{WF}}=\Theta_R(d^2/m_0^2),
\end{equation}
then $\lambda_\star^{\mathrm{WF}}=\Theta_R(d/m_0)$ and
\begin{equation}
        {\mathcal Q_{f,\epsilon}^{\mathrm{WF}}(\lambda_\star^{\mathrm{WF}})
        =O_R\left(
            \frac{\|f\|_{L^\infty(\Omega)}}{m_0^2\omega_0^2}
            \frac{d^3}{\epsilon^2}
        \right).}
    \label{eq:qhd-wf-optimized-query-scaling}
\end{equation}
These estimates count only potential-oracle queries.  The implementation of
$-\Delta_g^D$, discretization, state preparation, and measurement
are additional; in particular, coordinate-based implementation bounds may
depend on the face margin $\eta_\Omega^{-1}$.
\subsection{\texorpdfstring{Real-Space Adiabatic Algorithm on $\ell_p$ Balls}{Real-Space Adiabatic Algorithm on ell-p Balls}} \label{sec:rsaa-riemannian}
Recently, Herman et al.~\cite{herman2025mechanisms} introduced \emph{Real-Space Adiabatic Algorithm} (RSAA),
which is a digital implementation of the quantum adiabatic algorithm to prepare the ground state
of Schr\"odinger operators. Their algorithm uses the Dirichlet Laplacian as the kinetic operator and
evolves the Hamiltonian in Euclidean space according to a sufficiently slow schedule.  We extend this algorithm to a 
Riemannian manifolds by replacing the Euclidean Laplacian with the
Laplace--Beltrami operator corresponding to the underlying metric $g$. 

While RsAA can be seen as more general ground state preparation algorithm, we will consider optimization problems. Let $(\mathcal{M},g)$ be a compact connected Riemannian manifold, and let $f:\mathcal{M}\to\mathbb R$ be a continuous potential function. Without loss of generality we can assume $\min_{\mathcal{M}} f=0$ since shifting the potential by a constant only changes the evolution by a global phase. Let $F$ be the operator
\begin{equation*}
    (F\psi)(x)\coloneqq f(x)\psi(x)
\end{equation*}
on $L^2(\mathcal M)$.

Fix an initial parameter $\lambda_{\min}>0$ and a final parameter $\lambda_{\max}>0$ and choose a twice
continuously differentiable, nondecreasing schedule
$\alpha:[0,1]\to[\lambda_{\min}^2   ,\lambda_{\max}^2]$ with
$\alpha(0)=\lambda_{\min}^2 $ and $\alpha(1)=\lambda_{\max}^2$.  Define the
self-adjoint Hamiltonian
\begin{equation}
    H_g(s)\coloneqq-\Delta_g+\alpha(s)F,
    \qquad 0\leq s\leq1.
    \label{eq:riemannian-real-space-adiabatic-path}
\end{equation}
The goal is to prepare the ground state of this Hamiltonian at $s=1$ so that measurement yields a point $\widetilde{x}\in \mathcal{M}$ such that
\[
f(\widetilde{x}) - \min_{\mathcal{M}} f \leq \epsilon,
\]
for some small $\epsilon>0$ with high probability. 

The algorithm then sets the parameters $\lambda_{\min}^2$ and $\lambda_{\max}^2$ so that preparation of the ground state at $s=0$ is trivial and the ground state at $s=1$ is sufficiently concentrated around the optimum. The evolution from $s=0$ to $s=1$ is then determined by the schedule $\alpha$. The adiabatic theorem allows one to derive a simple schedule.

Let $E_0(s)$ and $E_1(s)$ be its two lowest eigenvalues, counted with
multiplicity, and define
\begin{equation}
    \gamma(s)\coloneqq E_1(s)-E_0(s),
    \qquad
    \gamma_{\min}\coloneqq\min_{s\in[0,1]}\gamma(s).
    \label{eq:riemannian-real-space-minimum-gap}
\end{equation}
Assume that $\gamma_{\min}>0$.  Choose a normalized ground state
$\phi_0(s)$ satisfying
\begin{equation*}
    H_g(s)\phi_0(s)=E_0(s)\phi_0(s),
    \qquad \norm{\phi_0(s)}=1.
\end{equation*}

\begin{theorem}[Adiabatic theorem]
\label{thm:riemannian-real-space-adiabatic}
Suppose $\gamma_{\min}>0$.  Define
\begin{align}
    \mathcal A_g
    \coloneqq &\frac{\norm{\partial_sH_g(0)}
                   +\norm{\partial_sH_g(1)}}
             {\gamma_{\min}^2} +\int_0^1\left(
          \frac{\norm{\partial_s^2H_g(s)}}{\gamma_{\min}^2}
          +\frac{\norm{\partial_sH_g(s)}^2}{\gamma_{\min}^3}
      \right)\dd s.
    \label{eq:riemannian-adiabatic-runtime-parameter}
\end{align}
For $T>0$, let $\psi(t)$ solve
\begin{equation*}
    i\partial_t\psi(t)=H_g(t/T)\psi(t),
    \qquad
    \psi(0)=\phi_0(0),
    \qquad 0\leq t\leq T.
\end{equation*}
There is a universal constant $C_{\mathrm{ad}}>0$ and a phase
$\vartheta_T\in\mathbb R$ such that
\begin{equation}
    \norm{\psi(T)-e^{i\vartheta_T}\phi_0(1)}
        \leq\frac{C_{\mathrm{ad}}\mathcal A_g}{T}.
    \label{eq:riemannian-adiabatic-tracking-error}
\end{equation}
\end{theorem}

\begin{proof}
Set
\begin{equation*}
    \overline H_g(s)\coloneqq H_g(s)-E_0(s)I.
\end{equation*}
The scalar shift changes the evolution only by a phase.  Since
$\partial_sH_g(s)=\alpha'(s)F$ is bounded, the Hellmann--Feynman theorem
gives $\norm{\partial_s\overline H_g(s)}\leq2\norm{\partial_sH_g(s)}$.
Thus the relative-bound hypothesis of Theorem~1 of
\cite{mozgunov2023quantum} holds with
$c_0(s)=4\norm{\partial_sH_g(s)}^2$ and $c_1(s)=0$.  Applying that theorem with
$2\Delta(s)=\gamma(s)$, proves \eqref{eq:riemannian-adiabatic-tracking-error}.
\end{proof}

For the RSAA schedule in
\eqref{eq:riemannian-real-space-adiabatic-path},
\begin{equation*}
    \partial_sH_g(s)=\alpha'(s)F,
    \qquad
    \partial_s^2H_g(s)=\alpha''(s)F,
    \qquad
    \norm F=\norm f_\infty.
\end{equation*}
Thus \eqref{eq:riemannian-adiabatic-runtime-parameter} is determined by
$\gamma_{\min}$, the schedule $\alpha$, and $\norm f_\infty$.

We now give an explicit digital implementation of the RSAA algorithm on $\ell_p$ ball.
The other manifolds can be treated similarly by applying appropriate simulation results in
previous sections with the appropriate boundary conditions and metrics.

For simplicity, we assume periodic boundary conditions, however our result can be generalized
to Dirichlet boundary conditions by smoothly reflecting the function across the boundary.
Since the Hamiltonian preserves the odd subspace, restricting this evolution to one odd sector recovers Dirichlet evolution.

Let $\mathcal{M}=\Omega_{p,\delta}^{\mathrm{ball}}$ be the projected Duffy domain
from Section~\ref{sec:global-lp-duffy}, equipped with the flat Duffy metric
and periodic boundary conditions.  Then
$-\Delta_g=-\Delta_\Omega^{\mathrm{per}}$ and the Riemannian volume measure
is $\dd\mu=\dd\mu_{p,\delta}^{\mathrm{ball}}$.

\begin{algorithm}[htbp]
\caption{Digital Implementation of RSAA optimization on an $\ell_p$ ball}
\label{alg:digital-riemannian-rsaa}
\begin{minipage}{\linewidth}
\textbf{Input:} Parameters $p\geq2$ and $0<\delta<1$, a potential oracle
for $f$, a schedule $\alpha$, an evolution time $T$, and a quadrature
degree $Q$ chosen according to
Corollary~\ref{cor:lp-duffy-periodic-discretization}.
\begin{enumerate}
    \item Set $M_{\mathrm{grid}}=2Q+1$ and prepare the sampled initial
    state
    \begin{equation*}
        \ket{\Psi(0)}
        =\left(\frac{2r_\delta}{M_{\mathrm{grid}}}\right)^{d/2}
          \sum_{u_j\in\mathcal G_{\widehat{\mathcal D}_p,Q}}
          \phi_0(0,u_j)\ket j.
    \end{equation*}
    \item Perform the finite-dimensional evolution in
    Corollary~\ref{cor:lp-duffy-periodic-discretization} with
    \begin{equation*}
        a(t)=1,
        \qquad
        b(t)=\alpha(t/T),
        \qquad 0\leq t\leq T.
    \end{equation*}
    \item Measure the grid register and return the corresponding point
    $\widetilde x\in \mathcal{M}$.
\end{enumerate}
\textbf{Output:} The point $\widetilde x$.
\end{minipage}
\end{algorithm}

\begin{theorem}[Optimization via RSAA on $\ell_p$ ball]
\label{cor:digital-riemannian-rsaa}
\label{cor:riemannian-rsaa-optimization}
Let $p\geq2$ and $0<\delta<1$.  Assume that $f$ is
$G_f$-Lipschitz in ambient Euclidean distance, that
$f\circ\widehat{\mathcal D}_p$ has matching values on opposite faces of
$C_\delta$, and that $\phi_0(0)\in\mathcal V_{L_0}^\Omega$.  Let $0<\eta\leq1$, let
$0<\underline\gamma\leq\gamma_{\min}$, choose
$\Lambda_f\geq\norm f_\infty$, and let $\epsilon>0$.  Define
$S_\rho\coloneqq\{x\in \mathcal{M}:f(x)\leq\rho\}$ and suppose that
\begin{equation}
    \int_{S_{\epsilon/2}}|\phi_0(1,x)|^2\dd\mu(x)
        \geq1-\frac{\eta}{3}.
    \label{eq:terminal-ground-state-concentration}
\end{equation}
There are choices of $T$ and $Q$ for which
Algorithm~\ref{alg:digital-riemannian-rsaa} returns
$\widetilde x\in S_\epsilon$ with probability at least $1-\eta$.
The algorithm uses $\mathcal O(d\log(2Q+1))$ qubits.
\end{theorem}

\begin{proof}
Choose
\begin{equation*}
    T=\frac{C}{\eta}\left[
        \frac{(2\norm{\alpha'}_\infty+\norm{\alpha''}_\infty)
              \Lambda_f}{\underline\gamma^2}
        +\frac{\norm{\alpha'}_\infty^2\Lambda_f^2}
              {\underline\gamma^3}
    \right],
\end{equation*}
where $C\geq3C_{\mathrm{ad}}$.  Then
$C_{\mathrm{ad}}\mathcal A_g/T\leq\eta/3$.

If $S_\epsilon=\mathcal{M}$, the claim is immediate.  Otherwise
$\norm{h_\epsilon}_\infty=1$ for
\begin{equation*}
    h_\epsilon(x)
        \coloneqq\min\!\left\{1,
            \frac{2}{\epsilon}\left(f(x)-\frac{\epsilon}{2}\right)_+
        \right\}.
\end{equation*}
This function vanishes on $S_{\epsilon/2}$, equals one on
$\mathcal{M}\setminus S_\epsilon$, is $2G_f/\epsilon$-Lipschitz, and has matching
values on opposite faces.  The concentration assumption therefore implies
\begin{equation*}
    \langle\phi_0(1),h_\epsilon\phi_0(1)\rangle\leq\frac{\eta}{3}.
\end{equation*}
Theorem~\ref{thm:riemannian-real-space-adiabatic} and the choice of
$T$ imply, for some phase $\vartheta$,
\begin{equation*}
    \norm{\psi(T)-e^{i\vartheta}\phi_0(1)}
        \leq\frac{\eta}{3}.
\end{equation*}
For $0\leq h_\epsilon\leq1$, the difference between its expectations in
two pure states is at most their trace distance, which is at most their
phase-adjusted vector distance.  Its expectation in the exact final state
is therefore at most $2\eta/3$.

Set $\epsilon_{\mathrm{disc}}=\eta/12$ and
$\overline\alpha=\int_0^1\alpha(s)\dd s$.  Choose $Q$ from
Corollary~\ref{cor:lp-duffy-periodic-discretization} with
\begin{equation*}
    L_0,
    \qquad B_1=\overline\alpha T,
    \qquad G_h\leq\frac{2G_f}{\epsilon},
    \qquad \epsilon=\epsilon_{\mathrm{disc}}.
\end{equation*}
Equivalently, $M_{\mathrm{grid}}=2Q+1$ may be chosen as
\begin{equation*}
    \mathcal O\!\left(
        L_0
        +\left[1+\frac{24r_\delta\kappa_{p,\delta}\sqrt d}{\eta}
            \max\!\left\{
                G_f\overline\alpha T,
                \frac{2G_f}{\epsilon}
            \right\}\right]
        \left[1+\overline\alpha T\Lambda_f
              +\log\!\left(\frac{12}{\eta}\right)\right]
    \right).
\end{equation*}
Corollary~\ref{cor:lp-duffy-periodic-discretization} changes the expectation
of $h_\epsilon$ by at most
$4\epsilon_{\mathrm{disc}}=\eta/3$.  Hence the grid expectation of
$h_\epsilon$ is at most $\eta$.  Since $h_\epsilon=1$ outside
$S_\epsilon$, the probability of returning a point outside $S_\epsilon$
is at most $\eta$.
\end{proof}

The authors in~\cite{herman2025mechanisms} gave a concrete class of non-convex objective functions for which the runtime of RsAA algorithm is polynomial.
Let $\{\mathcal I_i\}_{i=1}^k$ partition $[d]$ and define
\begin{equation}
    f_{\mathrm{sep}}(y):=\sum_{i=1}^k g_i(y_{\mathcal I_i}),
    \qquad
    b:=\max_i|\mathcal I_i|.
    \label{eq:duffy-block-separable-objective}
\end{equation}
For $R\in\mathrm{SO}(d)$ and $B>0$, define the rotated and rescaled box
objective
\begin{equation}
    f_{\mathrm{box}}(x)
        :=f_{\mathrm{sep}}\!\left(\frac{B}{r_\delta}Rx\right),
    \qquad x\in C_\delta,
    \label{eq:rotated-block-separable-box-objective}
\end{equation}
where $B$ is chosen so that the relevant minimizer is mapped into
$C_\delta$.  The rotation hides the separable coordinates from classical
algorithms~\cite{leng2023quantum,herman2025mechanisms}.  The corresponding
objective on $\Omega_{p,\delta}^{\mathrm{ball}}$ is
\begin{equation*}
    f_{p,\delta}
        :=f_{\mathrm{box}}\circ\widehat{\mathcal D}_p^{-1}.
\end{equation*}
If $f_{\mathrm{box}}$ has matching values on opposite faces of $C_\delta$,
then $f_{p,\delta}$ satisfies the periodic compatibility condition in
Theorem~\ref{cor:digital-riemannian-rsaa}.

To compare the adiabatic evolutions, define
\begin{align*}
    H_{\mathrm{box}}(s)
        &:=-\Delta_{C_\delta}^{\mathrm{per}}
            +\alpha(s)f_{\mathrm{box}},\\
    H_{p,\delta}(s)
        &:=-\Delta_{\mathcal D_p}^{\mathrm{per}}
            +\alpha(s)f_{p,\delta}.
\end{align*}
The definition of $-\Delta_{\mathcal D_p}^{\mathrm{per}}$ in
\eqref{eq:lp-periodic-duffy-kinetic-operator} and the identity
$f_{p,\delta}\circ\widehat{\mathcal D}_p=f_{\mathrm{box}}$ give
\begin{equation*}
    U_{p,\delta}H_{p,\delta}(s)U_{p,\delta}^\dagger 
        =H_{\mathrm{box}}(s).
\end{equation*}
Thus the two Hamiltonians have the same spectrum and the same spectral gap
for every $s\in[0,1]$.  Under the assumptions for rotated block-separable
objectives in \cite[Theorem~3.6]{herman2025mechanisms}, Herman et al. show
that this gap is inverse polynomial along the adiabatic path.  The
adiabatic runtime for $H_{\mathrm{box}}(s)$ is therefore polynomial, and
the same conclusion holds for its Duffy lift $H_{p,\delta}(s)$.
For known black-box optimization algorithms, the runtime is exponential for optimizing the same class of functions.

\section*{AI Use Disclosure}
Large language model tools were used interactively at all stages of the preparation of this manuscript: from the earliest exploratory research stages to the final exposition and writing steps. The final manuscript was substantially revised by the authors, who remain solely responsible for all mathematical claims, proofs, references, and conclusions. Most of the provable statements began as ideas from the author and GPT-5.6 Sol made substantial contributions to the analysis. GPT-5.6 Sol also generated some new ideas, notably, it suggested the use of the IMS Localization in the coherent-charts method.

\section*{Acknowledgment}
We thank Rob Otter and Ruslan Shaydulin for executive support and valuable feedback on this project. We also acknowledge our colleagues at the Global Technology Applied Research Center of JPMorganChase.

\section*{Disclaimer}
This paper was prepared for informational purposes by the Global Technology Applied Research center of JPMorgan Chase \& Co. This paper is not a product of the Research Department of JPMorgan Chase \& Co. or its affiliates. Neither JPMorgan Chase \& Co. nor any of its affiliates makes any explicit or implied representation or warranty and none of them accept any liability in connection with this paper, including, without limitation, with respect to the completeness, accuracy, or reliability of the information contained herein and the potential legal, compliance, tax, or accounting effects thereof. This document is not intended as investment research or investment advice, or as a recommendation, offer, or solicitation for the purchase or sale of any security, financial instrument, financial product or service, or to be used in any way for evaluating the merits of participating in any transaction.

\printbibliography

\end{document}